\documentclass{aa}  

\usepackage{graphicx}
\usepackage[varg]{txfonts}
\usepackage[]{hyperref}
\usepackage{amsmath}	% Advanced maths commands
\usepackage{amssymb}	% Extra maths symbols
\usepackage[normalem]{ulem}

\begin{document} 

 \title{Eclipse timing study of new hierarchical triple star candidates in the Northern Continuous Viewing Zone of \textit{TESS}}
\titlerunning{ETV study of triple stars in the NCVZ of \textit{TESS}}
   \author{T.~Mitnyan
          \inst{1,2}\fnmsep\thanks{Email: mtibor@titan.physx.u-szeged.hu}
          \and
          T.~Borkovits\inst{1,2,3,4,5}
          \and  
          D.~R.~Czavalinga\inst{1,2}
          \and
          S.~A.~Rappaport\inst{6,2}
          \and
          A.~P\'al\inst{3}
          \and
          B.~P.~Powell\inst{7}
          \and
          T.~Hajdu\inst{3,8}
          }

   \institute{Baja Astronomical Observatory of University of Szeged, H-6500 Baja, Szegedi út, Kt. 766, Hungary
   \and
   HUN-REN-SZTE Stellar Astrophysics Research Group, H-6500 Baja, Szegedi út, Kt.
766, Hungary
\and
Konkoly Observatory,HUN-REN Research Centre for Astronomy and Earth Sciences, H-1121 Budapest, Konkoly Thege Miklós út 15-17, Hungary
\and
ELTE Gothard Astrophysical Observatory, H-9700 Szombathely, Szent Imre h. u. 112, Hungary
\and
HUN-REN-ELTE Exoplanet Research Group, H-9700 Szombathely, Szent Imre h. u. 112, Hungary
         \and
Department of Physics, Kavli Institute for Astrophysics and Space Research, M.I.T., Cambridge, MA 02139, USA
\and
NASA Goddard Space Flight Center, 8800 Greenbelt Road, Greenbelt, MD 20771, USA
\and
Eszterh\'{a}zy K\'{a}roly Catholic University, Department of Physics, H-3300 Eszterh\'{a}zy t\'{e}r 1, Eger, Hungary 
}

   \date{Received ...; accepted ...}

% \abstract{}{}{}{}{} 
% 5 {} token are mandatory
 
  \abstract
   {}
       {We compiled a list of more than 3500 eclipsing binaries located in and near the Northern Continuous Viewing Zone (NCVZ) of the \textit{TESS} space telescope that have a sufficient amount of \textit{TESS} photometry to search for additional hidden components in these systems. Apart from discovering their hierarchical nature, we also planned to determine their orbital parameters and analyze their distributions.}
       {We obtained the \textit{TESS} light curves of all targets in an automated way applying convolution-aided differential photometry on the \textit{TESS} Full-Frame Images from all available sectors up to Sector 60. Using a new self-developed Python GUI, we visually vetted all of these light curves, determined the eclipsing periods of the objects and calculated their eclipse timing variations (ETVs). The ETV curves were used in order to search for nonlinear variations that could be attributed to a light travel time effect (LTTE) or dynamical perturbations caused by additional components in these systems. We pre-selected 351 such candidates and tried to model their ETVs with the analytic formulae of pure LTTE or the combination of LTTE and dynamical perturbations.}
       {In total we could fit a model solution for the ETVs of 135 hierarchical triple candidates in which 10 systems were already known in the literature and the remainder of the 125 systems are new discoveries. Among these systems, there are some more noteworthy ones, such as five tight triples very close to their dynamical stability limit with a period ratio of less than 20 and three newly discovered triply eclipsing triples. We point out that dynamical perturbations are occurring in GZ Dra, which turns out to be a triple and that the system is actually one of the most inclined one known in the literature with $i_\mathrm{m}=58\degr\pm7\degr$. We also made a comparison of the distributions of some orbital parameters coming from our solutions with those from the \textit{Kepler} sample derived by \citet{borkovits16}. Finally, we checked the correlations between the available parameters for systems that have \textit{Gaia} Non-Single Star orbital solutions with those from our ETV solutions.}
   {}

   \keywords{binaries: close --
                binaries: eclipsing --
                binaries: general
               }

   \maketitle

\section{Introduction}

In hierarchical triples both the inner binary and the outer, more distant third star are orbiting around the common center of mass of the system. They are especially interesting systems as their exact formation scenarios and their contributions to various evolutionary channels (e.~g. stellar mergers or Type Ia supernovae) are still under debate \citep[][]{toonen20,toonen22,tokovinin21}. They are also the most likely systems in which we can detect and study dynamical perturbations which can cause variations in their orbital elements \citep[][]{borkovits22}.

There are several ways to discover hierarchical triple systems using photometric, spectroscopic and/or astrometric observations. Photometric methods can be applied only for eclipsing binaries (EBs) in which one can detect: i) the presence of an extra light in the light curve (LC) of the inner binary; ii)  eclipse timing variations (ETVs) caused by light-travel time effect (LTTE); iii) eclipse depth variations (EDVs) caused by dynamically forced precession of the orbital planes; iv) extra eclipses of the components of the inner binary with the tertiary. Spectroscopy can either reveal directly the triple nature of an object by resolving three sets of atomic lines in the spectra or indirectly by detecting periodic variations in the systemic radial velocity (gamma velocity) of a binary star. Lastly, astrometric observations can detect changes in the position of a binary star in the plane of the sky while orbiting the common center of mass of the triple system. Of all of these options, photometry is the most convenient technique because it can cover larger areas on the sky than spectroscopic studies and with orders of magnitude more objects within them, while one also usually needs larger telescopes and more telescope time for spectroscopy than for photometry in case of the same object. Astrometry is also a bit problematic because currently the \textit{Gaia} space telescope is our only tool for the purpose of searching for hierarchical triples via this technique, and it just became such a tool very recently as \citet{czavalinga23} demonstrated after \textit{Gaia}'s 3rd Data Release \citep{gaia22b,babusiaux22}.
That means, currently, photometric techniques are our best option to discover new hierarchical triple systems. Apart from the first listed effect (extra light) which can be unreliable on its own and tells us only the flux ratio of the supposed tertiary relative to the total flux of the system, the other three (ETVs, EDVs, extra eclipses, or their combination) can be excellent indicators. The analysis of ETV (or more traditionally, O–C) curves has a century long history, nevertheless, the first triply eclipsing systems were discovered only very recently, in the last decade \citep{carter11,derekas11}, due to the \textit{Kepler} mission \citep{borucki10} and, EDVs in such objects are observed routinely only since the advent of the era of space photometry. The \textit{Kepler} space telescope, originally built to find transiting exoplanets, collected a 4 years long, almost uninterrupted photometric data set from a $\sim$100 square degree area of the sky with unprecedented precision that made it possible to detect the previously mentioned effects.  As its successor, \textit{TESS} \citep[Transiting Exoplanet Survey Satellite; ][]{ricker15} is conducting a nearly all-sky survey of exoplanets and its observations are also a treasure chest for almost all fields of stellar astronomy that should be exploited in many different ways, including searches for compact hierarchical triples.

There are a handful of ETV studies in the literature for large sky surveys such as \textit{CoRoT} \citep{hajdu17,hajdu22b}, \textit{Kepler} \citep{rappaport13,conroy14,borkovits15,borkovits16} or OGLE \citep{hajdu19,hajdu22a} that found hundreds of such compact hierarchical triples (CHTs) collectively. This number can be improved further significantly by the analysis of the whole collection of the available and future \textit{TESS} data. These systems could be of great help in searching for rare circumstances (e.g., super flat, super circular, super tight, super dynamically interacting, retrograde orbits, von Zeipel-Kozai-Lidov cycles in operation, etc), and fortuitous systems where we can make really precise measurements of the stars, their evolutionary state, and the orbits.

The \textit{TESS} space telescope during its 2-year-long primary mission performed a unique sky survey collecting a high-precision, almost uninterrupted, $\gtrsim 1$ month long photometric data set for the majority of the objects located both in the southern and northern hemispheres. Because of the observational strategy of \textit{TESS}, for the objects closer to the ecliptic poles, there are longer data sets available, as they can be involved in multiple sectors of observations with as long as a whole year for objects found in the Continuous Viewing Zones (CVZs). In its extended mission, \textit{TESS} had re-observed almost the same parts of the sky extending the length of these data sets. These temporal samples give us an opportunity to analyze the eclipse times of EBs with relatively short orbital periods (up to around a few tens of days) and search for the signal of additional components with short outer orbital periods (up to around a few hundreds of days) in the ETVs. In this way, we can discover new CHTs and also reanalyze previously known systems in order to obtain more accurate parameters for them.

In this paper, we carry out a comprehensive analysis of the ETV curves of those EBs which were observed by the \textit{TESS} space telescope in its Northern Continuous Viewing Zone (NCVZ) in between Sectors 14 and 60. As a close predecessor of this current study we refer to the work of \citet{borkovits16} who analyzed the four-year-long data set of the prime \textit{Kepler}-mission to identify and study CHTs amongst more than 2600 EBs via their ETV curves. In this work we follow a similar (but not identical) method for the identification and ETV analyses of CHT candidates amongst \textit{TESS} EBs.

The structure of the paper is the following. In Sect.~\ref{sec:obsdata} we describe briefly the methods and data sets to identify EBs in the \textit{TESS} NCVZ sample, then the data reduction and the formation of the ETV curves are discussed. The basics of the mathematical modeling of the ETVs are summarized in Sect.~\ref{sec:analysis}, while the results of our comprehensive, detailed analysis are discussed and tabulated in Sect.~\ref{sec:discussion} and then, our work is concluded in Sect.~\ref{Sect:Summary}.

\section{Observational data and data reduction}
\label{sec:obsdata}

We used a list of objects showing eclipse-like variations during the Year 2 observations of \textit{TESS} found in the \textit{TESS} data using a machine learning approach further described by \citet{powell21}. From this list, we selected those objects that had at least 8 Sectors of observations available during Year 2, i.~e. those that are located in or near the NCVZ of \textit{TESS}. That yielded 5139 such targets which had long enough and almost uninterrupted data sets that could be useful for our survey. Moreover, we have added one more object (TIC 377105433) to our list from the list of \citet{czavalinga23} which, although located in the NCVZ, was too faint to be found by the search algorithm used to construct the GSFC EB Catalog. After filtering out duplicate sources, 3553 objects have made the final set of targets for further analysis.

We obtained all \textit{TESS} light curves of the objects in our sample up to Sector 60 in an automated way using a convolution-aided differential photometric pipeline based on the FITSH software package \citep{pal12}. We removed the remaining trends using the WOTAN \citep{hippke19} Python package or using the Principal Component Analysis method implemented in the lightkurve \citep{lightkurve18} Python package in some more complicated cases. In order to determine the eclipsing periods of the systems, first we ran an initial Box Least Squares \citep[BLS;][]{kovacs02} search on the full light curves. Then, using the best BLS period, we transformed the light curves to the orbital phase domain and visually checked if the initial period found by the BLS is correct. As a final step, we used the Phase Dispersion Minimization \citep[PDM;][]{stellingwerf78} period search method to obtain more accurate eclipsing periods.

After we determined the periods of the systems, as a next step, we calculated eclipse timing variations in a similar way as \citep{borkovits15} using the following equation:
\begin{equation}
\Delta=T(E)-T_{0}-P_{\mathrm{s}}E,
\end{equation}
where $\Delta$ is the time difference of the observed and calculated mid-eclipse times (O--C) for the given orbital cycle, $T(E)$ is the mid-eclipse time of the $E$th orbital cycle, $T_{0}$ is the mid-eclipse time of the zeroth orbital cycle (reference epoch), $P_{\mathrm{s}}$ is the reference period and $E$ is the cycle number.  Finally, we visually vetted all ETV curves in order to select those which show clear non-linear variations for further analysis.

\begin{figure*}
\includegraphics[width=\textwidth]{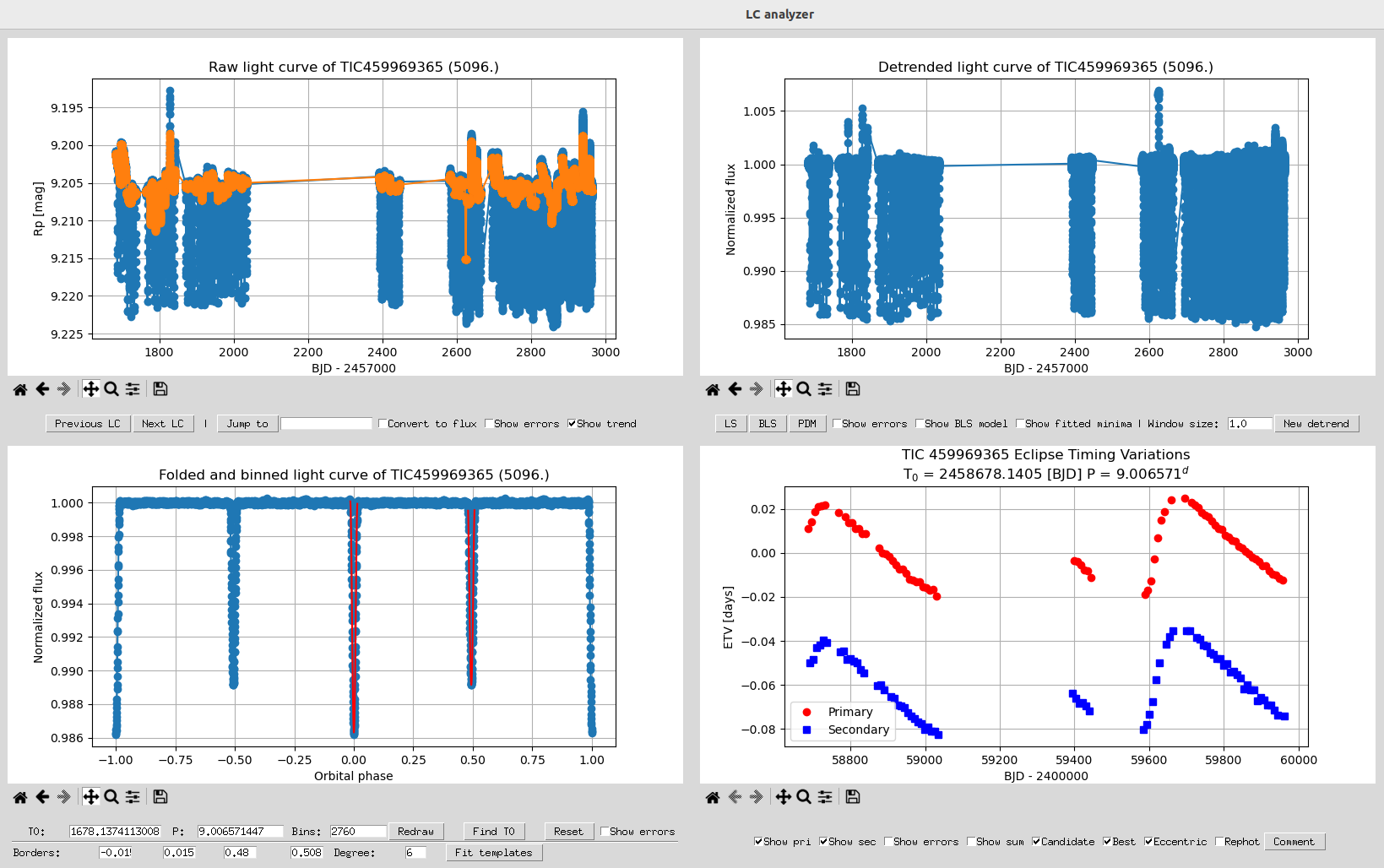}
\caption{Current look of our self-developed, unique light and ETV curve analyzer GUI. On the top left corner, the raw lightcurve can be seen (blue) alongside the fitted trends (orange). On the top right corner, the detrended lightcurve is plotted after the normalization by the previously fitted trends. On the lower left corner, the previously detrended and normalized lightcurve binned and folded with the specified bin numbers and ephemeris is plotted alongside the polinomial templates fitted on the two types of eclipses. Finally, on the lower right corner, the ETVs calculated from the primary (red) and secondary (blue) mid-eclipse times using the indicated ephemeris on the top are plotted. Below each panel there are several useful features (see the text for details) which enable a fast interactive analysis of any set of lightcurves.}
 \label{fig:gui}
\end{figure*}

For this purpose, we developed an interactive program with a graphical user interface (GUI) utilizing the Tkinter module of Python which can be seen in Fig.~\ref{fig:gui}. This program allows the user to quickly load and view any raw LC with several useful interactive features that help to analyze the LC and quickly calculate the ETV curve of an object using the algorithm described above. The already implemented features are: i) interactive detrending of the raw LC; ii) multiple period searching methods, e.g. BLS Periodogram and PDM in order to determine the eclipsing period of the system automatically; iii) reference epoch determination with the calculation of the (folded) orbital phase curve; iv) determination of the times of eclipsing minima or out-of-eclipse maxima with the fitting of higher-order polynomials; v) displaying the ETV curve of the object calculated from the previously determined minima/maxima using the previously found orbital period and reference epoch; vi) save the latest results in a database that allows one to do more complex external analyses on the candidate systems, and also ensure reproducibility.

\section{Analysis}
\label{sec:analysis}

\subsection{Theoretical basics of the ETV analysis}

The ETVs were modeled in a very similar manner as described in \citet{borkovits15,borkovits16}. The full analytic model takes the form:
\begin{equation}
\Delta=\sum_{\mathrm{i}=0}^{3}c_{\mathrm{i}}E^{\mathrm{i}}+[\Delta_{\mathrm{LTTE}}+\Delta_{\mathrm{dyn}}+\Delta{\mathrm{apse}}]_{0}^{E},
\end{equation}
where the polynomial term contains corrections for the calculated eclipse times (in the case of $i=0$, 1), constant-rate and, linear period variations (in the case of $i=2$, 3). The constant and linear ($i=0$, 1) terms were used for all the sources in our analyses. In contrast to this, the $i=2, 3$ polynomial terms were fitted only in a few cases, where the presence of a quadratic or even more complicated additional variations in the ETV curves were evident. 

Moreover, the ($\Delta_\mathrm{LTTE,dyn,apse}$) terms represent the contributions of LTTE, $P_2$-timescale dynamical perturbations and apsidal motion, respectively, while the square bracket denotes the difference of each terms measured in the $E$th and zeroth (i.\,e. reference) cycle. Here, for consistency, we give only very brief descriptions and the formulae of these latter terms as they were used, while their detailed explanation can be found in \citet{borkovits15,borkovits16}.

\subsubsection{Light-travel time effect (LTTE)}
\label{LTTE}
The classical LTTE (or R\o mer-delay) contribution comes from the periodically varying distance of the inner binary during its orbital revolution around the common center of mass of the triple system that causes a periodic variation in the light travel time towards the observer detected as periodic eclipse timing variation. Based on \citet{irwin52}, it was taken into account in the form of
\begin{equation}
\Delta_\mathrm{LTTE}=-\frac{a_\mathrm{AB}\sin i_2}{c}\frac{\left(1-e_2\right)^2\sin(v_2+\omega_2)}{(1+e_2\cos v_2)},
\label{Eq:LTTEdef}
\end{equation}
where $i_2$, $e_2$, $v_2$ and $\omega_2$ refer to the inclination, eccentricity, true anomaly and the argument of the periastron of the outer orbit\footnote{From now on, the orbital parameters with subscript 1 and 2 refer to the inner and the outer orbit, respectively.} (i.e. the relative orbit, on which the tertiary star orbits around the center of mass of the inner binary), respectively, while $a_\mathrm{AB}$ stands for the semi-major axis of the (physical) orbit of the center of mass of the inner pair around the center of mass of the whole triple system. Finally, $c$ denotes the speed of light.

It can be shown easily that, as far as the orbital elements of the outer orbit remain constant, this effect can be described with a pure sine in the eccentric anomaly domain, of which the amplitude is
\begin{equation}
\mathcal{A}_\mathrm{LTTE}=\frac{m_\mathrm{C}}{m_\mathrm{ABC}}\frac{a_2\sin i_2}{c}\sqrt{1-e_2^2\cos^2\omega_2}.
\end{equation}
By analogy with the mass function used for single-lined spectroscopic binaries (SB1), it is usual to introduce the mass function as
\begin{equation}
f(m_\mathrm{C})=\frac{m_\mathrm{C}^3\sin^3i_2}{m_\mathrm{ABC}^2}\approx0.75\times10^{12}\frac{\mathcal{A}_\mathrm{LTTE}^3}{P_2^2}\left(1-e_2^2\cos^2\omega_2\right)^{-3/2},
\end{equation}
where, in the last form, the numerical constant is correct when time is expressed in days and masses in solar mass. (Note, here and hereafter, $m_\mathrm{A,B,C}$ denote the individual masses of the three constituent stars, while $m_\mathrm{ABC}$ stands for the total mass of the triple system.)

In regard to the wide, or outer orbit, an LTTE solution carries exactly the same information that can be extracted from the radial velocity measurements of a single-line spectroscopic binary (SB1). Therefore, besides the orbital period of the outer orbit ($P_2$) one can determine three of the six orbital elements of the outer orbit ($e_2$, $\omega_2$, $\tau_2$) and, moreover, the projected semi-major axis of the inner pair's orbit around the common centre of mass of the triple ($a_\mathrm{AB}\sin i_2$) from which, in the absence of any external information, one can determine the mass of the third companion as a unique function of two additional parameters, namely, the total mass of the inner binary $m_\mathrm{AB}$ and the orbital inclination ($i_2$). Such external information might have come from the observations of radial velocities of the third star (producing a quasi SB2-like system) and/or from detection of third-body eclipses \citep[a good example for both situations is e.~g., HD~181068,][]{borkovits13} or from the detection of dynamically driven ETVs in tight triple systems. Note, when such external information was not available for the total mass of the inner pair, in the case of W~UMa-type, i.~e., overcontact binaries, we estimated the binary's mass with the use of the empirical mass-period relations of \citet{2008MNRAS.390.1577G}, while for other systems we simply set $m_\mathrm{AB}=2\,\mathrm{M}_{\sun}$.

\subsubsection{$P_2$-timescale dynamical perturbations}
\label{subsect:ETV_dyn}

The contribution of dynamical perturbations comes from the possible gravitational interaction between the outer component and the inner binary which can result in the change of the orbital parameters of the inner binary and hence period variation. While LTTE terms were fitted by default during our analysis, the $P_2$-timescale dynamical contributions were taken into account only when the tightness of the system under investigation made it necessary. The most elaborate presentation of the theory of the $P_2-$ (medium-) timescale perturbations and their manifestations in the ETV of a tight EB was presented and discussed in \citet{borkovits15}. For the sake of brevity, here we repeat only the lowest order (i.e. most significant) quadrupole terms (modifying slightly Eqs.~5--11 of \citealt{borkovits15}), while higher order (and more lengthy) formulae can be found in the appendices of \citet{borkovits15}. According to this theory, the quadrupole-level dynamical contribution to an ETV curve in a CHT is the following:
\begin{eqnarray}
\Delta_\mathrm{dyn}&=&\mathcal{A}_\mathrm{dyn}\left(1-e_1^2\right)^{1/2}\left\{\left[f_1+\frac{3}{2}K_1\right]\mathcal{M}\right.  \nonumber \\
&&+\left(1+I\right)\left[K_{11}\mathcal{S}(2u_2-2\alpha)-K_{12}\mathcal{C}(2u_2-2\alpha)]\right] \nonumber \\
&&+\left(1-I\right)\left[K_{11}\mathcal{S}(2u_2-2\beta)+K_{12}\mathcal{C}(2u_2-2\beta)]\right] \nonumber \\
&&+\sin^2i_\mathrm{m}\left(K_{11}\cos2n_1+K_{12}\sin2n_1\right. \nonumber \\
&&\left.\left.-\frac{4}{3}f_1-2K_1\right)\left[2\mathcal{M}-\mathcal{S}(2u_2-2n_2)\right]\right\} \nonumber \\
&&+\Delta_1^*(\sin i_\mathrm{m}\cot i_1),
\label{Eq:dyndef}
\end{eqnarray}
where
\begin{equation}
\mathcal{A}_\mathrm{dyn}=\frac{1}{2\pi}\frac{m_\mathrm{C}}{m_\mathrm{ABC}}\frac{P_1^2}{P_2}\left(1-e_2^2\right)^{-3/2}
\end{equation}
stands for the characteristic amplitude of the dynamical effects on a timescale equal to the outer period ($P_2$). Furthermore, $I=\cos i_\mathrm{m}$ denotes the cosine of the mutual or relative inclination ($i_\mathrm{m}$), i.~e., the angle between the inner and outer orbital planes, while $u_2=v_2+\omega_2$ stands for the true longitude of the third star measured from the ascending node of the wide orbit with respect to the tangential plane of the sky. Moreover, the node-like quantities $n_{1,2}$ represent the angular distances of the intersection of the two orbital planes (i.e. the dynamical node) from the intersections of the respective orbital planes with the tangential plane of the sky (i.e. the observable nodes), while $\alpha=n_2-n_1$ and $\beta=n_2+n_1$ \citep[See Fig.~1 of][for an easy visualization]{borkovits15}.
Furthermore, the trigonometric terms, depending on the parameters of the outer orbit (and the current position of the third star) are as follows:
\begin{eqnarray}
\mathcal{M}&=&v_2-l_2+e_2\sin v_2 \nonumber \\
&=&3e_2\sin v_2-\frac{3}{4}e_2^2\sin2v_2+\frac{1}{3}e_2^3\sin3v_2+\mathcal{O}(e_2^4), \\
\mathcal{S}(2u_2)&=&\sin2u_2+e_2\left[\sin(u_2+\omega_2)+\frac{1}{3}\sin(3u_2-\omega_2)\right], \\
\mathcal{C}(2u_2)&=&\cos2u_2+e_2\left[\cos(u_2+\omega_2)+\frac{1}{3}\cos(3u_2-\omega_2)\right],
\end{eqnarray}
where $l_2$ is the mean anomaly of the outer orbit.
Moreover, the inner orbit dependent quantities are
\begin{eqnarray}
f_1&=&1+\frac{25}{8}e_1^2+\frac{15}{8}e_1^4+\frac{95}{64}e_1^6+\mathcal{O}(e_1^8), \\
K_1&=&\mp e_1\sin\omega_1+\frac{3}{4}e_1^2\cos2\omega_1\pm\frac{1}{2}e_1^3\sin3\omega_1+\mathcal{O}(e_1^4), \\
K_{11}&=&\frac{45}{32}e_1^2\pm\frac{15}{8}\left(e_1+\frac{1}{2}e_1^3\right)\sin\omega_1+\frac{153}{64}e_1^2\cos2\omega_1 \nonumber \\
&&\mp\frac{45}{128}e_1^3\sin3\omega_1+\mathcal{O}(e_1^4), \\
K_{12}&=&\mp\frac{15}{8}\left(e_1-\frac{1}{2}e_1^3\right)\cos\omega_1+\frac{153}{64}e_1^2\sin2\omega_1 \nonumber \\
&&\mp\frac{45}{128}e_1^3\cos3\omega_1+\mathcal{O}(e_1^4),
\end{eqnarray}
and, finally, $\Delta_1^*(\sin i_\mathrm{m}\cot i_1)$ stands for the usually negligibly small precession terms which are not shown here, however, they are taken into account during the analytic fitting process. (Note, that the upper signs refer to the primary eclipses, while the lower ones for the secondary events.)

As one can see, in the case of a circular inner orbit (which occurs frequently for close binary stars and, hence, for most of the EBs), all the $K$ functions disappear and, moreover, $f_1$ reduces to unity, resulting in a much simpler equation for $\Delta_\mathrm{dyn}$.

The $\Delta_\mathrm{dyn}$ term was included into the ETV analysis only when the tightness (i.e. the ratio $P_2/P_1$) of the triple system made it necessary. In several cases, typically for the tightest, eccentric and/or inclined triples, the shapes of the ETV curves have shown already at a first inspection that the dynamical effects are important/dominant. For some other cases, however, the need to include dynamical effects was far from evident. For this reason, in the case of all the LTTE-only fitted ETVs the software automatically calculated the nominal amplitude ratio of $\mathcal{A}_\mathrm{dyn}/\mathcal{A}_\mathrm{LTTE}$ under the assumption that $m_\mathrm{AB} = 2$ (discussed above) and $i_2=90\degr$, the latter of which yields the minimum value of the amplitude ratio.  If this ratio was found to be higher than 0.25, we repeated the fitting process, switching on the dynamical contribution.

The dynamical ETV terms, in theory, give much more orbital, geometric and other dynamical parameters, than does a pure LTTE-dominated ETV solution. This is so because the dynamical terms depend at the same time not only upon the orientations of both orbits relative to the observer, but also upon the orientations of the two orbits relative to each other (as the gravitational perturbations are driven by the relative positions of the three stars with respect to each other) and, hence, the dynamical terms give the full, three-dimensional, spatial configuration of the triple system, including such a key-parameter as the mutual (or, relative) inclination ($i_\mathrm{m}$) of the inner and outer orbits. Moreover, in the amplitude of the dynamical ETV the outer mass ratio ($q_2=m_\mathrm{C}/m_\mathrm{AB}$) also occurs, and moreover, the outer inclination $i_2$ can also be obtained. Thus, with the combination of the LTTE and dynamical terms, $m_\mathrm{C}$ and $m_\mathrm{AB}$ can also be determined directly. Unfortunately, however, the parameter space of a combined LTTE and dynamical ETV solution is so complex and multiply degenerate, that in the absence of any other, external information, one cannot expect accurate mass determination from such a combination of contributions to the ETV. A detailed discussion of the parameters which can be mined out from a dynamical LTTE solution can be found in \citet{borkovits15}, while the degeneracies of such combined ETV solutions are discussed in Sect. 6.3 of \citet{rappaport13} and Sect.~5.2 of \citet{borkovits15}.
%Hence, the dynamical ETV depends on both eccentricities ($e_{1,2}$), both observable arguments of periastrons ($\omega_{1,2}$) and also on both dynamical arguments of periastrons ($g_{1,2}=\omega_{1,2}-n_{1,2}$

\subsubsection{Apsidal motion}

Apsidal motion (i.e. the rotation of the orbital ellipse within the orbital plane) may have several different origins. The most frequent and well known variations are the relativistic, classical tidal (arising from the non-spherical mass distribution of the tidally deformed bodies) and, the dynamical (third-body perturbed) apsidal motions \citep[see, e.g.,][for a short review]{borkovits19}. Irrespective on the origin of this phenomena, its effect on the ETV curves can be described mathematically as follows:
\begin{eqnarray}
\Delta_\mathrm{apse}&=&\frac{P_1}{2\pi}\left[2\arctan\left(\frac{\pm e_1\cos\omega_1}{1+\sqrt{1-e_1^2}\mp e_1\sin\omega_1}\right)\right. \nonumber \\
&&\left.\pm\sqrt{1-e_1^2}\frac{e_1\cos\omega_1}{1\mp e_1\sin\omega_1}\right]. 
\label{Eq:apse-def}
\end{eqnarray}
(The form of this equation in widespread use is given as a trigonometric series in $\omega_1$. The slight inclination angle dependence is also taken into account in \citealt{gimenez83}.)

In general, for tight triples and, especially, for such triples with less compact inner pairs, the dynamically forced apsidal motion substantially dominates over the other two contributions. We will show in Sect.~\ref{sec:discussion}, that for all of the currently investigated triples with eccentric inner binaries, this is the actual scenario. Hence, instead of fitting a (linear) apsidal advance rate ($\Delta\omega_1$), hidden in Eq.~\ref{Eq:apse-def}, as an additional free parameter, we calculated the current values of $\omega_1$ (and also of $e_1$) inherently for each time during the fitting process, with the use of the perturbation equations, and in the manner, as described in \citet{borkovits15}.

\subsection{Methods of the ETV analysis}

The software package which realizes the mathematical model discussed above is described in detail in \citet{borkovits15}, and it has been applied previously in several studies analyzing ETV curves of several \textit{Kepler} \citep{borkovits15,borkovits16}, \textit{K2} \citep{2019MNRAS.483.1934B}, \textit{CoRoT} \citep{hajdu17}, OGLE \citep{hajdu19} and \textit{TESS} \citep{2021MNRAS.503.3759B,2023MNRAS.524.4220P} EBs. For those systems where there was no need to include the medium-period dynamical ETV contribution (Sect.~\ref{subsect:ETV_dyn}) we applied a simple Levenberg-Marquardt-type (LM) differential least squares parameter optimization to find the most probable solution. (The pros and cons of the use of this method for the current problem, and some further details are discussed in Sect.~5.2 of \citealt{borkovits15}.)

When the inclusion of the dynamical contribution was necessary, we followed a different parameter optimization technique. Because of the large number of parameters to be fitted, as well as their highly non-linear correlations, this is not an ideal situation for the LM method.  Therefore, we switched to Markov-Chain Monte Carlo (MCMC) approach. This is a novelty of the software package relative to its formerly used versions. In such a manner we were able to explore the whole (or, at least, physically realistic) part of the parameters and obtain more realistic uncertainties.

%Finally, experiencing the effectiveness of the MCMC-based parameter optimization on the dynamically effected ETV curves, we made a second, MCMC-based fitting process on the pure LTTE ETV curves, too. This was carried out with an independent software package and, hence, can be used as an independent check of the LM-based former ETV results.
%
%{\blue For alternative methods of fitting the LTTE, we employ the same ETVs detailed earlier. This approach involves initially selecting ETVs containing sufficient data to appropriately accommodate the LTTE curve. Subsequently, we manually fit the LTTE curve using a custom Python3 program with a Graphical User Interface (GUI) built using PyQt5\footnote{\href{https://pypi.org/project/PyQt5/}{https://pypi.org/project/PyQt5/}}. To identify a suitable starting point for the parameters in the MCMC method, we utilize a Nelder-Mead optimizer. We then derive the parameters and their corresponding errors from the MCMC runs. Finally, we assess the parameters to determine the potential dynamical contribution of the ETVs. ETVs with a dynamical-to-LTTE contribution ratio below 0.1 are retained in our sample.}
%{\red\bf Description of Don\'at's method...}

\section{Results and discussion}
\label{sec:discussion}

\begin{figure*}
\includegraphics[width=\textwidth]{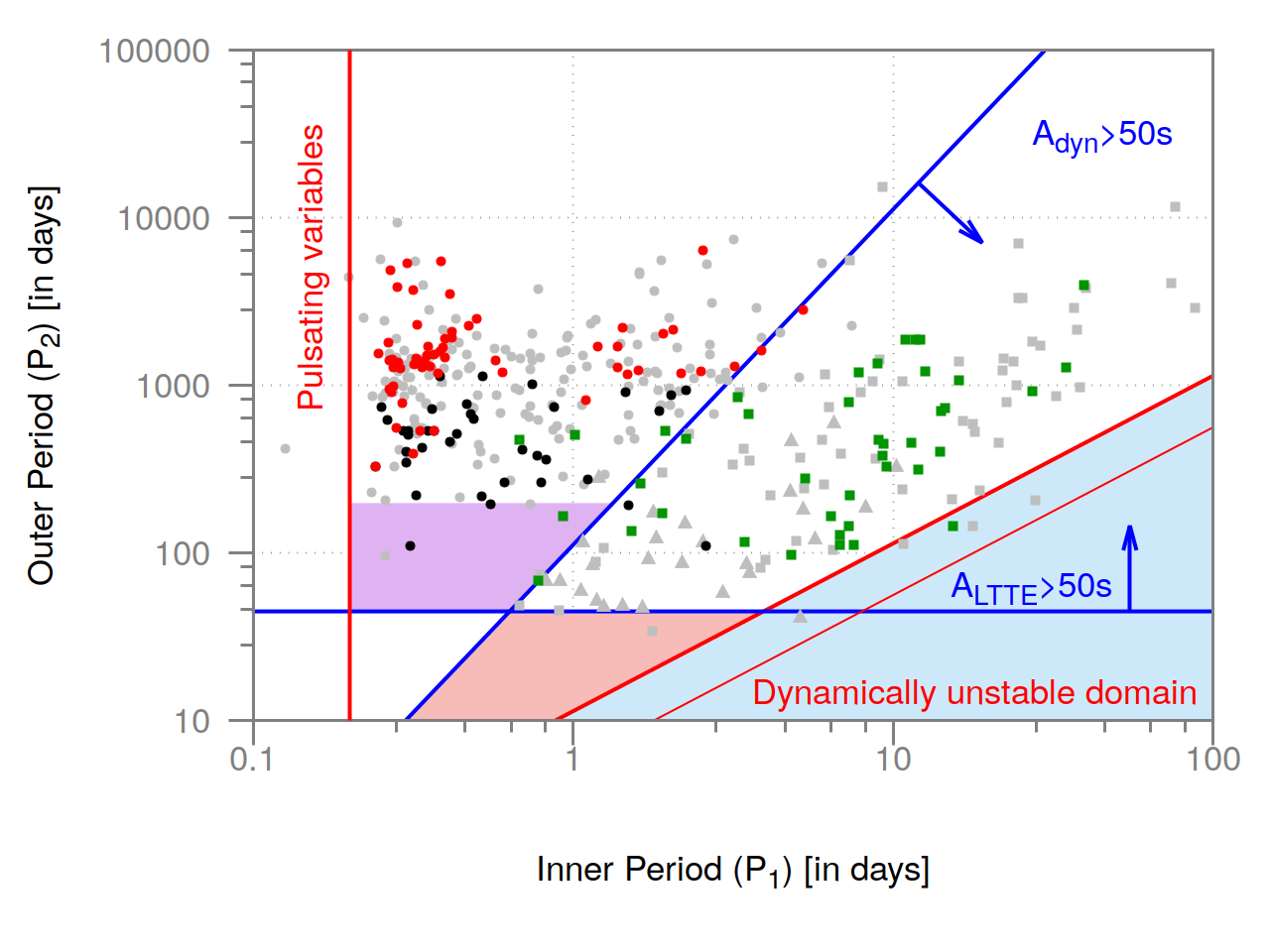}
\caption{Locations of the analyzed hierarchical triple star candidates together with other triples, detected with space telescopes \textit{Kepler} and \textit{TESS} in the $P_1-P_2$ plane. Black and red dots and green squares represent the current NCVZ candidates. For the black and red systems simple LTTE solutions were satisfactory, while for the green ones the dynamical effects were also taken into account. (The red symbols, however, denote the most uncertain solutions which were omitted from the statistical investigations.) Gray dots and squares denote the same classes of those hierarchical triple star candidates in the original \textit{Kepler}-field, which were analyzed in \citet{borkovits16}. Gray triangles represent recently discovered \textit{K2} and \textit{TESS} systems with accurately known photodynamical solutions. (These are mostly triply eclipsing triple systems.) The vertical red line at the left shows the  lower limit of the period of overcontact binaries. The horizontal and sloped blue lines are boundaries that roughly separate detectable ETVs from the undetectable ones. The detection limits again, were set to 50\,sec. These amplitudes were calculated following the same assumptions as in Fig.~8 of \citet{borkovits16}, i.e., $m_\mathrm{A}=m_\mathrm{B}=m_\mathrm{C}=1\,\mathrm{M}_\odot$, $e_2=0.35$, $i_2=60\degr$, $\omega_2=90\degr$. The arrows indicate the direction of the increasing of the respective amplitudes. The shaded regions from left to right represent $(i)$ the W~UMa desert, i.e. the (almost) empty domain where a tight third companion of a short-period EB would be certainly detectable through its LTTE, even in the absence of a measurable dynamical delay; $(ii)$ the purely dynamical region, i.~e., where the dynamical effect should be detectable, while the LTTE would not and; $(iii)$ the dynamically unstable region in the sense of the \citet{2001MNRAS.321..398M} formula. Note, that while the border of this latter shaded area was also calculated with $e_2=0.35$, we also give the limit for $e_2=0.1$, as the thinner red line within this (light blue) region.}
 \label{fig:P1vsP2}
\end{figure*}   

As was mentioned in Sect.~\ref{sec:obsdata} in total we identified 351 EBs of which the ETVs exhibited promising non-linear variations.
In order to identify already known triple systems or candidates in our sample, we thoroughly checked the literature of these 351 EBs. We found five formerly known EBs, where previous ETV studies (based on former, ground-based observed eclipse times) indicate the presence of an additional, more distant component. These are TIC~219109908 \citep[EF Dra,][]{yang12}, TIC~219738202 \citep[BX Dra,][]{parketal13}, TIC~233532554 \citep[RR Dra,][]{2021MNRAS.507.2804W,2022AcA....72...31S}, TIC~298734307 \citep[HL Dra,][]{2021MNRAS.505.6166S} and TIC~288734990 \citep[NSVS 01286630,][]{2016A&A...587A..82W,2018RAA....18..116Z}. In the first four cases, however, the previously found third-body periods are in the range of decades and, hence, the currently found, short-term non-linear ETV behaviors cannot be accounted for by those putative companions, while in the last system although the outer orbital period is on the same order as in our solution, the \textit{TESS} ETV data and, hence, our solution are not in accord with the former results.
%(see below in Sect.~\ref{subsect:individual_systems}).\footnote{\red\bf Ett\H ol f\"uggetlen\"ul, ezekn\'el a csillagokn\'al ki kell majd gy\H ujteni a teljes minimumlist\'at, \'es azzal egy\"utt elemezni!!!!} 
For 9 other objects in our candidate sample we found some external archival eclipse times, but there were no actual earlier ETV studies of these objects.

In conclusion, out of the promising 351 EBs, we identified 135 hierarchical triple system candidates, of which the CHT nature looks almost certain, and another 77 systems which are at least moderately likely CHT candidates. The remaining 58 EBs are less likely cases. In Fig.~\ref{fig:P1vsP2} we plot the locations of these 77 likely CHT candidates in the $P_1$ vs $P_2$ plane where, for a comparison, we also plot some previously known CHTs from some other surveys. Moreover, we plot the ETV curves together with the best analytic LTTE or, combined LTTE+dynamical fits for all the 135 systems in Appendix~\ref{Sect:ETV_plots}.

Here, in what follows, first we present our findings both for the pure LTTE and the combined LTTE+dynamical solution systems in general, and then we discuss some specific subgroups of the newly discovered CHT candidates.

\subsection{Triple system candidates with pure LTTE solutions}

From the ETVs of the narrowed sample of the 135 good CHT candidate systems, we identified 94 EBs in which case we were able to find pure LTTE solutions. According to the reliability of the solutions we divided these systems into three groups. Into the first group we ranked those 19 EBs where we judge our LTTE solutions to be certain or, at least highly likely. These are typically such close likely triple systems, where the period of the third component (i.e. the outer period) does not exceed half of the nominal $\sim1278$-day length of the complete \textit{TESS} NCVZ data set, the amplitude of the ETV curve(s) is substantially higher than the scatter of the individual ETV points and, finally, all the sensitive sections of the light-time orbits are well covered. Note, while the first two criteria look trivial, we had to introduce the third one because of the large intermediate gaps in the observations between Sectors 26 and 40, and Sectors 41 and 47. An example of those CHT candidates where this last criterion was applied is TIC~159398028. This system, despite its outer period of $P_2=671\pm3$\,d, was ranked only in the second (less certain) group because none of the lower ETV extrema -- and only one of the two upper extrema -- was covered by the observations. The outer $P_2$ periods of the third companions are in the range of 109 and 628 days. All these compact triples have very close inner EBs (there are only three inner pairs with $P_1>1$\,days). This is naturally not an unexpected fact, since for longer inner periods such compact systems would have dynamically dominated ETVs. We tabulate the main parameters of the LTTE solutions of this group in Table~\ref{Tab:Orbelem_LTTE1}. The uncertainties of each parameters are given in parentheses and refer to the last digit(s). As the pure LTTE solutions were calculated with non-linear LM optimization, these are formal uncertainties obtained via matrix-inversion and the resultant correlation matrices after reducing the $\chi^2$-s to unity. In the last three columns we give the minimum third-body mass $(m_\mathrm{C})_\mathrm{min}$, the nominal dynamical to LTTE amplitude ratio, $\frac{{\cal{A}}_\mathrm{dyn}}{{\cal{A}}_\mathrm{LTTE}}$, calculated with the use of this minimum mass and, finally, the assumed mass of the binary, $m_\mathrm{AB}$, which was used for both the calculation of the minimum of the outer mass, and the amplitude ratio. When the total mass of the inner binary is known from the literature, then that value is given together with the current references, but in most cases we give only some realistic estimate (which we denote with : after the numerical value).

    \setlength{\tabcolsep}{2.0pt}
\begin{table*}
\begin{center}
\caption{Triple system candidates with certain or very likely LTTE solutions.} 
\label{Tab:Orbelem_LTTE1}  
\begin{tabular}{lccccccccccc} 
\hline
TIC No. & $P_1$ & $\Delta P_1$ & $P_2$ & $a_\mathrm{AB}\sin i_2$ & $e_2$ & $\omega_2$ & $\tau_2$ & $f(m_\mathrm{C})$ & $(m_\mathrm{C})_\mathrm{min}$ & $\frac{{\cal{A}}_\mathrm{dyn}}{{\cal{A}}_\mathrm{LTTE}}$ & $m_\mathrm{AB}$\\
        & (day) &$\times10^{-10}$ (d/c)&(day)&(R$_\odot$)  &       &   (deg)    &   (MBJD) & (M$_\odot$)       & (M$_\odot$)            & &  (M$_\odot$)    \\
\hline
165550395$^a$ &0.32253796(6)&$-8.39$(31)&221.1(3)&60.0(8)&0.13(3)&356(12)&58733(8)&0.059(3)  & 0.60 & 0.02 & 1.30\\
198280388     &0.488399845(7)&$-$ &628.9(4)& 93.0(3) &0.146(7)&354(2) & 58663(3) & 0.0273(3) & 0.56 & 0.006& $2.00^*$ \\
229711743     &0.30085045(1)& $-$ &402.9(5)& 104(1)  & 0.34(2)& 290(3)& 58632(4) & 0.092(3)  & 0.77 & 0.005& $1.46^*$ \\
229751802     &0.51625518(3)&$-12.7$(2)&217.28(3)&112.7(4)&0.491(4)&193.5(4)&58593.7(3)&0.406(4)&1.80&0.05 & 2: \\
229762991     & 0.338461410(8)&$-$&425.0(5)& 75.5(5) & 0.15(2)& 256(5)& 58531(6) & 0.0319(7) & 0.52 & 0.005& $1.57^*$ \\
229771231     &0.82095399(2)& $-$ &358.6(2)& 84.8(8) &0.50(1) &146(1) &58854(1)  & 0.063(2)  & 0.79 & 0.06 & 2: \\
229787617     &0.41265202(3)& $-$ & 456(1) & 79(2)   & 0.12(4)&133(24)&58724(30) & 0.032(2)  & 0.56 & 0.007& $1.79^*$ \\ 
232606864$^b$ &0.29430400(4)&12.9(1)&533.9(4)&97.4(5)&0.411(7)& 48(1) & 58775(2) & 0.0435(7) & 0.56 & 0.004& $1.44^*$ \\ 
233050384     &0.552454498(6)&$-$ &193.81(4)&83.0(2) &0.027(6)&291(12)& 58602(6) & 0.204(2)  & 1.37 & 0.04 & $2.17^*$ \\  
233130175     &1.4940862(2) & $-$ &191.0(7)& 67(3)   &0.07(10)&317(88)&58711(47) &  0.11(2)  & 1.00 & 0.06 & 2: \\
233496533     &2.6064780(1) & $-$ &109.79(8)& 55(1)  &0.002(39)&108(360)&58755(110)& 0.19(1) & 1.25 & 0.01 & 2: \\
233574062     &0.299538793(4)&$-$ &346.5(2) & 91.9(3)&0.171(7)&100(2) & 58719(2) & 0.0866(8) & 0.75 & 0.006& $1.46^*$ \\
233611409$^c$ &0.69617520(2)& $-$ &408.9(4) & 86.7(6)&0.23(1) & 20(3) & 58631(3) & 0.052(1)  & 0.84 & 0.02 & $2.53^*$ \\
236760823     &1.11377193(2)& $-$ & 274.8(1)& 68.3(4)&0.03(1) &106(24)& 58804(18)& 0.0565(9) & 0.75 & 0.11 & 2: \\
237202126     &0.30552307(1)& $-$ & 531.2(5)&174.9(9)&0.138(9)& 207(4)& 58424(5) & 0.254(4)  & 1.23 & 0.003& $1.47^*$ \\
258351350$^d$ &0.772935949(8)&$-$ & 380.7(1)&121.1(3)&0.284(4)&196.5(8)&58836.5(9)& 0.164(1) & 1.61 & 0.03 & 3.43\\  
264218477     &0.308556266(6)& $-$&109.09(4)& 45.1(5)&0.38(2) & 294(3)& 58695(1) & 0.103(4)  & 0.85 & 0.06 & 1.6: \\
288430645$^e$ &0.7915110(3) & $-$ & 261(3) & 93(7)   &0.14(14)&146(59)&58584(44) &  0.16(3)  & 1.18 & 0.02 & 2: \\
320324245     &0.35113308(3)&5.49(18)&532.9(4)&111.0(4)&0.174(5)&37(2)& 58900(3) & 0.0645(8) & 0.70 & 0.004& $1.61^*$ \\ 
\hline
\end{tabular}
\end{center}
{\bf Notes.} {In the last column ($m_\mathrm{AB}$) EB masses, followed by asterisk ($^*$) are estimated from the empirical Period-Mass relations of overcontact binaries \citep{2008MNRAS.390.1577G}; while values followed by semi-colon (:) are our own estimations; \\ 
$a$: $m_\mathrm{AB}$ was taken from \citet{2020ApJS..247...50S}; $b$: V504 Dra; $c$: V411 Dra; $d$: HZ Dra -- $m_\mathrm{AB}$ was taken from \citet{2017MNRAS.465.1181L}; $e$: five third-body eclipses.}
\end{table*}

We classified 17 additional systems with less certain but likely LTTE solutions. These systems are CHTs with a high certainty, but do not fulfill all the criteria which were required for the first group. Their outer periods are in between 507 and 1011 days, i.e. more than one full orbit is covered at least partly, in all cases. All but four systems have inner periods less than 1 days. These are listed in Table~\ref{Tab:Orbelem_LTTE2}.

\setlength{\tabcolsep}{2.0pt}
\begin{table*}
\begin{center}
\caption{Triple system candidates with less certain but likely LTTE solutions.} 
\label{Tab:Orbelem_LTTE2}  
\begin{tabular}{lccccccccccc} 
\hline
TIC No. & $P_1$ & $\Delta P_1$ & $P_2$ & $a_\mathrm{AB}\sin i_2$ & $e_2$ & $\omega_2$ & $\tau_2$ & $f(m_\mathrm{C})$ & $(m_\mathrm{C})_\mathrm{min}$ & $\frac{{\cal{A}}_\mathrm{dyn}}{{\cal{A}}_\mathrm{LTTE}}$ & $m_\mathrm{AB}$\\
        & (day) &$\times10^{-10}$ (d/c)&(day)&(R$_\odot$)  &       &   (deg)    &   (MBJD) & (M$_\odot$)       & (M$_\odot$)            & &  (M$_\odot$)    \\
\hline
159398028     &0.48007761(1)& $-$ & 671(3) & 27.3(8) & 0.10(4)&296(21)&58866(39) &0.00610(5) & 0.14 & 0.005& 2: \\
159465833     & 1.852848(2) & $-$ & 696(15)& 194(29) & 0.3(2) &181(32)&58721(64) & 0.20(9)   & 2.17 & 0.05 & 2: \\
159509734     &0.262377357(4)&$-$ & 619(5) & 14.0(6) & 0.63(6)&240(5) & 58919(11)&0.000095(13)&0.06 & 0.004&$1.34^*$ \\
160518449     &2.25604658(5)& $-$ & 926(1) &103.7(5) &0.25(1) & 94(1) & 58763(3) & 0.0174(3) & 0.47 & 0.07 & 2: \\
219809405     &0.304936731(6)&$-$ & 507(3) & 12.1(5) & 0.06(9)&128(76)&58863(107)&0.00009(1) & 0.06 & 0.004&$1.47^*$ \\
229412530     &0.36690442(2)& $-$ & 530(3) & 112(4)  &0.82(2) & 27(1) & 58585(5) & 0.066(8)  & 0.72 & 0.03 & $1.66^*$ \\
229789945     &0.36334623(3)& $-$ & 720(4) & 64(2)   &0.13(5) &157(18)& 58961(36)& 0.0069(6) & 0.30 & 0.003& $1.65^*$ \\
229972643     &0.87325531(8)& $-$ & 740(5) & 72(2)   &0.19(3) &323(14)& 58869(30)& 0.0091(7) & 0.37 & 0.01 & 2: \\
258875507     &0.60881129(1)& $-$ &262.3(8)& 18.4(7) &0.39(6) &334(9) & 58676(7) & 0.0012(1) & 0.17 & 0.05 & 1.8: \\
259038777     &0.43327169(3)& $-$ & 510(3) & 59(1)   & 0.04(4)&264(76)&58639(107)& 0.0108(7) & 0.38 & 0.006& $1.85^*$ \\
259264759     &0.5204445(1) & $-$ &1126(13)& 151(3)  &0.26(2) &290(4) & 58658(15)& 0.036(2)  & 0.65  &0.002& $2.09^*$ \\
288510753     &2.0198571(2) & $-$ & 867(4) & 113(1)  &0.29(2) &268(3) & 59060(9) & 0.0254(8) & 0.55 & 0.06 & 2: \\
288734990 &0.383928424(2)&$-$ &1121(5) & 34(2)   &0.00(2) & 89(5) & 54529(8) &0.00044(6) & 0.09 & 0.002& 1.2: \\
357596894     & 0.4646877(2)& $-$ & 766(15)& 165(9)  &0.38(6) &191(7) & 58435(27)& 0.10(2)   & 0.82 & 0.005& 1.5: \\
377105433     &0.25140325(1)& $-$ & 734(3) & 57.4(9) & 0.41(3)&163(3) & 58934(7) & 0.0047(2) & 0.22 & 0.002& $1.30^*$\\
424461577     &0.74483611(2)& $-$ &1011(2) & 185(3)  &0.38(2) & 124(1)& 58295(3) & 0.083(5)  & 0.88 & 0.007& 2: \\
441788515     & 1.4548748(5)& $-$ & 911(13)& 169(10) &0.32(4) & 79(15)& 58637(39)& 0.078(14) & 0.86 & 0.029& 2: \\
\hline
\end{tabular}
\end{center}
%{\bf Notes.} {$a$: See Sect.~\ref{subsect:individual_systems} for details.}
\end{table*}

Finally, in Table~\ref{Tab:Orbelem_LTTE3} we tabulate the parameters of the LTTE `solutions' of the remaining 58 systems. This is a very inhomogeneous group. In the case of some EBs the LTTE solution looks quite reasonable, however, when the period is certainly longer than the duration of the dataset, the parameters we obtained are likely not very robust. In other cases, however, even the interpretation of the non-linear ETVs as LTTE signals seem suspect. Strictly speaking, from the fact, that we were able to obtain any LTTE solutions do not follow that there is any third component in the given systems. Therefore, this third set should be considered only with great caution, and we do not recommend using the numerical results of this subgroup for any statistical studies. Instead, the group of these systems should be considered only as EBs which display clearly non-linear ETVs and hence, are worthy of further attention.

\setlength{\tabcolsep}{0.75pt}
\begin{table*}
\begin{center}
\caption{Triple system candidates with very uncertain, questionable LTTE solutions.} 
\label{Tab:Orbelem_LTTE3}  
\begin{tabular}{lccccccccccc} 
\hline
TIC No. & $P_1$ & $\Delta P_1$ & $P_2$ & $a_\mathrm{AB}\sin i_2$ & $e_2$ & $\omega_2$ & $\tau_2$ & $f(m_\mathrm{C})$ & $(m_\mathrm{C})_\mathrm{min}$ & $\frac{{\cal{A}}_\mathrm{dyn}}{{\cal{A}}_\mathrm{LTTE}}$ & $m_\mathrm{AB}$\\
        & (day) &$\times10^{-10}$ (d/c)&(day)&(R$_\odot$)  &       &   (deg)    &   (MBJD) & (M$_\odot$)       & (M$_\odot$)            & &  (M$_\odot$)    \\
\hline
159510983     &0.32427244(4)& $-$ &1441(25)& 49(2)   & 0.66(3)&267(3) & 58433(22)&0.00075(8) & 0.13 & 0.002&$1.53^*$ \\
160396455     & 1.3800761(3)& $-$ &1689(873)& 25(15) &0.63(54)& 59(38)&59069(369)&0.00007(15)& 0.07 & 0.02 & 2: \\
160418633     &0.241149471(7)&$-$ & 327(2) & 17(2)   &0.68(9) &330(7) & 58615(9) & 0.0006(2) & 0.10 & 0.02 & $1.27^*$ \\
160492808     &0.315173247(3)&$-$ & 391(3) & 6.2(6)  & 0.7(1) &145(10)& 58671(13)&0.000021(7)& 0.04 & 0.02 &$1.50^*$ \\
160520144     &2.17965944(5)& $-$ &1173(26)& 85(4)   & 0.7(1) & 91(11)& 58750(44)& 0.006(1)  & 0.27 & 0.11 & 1.5:\\
165500662     & 0.3668363(1)& $-$ &1531(145)& 64(8)  & 0.8(2) & 149(3)& 58677(20)& 0.0015(6) & 0.17 & 0.005& $1.66^*$ \\
165527014     &0.47028292(4)& $-$ &2255(382)& 35(3)  & 0.8(2) & 30(7) & 58948(95)&0.00011(5) & 0.07 & 0.004& 1.5: \\ 
165527500     &0.3962501(2) & $-$ &1451(137)&35(9)   & 0.57(8)&260(6) & 58638(50)& 0.0003(2) & 0.10 & 0.002& $1.74^*$ \\
165529344     &0.28017833(4)&$-4.3(2)$&553(4)&26.7(8)& 0.52(4)&260(4) & 58417(10)&0.00083(7) & 0.12 & 0.004&$1.39^*$ \\
198158617     &0.28314100(6)& $-$ &3845(408)&133(7)  & 0.8(2) & 34(2) & 59101(61)& 0.0021(6) & 0.17 & 0.0006& $1.40^*$ \\
198183185$^a$ &0.28907245(4)& $-$ &1251(21)& 82(1)   & 0.33(2)&154(5) & 58501(22)& 0.0047(3) & 0.23 &0.0009& $1.42^*$ \\   
198280146     &0.26745252(2)& $-$ &1402(12)& 197(12) & 0.91(1)& 16(1) & 58952(14)& 0.052(10) & 0.58 & 0.01 & $1.35^*$ \\
198355086     & 0.333710(1) & $-$ &1380(50)& 267(83) & 0.8(1) &246(16)& 58705(97)& 0.13(13)  & 0.94 & 0.003& $1.56^*$ \\
199632809     & 1.5980418(6)& $-$ &1229(34)& 148(4)  & 0.38(4)&175(9) & 58562(36)& 0.029(3)  & 0.57 & 0.03 & 2: \\
219761337$^b$ & 0.3174737(9)& $-$ &3662(1520)&211(97)& 0.71(5)&294(4) & 58789(99)& 0.009(15) & 0.32 &0.0004& $1.51^*$ \\
219787718$^c$ &0.35259030(3)& $-$ &1700(100)&134(2)  &0.281(5)&278(2) & 58066(38)& 0.011(1)  & 0.35 &0.0007& $1.62^*$ \\
219856620     &0.37941453(2)& $-$ &1171(16)& 74(2)   & 0.21(4)&294(9) & 58764(30)& 0.0039(3) & 0.25 & 0.001&$1.70^*$ \\
219859424$^d$ &0.34722214(2)& $-$ &1319(30)& 30.8(5) & 0.30(2)&334(4) & 58131(26)& 0.00023(2)& 0.09 & 0.001&$1.60^*$ \\
219876159     &0.28160414(6)& $-$ &1265(52)& 49(3)   & 0.59(9)&154(5) & 58741(31)& 0.0010(2) & 0.13 & 0.001&$1.40^*$ \\
229437073     &0.33789064(3)& $-$ &1270(14)& 120(1)  & 0.40(2)&316(2) & 58742(10)& 0.0143(6) & 0.38 &0.001 & $1.57^*$ \\
229451028     &0.3182817(1) & $-$ &1325(36)& 138(7)  & 0.70(2)&107(1) & 58677(16)& 0.020(3)  & 0.42 &0.002 & $1.51^*$ \\
229500406     &0.3834058(6) & $-$ &1591(152)&255(23) & 0.16(5)&113(28)&58010(151)& 0.087(29) & 0.82 &0.0007& $1.71^*$ \\
229507536     &1.0926249(1) & $-$ & 808(5) & 84(3)   & 0.51(5)& 13(5) &58989(12) & 0.012(1)  & 0.41 & 0.03 & 2: \\
229584508     &0.35845415(3)& $-$ &1296(17)& 76(1)   & 0.52(2)& 276(1)&58656(12) & 0.0034(2) & 0.23 & 0.002& $1.63^*$ \\
229592690     &0.27497593(3)& $-$ &1277(7) &104(2)   & 0.82(1)& 68(1) &58682(7)  & 0.0092(5) & 0.30 & 0.003& $1.38^*$ \\
229605410     & 3.193544(5) & $-$ &1288(92)& 211(20) &0.25(13)&135(13)&58710(71) & 0.076(24) & 0.85 & 0.07 & 2: \\
229651225     &0.34981401(4)& $-$ &1500(200)& 157(2) & 0.27(1)& 195(4)&58007(96) & 0.023(6)  & 0.46 &0.0008& $1.61^*$ \\ 
229773372     & 0.2638422(1)& $-$ &1800(100)& 169(6) & 0.07(1)&117(27)&58949(137)& 0.020(3)  & 0.39 &0.0003& $1.34^*$ \\
229791016     & 0.5011304(9)& $-$ &2479(352)& 202(33)& 0.57(5)& 40(7) &57452(251)& 0.018(10) & 0.49 & 0.001& $2.03^*$ \\
229799471     & 0.398328(5) & $-$ &1902(757)&388(231)&0.23(10)&229(28)&59609(499)& 0.22(42)  & 1.25 &0.0006& $1.75^*$ \\
229800815     & 0.4167940(1)& $-$ &2094(280)& 38(5)  & 0.56(3)& 151(5)&58900(55) & 0.00017(9)& 0.09 & 0.001& $1.80^*$ \\
229910746     & 2.506324(1) & $-$ &1199(64) & 40(5)  & 0.37(5)&131(5) &58636(46) &0.00058(22)& 0.14 & 0.07 & 2: \\
229952080     & 0.385382(9) & $-$ &5498(15210)&532(1332)&0.62(62)&198(14)&59246(2104)&0.07(62)&0.74 &0.0002& $1.71^*$ \\
230003294     & 0.4165119(3)& $-$ &1916(466)& 41(11) & 0.65(7)& 137(8)&58758(60) & 0.0003(2) & 0.10 & 0.002& $1.80^*$ \\
230007820$^e$ & 5.253221(5) & $-$ &2815(50) &380(118)&0.63(12)& 203(2)&57355(31) & 0.093(97) & 0.93 & 0.13 & 2: \\
233056681     &0.60112646(6)& $-$ &1184(60) & 89(127)& 0.99(5)&153(41)&59173(191)& 0.007(29) & 0.33 & 1.43 & 2: \\
233680160     &0.24739890(3)& $-$ &1535(25) &222(18) & 0.73(6)& 128(3)&59258(21) & 0.06(2)   & 0.61 & 0.001& $1.29^*$ \\
233689972$^f$ & 1.3807321(3)& $-$ &1276(104)& 40(4)  & 0.6(1) & 274(6)&58684(47) & 0.0005(2) & 0.21 & 0.02 & $4.02^*$ \\
233719825     & 2.05091(2)  & $-$ &2141(620)&514(194)& 0.41(7)& 141(9)&58835(103)& 0.40(50)  & 1.78 & 0.01 & 2: \\
258920306     & 3.882199(8) & $-$ &1600(194)& 153(35)&0.21(8) & 170(9)&58643(99) & 0.019(14) & 0.49 & 0.08 & 2: \\  
258921745     & 1.907185(6) & $-$ &2027(892)&366(168)&0.06(24)&118(126)&58434(775)&0.16(26)  & 1.17 & 0.01 & 2: \\
259168350     & 1.4770090(5)& $-$ & 1158(13)& 227(7) & 0.58(5)& 123(3)&58808(14) & 0.12(1)   & 1.02 & 0.04 & 2: \\
275692690     & 1.429690(3) & $-$ &2208(482)&334(59) & 0.4(1) & 230(5)&59177(155)& 0.10(7)   & 0.97 & 0.007& 2: \\
298600301     & 2.54142(1)  & $-$ &6326(16993)&297(607)&0.9(2)& 20(12)&59380(2523)&0.009(72) & 0.37 & 0.07 & 2: \\  
353990339     & 0.268689(4) & $-$ &4868(3932)&662(641)&0.4(2) & 6(4)  &59429(815)& 0.16(54)  & 0.96 &0.00007& $1.36^*$ \\
359629786     &0.39148653(6)& $-$ &1678(162)& 47(3)  & 0.88(5)& 259(6)&59005(66) &0.00050(13)& 0.13 & 0.006& 2: \\
362227092     &0.2839931(3) & $-$ &1359(147)& 60(13) & 0.45(4)&268(6) &58800(38) & 0.0015(11)& 0.16 &0.0008& $1.41^*$ \\
362259979     &0.29305558(1)& $-$ & 785(12)& 14.5(5) & 0.43(6)&237(15)&58954(37) &0.000066(7)& 0.05 & 0.002& $1.44^*$ \\ 
367853009     &  0.410870(1)& $-$ &3471(959)&425(113)&0.34(10)& 19(15)&58914(197)& 0.086(83) & 0.84 &0.0003& $1.79^*$ \\
377054541$^g$ & 0.3255584(1)& $-$ & 2300   & 382(8)  &0.329(8)& 83(1) & 58865(10)& 0.141(9)  & 0.96 &0.0003& $1.54^*$\\    
377251803$^h$ & 0.5713695(2)& $-$ &1394(76)& 56(4)   & 0.26(2)&286(16)& 58257(72)& 0.0012(3) & 0.26 & 0.002 & 3.5: \\
377307592$^i$ &1.19283483(3)& $-$ &1706(5) & 267(14) & 0.80(3)& 49(4) & 58635(19)& 0.088(14) & 0.90 & 0.03 & 2: \\
389962894     &0.273380963(8)&$-$ &1422(54)& 49(2)   &0.82(2) & 27(2) & 58944(19)& 0.0008(1) & 0.12 & 0.004& $1.37^*$ \\
389966320     &0.3039930(5) & $-$ &5328(748)&557(60) & 0.8(2) & 344(2)&59184(101)& 0.082(35) & 0.74 &0.0005& $1.47^*$ \\ 
389968719     &0.33178698(3)&12.3(2)&530(1)& 26.4(4) & 0.48(2)& 341(3)& 58772(4) & 0.00088(4)& 0.13 & 0.006& 1.5: \\ 
394089883     & 0.2669306(2)& $-$ & 938(26)& 241(24) &0.49(10)& 201(5)& 58424(29)& 0.21(6)   & 1.08 &0.001 & $1.35^*$\\
441803530     &0.275244907(9)&$-$ & 982(10)& 27.0(6) & 0.01(7)&326(268)&59102(731)&0.00027(2)& 0.08 & 0.001& $1.38^*$ \\
441806190     &0.270233738(9)&$-$ & 912(3) & 113(7)  & 0.69(7)& 288(3)& 58585(8) & 0.023(5)  & 0.42 & 0.003& $1.36^*$ \\
\hline
\end{tabular}
\end{center}
{\bf Notes.} {$a$: V489 Dra; $b$: V516 Dra; $^c$: V377 Dra; $^d$: V362 Dra; $e$: SWASP data used; $f$: V402 Dra. One former mid-eclipse time is available in \citet{nelson16}, which does not fit to the current solution; $g$: Outer period is fixed to get realistic third-body mass; $h$: DG Dra; $i$: KK Dra. SWASP and other ground-based eclipse times were used.}
\end{table*}

\subsection{Triple system candidates with combined LTTE and dynamical solutions}

For 41 EBs we obtained combined LTTE and dynamical ETV solutions. The inferred outer periods in this group range from 68 days to $\sim 4000$ days. Similar to the pure LTTE systems, again, we consider the solutions of those 27 triples certain where the outer period does not exceed half the length of the given data set (see Table~\ref{Tab:Orbelemdyn1}). (Naturally, the same caution that was discussed in case of the LTTE systems, must be taken into account here as well.) The nominally less certain solutions of the remaining 14 systems are tabulated in Table~\ref{Tab:Orbelemdyn2}. 

The structure of these Tables~\ref{Tab:Orbelemdyn1},~\ref{Tab:Orbelemdyn2} departs slightly from that of the three tables of LTTE triples (Tables~\ref{Tab:Orbelem_LTTE1}--\ref{Tab:Orbelem_LTTE3}). This is in accord with the fact that from a combined solution one can determine individual masses (at least $m_\mathrm{C}$ and $m_\mathrm{AB}$), and, hence the semi-major axis of the outer orbit ($a_2$) can also be readily calculated. Moreover, in these tables instead of a theoretically calculated, nominal dynamical to LTTE contribution amplitude ratio, we give the true, `measured' ratio of $\mathcal{A}_\mathrm{dyn}/\mathcal{A}_\mathrm{LTTE}$, which can be determined directly from the analytic ETV solution curves. Because of the large number of the parameters that can be obtained from such a dynamical solution, we also introduce Table~\ref{Tab:AMEparam} where we give additional parameters for the combined solution systems. These are mainly the apsidal motion and/or orientation and nodal precession parameters of these triples. In this regard, however, some caution should be taken as follows.

First, as was discussed in Sect.~5.2 of \citet{borkovits16} in the context of the mutual inclination ($i_\mathrm{m}$) parameter, practically one cannot distinguish between, e.g. an $i_\mathrm{m}=0\degr$ and $i_\mathrm{m}=10\degr$ ETV solution, at least not on the timescale of \textit{TESS} observations. From an observational point of view, however, there may be strong differences between the two solutions, as the second one may result in rapid inner inclination ($i_1$) and, hence eclipse depth variations and, therefore, they should be clearly distinguishable. Hence, in all cases, when our solution resulted in such a mutual inclination value, which was not in accord with the rate of the observed EDVs (including non-EDV cases), we did not accept that solution, and forced a coplanar ($i_\mathrm{m}=0\degr$) solution.

Second, the tabulated inner inclinations ($i_1$) should also be considered with great caution. The dynamical ETV solution, especially for coplanar cases, is almost insensitive to this parameter. Hence, it is better to fix its value a priori. Since for the vast majority of the combined LTTE+dyn systems no light curve solution is available, generally $i_1$ is an unknown parameter. Fortunately, however, most of the dynamically dominated systems have longer inner periods (see below) and, hence, one should not get large errors by assuming inclinations of $i_1\sim87\degr-89\degr$, as we have done.

Comparing these triple systems with the pure LTTE triples, one can see, that in general they contain less close inner binaries. (The mean and the median of the inner periods are $\overline{P}_1=9\fd341$ and $P_1^\mathrm{med}=7\fd539$, respectively.) This property may come about for different reasons. For example, the amplitude of the dynamical ETV contribution is proportional to $P_1^2/P_2$, i.e., besides the tightness ($P_1/P_2$) of the system, it also scales with the inner period and, hence, amongst two similarly tight triples, the longer period one produces larger amplitude, better detectable ETVs (assuming, of course, that the other orbital and dynamical parameters are similar). This fact may explain, at least in part, the lack of tight triple systems containing the closest binaries. Moreover, in the case of the closest binaries, tidal effects might be effective enough to circularize the inner orbits and, hence, most of the observed closest binaries revolve on circular, or very low eccentricity orbits. Again, this can reduce the amplitudes of the dynamical terms effectively \citep{2011A&A...528A..53B}, which may explain, again, the lack of the detection of triples (at least, via ETV) in the light red region of Fig.~\ref{fig:P1vsP2}. On the other hand, for somewhat wider inner EBs, those more distant third companions which would produce pure LTTE-generated ETVs, must have orbital periods in the range of at least a few years and, hence, they cannot be detected with a high confidence during the 3-year-long, and non-continuous observations of the NCVZ by \textit{TESS}. Consequently, the pure LTTE triple star candidates should be offset towards the shorter inner period, i.e., closest systems.

\subsection{Systems with special interests}

In what follows, we discuss some small subgroups of the identified triple star candidates which are worthy of special interest for different reasons.

\subsubsection{Tightest triples}
\label{Sect:tightest_triples}

We found 5 systems with $P_2/P_1\lesssim20$. These are TICs~236774836 (with $P_2/P_1\sim9.35$), 219885468 (14.7), 389966039 (16.3), 235934882 (18.6), 233684019 (19.9). (In Fig.~\ref{fig:P1vsP2} these systems are represented with the five green rectangles which are the closest to the border of the dynamically unstable domain at the lower right part of the plot.)  Despite the weakly hierarchical nature of these triples, our analytic, hierarchical three-body models fit and describe these ETVs fairly well. However, according to our prior experiences with \textit{Kepler} systems, for such tight triples, our analytic hierarchical triple star model becomes less accurate on a somewhat longer timescale of a decade (which, in general, is about the timescale of the ``apse-node'' type or, long-period perturbations of these triples), suggesting that the analytic model of this latter class should be improved). In this regard, TIC~219885468 serves as a good case study of that effect, because recently \citet{borkovitsmitnyan23} had carried out a more complex photodynamical analysis of this triple star system. During such an analysis the motion of the triple star system is integrated numerically and, hence, we can compare our analytical results with the outputs of the more exact numerical integration. Regarding the masses, \citet{borkovitsmitnyan23} obtained $m^\mathrm{photdyn}_\mathrm{AB}=2.37\pm0.04\,\mathrm{M}_{\sun}$ and $m^\mathrm{photdyn}_\mathrm{C}=0.76\pm0.02\,\mathrm{M}_{\sun}$ (while the analytical model resulted in $m^\mathrm{analytical}_\mathrm{AB}=2.45\pm0.13\,\mathrm{M}_{\sun}$ and $m^\mathrm{analytical}_\mathrm{C}=0.70\pm0.16\,\mathrm{M}_{\sun}$, respectively). The eccentricities\footnote{Comparing, however, the orbital elements, one should keep in mind, that the definition and, hence, the meaning of the orbital elements in a photodynamical and an analytical model are different. In the former one, these are instantaneous, osculating orbital elements at a given epoch, while in the analytic model, these are some kind of averaged orbital elements for not necessarily the same epoch.} are also in good accord ($e^\mathrm{photdyn}_1=0.0423\pm0.0006$ vs. $e^\mathrm{analytical}_1=0.052\pm0.006$) and ($e^\mathrm{photdyn}_2=0.3903\pm0.0007$ vs. $e^\mathrm{analytical}_2=0.41\pm0.06$). Similarly, the argument of periastron of the inner orbit is quite close to each other in the two solutions ($\omega^\mathrm{photdyn}_1=259\fdg3\pm0\fdg2$ vs. $\omega^\mathrm{anal}_1=266\degr\pm1\degr$). In contrast to this, at first sight, there is remarkable discrepancy in the arguments of periastron of the outer orbit ($\omega^\mathrm{photdyn}_2=256\fdg0\pm0\fdg7$ vs. $\omega^\mathrm{analytical}_2=83\degr\pm8\degr$). However, this difference is close to $180\degr$, and can be understood from that point of view, that the analytic perturbation formulae depend upon only $2\omega_\mathrm{out}$ and, hence, they are insensitive to a $180\degr$ difference in $\omega_2$. On the other hand, the LTTE-term depends on $\omega_2$, which fact, in general, could have resolved this ambiguity, but in the current situation the LTTE contribution is only about $2-3\%$ of the dynamical one in the ETV and, thus, our analytical solution still suffers from this ambiguity. In conclusion, we can say that our approximate, analytical ETV solutions look quite acceptable even for such tight triples. On the other hand, however, turning to the some decade-long evolution (i.~e. apsidal motion) of the current (and also, other, similarly tight) systems, we again refer to \citet{borkovitsmitnyan23} who pointed out, that there are some discrepancies between the analytically predicted and numerical integrated apsidal motion periods (for the current system $P_\mathrm{apse,1}^\mathrm{analytical}=20$\,yr versus $P_\mathrm{apse,1}^\mathrm{photdyn}=13.6$\,yr).
%On the other hand, one should keep in mind, that the inner, eclipsing pair of TIC~219885468 is formed by two very similar stars (according to \citealt{borkovitsmitnyan23} the inner mass ratio is $q_\mathrm{in}=0.99\pm0.02$) and thus, the octupole order perturbations should be negligible in this system. So, our conclusion does not remain necessarily true for such tight triples, where octupole perturbations may be non-negligible.} 
Hence, future follow-up observations and then numerical integration and modeling of the triple star motion of these tight systems would be especially important.

\subsubsection{Triple star candidates with eclipse depth variations (EDVs)}
\label{Sect:EDVtriples}

We identified eight triple star candidates where the inner EBs exhibit EDVs. These systems are: TICs~233684019, 233729038, 236774836, 259004910, 288611133, 357686232, 417701432, 441768471. Amongst them, the most dramatic variation can be found in TIC 236774836, which is the tightest triple system candidate in our entire sample (with $\mathbf{P_2/P_1\sim9.35}$). While during the Year 2 observations the eclipse depths int his source remain constant, a dramatic decrease in the eclipse depths takes place in the Year 4 data and, the eclipses disappear completely near the beginning of Year 5 (Sector 56). The most likely explanation for such a rapid EDV is a non-coplanar configuration of the given triple system which forces precession of the orbital planes and, hence, variation of the inclination of the EB. As the timescale of such an effect is proportional to $P_2^2/P_1$, one can expect the detection of such a rapid effect in the tightest and most compact triple systems.
Finally, note that TIC~230002837 also exhibits slight, suspicious EDVs, but we were unable to find any reliable (or, at least, uncertain) third-body solution for its ETVs.

\subsubsection{Eccentric EBs (eEBs) with detectable apsidal motion}

We found eccentric EB orbits in 27 candidate systems (see Table~\ref{Tab:AMEparam}). All of them belong to the systems with combined LTTE+dynamical ETV solution. In an eEB one can expect apsidal motion.
%. The three main sources of apsidal motion are i) the classic tidal perturbations due to the non-spherical mass distributions of the tidally and rotationally oblated binary components; ii) the general relativistic effects and; iii) the third-body forced dynamical apsidal motions. While the tidal and relativistic effects, in general, result very long period apsidal motion (typically from several decades to millennia), the third-body effects may lead to very rapid apsidal motion on timescales of a few years in the most compact and tightest triples (as the dynamical apsidal motion period, again, is proportional to $P_2^2/P_1$, see, e.~g., \citealt{borkovits15}).}
While the eccentricity of an EB's orbit can be discerned, in general, simply by a visual inspection of a possibly unequal time interval between the odd-even and even-odd eclipse events and/or the different durations of the even and odd eclipses, the apsidal motion most readily reveals itself via the divergence or convergence of the primary and secondary ETV curves (which is the manifestation of the anticorrelated behaviour of the primary and secondary ETV curves on a time interval, that is substantially shorter than the period of the apsidal precession). Because the duration of the nominally $\sim1300$\,d \textit{TESS} observations is very short relative to the decades to millennia long tidal and relativistic apsidal motion periods, one cannot expect the detection of apsidal motion caused primarily by these effects in our ETV curves. Thus, where we clearly detect apsidal motion in our short-term ETVs, we can be certain that this effect is due to rapid, dynamically forced apsidal motion.

We identified clear apsidal motion in 10 systems. These are TICs~199616648, 219771659, 219885468, 233684019, 233729038, 233738966, 236774836, 256514937, 357686232, 417701432. Note, there are some other systems, where the ETVs suggest apsidal motion but the detection is less certain since the $P_2$-period dynamically forced ETVs may hide the convergence or divergence of the two ETVs. These 7 systems are TICs~236769201, 243337122, 259271740, 272705788, 288611133, 376976908, 441768471.

In Table~\ref{Tab:AMEparam} we give the theoretical dynamical apsidal motion periods ($P_\mathrm{apse}$) which are calculated by our software using the quadrupole order perturbation theory of the hierarchical three-body problem \citep[see][Appendix C, for a detailed description]{borkovits15}. As one can see, the shortest theoretical apsidal motion period is $P_\mathrm{apse}=9.6$\,yr. It was detected in TIC~236774836, i.~e., in that notable system where the eclipses completely disappeared for 2022 (see in Sect.~\ref{Sect:EDVtriples} above).

Note also, that our software package calculates the theoretical, dynamical apsidal motion for all non-circular, dynamically dominated ETVs and, hence, $P_\mathrm{apse}$ is tabulated in Table~\ref{Tab:AMEparam} for all eEBs, irrespective of whether the apsidal motion manifests itself during the short interval of \textit{TESS} observations. Hence, one can realize that there are four systems, for which the given $P_\mathrm{apse}$ value is negative. Here, the negative sign denotes retrograde apsidal motion. While the tidal and relativistic apsidal motions are always prograde (i.e., the orbital ellipse revolves in the same direction as that of the orbiting bodies), dynamically forced apsidal motion can be retrograde either for dynamical or geometric reasons \citep[see, e.~g.][]{borkovits22a}.

\subsubsection{Triply eclipsing triples}

We found third-body eclipses in four systems. These are TICs~229785001, 233684019, 288430645, 441738417. Their third-body periods range from 145 to 472 days. The periods deduced from the ETV curves are in agreement with the locations of the extra eclipses in all the four systems, confirming that the same objects are the origin of the extra eclipses and the timing variations. Note, a detailed photodynamical analysis of TIC~229785001 was published earlier by \citet{rappaport23}. The results of our analytic ETV solution are in good accord with their results. We plan complex photodynamical investigations of the remaining three systems in the near future.

%\subsubsection{Notes on some individual systems}
%\label{subsect:individual_systems}

\subsubsection{Interpretation of the peculiar ETV curve of GZ Draconis}

\begin{figure}
    \centering
    \includegraphics[width=0.48\textwidth]{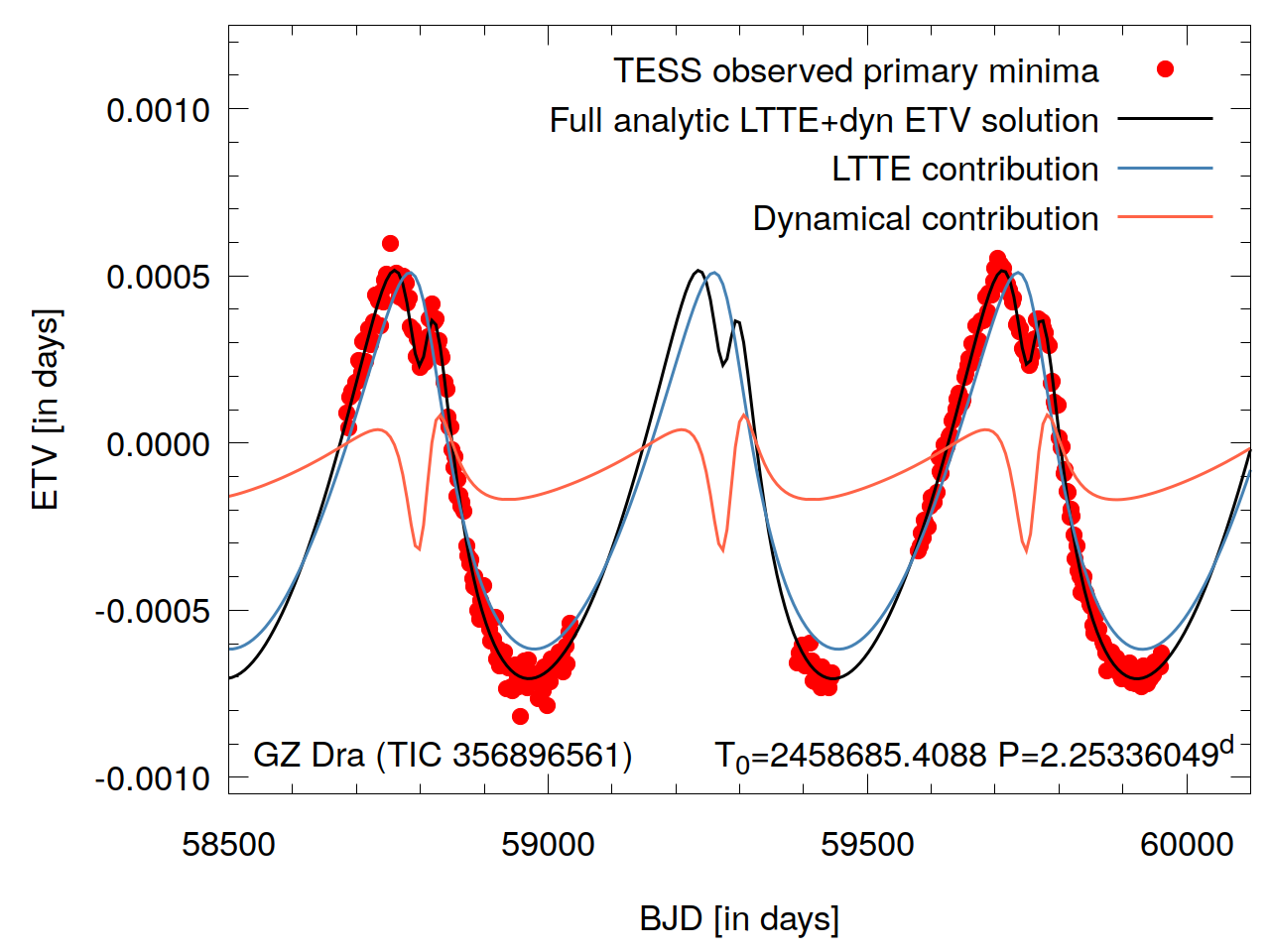}
    \caption{Primary ETV curve of GZ Draconis, formed from the 2-min \textit{TESS} measurements (red dots), together with the combined LTTE+dyn solution (black curve). We plot separately both the LTTE (light blue) and quadrupole-order dynamical (light red) contributions. The unusual bumps near to the upper extrema of the observed curves are clearly visible and, indicate a significant, non-coplanar dynamical contribution. See text for further details.}
    \label{fig_GZDra-ETVbumps}
\end{figure}

While GZ Draconis does not appear to be an extremely interesting, peculiar triple system; nonetheless we feel it worthwhile to discuss the cautionary tale of this quite average EB. This relatively bright (V=9.4) object first was known as a visual double star from the 1980's \citep[see, e.~g.][]{1984A&AS...57..467M,1987ApJS...65..161H} and then, the EB nature of the brighter component was discovered by the \textit{Hipparcos} satellite \citep{1999IBVS.4659....1K,2000A&A...356..141F}. The system was observed continuously in 2-min cadence mode during Years 2, 4 and 5 of \textit{TESS} observations. From our point of view, only the ETV curves calculated from the \textit{TESS}-observed primary and secondary eclipses have importance. It was clear for us already after the Year 2 observations that the timing diagrams show clear, third-body induced ETVs. By the end of the Year 2 observations, and using some former, less accurate times of primary eclipses, calculated from the publicly available SuperWASP observations, we were able to find reasonably good LTTE solution with a period of $P_2=476\pm1$\,d. We realized, however, that there was an unexpected dip on the primary ETV curve around BJD $\sim2\,458800$ with an amplitude of $\sim0\fd0003$ and duration of $\sim40$\,days, but did not pay careful attention to this feature, assuming that it was caused by some light curve distortions (for example, spottedness). Note, this dip cannot be seen on the much less accurate secondary ETV curve. Then the new NCVZ monitoring of \textit{TESS} began during the second year of the first extended mission, and the new eclipse times confirmed the former ETV solution sectors by sectors. The uninteresting characteristics of such a straightforward  ETV curve changed immediately when the Sector 53 observations became available, and we calculated the new set of eclipse times. The same dip in the ETV curve as in the Year 2 data occurred again and, at the very same phase of the third-body orbit as two outer orbital cycles earlier (Fig.~\ref{fig_GZDra-ETVbumps}). By this time we realized that even though the P2/P1 ratio of $\sim$211 is large (i.e., not a very tight triple) the dynamical effect might still actually produce a non-negligible contribution to the ETV curve and explain the unexpected "extra" dips.\footnote{Similar dips can be seen in the dynamically contributed ETV curves of several other, mostly \textit{Kepler}-discovered triple systems \citep[see, e.~g.,][]{borkovits15,borkovits16,gaulmeetal22}.} A further, more in-depth consideration of the shape of the ETV curve makes this possibility even more exciting. Taking a look at on the other (bottom) extrema of the ETV curves, one realizes that during these phases both in the Year 2 (in between cca. JDs 2\,458\,930 and 2\,459\,000) and the Year 5 (JDs 2\,459\,880 -- 2\,459\,950) the curvatures of the ETVs practically disappear, which suggests the presence of a second bump, which counteracts the LTTE. Therefore, we may assume two bumps within one outer period (for the dynamical curve) which, in the case of a circular inner orbit (as is the case in GZ~Dra), may occur only if the two orbital planes are substantially inclined to each other \citep[see Eq.~46 of][]{2003A&A...398.1091B}. Thus, the shape of the ETV suggests that there is an exceptional chance in the case of GZ Draconis for determining the mutual inclination purely from the ETV for such a non-tight triple star system. Then, our search for an inclined, combined LTTE and dynamical ETV solution became quite fruitful, and we realized that this triple star must be one of the currently most inclined triple star systems known with $i_\mathrm{m}=58\degr\pm7\degr$.

Finally, in this regard, we make two additional notes. First, we refer to the fact that that this mutual inclination is well within the high mutual inclination domain of the original von Zeipel-Lidov-Kozai phenomena \citep[see, e.g.][and further references therein]{naoz16}, where large amplitude eccentricity cycles may occur. Despite this fact, one cannot expect such large amplitude eccentricity variations in the currently circular inner pair, because of the tidal interactions between the two components, which cancels the ZLK cycles.

Second, such a large mutual inclination implies very large amplitude orbital plane precession cycles, which would lead to substantial EDVs and also the disappearance (and later, reappearance) of the eclipses. While this may be true, as one can see in Table~\ref{Tab:AMEparam}, the theoretical period of such variations in GZ Draconis (with the measured parameters) is about $P_\mathrm{node}\sim3\,800$\,yr, which explains the absence of EDV during the few-year-long \textit{TESS} observations.

~

%\noindent
%\textit{Comparison of the former and the current ETV solution of TIC~288734990}

\setlength{\tabcolsep}{1.0pt}
\begin{table*}
\begin{center}
\caption{Orbital Elements from combined dynamical and LTTE solutions with outer period shorter than half of the length of the data sets.} 
\label{Tab:Orbelemdyn1}  
\begin{tabular}{lccccccccccc} 
\hline
TIC No. & $P_1$ & $P_2$ & $a_2$ & $e_2$ & $\omega_2$ & $\tau_2$ & $f(m_\mathrm{C})$ & $\frac{m_\mathrm{C}}{m_\mathrm{ABC}}$ & $m_\mathrm{AB}$ & $m_\mathrm{C}$ & $\frac{{\cal{A}}^\mathrm{meas}_\mathrm{dyn}}{{\cal{A}}_\mathrm{LTTE}}$\\
        & (day) & (day) &(R$_\odot$)  &       &   (deg)    &   (MBJD) & (M$_\odot$)       & (M$_\odot$)            &  (M$_\odot$)  &   \\
\hline
198241524$^a$&0.77517979(7)& 68.34(2) & 107(1)& 0.09(2)&  68(8) & 58684(2) & 0.075(6)  & 0.28(1) & 2.53(9)    & 0.99(4) & 0.14\\
199616648 &14.074093(3) & 399.4(9) & 265(39) & 0.14(1) &  88(3) & 58575(4) & 0.06(8)   & 0.34(19)& 1.04(60)   & 0.54(34) & 9.2\\
219771659 &14.494786(2) & 728.8(6) & 523(3)  &0.0000(9)&  77(3) & 58436(6) & 0.122(8)  & 0.324(8)& 2.45(4)    & 1.17(4)  & 3.28\\
219885468 & 7.539395(7) & 110.8(2) & 142(3)  & 0.41(6) &  83(8) & 58816(1) & 0.034(14) & 0.22(4) & 2.45(13)   & 0.70(16) & 42 \\ 
219892913 & 3.5444107(4)& 669.6(7) & 432(56) &0.763(8) & 218(3) & 58927(8) & 0.15(14)  & 0.39(17)& 1.46(86)   & 0.95(37) & 2.79 \\
229785001$^b$&0.92976(2)& 165.37(5)& 200(1)  & 0.46(3) & 31(3)  &58730.7(7)& 0.29()    & 0.42(2) & 2.28(3)    & 1.65(3)  & \\
          &             &3254(160) & 1573(60)& 0.14(4) & 121(10)&57817(60) & 0.040()   & 0.20(10)& 3.93(6)    & 1.00(10) & \\
233073872 &14.139796(9) & 703(2)   & 633(18) & 0.52(2) & 61(7)  & 59029(2) & 0.21(6)   & 0.31(4) & 4.73(51)   & 2.17(30) & 7.19 \\
233530543 & 1.6215399(4)& 260(1)   & 280(12) & 0.35(7) & 83(5)  &58767.3(7)& 0.16(5)   & 0.33(4) & 2.91(47)   & 1.45(34) & 0.27 \\
233684019 & 7.2814918(8)& 145.04(6)& 159(6)  & 0.30(2) & 88(2)  &58742.2(5)& 0.0066(14)& 0.14(1) & 2.23(29)   & 0.35(5)  & 14 \\
233729038 &11.402383(1) & 453.1(3) & 425(22) & 0.17(4) & 140(8) & 58529(11)& 0.30(16)  & 0.39(8) & 3.04(49)   & 1.98(61) & 2.05 \\
233738966 & 9.2705839(4)& 379.4(1) & 347(9)  & 0.104(2)& 184(1) &58750.9(7)& 0.059(6)  & 0.250(4)& 2.93(28)   & 0.98(10) & 3.12 \\
235687051 & 3.4327181(3)& 116.2(3) & 170(9)  & 0.18(7) & 138(5) & 58716(3) & 0.0016(12)&0.069(23)& 4.55(77)   & 0.34(10) & 2.06 \\ 
235934882 & 6.8382493(8)& 127.09(2)& 158(3)  & 0.327(6)& 3.8(6) &58630.6(3)& 0.050(9)  & 0.25(2) & 2.44(17)   & 0.81(6)  & 17.6 \\
236769201 & 6.3754483(6)& 165.76(3)& 198(4)  & 0.44(1) & 146(2) & 58718(1) & 0.075(13) & 0.27(2) & 2.76(24)   & 1.03(7)  & 14.8 \\
236774836 & 15.42852(3) & 144.2(1) & 202(8)  & 0.021(2)& 31(5)  & 58701(2) & 0.78(21)  & 0.53(6) & 2.51(43)   & 2.82(43) & 12.9 \\
237234024 &1.89581975(6)& 172.5(1) & 202(10) & 0.21(2) & 175(7) & 58653(3) & 0.070(26) & 0.27(5) & 2.73(51)   & 0.99(14) & 0.43 \\
256514937  &5.3289107(2)& 277.75(7)& 328(5)  & 0.668(4)& 205(1) &58834.4(3)& 0.312(22) & 0.394(3)& 3.72(25)   & 2.42(16) & 9.56 \\ 
356014478 & 9.531384(6) & 328(1)   & 364(17) & 0.20(2) & 172(13)& 58659(16)& 3.13(71)  & 0.80(7) & 1.18(51)   & 4.85(67) & 2.95 \\
356896561$^c$&2.25336049(2)&476(2) & 382(30) & 0.53(4) & 307(6) & 58806(6) & 0.0006(4) & 0.08(3) & 3.03(78)   & 0.27(7)  & 0.35 \\
357686232 & 7.3089436(3)& 221.2(2) & 201(5)  & 0.059(3)& 320(3) & 58609(2) & 0.0014(2) & 0.087(4)& 2.04(15)   & 0.19(2)  & 3.02 \\
389966039 & 6.833337(8) & 111.3(1) & 161(8)  & 0.38(3) & 84(6)  &58701.1(9)& 0.029(15) & 0.19(4) & 3.65(63)   & 0.84(20) & 37.7 \\
402656423 & 1.5192997(2)& 135.0(4) & 177(12) & 0.45(5) & 214(9) & 58854(3) & 0.16(9)   & 0.34(9) & 2.66(71)   & 1.40(37) & 1.19 \\
417701432 & 9.366182(9) & 445(1)   & 413(24) & 0.45(2) & 340(3) & 58910(2) & 0.06(3)   & 0.23(6) & 3.67(78)   & 1.10(27) & 9.14 \\
420263314 & 12.010564(4)& 314(1)   & 327(10) & 0.53(5) & 300(3) &58725(4)  &0.0000019(6)&0.009(1)& 4.75(43)   &0.045(7)  & 54.6 \\
423927178$^d$&4.79972(2)& 97.0(1)  & 167(15) & 0.07(2) & 170(6) & 58716(2) & 1.44(74)  & 0.60(9) & 2.68(57)   &4.02(1.69)& 2.11 \\
441738417 &0.67658321(2)& 471.7(7) & 378(6)  & 0.833(8)& 1(4)   & 58807(7) & 1.06(9)   & 0.69(2) & 1.01(9)    & 2.24(13) & 0.41 \\
459969365 & 9.006622(2) & 473(2)   & 388(63) & 0.53(3) & 319(8) & 58677(17)& 0.17(22)  & 0.36(21) & 2.23(1.51) & 1.28(77) & 7.6 \\ 
\hline
\end{tabular}
\end{center}
{\bf Notes.} {$a$: Cubic ephemeris: $\Delta P=-67(2)\times10^{-10}$\,d/c$^2$, $c_3=0.68(4)\times10^{-12}$\,d/c$^3$; $^b$: 2+1+1 configuration; photodynamical solution taken from \citet{rappaport23}; $^c$: GZ Dra -- WASP data used; $^d$: quadratic ephemeris: $\Delta P=10(1)\times10^{-7}$\,d/c$^2$. }
\end{table*}

\setlength{\tabcolsep}{2.5pt}
\begin{table*}
\begin{center}
\caption{Orbital elements from less certain combined dynamical and LTTE solutions. (The outer period is longer than half of the length of the data sets or, critical section(s) are missing.)} 
\label{Tab:Orbelemdyn2}  
\begin{tabular}{lccccccccccc} 
\hline
TIC No. & $P_1$ & $P_2$ & $a_2$ & $e_2$ & $\omega_2$ & $\tau_2$ & $f(m_\mathrm{C})$ & $\frac{m_\mathrm{C}}{m_\mathrm{ABC}}$ & $m_\mathrm{AB}$ & $m_\mathrm{C}$ & $\frac{{\cal{A}}^\mathrm{meas}_\mathrm{dyn}}{{\cal{A}}_\mathrm{LTTE}}$\\
        & (day) & (day) &(R$_\odot$)  &       &   (deg)    &   (MBJD) & (M$_\odot$)       & (M$_\odot$)            &  (M$_\odot$)  &   \\
\hline
219788156 & 7.8244925(8)& 1187(96) & 614(99) & 0.15(8) & 197(47)&58864(164)& 0.0003(3) & 0.05(1) & 2.09(1.00) & 0.12(7)  & 0.24 \\
229594479 & 1.9375364(2)& 535(6)   & 359(27) & 0.34(9) & 152(11)& 58948(19)& 0.020(6)  & 0.21(1) & 1.72(46)   & 0.45(13) & 0.15 \\
230012179 & 8.9234907(4)& 1348(7)  & 746(52) & 0.19(1) & 188(4) & 58978(14)& 0.008(3)  & 0.14(2) & 2.65(64)   & 0.42(12) & 0.32 \\
243337122 &10.937295(1) & 1853(70) & 992(51) & 0.51(1) & 270(3) & 59012(5) & 0.033(11) & 0.22(4) & 2.80(50)   & 0.84(10) & 1.73 \\
259004910 &12.607407(2) & 1215(119)& 621(52) & 0.12(7) & 222(6) & 59196(35)& 0.039(9)  & 0.26(1) & 1.61(32)   & 0.57(12) & 0.76 \\
259271740 &11.751608(1) & 1865(74) &1183(49) & 0.35(2) & 310(5) & 59488(26)& 0.17(2)   & 0.300(7)& 4.48(56)   & 1.91(25) & 0.95 \\
272705788 & 7.2566223(2)& 795.4(8) & 489(5)  & 0.273(7)&175.0(9)& 58847(2) & 0.043(3)  & 0.262(6)& 1.84(7)    & 0.65(3)  & 0.77 \\
288611133 & 16.130427(5)& 1070(9)  & 760(38) & 0.073(5)& 69(3)  & 59078(20)& 0.22(9)   & 0.38(7) & 3.19(64)   & 1.95(42) & 1.81 \\
288611883 & 1.0104126(4)&  507(8)  & 354(14) & 0.75(4) & 66(8)  & 58857(14)& 0.15(5)   & 0.41(5) & 1.38(11)   & 0.94(22) & 0.27 \\ 
376976908 & 34.77081(3) & 1271(19) & 707(45) & 0.31(1) & 53(2)  & 59440(2) & 0.12(6)   & 0.37(9) & 1.86(47)   & 1.08(29) & 8.55 \\
392569978 & 39.44393(2) & 3961(24) & 1934(27)& 0.423(3)& 129(3) & 59411(15)& 0.16(2)   & 0.30(2) & 4.36(22)   & 1.82(10) & 3.65 \\
392572173 &12.151204(2) & 1856(38) & 1184(68)& 0.06(2) & 59(9)  & 59929(44)& 0.055(22) & 0.26(5) & 4.81(1.06) & 1.68(22) & 1.13 \\
420177014 &3.25857999(3)& 846(2)   & 515(32) & 0.38(1) & 153(1) & 58277(4) & 0.10(3)   & 0.34(1) & 1.70(42)   & 0.86(22) & 0.22 \\
441768471 &27.295906(5) & 917.2(3) & 577(9)  & 0.262(1)&267.3(3)& 58806(1) & 0.14(2)   & 0.36(2) & 1.96(13)   & 1.11(7)  & 8.31 \\
\hline
\end{tabular}
\end{center}
\end{table*}

\begin{table*}
\begin{center}
\caption{Apsidal motion and/or orientation parameters from AME and dynamical fits.} 
\label{Tab:AMEparam}
\begin{tabular}{lccccccccccc} 
\hline
TIC No. & $P_\mathrm{anom}$ & $a_1$      & $e_1$ & $\omega_1$ & $\tau_1$ & $P_\mathrm{apse}$ & $i_\mathrm{m}$ & $i_1$ & $i_2$ & $\Delta\Omega$ & $P_\mathrm{node}$\\
        & (days)            &(R$_\odot$) &       & (deg)      & (MJD)    &   (years)         & (deg) & (deg) & (deg) &   (deg)   &  (years) \\
\hline
198241524    &0.77515141(7)&4.84(5)& 0        & $-$     & $-$        & $-$         & 0      & 80 & 80 &   0     & $-67$\\
199616648    &14.078392(3) & 25(5) & 0.088(5) & 107(1)  & 58707.87(6)& 122         & 8(2)   & 89 & 93 &   6(4)  & $-98$\\
219771659    &14.494829(2) &33.7(2)&0.2716(4)&   2(2)  & 58675.77(8) & 902         & 21(1)  & 89 & 89 &  21(1)  & $-288$\\
219788156$^a$& 7.8244736(8)& 21(3) & 0        & $-$     & $-$        & $-$         & 0      &89(2)&89 &   0     & $-6373$ \\
219885468$^b$& 7.547063(7) &21.8(4)& 0.052(6) & 266(1)  & 58686.75(2)&  20         & 0      & 88 & 88 &   0     & $-14$\\
219892913    & 3.5445656(4)& 11(2) & 0.014(1) & 205(7)  & 58679.39(7)& 317         & 2(4)   & 88 &86.8&  $-$1(3)& $-285$\\
229594479    & 1.9375300(2)& 7.8(7)& 0        & $-$     & $-$        & $-$         & 0      & 87 & 87 &   0     & $-1831$ \\
229785001$^c$&             &5.28(2)&0.0005(3) & 84(60)  & 58683.0(1) &             & 3(1)   &87.7(6)&89.23(4)& 3(1)& \\
230012179    & 8.9235427(4)& 25(2) & 0.031(5) & 91.0(3) &58670.490(9)& 5103        & 0      & 89 & 89 &   0     & $-3880$ \\
233073872$^d$&14.142051(9) & 41(1) & 0.039(2) & 139(3)  & 58665.5(1) & 254         & 0      & 89 & 89 &   0     & $-213$ \\ 
233530543    &1.6215143(4) & 8.3(4)& 0        & $-$     & $-$        & $-$         & 0      & 86 & 86 &   0     & $-339$ \\ 
233684019$^e$&7.2836669(8) &20.7(9)&0.0074(5) & 84.2(4) &58680.575(8)& 67          & 8(2)   & 92 & 89 &   8(2)  & $-41$ \\
233729038$^f$&11.404789(1) & 31(2) & 0.013(2) &  72(4)  & 58706.6(1) & 148         & 18(7)  & 89 & 97 &  16(7)  & $-149$ \\
233738966    & 9.2715706(4)&26.6(8)& 0.134(2) & 298.9(6)& 58677.71(1)& 230         & 8(3)   & 89 & 82 &  2(2)   & $-177$ \\
235687051    & 3.4324994(3)&15.9(9)& 0        & $-$     & $-$        & $-$         & 0      & 89 & 89 &   0     & $-95$ \\
235934882    & 6.8382380(8)&20.4(5)&0.00250(3)& 351(2)  & 58680.93(3)& 29          & 0      & 89 & 89 &   0     & $-22$ \\
236769201$^g$& 6.3781832(6)&20.3(6)& 0.009(5) & 5(5)    & 58679.60(8)& 42          & 0      & 87 & 87 &   0     & $-33$ \\
236774836$^h$& 15.49633(3) & 36(3) & 0.194(5) & 236(1)  & 58683.12(4)& 9.6         & 5(2)   & 89 & 84 &   3(2)  & $-8.2$ \\
237234024    &1.89575417(6)&9.0(6) & 0        & $-$     & $-$        & $-$         & 0      & 86 & 86 &   0     & $-171$ \\
243337122    &10.937381(1) & 30(2) & 0.26(2)  & 252(2)  & 58677.21(5)& 3861        & 26(4)  & 89 & 69 & $-17$(5)& $-2969$ \\
256514937    & 5.3294342(2)&19.9(4)&0.0260(8) & 56(2)   & 58682.38(2)& 142         & 23(1)  & 89 & 110& $-10$(1)& $-52$ \\
259004910$^f$&12.607183(2) & 27(2) &0.14758(8)&357.5(6) & 58678.37(2)& $-1640$     & 45(4)  & 93 & 94 & 45(4)   & $-1824$ \\
259271740    &11.751674(1) & 36(2) & 0.316(8) & 230(1)  & 58674.09(2)& 3950        & 31(5)  & 92 & 90 & $-31$(5)& $-2778$ \\
272705788    & 7.2566190(2)&19.3(2)& 0.151(2) & 181(1)  & 58678.68(2)& $-1468$     & 43(1)  & 89 & 96 & 42(1)   & $-1191$ \\
288611133$^f$& 16.129992(5)& 40(3) & 0.21(1)  & 38(17)  & 58678.1(9) & $-1531$     & 37(4)  & 89 &113 & $-29$(5)& $-651$ \\
288611883    & 1.0104071(4)& 4.7(1)& 0        & $-$     &  $-$       & $-$         & 0      & 89 & 89 &   0     & $-625$ \\
356014478$^i$& 9.524509(6) & 20(3) & 0        & $-$     &  $-$       &  $-$        & 0      & 89 & 89 &   0     & $-47$ \\ 
356896561    &2.25336107(2)&10.5(9)& 0        & $-$     &  $-$       &  $-$        & 58(7)  &85.3&135.7&$-33$(6)& $-3825$ \\
357686232    & 7.3089470(3)&20.1(5)& 0.0654(6)& 333(1)  & 58680.54(2)& 284         & 3(3)   &89.7&87.9& -3(3)   & $-147$ \\
376976908    & 34.77571(3) & 55(5) & 0.092(4) & 107.2(8)& 58646.09(9)& 886         & 22(1)  & 89 & 111&   4(2)  & $-358$ \\
389966039$^j$& 6.836605(8) & 23(1) & 0.13(2)  & 67(3)   & 58677.43(9)& 28          & 0      & 88 & 88 &   0     & $-19$ \\
392569978    & 39.44786(2) & 80(1) & 0.28(1)  & 228(2)  & 58662.4(2) & 4506        & 15(2)  & 89 & 92 & $-15$(2)& $-2698$ \\
392572173    & 12.151180(2)& 38(2) & 0.39(2)  & 322(6)  & 58710.1(1) & 2680        & 51(6)  & 89 & 52 &  38(5)  & $-6275$ \\
402656423    & 1.5192070(2)& 7.7(7)& 0        & $-$     & $-$        & $-$         & 0      & 80 & 80 &   0     & $-80$ \\
417701432$^f$& 9.366882(9) & 29(2) & 0.22(2)  & 241(3)  & 58676.06(6)& 312         & 14(4)  & 89 & 78 &   9(6)  & $-158$ \\
420177014    &3.25855942(3)&11.0(9)& 0        & 0       &  $-$       &  $-$        & 0      & 88 & 88 &   0     & $-1719$\\ 
420263614    &12.010766(4) & 37(1) & 0.3908(6)& 190.5(4)&58732.908(8)& 2055        & 56(4)  &86.4&129.4& 39(3)  & $-314$ \\
423927178    & 4.79262(2)  & 17(1) & 0        & $-$     &  $-$       &  $-$        & 0      & 88 & 88 &   0     & $-11$ \\
441738417$^k$& 0.67657755(2)&3.3(1)& 0        & $-$     &  $-$       &  $-$        & 0      & 89 & 89 &   0     & $-288$\\
441768471$^f$&27.296093(5) & 48(1) & 0.0687(3)& 46.7(2) & 58674.99(2)& 492         & 20.8(2)& 89 &98.1&$-$18.8(2)& $-254$ \\
459969365    & 9.008075(2) & 24(5) & 0.012(2) &  28(17) & 58681.1(4) & 153         & 4(5)   &88.6&89.5&  4(5) & $-131$ \\     
\hline
\end{tabular}
\end{center}
{\bf Notes.} {$^a$: Allowing non-coplanar solution resulted in $i_\mathrm{m}=43\degr\pm10\degr$, but we do not trust in it; $^b$: Adjusted mutual inclination resulted in $i_\mathrm{m}=23\degr\pm7\degr$ which would lead to $\Delta i_1\sim 27\degr$ during \textit{TESS} observations and hence, certainly overestimated; $^c$: Third-body eclipses; 2+1+1 configuration; photodynamical solution from \citet{rappaport23}; $^d$: Adjusted mutual inclination resulted in $i_\mathrm{m}=22\degr\pm7\degr$ which would lead to $\Delta i_1\sim 4\degr$ during \textit{TESS} observations and hence, certainly overestimated; $^e$: Grazing third-body eclipses + EDV $^f$: Eclipse depth variation; $^g$: Adjusted mutual inclination resulted in $i_\mathrm{m}=11\degr\pm2\degr$ which would lead to $\Delta i_1\sim 6\degr$ during \textit{TESS} observations and hence, certainly overestimated; $^h$: Disappearing eclipses; $^i$: Adjusted mutual inclination resulted in $i_\mathrm{m}=13\degr\pm7\degr$ which would lead to $\Delta i_1\sim 9\degr$ during \textit{TESS} observations and hence, certainly overestimated; $^j$: No secondary ETV curve; $^k$: Third-body eclipses}
\end{table*}

\subsection{Statistics}

\citet{borkovits16} published the largest collection of CHTs with known orbital periods using a sample of 222 objects found in the \textit{Kepler} field. We have about a third of that sample size with 77 objects that are likely or very likely to be triples, but since we applied similar detection and analysis techniques, it may be worthwhile to perform a similar statistical analysis on our newly found systems in the \textit{TESS} NCVZ and make a comparison.

\begin{figure}
\centering
\includegraphics[width=0.5\textwidth]{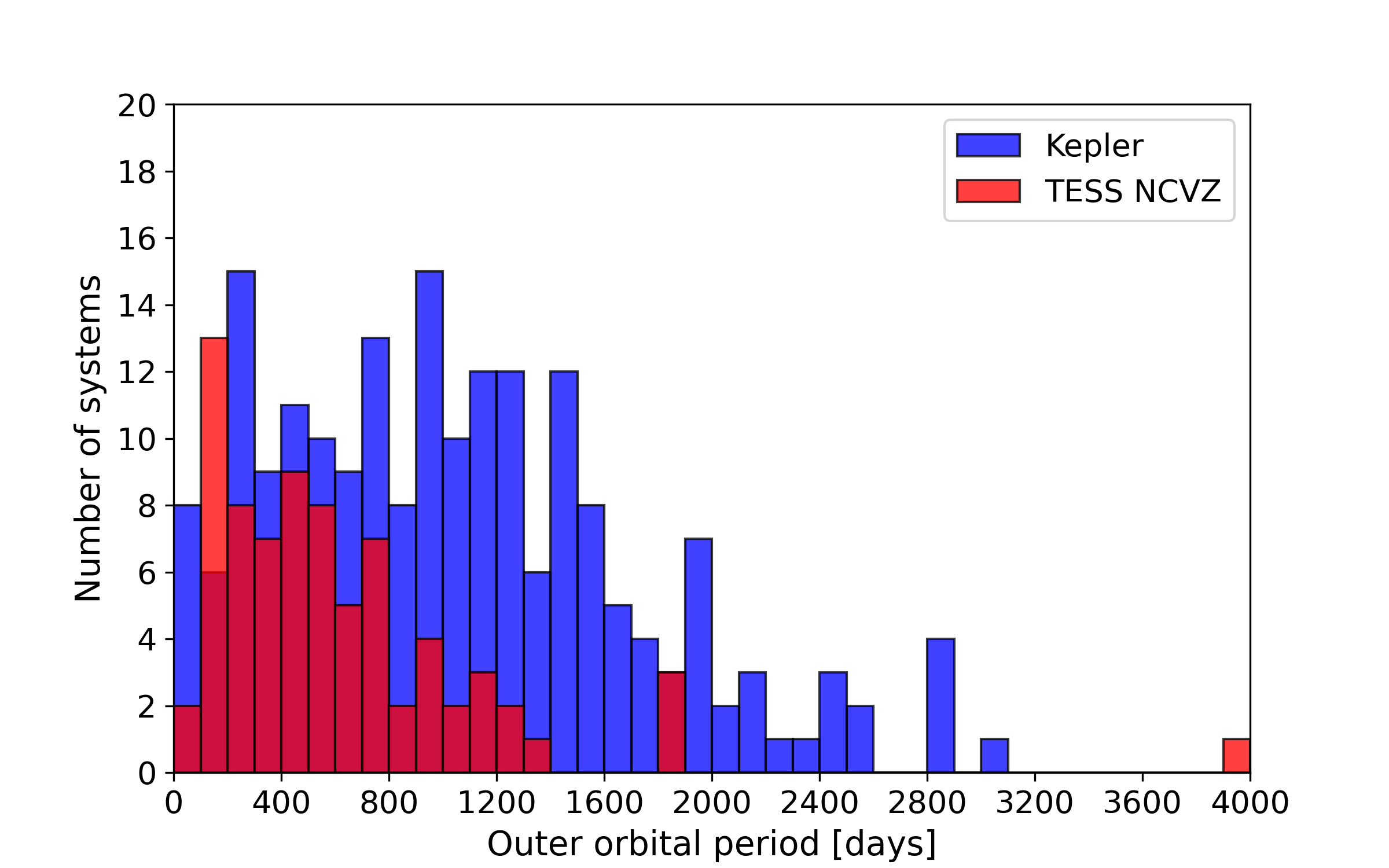}
\caption{Distribution of outer orbital periods for the \textit{Kepler} (blue) and \textit{TESS} NCVZ (red) samples.}
 \label{fig:P2dist}
\end{figure}

In Fig. \ref{fig:P2dist}, we show the distribution of the outer orbital periods in our sample. It can be seen that most systems have outer orbital periods of 100-800 days. In our photometric data sets there are smaller and larger gaps due to the observing strategy of \textit{TESS}: the former are caused by the fact that the telescope changes its observational field of view in every $\sim$27 days and it is not guaranteed that the object will continue to be observed in all Sectors; while the latter are simply caused by the fact that \textit{TESS} observes different hemispheres during alternating years. This observational strategy could be responsible for the slightly elevated number of systems with 100-800 day outer orbital periods as they are easier to detect with such data sets. The total temporal coverage of our data sets is about 1300 days long and one can see that the number of systems with longer orbital periods are decreasing more or less with $P_{2}^{-1} $ as suggested by \citet{borkovits16}.

\begin{figure}
\centering
\includegraphics[width=0.5\textwidth]{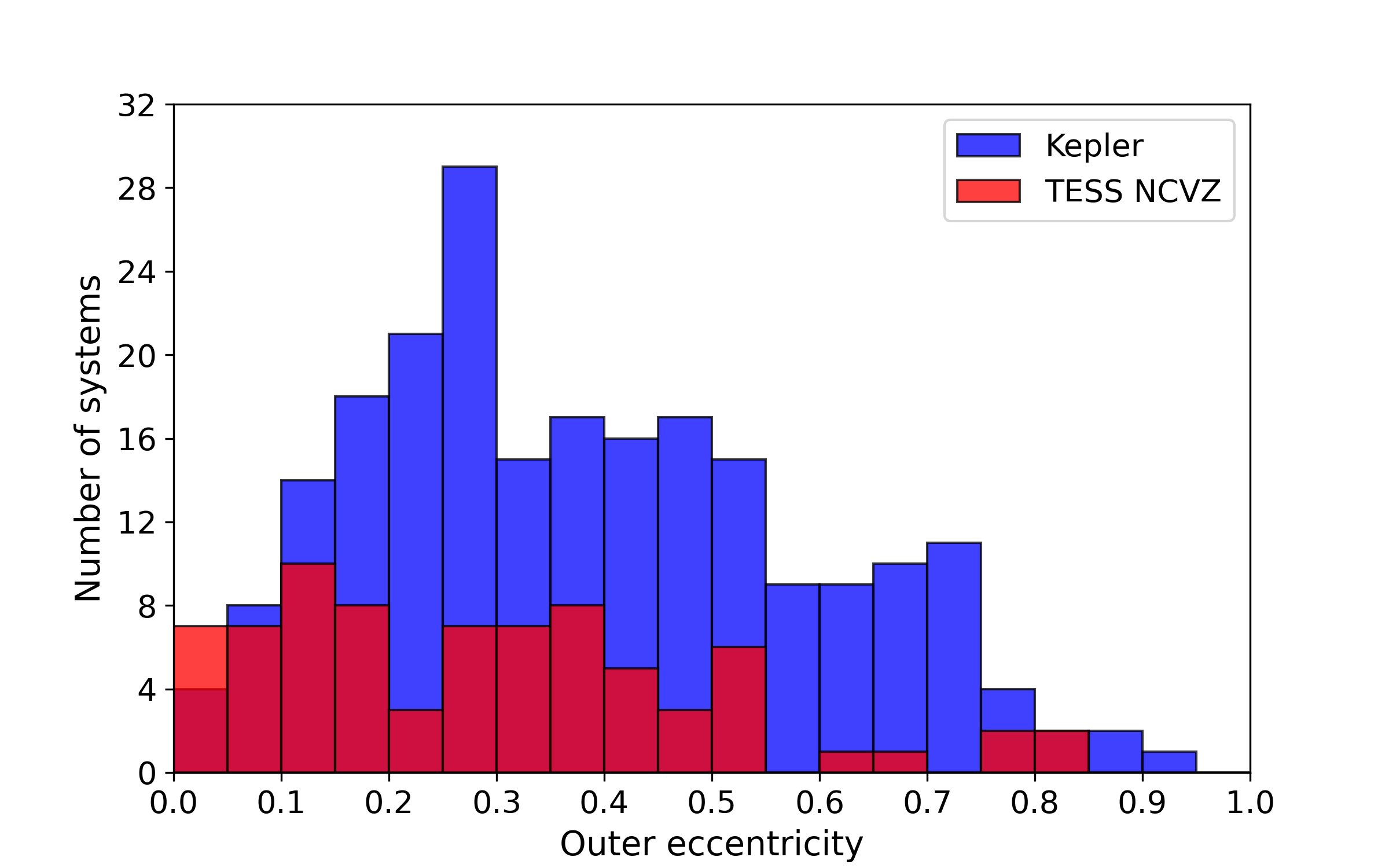}
\caption{Distribution of outer orbital eccentricities for the \textit{Kepler} (blue) and \textit{TESS} NCVZ (red) samples.}
 \label{fig:e2dist}
\end{figure}

\begin{figure}
\centering
\includegraphics[width=0.5\textwidth]{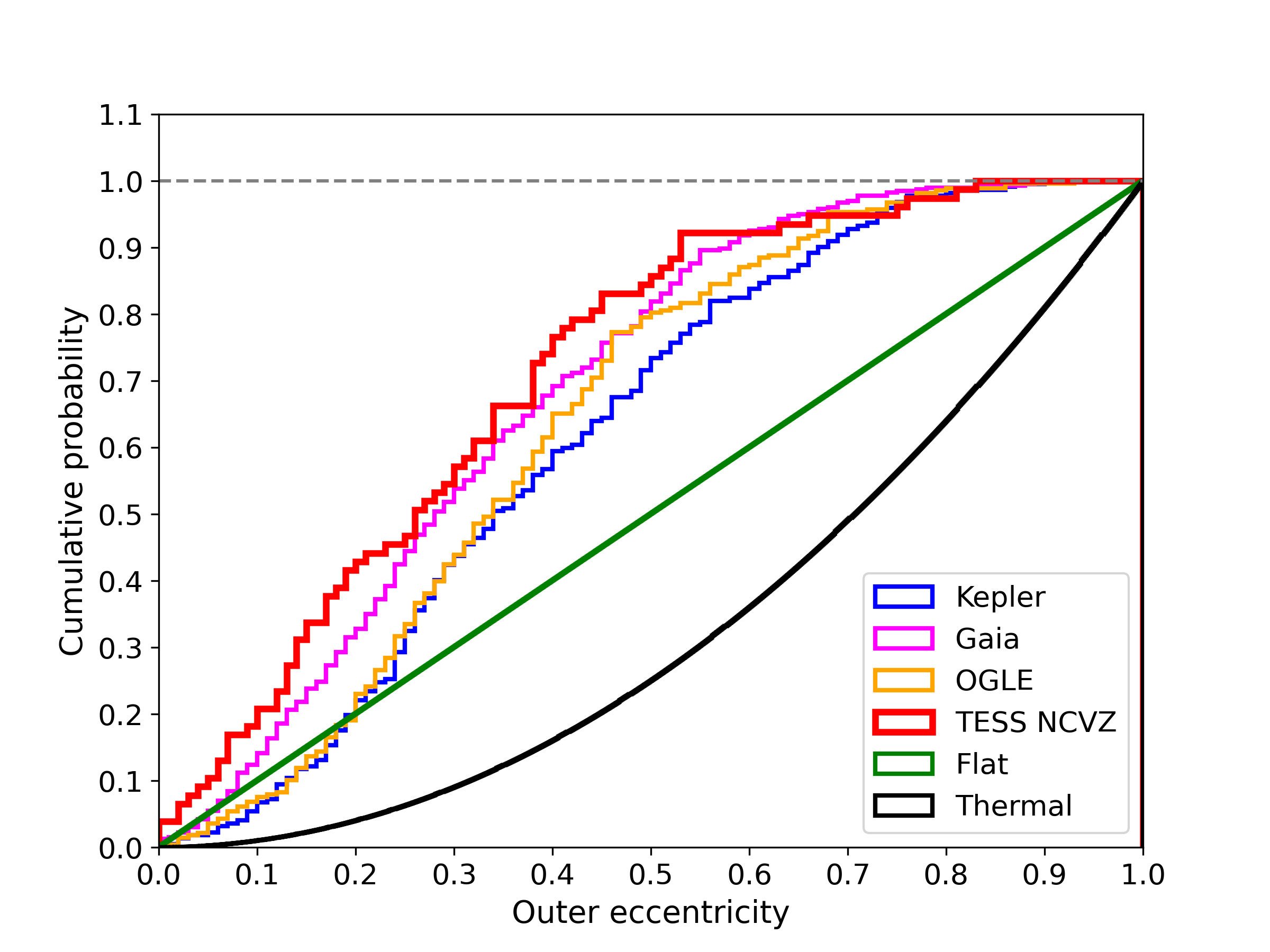}
\caption{Cumulative distribution of outer orbital eccentricities for our \textit{TESS} NCVZ (red) systems along with samples from earlier studies as \textit{Kepler} (blue), \textit{Gaia} (magenta), OGLE (orange) and theories as flat (green) or thermal (black) distributions.}
 \label{fig:e2cumuldist}
\end{figure}

In Fig. \ref{fig:e2dist}., we plot the distribution of the outer eccentricities of systems in our sample. Except six objects, all of them have an eccentricity value smaller than 0.55 and the distribution looks more or less flat with some small fluctuations. It does not have similar characteristics to those of the distribution from the \textit{Kepler} sample. Nevertheless, if we compare the cumulative eccentricity distribution (Fig. \ref{fig:e2cumuldist}.) with those from earlier studies, they are very similar to each other\footnote{This may come from that cumulative distributions are better at hiding statistical fluctuations which can dominate the appearance of a regular distribution.} and completely different from simple linear or thermal distributions from theory. It resembles more the distribution that of found by \citet{duchene13} for populations of field EBs. 

\begin{figure}
\centering
\includegraphics[width=0.5\textwidth]{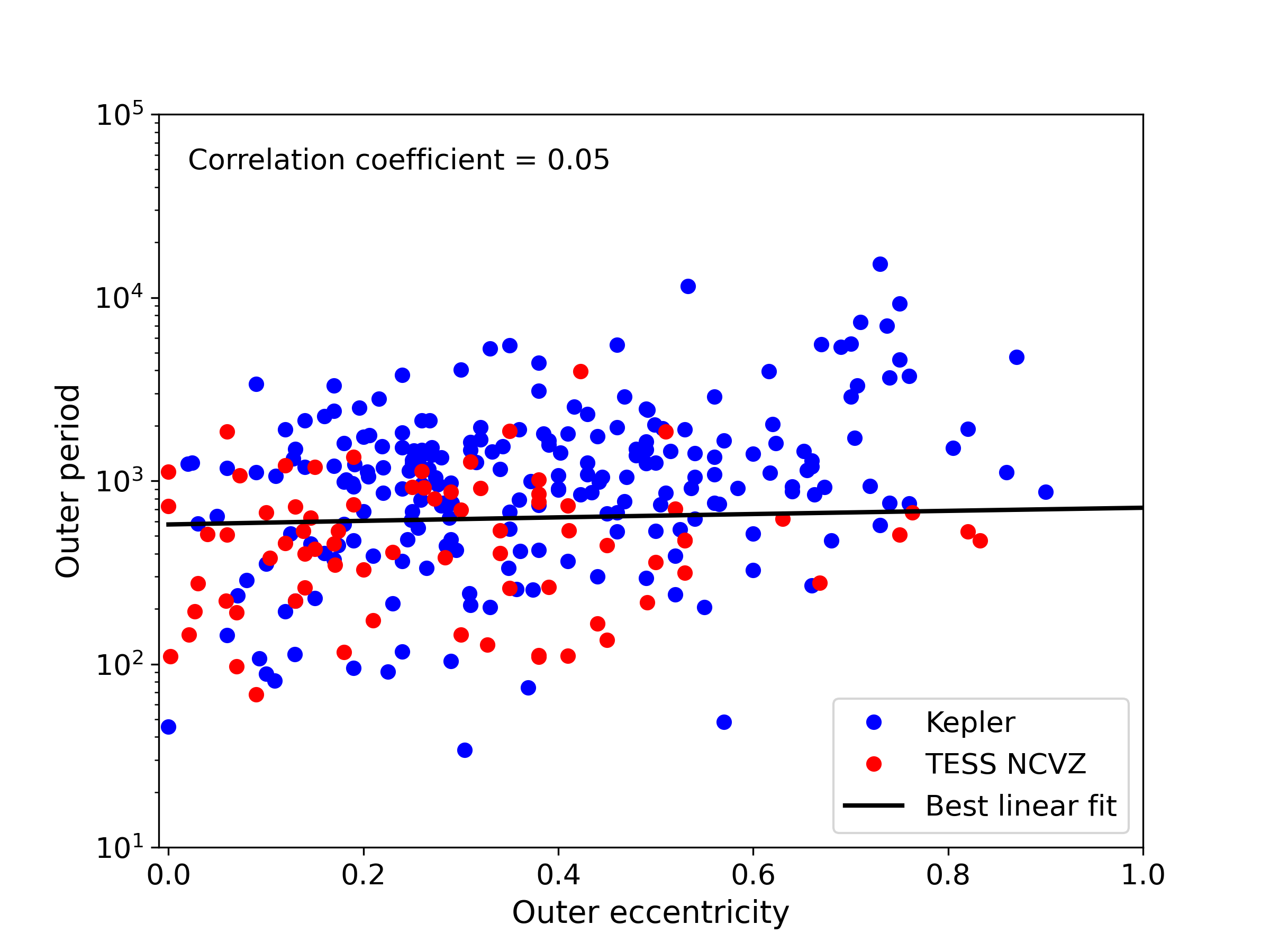}
\caption{Correlation of outer orbital periods and eccentricities for the \textit{Kepler} (blue) and \textit{TESS} NCVZ (red) samples. Note, the correlation value pertains to the NCVZ set only.}
 \label{fig:e2P2corr}
\end{figure}

Fig. \ref{fig:e2P2corr}. shows the relation between the outer orbital period and the outer eccentricity for both studies. The black line denotes a fit for a linear relation to the \textit{TESS} NCVZ sample with a correlation of 0.05 (i.e., not very correlated) which is significantly lower than the value of 0.34 found by \citet{borkovits16} for the \textit{Kepler} sample. We emphasize again, however, that our \textit{TESS} sample is much smaller than the \textit{Kepler} sample of \citet{borkovits16} and, this is more expressively true for the longer outer period systems, where one would expect higher outer eccentricities, as they cannot be detected certainly in the current stage of the \textit{TESS} sample.

\begin{figure}
\centering
\includegraphics[width=0.5\textwidth]{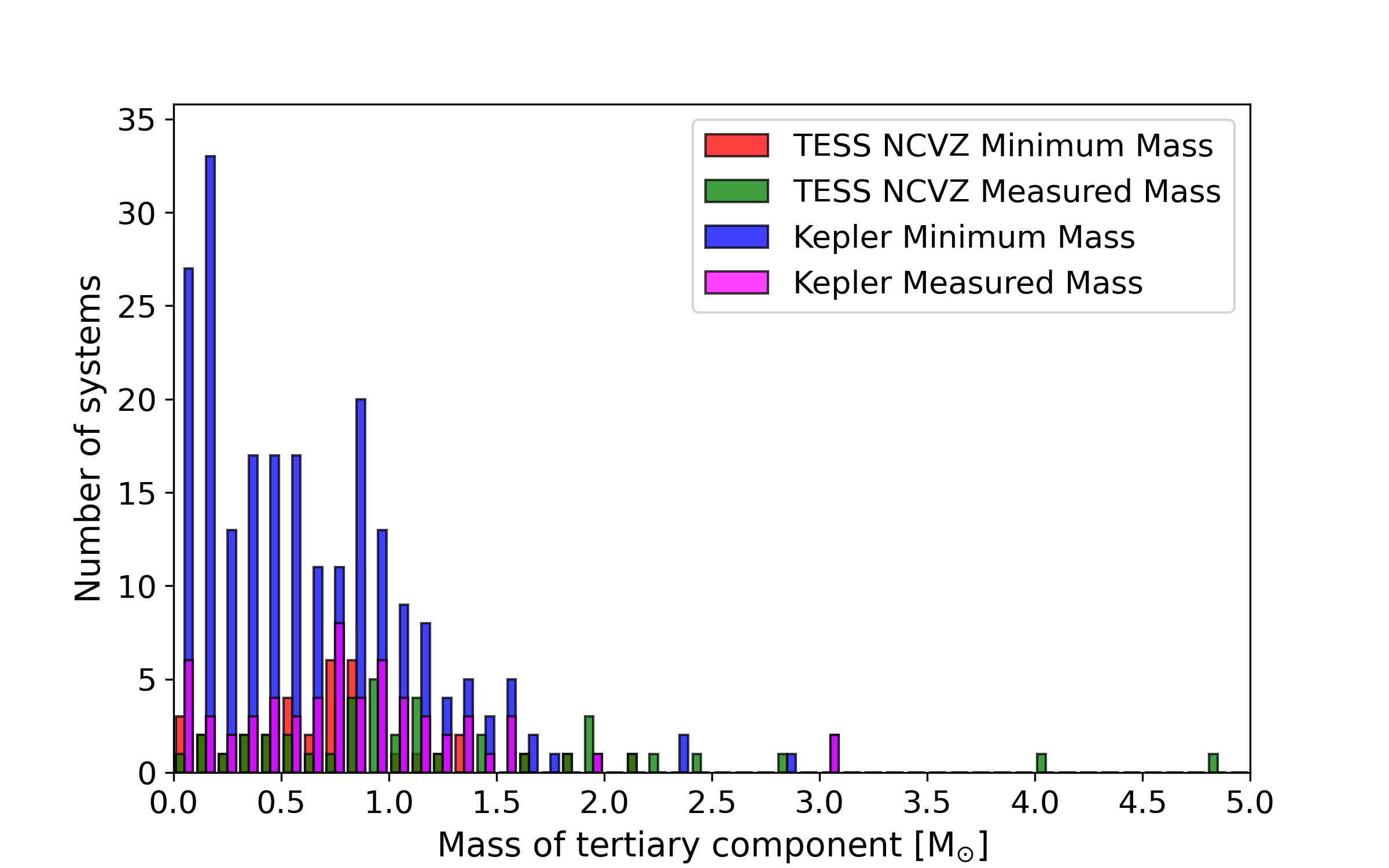}
\caption{Distribution of tertiary masses for the \textit{Kepler} and \textit{TESS} NCVZ samples directly measured from combined dynamical+LTTE ETV solutions (magenta and green, respectively) and lower limits estimated from pure LTTE solutions (blue and red, respectively).}
 \label{fig:Mcdist}
\end{figure}

In Fig. \ref{fig:Mcdist}., we plot the distribution of the tertiary component masses ($\mathrm{M_{C}}$) for both our and the \textit{Kepler} sample. For systems with measurable dynamical perturbations, exact masses can be determined, nevertheless for systems with LTTE only minimum masses can be estimated from the mass function using the constraints on the inner binary mass explained in Sect. \ref{LTTE}. Similarly to the \textit{Kepler} sample, the majority of the systems with measured masses have tertiaries with less than 1.8\,$\mathrm{M_{\odot}}$.

\begin{figure}
\centering
\includegraphics[width=0.5\textwidth]{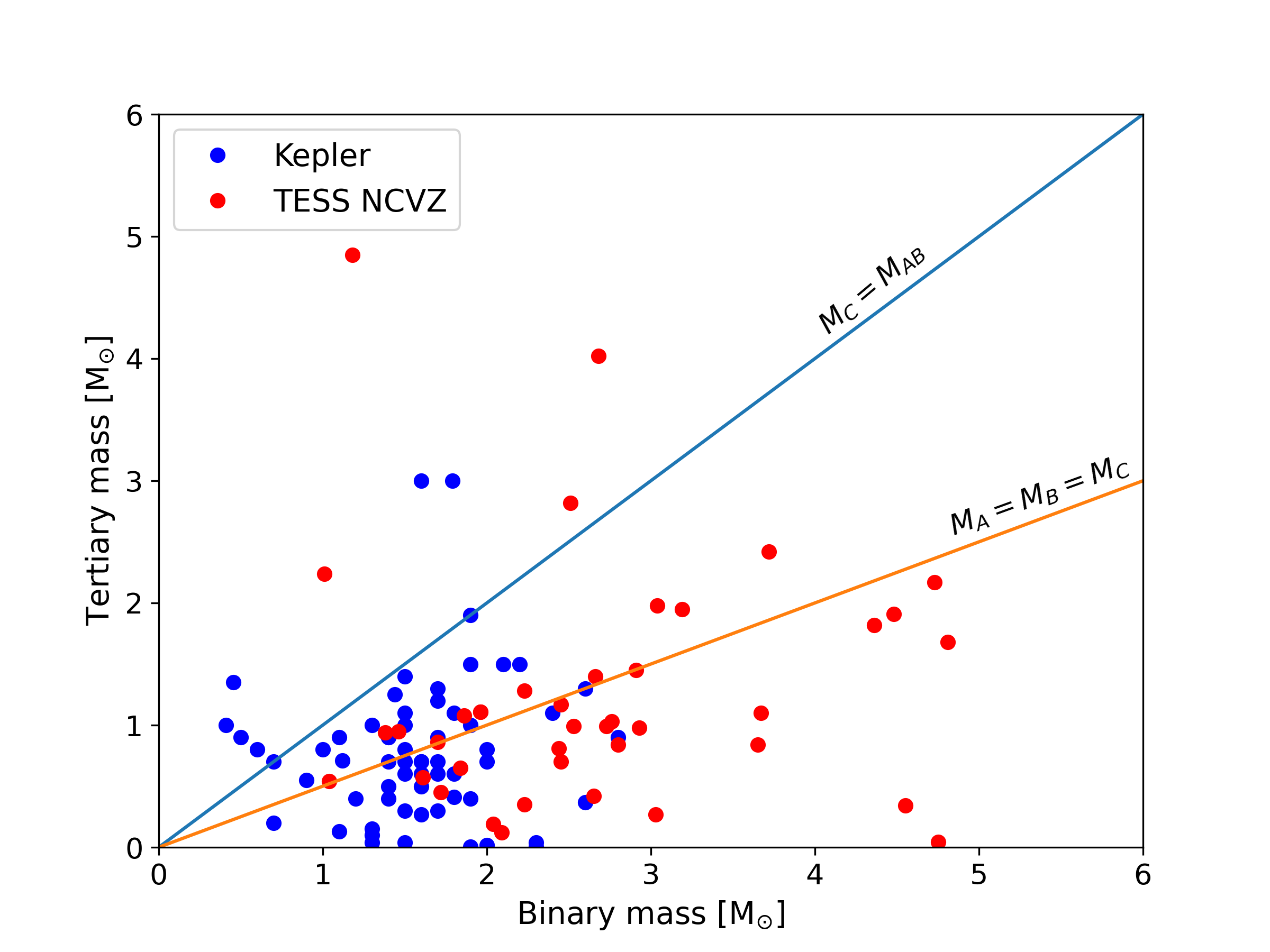}
\caption{Correlation of tertiary and inner binary masses for the \textit{Kepler} (blue) and \textit{TESS} NCVZ (red) samples with combined dynamical + LTTE ETV solutions.}
 \label{fig:McMabcorr}
\end{figure}

Fig. \ref{fig:McMabcorr}. shows the correlation between the exact masses of the binary and the tertiary components for those systems in which measurable dynamical perturbations occur for both our \textit{TESS} NCVZ and the \textit{Kepler} sample. The orange and blue lines represent the cases where all three components have the same mass, and where the mass of the tertiary equals the mass of the inner binary, respectively. The majority of the systems are located under or close to the orange line, while only 4 systems are above the blue line indicating that the tertiary has a higher mass than the inner binary components. It may also be worthwhile pointing out that in the \textit{Kepler} sample, the majority of the systems have an inner binary mass of less than 2\,$\mathrm{M_{\odot}}$, while in our \textit{TESS} NCVZ sample, there are only a handful of systems with such inner binary masses and the majority of the systems are above this value. Hence these systems may serve as an extension to the regime of higher binary masses.

\begin{figure}
\centering
\includegraphics[width=0.5\textwidth]{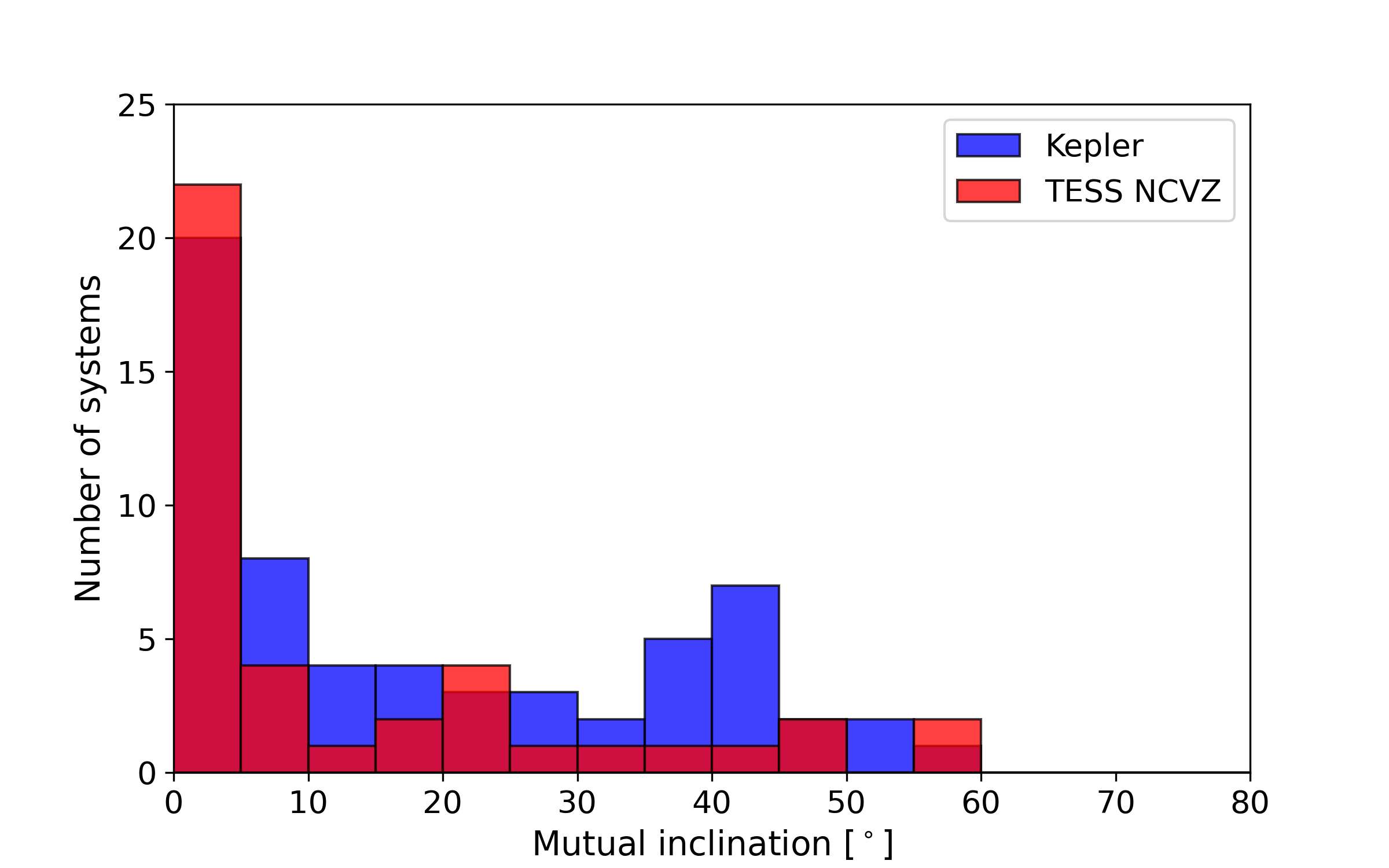}
\caption{Distribution of mutual orbital inclinations for the \textit{Kepler} (blue) and \textit{TESS} NCVZ (red) samples with combined dynamical + LTTE ETV solutions.}
 \label{fig:imutdist}
\end{figure}

On Fig. \ref{fig:imutdist}., we plot the distribution of the mutual inclination for systems in which the measurable dynamical perturbations allowed us to determine this quantity for both samples. The vast majority of the systems in our \textit{TESS} NCVZ sample have a (nearly) coplanar configuration and only a few systems can be found with higher mutual inclination values. This is different from the Kepler results which showed a small peak centered at $\mathrm{i_{mut}}=40^{\circ}$. However, in both cases the results are somewhat hampered by small number statistics.

\begin{figure}
\centering
\includegraphics[width=0.5\textwidth]{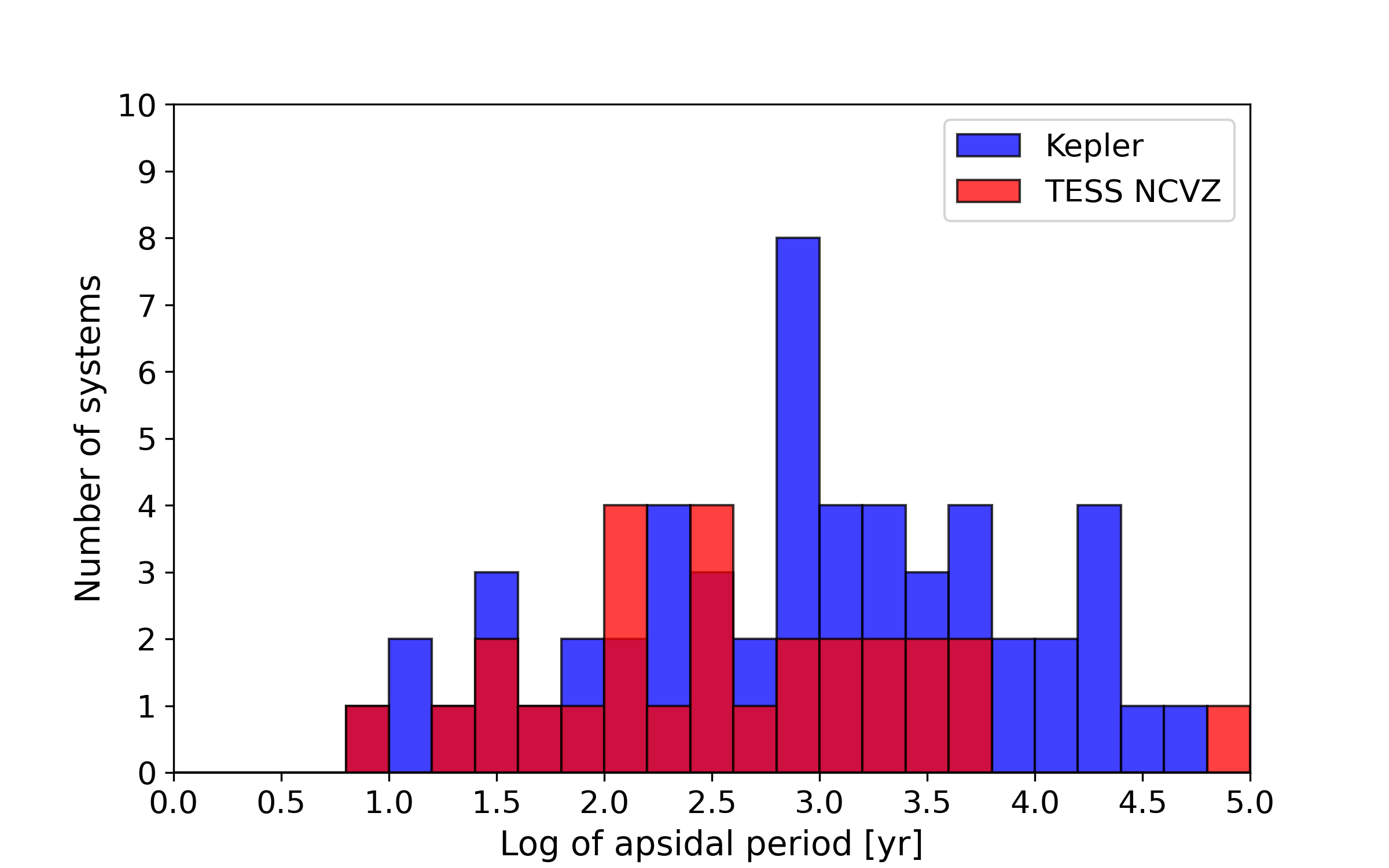}
\caption{Distribution of apsidal motion periods for the \textit{Kepler} (blue) and \textit{TESS} NCVZ (red) samples with combined dynamical + LTTE ETV solutions.}
 \label{fig:Papsedist}
\end{figure}

Finally, in Fig. \ref{fig:Papsedist}., we show the distribution of apsidal motion timescales for both samples. 

\subsection{Comparison with \textit{Gaia} NSS candidates}

\begin{figure*}
\caption{Correlations between the orbital periods (upper left panel), projected semi-major axes (upper right panel), eccentricities (lower left panel) and argument of periastrons (lower right panel) coming from the \textit{Gaia} NSS solutions and fitting LTTE models to the \textit{TESS} ETVs of the NCVZ sample. Different colors show different types of \textit{Gaia} NSS orbital solutions. The yellow dashed lines represent a ratio of unity, while for the bottom right panel the green dashed line shows the $\pm180\degr$ difference between the argument of periastrons coming from the two kinds of solutions. In order to avoid breaking the dashed lines in the bottom right panel, we employ a $\pm360\degr$ shift for some of the $\omega_{\mathrm{out,NSS}}$ values which is equivalent because of the periodicity of the argument of periastron. The faded dots are additional previously known systems from \citet{czavalinga23}}             % title of figure
\label{gaiacomparison}
    %\centering
    \includegraphics[scale = 0.6]{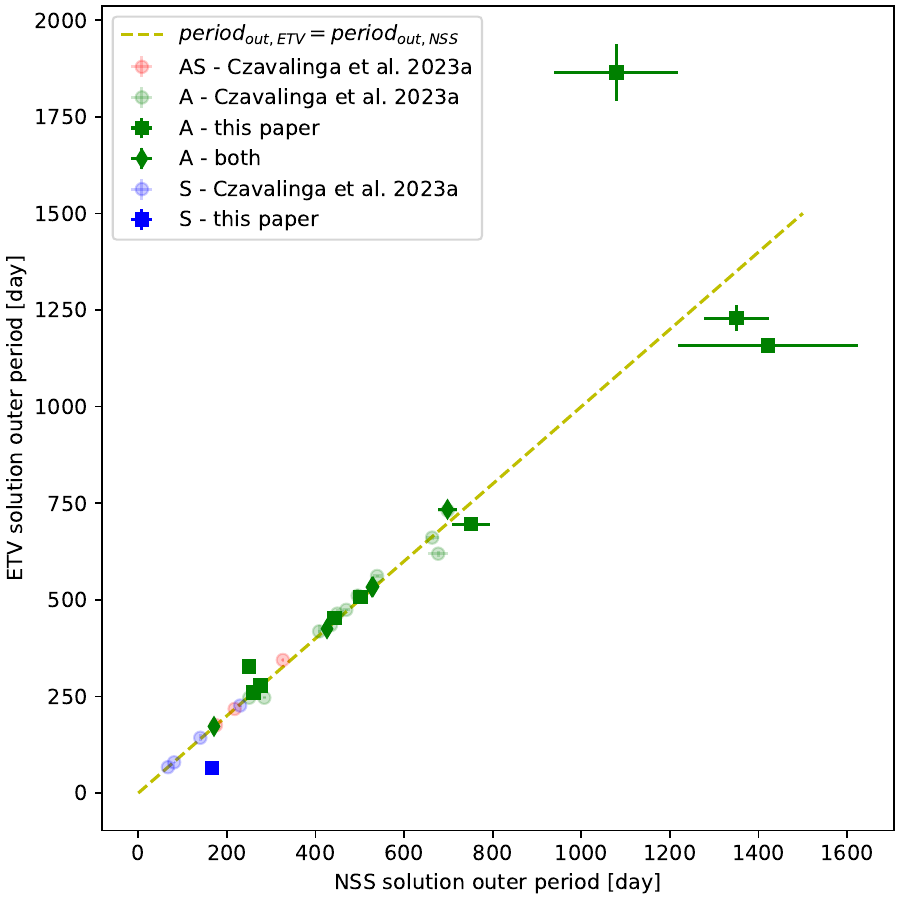}
    \includegraphics[scale = 0.6]{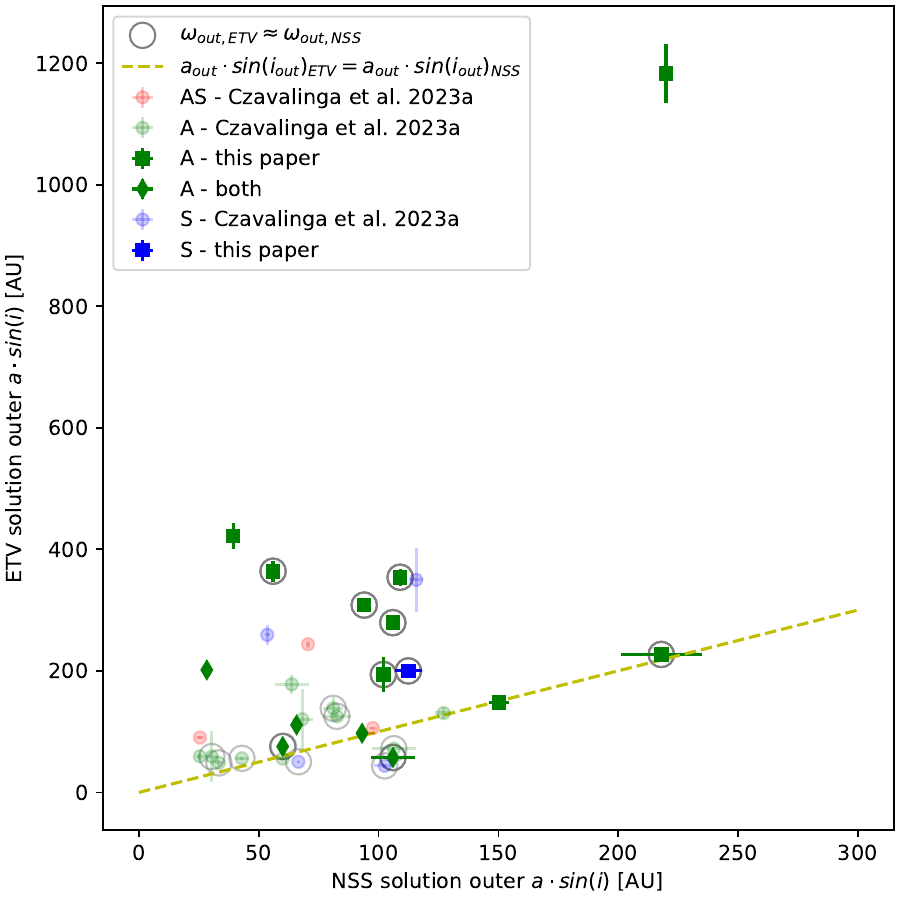}
    \includegraphics[scale = 0.6]{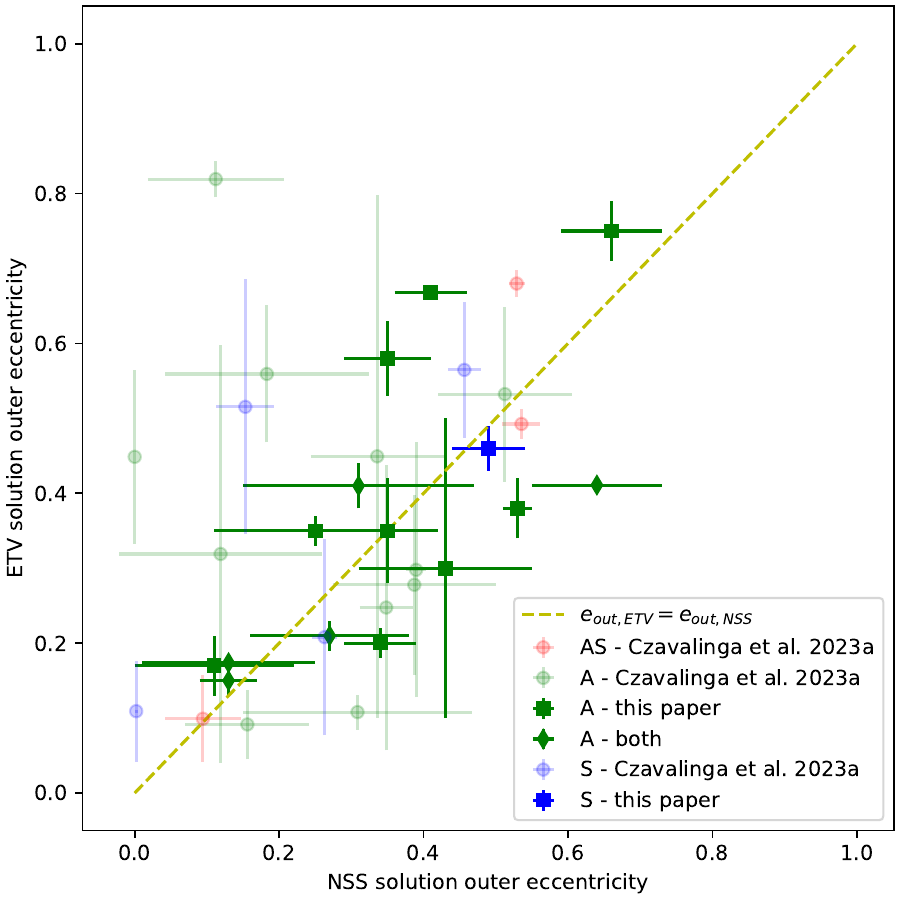}
    \includegraphics[scale = 0.6]{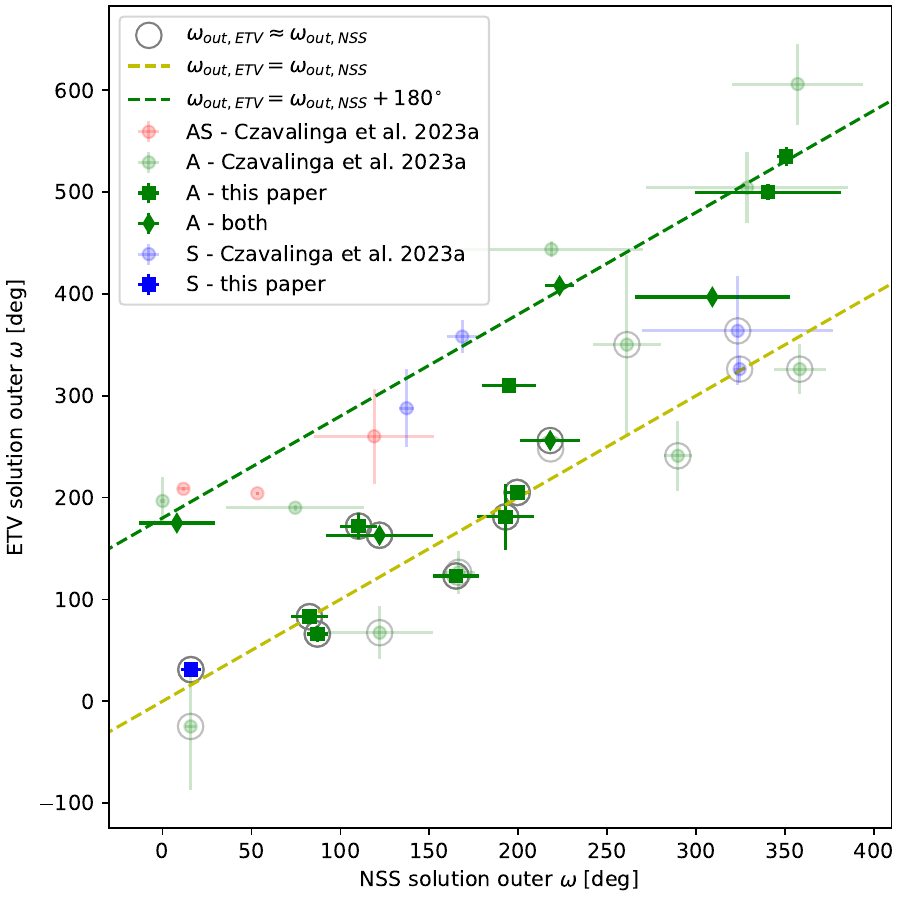}
\end{figure*}

As it was recently demonstrated in \citet{czavalinga23}, it is also possible to identify new hierarchical triple candidates using \textit{Gaia} data comparing the orbital periods from the astrometric or spectroscopic \textit{Gaia} NSS orbital solutions with the eclipsing periods from lightcurves and finding a difference of at least 5 times. For this purpose, we also crossmatched our full sample of 5139 objects with the \textit{Gaia} DR3 Non-Single Star catalog \citep{gaia22a}. That yielded 32 and 116 matches with \textit{Gaia} astrometric and spectroscopic orbital solutions, respectively. After the comparison of the orbital periods of the \textit{Gaia} orbital solutions with the binary periods from the \textit{TESS} light curves, we excluded 8 and 115 objects as the two types of periods were identical or exactly integer or half integer multiples of each other (i.~e \textit{Gaia} picked up the EB period or its possible aliases because of limited \textit{Gaia} data). That left a total of 25 triple candidates in our sample with available \textit{Gaia} orbital solution for the outer orbit of a possible tertiary component. As a final step, we analyzed the \textit{TESS} light and ETV curves of these systems one-by-one and excluded 10 further systems for various other reasons (e.g. the source was not an eclipsing binary, or the light curve and the \textit{Gaia} NSS solution actually belonged to another close star without a \textit{Gaia} NSS solution). After all this, in total we found 15 systems that have a \textit{Gaia} NSS orbital solution, and, moreover, the ETVs also confirm the validity of these solutions for all of them. For these objects, we tabulated the orbital parameters from the two types of orbital solutions in Table \ref{gaiapars}. Five of these objects (TICs 229762991, 232606864, 237234024, 320324245, 377105433) were already found and published recently by \citet{czavalinga23} and our ETV solutions for these triples are in a good agreement with theirs, while the other 10 systems are new discoveries that also demonstrate the potential of the \textit{Gaia} NSS catalog in the field of discovering new hierarchical triples. In Fig. \ref{gaiacomparison}. we plotted the correlations of different orbital parameters determined from the two types of orbital solutions (\textit{Gaia} NSS vs. ETVs) for the 22 systems found by \citet{czavalinga23} and the 10 systems newly found in our NCVZ analysis in a similar way as \citet{czavalinga23}. From these figures, it can be seen that the 10 new systems found by us can further confirm the conclusions made by \citet{czavalinga23}, i.~e. \textit{Gaia} does a good job picking up the correct outer orbital period, but the ETV results are often more accurate, especially in case of the outer eccentricity, projected semi-major axis and argument of periastron.

\begin{table*}[!h]
\caption{Outer orbital parameters from the \textit{Gaia} NSS and our \textit{TESS} ETV solutions for those 15 triple candidates that have both. For each object, the upper values belong to the former, while the lower values correspond to the latter solution, respectively.}             % title of Table
\label{gaiapars}      % is used to refer this table in the text
\centering                          % used for centering table
\begin{tabular}{c | c | c | c | c | c | c} 
\hline\hline 
\begin{tabular}{@{}c@{}} Common name \\ \textit{Gaia} DR3 source ID \end{tabular} & $\mathrm{P_{out} / P_{in}}$ & \begin{tabular}{@{}c@{}} $\mathrm{P_{out}}$ \\ $\mathrm{[days]}$ \end{tabular} & \begin{tabular}{@{}c@{}} $\mathrm{a_{out}\sin{i_{out}}}$ \\ $\mathrm{[R_{\odot}]}$ \end{tabular} & \begin{tabular}{@{}c@{}} $\mathrm{e_{out}}$ \\  \end{tabular} & \begin{tabular}{@{}c@{}} $\mathrm{\omega_{out}}$ \\ $[^\circ]$ \end{tabular} & \begin{tabular}{@{}c@{}} $\mathrm{\tau_{out}}$ \\ $\mathrm{[days]}$ \end{tabular}\\
\hline
\begin{tabular}{@{}c@{}} *\object{TIC 159465833} \\ 1707677283198785152 (A) \end{tabular} & 376 & \begin{tabular}{@{}c@{}} 751 (43) \\ 696 (15) \end{tabular} & \begin{tabular}{@{}c@{}} 102 (1) \\ 194 (29) \end{tabular} & \begin{tabular}{@{}c@{}} 0.43 (12) \\ 0.30 (20) \end{tabular} & \begin{tabular}{@{}c@{}} 193 (16) \\ 181 (32) \end{tabular} & \begin{tabular}{@{}c@{}} 58849 (28) \\ 58721 (64) \end{tabular}\\
\hline
\begin{tabular}{@{}c@{}} \object{TIC 199632809} \\ 1431986081247699200 (A) \end{tabular} & 769 & \begin{tabular}{@{}c@{}} 1350 (73) \\ 1229 (34) \end{tabular} & \begin{tabular}{@{}c@{}} 150 (4) \\ 148 (4) \end{tabular} & \begin{tabular}{@{}c@{}} 0.53 (2) \\ 0.38 (4) \end{tabular} & \begin{tabular}{@{}c@{}} 351 (5) \\ 175 (9) \end{tabular} & \begin{tabular}{@{}c@{}} 58585 (6) \\ 58562 (36) \end{tabular}\\
\hline
\begin{tabular}{@{}c@{}} *\object{TIC 229762991} \\ 2259293058444658432 (A) \end{tabular} & 1256 & \begin{tabular}{@{}c@{}} 426 (3) \\ 425.0 (5) \end{tabular} & \begin{tabular}{@{}c@{}} 60 (2) \\ 75.5 (5) \end{tabular} & \begin{tabular}{@{}c@{}} 0.13 (4) \\ 0.15 (2) \end{tabular} & \begin{tabular}{@{}c@{}} 218 (17) \\ 256 (5) \end{tabular} & \begin{tabular}{@{}c@{}} 58506 (20) \\ 58531 (6) \end{tabular}\\
\hline 
\begin{tabular}{@{}c@{}} *\object{TIC 229785001} \\ 2256269710706835712 (S) \end{tabular} & 178 & \begin{tabular}{@{}c@{}} 166.0 (5) \\ 165.37 (5) \end{tabular} & \begin{tabular}{@{}c@{}} 112 (6) \\ 200 (1)  \end{tabular} & \begin{tabular}{@{}c@{}} 0.49 (5) \\ 0.46 (3) \end{tabular} & \begin{tabular}{@{}c@{}} 16 (5) \\ 31 (3) \end{tabular} & \begin{tabular}{@{}c@{}} 58731 (1) \\ 58730.7 (7) \end{tabular}\\
\hline 
\begin{tabular}{@{}c@{}} \object{TIC 232606864} \\ 1438455538945959680 (A) \end{tabular} & 1814 & \begin{tabular}{@{}c@{}} 530 (7) \\ 533.9 (4) \end{tabular} & \begin{tabular}{@{}c@{}} 93 (1) \\ 97.4 (5) \end{tabular} & \begin{tabular}{@{}c@{}} 0.64 (9) \\ 0.411 (7) \end{tabular} & \begin{tabular}{@{}c@{}} 223 (8) \\ 48 (1) \end{tabular} & \begin{tabular}{@{}c@{}} 58805 (8) \\ 58775 (2) \end{tabular}\\
\hline 
\begin{tabular}{@{}c@{}} *\object{TIC 233530543} \\ 2256492980286409728 (A) \end{tabular} & 160 & \begin{tabular}{@{}c@{}} 260 (1) \\ 260 (1) \end{tabular} & \begin{tabular}{@{}c@{}} 105.95 (18) \\ 279 (12) \end{tabular} & \begin{tabular}{@{}c@{}} 0.35 (7) \\ 0.35 (7) \end{tabular} & \begin{tabular}{@{}c@{}} 83 (10) \\ 83 (5) \end{tabular} & \begin{tabular}{@{}c@{}} 58773 (8) \\ 58767.3 (7) \end{tabular}\\
\hline 
\begin{tabular}{@{}c@{}} \object{TIC 233729038} \\ 2159794883992807680 (A) \end{tabular} & 140 & \begin{tabular}{@{}c@{}} 443 (6) \\ 453.1 (3) \end{tabular} & \begin{tabular}{@{}c@{}} 39.29 (26) \\ 422 (22) \end{tabular} & \begin{tabular}{@{}c@{}} 0.11 (11) \\ 0.17 (4) \end{tabular} & \begin{tabular}{@{}c@{}} 340 (41) \\ 140 (8) \end{tabular} & \begin{tabular}{@{}c@{}} 58535 (52) \\ 58529 (11) \end{tabular}\\
\hline 
\begin{tabular}{@{}c@{}} \object{TIC 237234024} \\ 2288385277122496896 (A) \end{tabular} & 91 & \begin{tabular}{@{}c@{}} 171 (1) \\ 172.5 (1) \end{tabular} & \begin{tabular}{@{}c@{}} 28.33 (6) \\ 202 (10) \end{tabular} & \begin{tabular}{@{}c@{}} 0.27 (11) \\ 0.21 (2) \end{tabular} & \begin{tabular}{@{}c@{}} 8 (21) \\ 175 (7) \end{tabular} & \begin{tabular}{@{}c@{}} 58651 (10) \\ 58653 (3) \end{tabular}\\
\hline 
\begin{tabular}{@{}c@{}} *\object{TIC 256514937} \\ 2241281133839154304 (A) \end{tabular} & 52 & \begin{tabular}{@{}c@{}} 276.02 (62) \\ 277.75 (7) \end{tabular} & \begin{tabular}{@{}c@{}} 94.01 (17) \\ 308 (5) \end{tabular} & \begin{tabular}{@{}c@{}} 0.41 (5) \\ 0.668 (4) \end{tabular} & \begin{tabular}{@{}c@{}} 199 (6) \\ 205 (1) \end{tabular} & \begin{tabular}{@{}c@{}} 58834 (5) \\ 58834.4 (3) \end{tabular}\\
\hline
\begin{tabular}{@{}c@{}} *\object{TIC 259168350} \\ 2254791692199089536 (A) \end{tabular} & 784 & \begin{tabular}{@{}c@{}} 1421 (203) \\ 1158 (13) \end{tabular} & \begin{tabular}{@{}c@{}} 218 (17) \\ 227 (7) \end{tabular} & \begin{tabular}{@{}c@{}} 0.35 (6) \\ 0.58 (5) \end{tabular} & \begin{tabular}{@{}c@{}} 165 (13) \\ 123 (3) \end{tabular} & \begin{tabular}{@{}c@{}} 58908 (16) \\ 58808 (14) \end{tabular}\\
\hline 
\begin{tabular}{@{}c@{}} \object{TIC 259271740} \\ 2263535592780194688 (A) \end{tabular} & 159 & \begin{tabular}{@{}c@{}} 1079 (140) \\ 1865 (74) \end{tabular} & \begin{tabular}{@{}c@{}} 219.93 (17) \\ 1183 (49)  \end{tabular} & \begin{tabular}{@{}c@{}} 0.25 (14) \\ 0.35 (2) \end{tabular} & \begin{tabular}{@{}c@{}} 195 (15) \\ 310 (5) \end{tabular} & \begin{tabular}{@{}c@{}} 59034 (42) \\ 59488 (26) \end{tabular}\\
\hline
\begin{tabular}{@{}c@{}} *\object{TIC 288611883} \\ 2294298691174194048 (A) \end{tabular} & 502 & \begin{tabular}{@{}c@{}} 502 (8) \\ 507 (8) \end{tabular} & \begin{tabular}{@{}c@{}} 109 (1) \\ 354 (14)  \end{tabular} & \begin{tabular}{@{}c@{}} 0.66 (7) \\ 0.75 (4) \end{tabular} & \begin{tabular}{@{}c@{}} 87 (6) \\ 66 (8) \end{tabular} & \begin{tabular}{@{}c@{}} 58837 (9) \\ 58857 (14) \end{tabular}\\
\hline 
\begin{tabular}{@{}c@{}} \object{TIC 320324245} \\ 1703274357605001728 (A) \end{tabular} & 1518 & \begin{tabular}{@{}c@{}} 527 (8) \\ 532.9 (4) \end{tabular} & \begin{tabular}{@{}c@{}} 66 (1) \\ 111.0 (4)  \end{tabular} & \begin{tabular}{@{}c@{}} 0.13 (12) \\ 0.174 (5) \end{tabular} & \begin{tabular}{@{}c@{}} 309 (44) \\ 37 (2) \end{tabular} & \begin{tabular}{@{}c@{}} 58770 (64) \\ 58900 (3) \end{tabular}\\
\hline 
\begin{tabular}{@{}c@{}} *\object{TIC 356014478} \\ 1636535307469148544 (A) \end{tabular} & 34 & \begin{tabular}{@{}c@{}} 250 (1) \\ 328 (1) \end{tabular} & \begin{tabular}{@{}c@{}} 55.97 (46) \\ 364 (17)  \end{tabular} & \begin{tabular}{@{}c@{}} 0.34 (5) \\ 0.20 (2) \end{tabular} & \begin{tabular}{@{}c@{}} 110 (10) \\ 172 (13) \end{tabular} & \begin{tabular}{@{}c@{}} 58686 (9) \\ 58659 (16) \end{tabular}\\
\hline
\begin{tabular}{@{}c@{}} *\object{TIC 377105433} \\ 2155105088944426624 (A) \end{tabular} & 2920 & \begin{tabular}{@{}c@{}} 698 (21) \\ 734 (3) \end{tabular} & \begin{tabular}{@{}c@{}} 106 (9) \\ 57.4 (9)  \end{tabular} & \begin{tabular}{@{}c@{}} 0.31 (16) \\ 0.41 (3) \end{tabular} & \begin{tabular}{@{}c@{}} 122 (30) \\ 163 (3) \end{tabular} & \begin{tabular}{@{}c@{}} 59046 (59) \\ 58934 (7) \end{tabular}\\
\hline
\end{tabular}
\tablefoot{Those systems in which \textit{Gaia} tracks the same component as the ETVs are indicated with an asterisk before their names. The \textit{Gaia} NSS solution type of the objects are noted after their DR3 source IDs (A: astrometric solution, S: spectroscopic solution). $\mathrm{\tau_{\mathrm{out}}}$ is the epoch of periastron passage in MBJD.}
\end{table*}

\section{Summary}
\label{Sect:Summary}

In the hope of discovering new compact hierarchical triples, we have conducted a survey of eclipsing binaries based on \textit{TESS} data. We have selected 3553 objects from a list of EBs found in the \textit{TESS} data using a machine learning approach which are in or near the NCVZ of \textit{TESS} and had a sufficient amount of photometric data. We have obtained the light curves of every source from all available \textit{TESS} sectors up to Sector 60 in an automated way. After that, using our newly developed Python GUI, we have visually vetted all of the light curves of our targets along with determining their eclipsing periods and calculating their ETV curves. In the end of this process, we have selected 351 objects for further detailed analysis that seemed to have non-linear ETVs which could be caused by previously unknown third bodies in these systems. Finally, we tried to fit these ETVs with analytical formulae of LTTE or LTTE + dynamical perturbations. 

In total we were able to derive models for 94 objects with pure LTTE and 41 objects with LTTE + dynamical solutions which we divided into three and two subgroups based on the quality of the models, respectively. The first subgroup contains the most reliable, highly certain solutions for 19 and 27 objects with the two types of ETV solutions, respectively. The second subgroup has 17 and 14 objects with likely valid solutions, but with probably higher uncertainties in their orbital parameters. For the pure LTTE systems only, the third subgroup consists of 58 objects having very uncertain solutions that are still likely triple systems, but their parameters should be used with considerable caution. That makes a total of 135 hierarchical triple candidates in our NCVZ sample of which 10 objects were already known from previous studies, while the rest of the 125 systems are new discoveries. Among these systems there are a number that are individually interesting. First, there are five very tight systems with a period ratio of less than 20 which are the closest to the dynamical stability limit. It is worth noting that, although our analytic models describe these systems well on a shorter timescale of the available \textit{TESS} data set, they are inaccurate on decade-long timescales and numerical integration of the orbits should be applied to model these systems correctly. There are also four triply eclipsing triples in our sample, out of which one (TIC\,29785001) was already known and for that one our ETV solution is consistent with the previous one of \citet{rappaport23}. The other three systems are new discoveries and we plan a separate, more detailed comprehensive analysis in the near future for them. 

We have outlined the case of GZ Dra in which we have recognized and successfully modelled the previously unknown effect of dynamical perturbations. This analysis showed that GZ Dra is actually one of the most inclined triple systems known with $i_\mathrm{m}=58\degr\pm7\degr$. Although this inclination is well within the range of the original von Zeipel-Lidov-Kozai effect, we cannot expect large amplitude eccentricity cycles because of the presence of strong tidal interactions between the components of the highly circularized inner pair. Nevertheless, one would expect large amplitude orbital plane precession in the system which would cause EDVs in its light curve, but the timescale of this variation is on the order of a thousand years and that is the reason we cannot see its effect in the few years long \textit{TESS} data set. 

Because we have also determined all the same parameters for every triple candidates in our sample, we are able to make a comparison with the distributions of \citet{borkovits16} found for the full \textit{Kepler} sample. For this purpose, we used only those 77 systems that have a certain or likely valid ETV solutions (i.~e. the first two subgroups). In this sample of ours, there is a slightly elevated number of systems with an outer orbital period of 100-800\,days that is most likely caused by the observing strategy of \textit{TESS}. In the case of outer eccentricities we have found that their cumulative distribution is similar to those found by earlier surveys (\textit{Kepler}, OGLE, \textit{Gaia}), but very different from theoretical (flat and thermal) distributions. We have not found any correlation between the outer periods and eccentricities for systems in our sample. Regarding the masses of the outer components in our systems, we have found that the majority of them are below 1.8\,$\mathrm{M_{\odot}}$ which is similar to the \textit{Kepler} sample. Among those 41 systems in which dynamical perturbations occur, and exact masses can be derived for the components, we have found that only four have a more massive tertiary component than their inner binaries. In the \textit{Kepler} sample the majority of the systems have an inner binary mass lower than 2\,$\mathrm{M_{\odot}}$, while in our sample it is the opposite complementing the regime of triples with higher inner binary masses. Similarly to what our other studies of \textit{TESS} binaries have shown \citep{rappaport23}, the vast majority of these systems also have negligible mutual inclination (i.~e. coplanar configuration) and only a few systems have distinctly non-zero mutual inclination. The latter also implies that we cannot see any peak at around 40$^{\circ}$ as found by \citet{borkovits16} for the \textit{Kepler} sample.

Finally, we have crossmatched our list of targets with the \textit{Gaia} NSS catalog and found 15 systems that have either astrometric or spectroscopic orbital solutions with substantially different orbital period than their eclipsing period. For all of these objects, our ETV solutions can confirm that they are triple systems, nevertheless 5 of them were already found and published by \citet{czavalinga23}. We have also checked the same correlations as \citet{czavalinga23} for the orbital parameters of these systems derived from the two different methods and we can draw the same conclusions as they did. The \textit{Gaia} NSS catalog has great potential in discovering new hierarchical triples, the most reliable parameter in these solutions is the outer orbital period, while the other parameters can have larger uncertainties than found via the ETV fits.

In this study, we have demonstrated that \textit{TESS} data -- in addition to many other applications -- is an excellent tool for finding and analyzing new hierarchical triple stars. Since a lot of systems in our sample have uncertain solutions and more observations would be needed in order to improve their quality, we hope that \textit{TESS} will continue monitoring these objects and their reanalysis will become possible in the not too distant future.

\begin{acknowledgements}

This project has received funding from the HUN-REN Hungarian Research Network.

T.\,M., T.\,B., T.\,H. and A.\,P. acknowledge the financial support of the Hungarian National Research, Development and Innovation Office -- NKFIH Grants K-147131 and K-138962.

This paper includes data collected by the \textit{TESS} mission. Funding for the \textit{TESS} mission is provided by the NASA Science Mission directorate. Some of the data presented in this paper were obtained from the Mikulski Archive for Space Telescopes (MAST). STScI is operated by the Association of Universities for Research in Astronomy, Inc., under NASA contract NAS5-26555. Support for MAST for non-HST data is provided by the NASA Office of Space Science via grant NNX09AF08G and by other grants and contracts.

This work has made use  of data  from the European  Space Agency (ESA)  mission {\it Gaia}\footnote{\url{https://www.cosmos.esa.int/gaia}},  processed  by  the {\it   Gaia}   Data   Processing   and  Analysis   Consortium   (DPAC)\footnote{\url{https://www.cosmos.esa.int/web/gaia/dpac/consortium}}.  Funding for the DPAC  has been provided  by national  institutions, in  particular the institutions participating in the {\it Gaia} Multilateral Agreement.

This publication makes use of data products from the Wide-field Infrared Survey Explorer, which is a joint project of the University of California, Los Angeles, and the Jet Propulsion Laboratory/California Institute of Technology, funded by the National Aeronautics and Space Administration. 

This publication makes use of data products from the Two Micron All Sky Survey, which is a joint project of the University of Massachusetts and the Infrared Processing and Analysis Center/California Institute of Technology, funded by the National Aeronautics and Space Administration and the National Science Foundation.

We  used the  Simbad  service  operated by  the  Centre des  Donn\'ees Stellaires (Strasbourg,  France) and the ESO  Science Archive Facility services (data  obtained under request number 396301).   
\end{acknowledgements}

%%%%%%%%%%%%%%%%%%%%%%%%%%%%%%%%%%%%%%%%%%%%%%%%%%

\bibliographystyle{aa}
\bibliography{example} % if your bibtex file is called example.bib

%%%%%%%%%%%%%%%%%%%%%%%%%%%%%%%%%%%%%%%%%%%%%%%%%%

% Don't change these lines
%\bsp	% typesetting comment
%\label{lastpage}
%\end{document}

\begin{appendix}

\section{ETV plots}
\label{Sect:ETV_plots}

\begin{figure*}
\includegraphics[width=60mm]{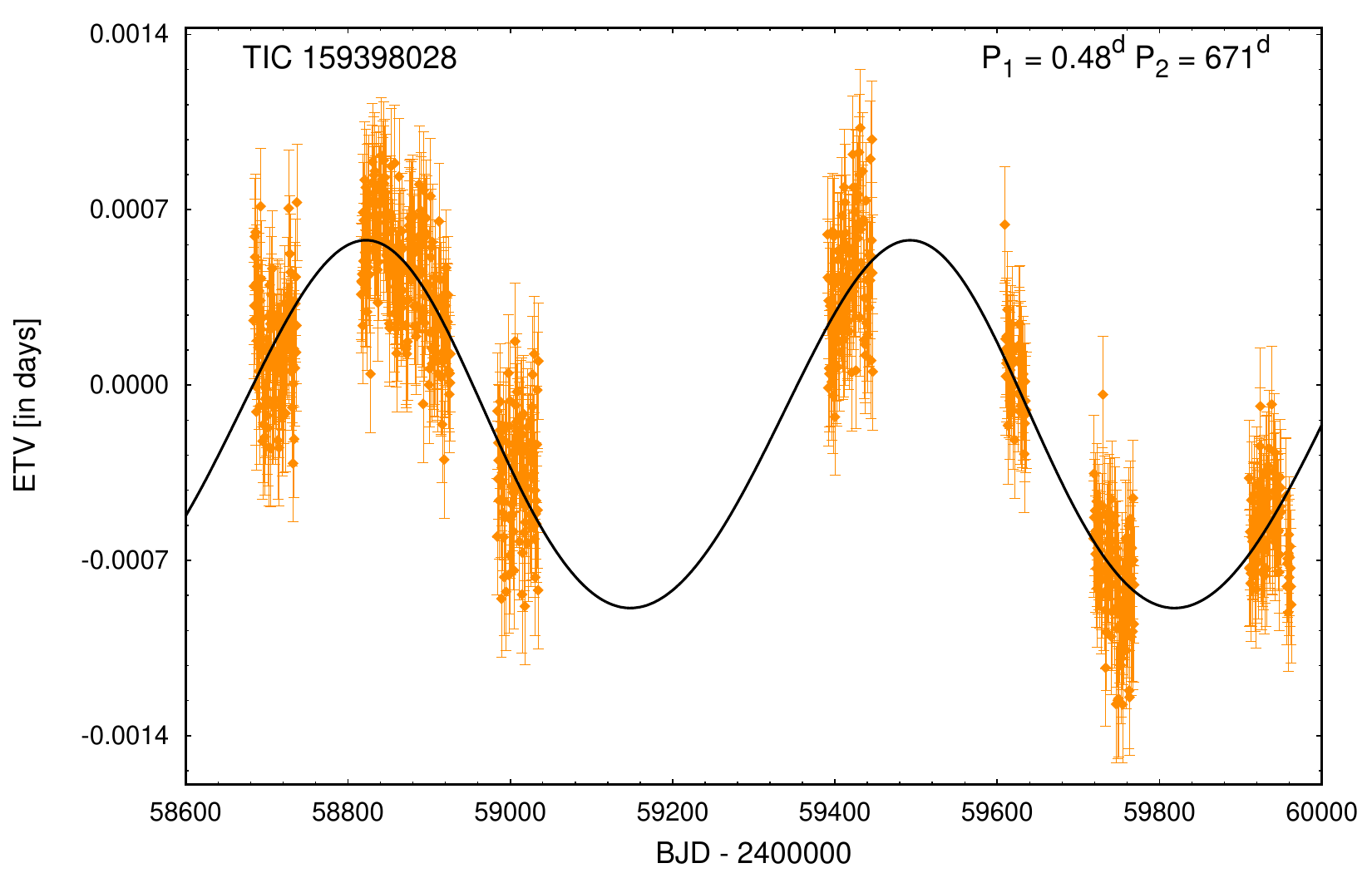}\includegraphics[width=60mm]{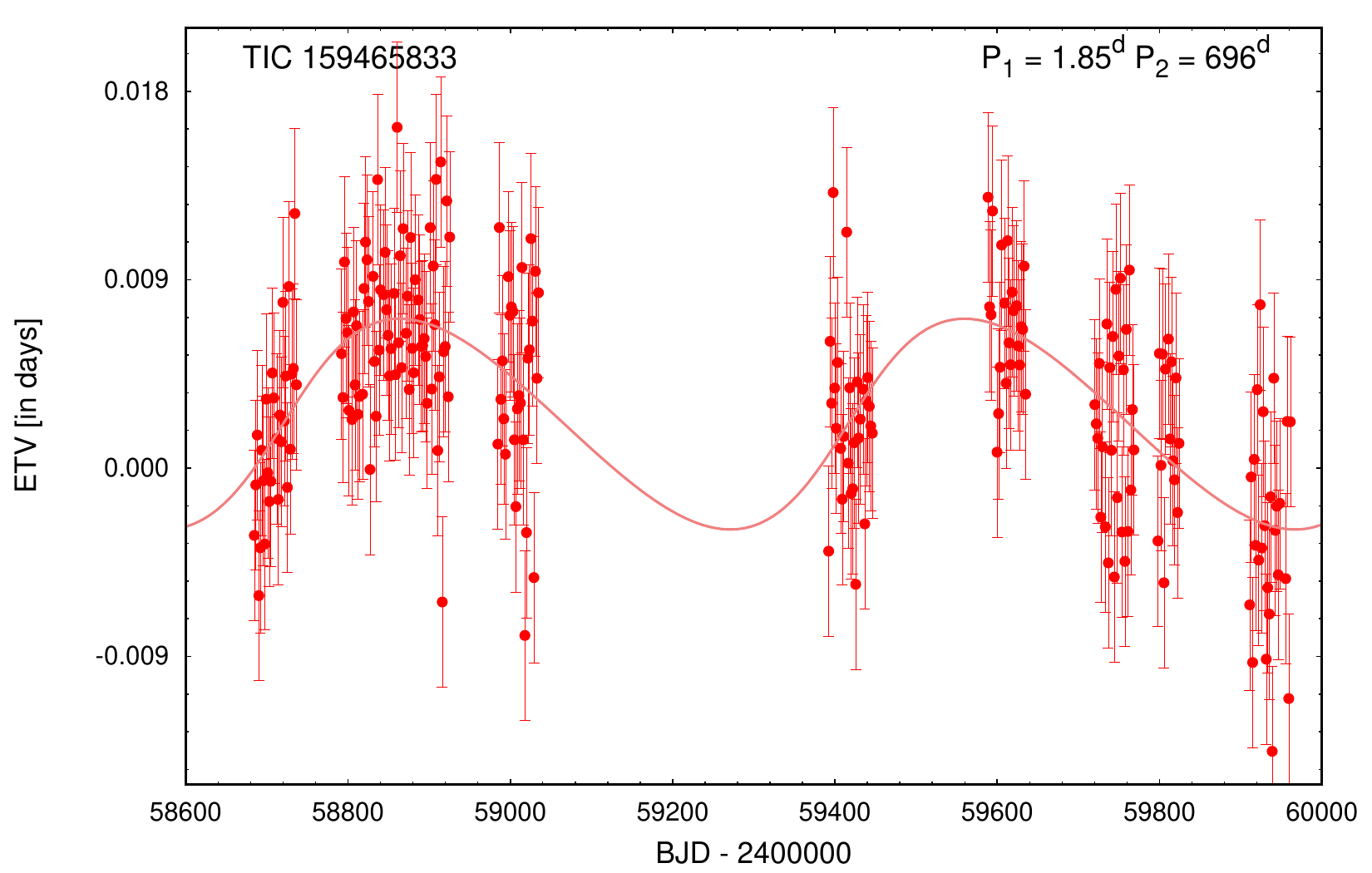}\includegraphics[width=60mm]{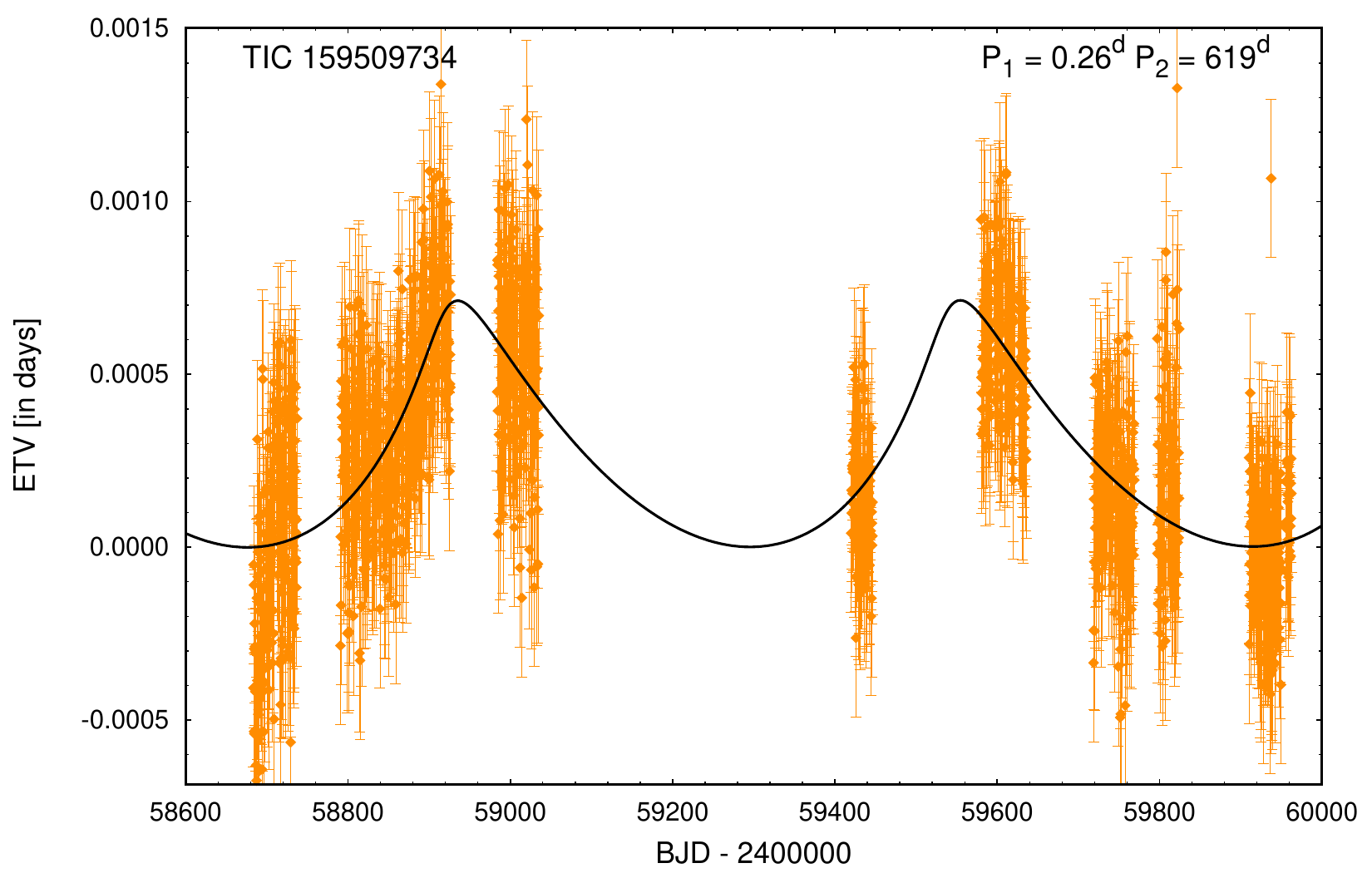}
\includegraphics[width=60mm]{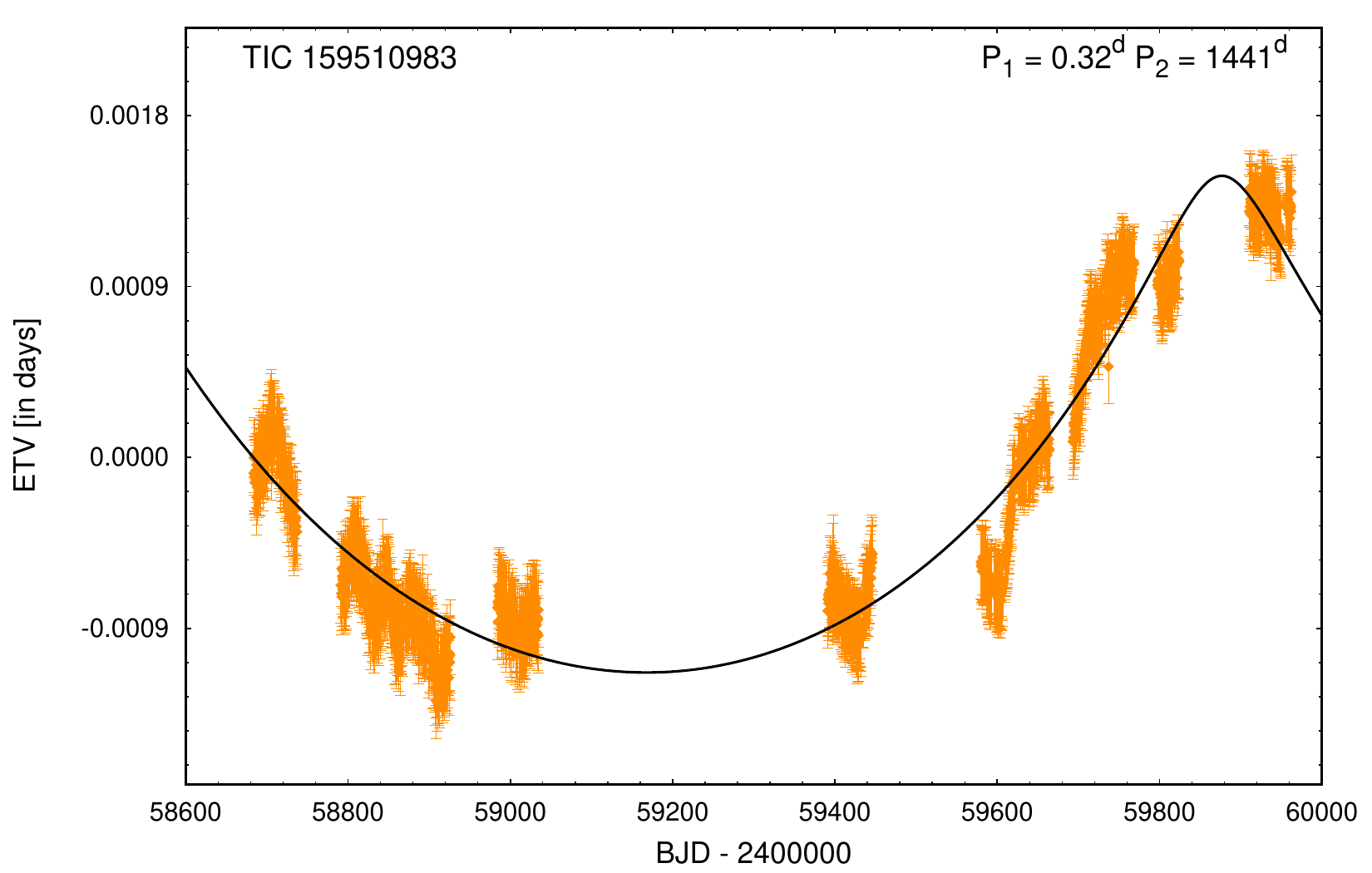}\includegraphics[width=60mm]{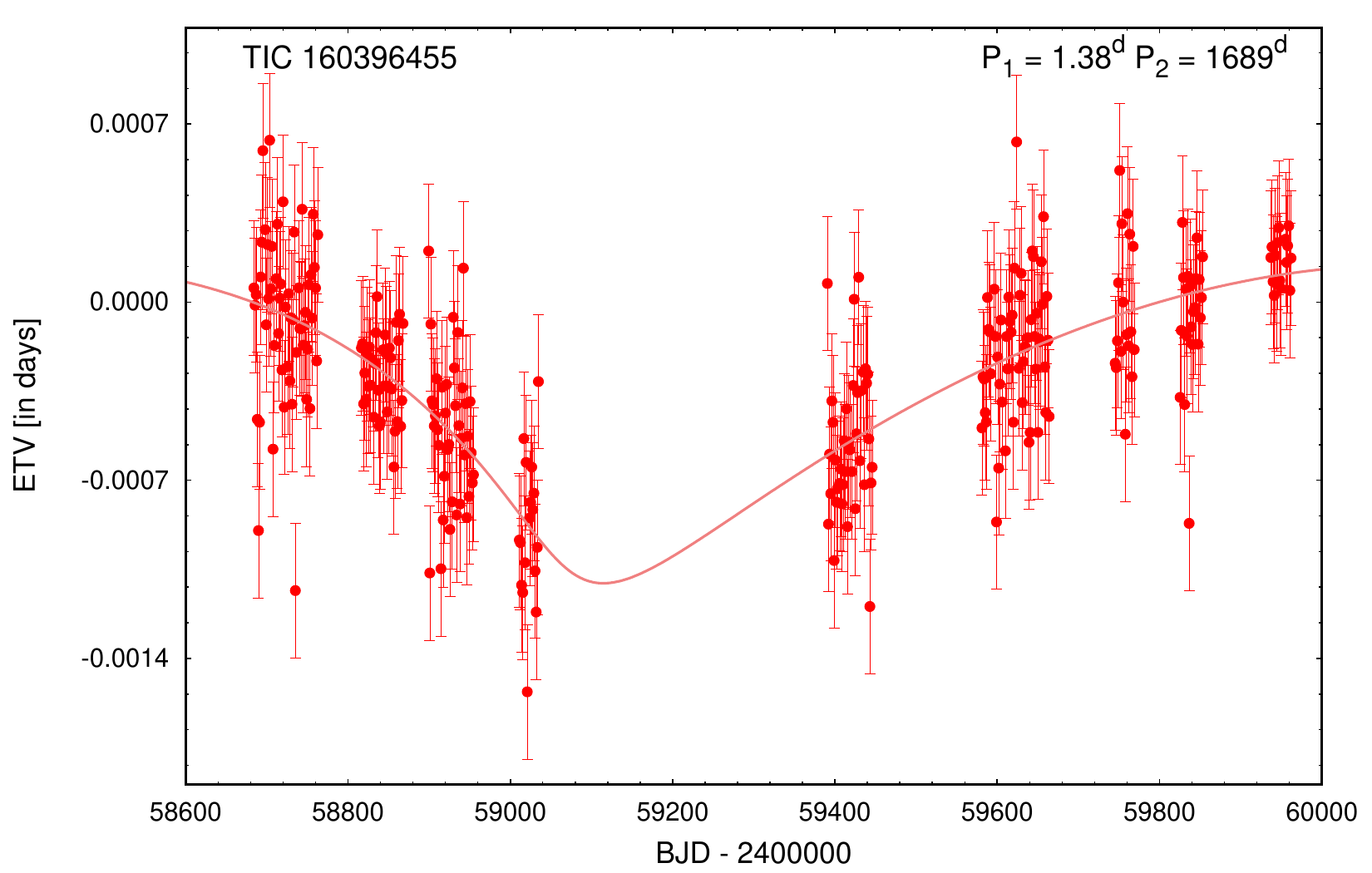}\includegraphics[width=60mm]{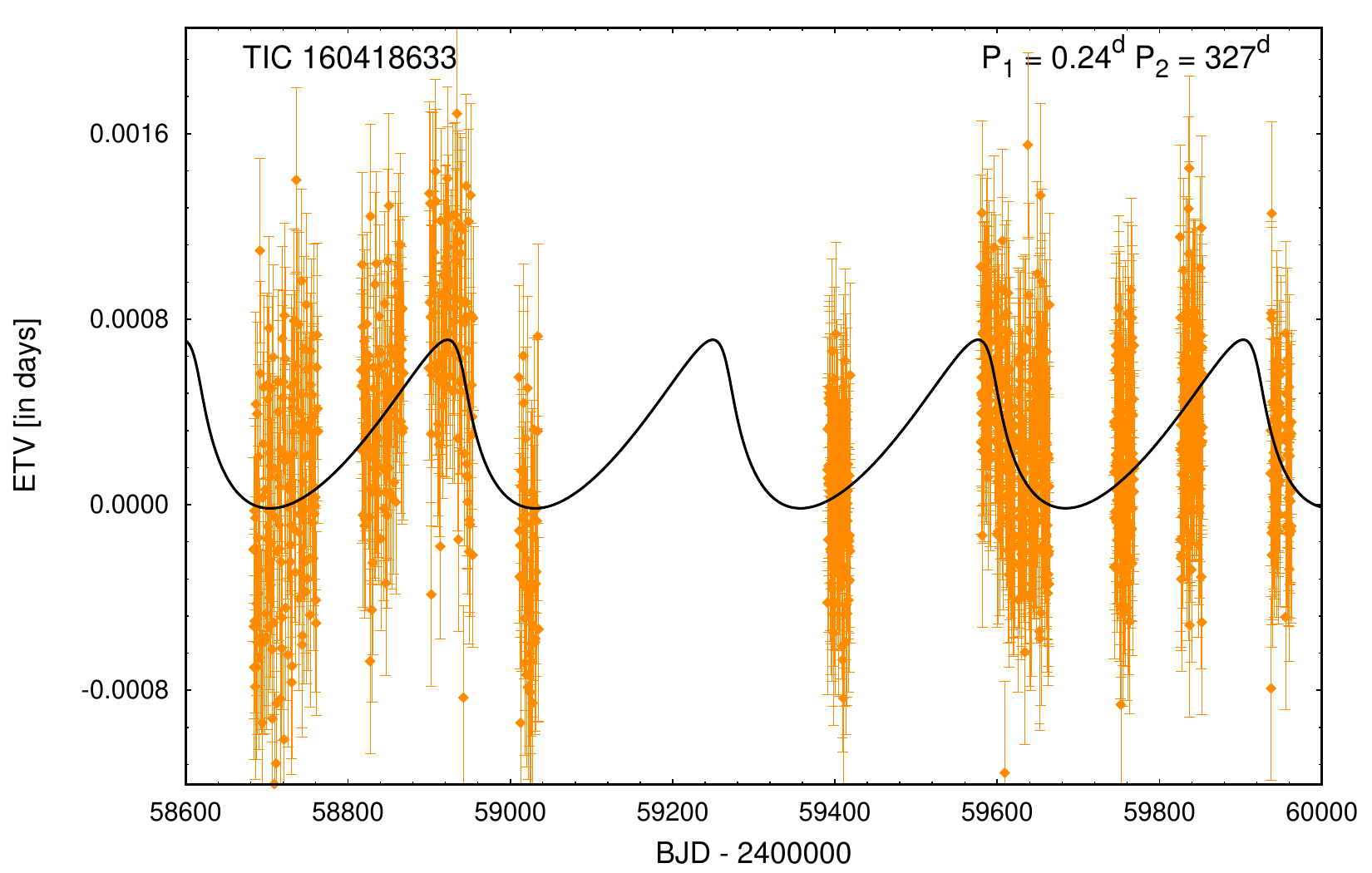}
\includegraphics[width=60mm]{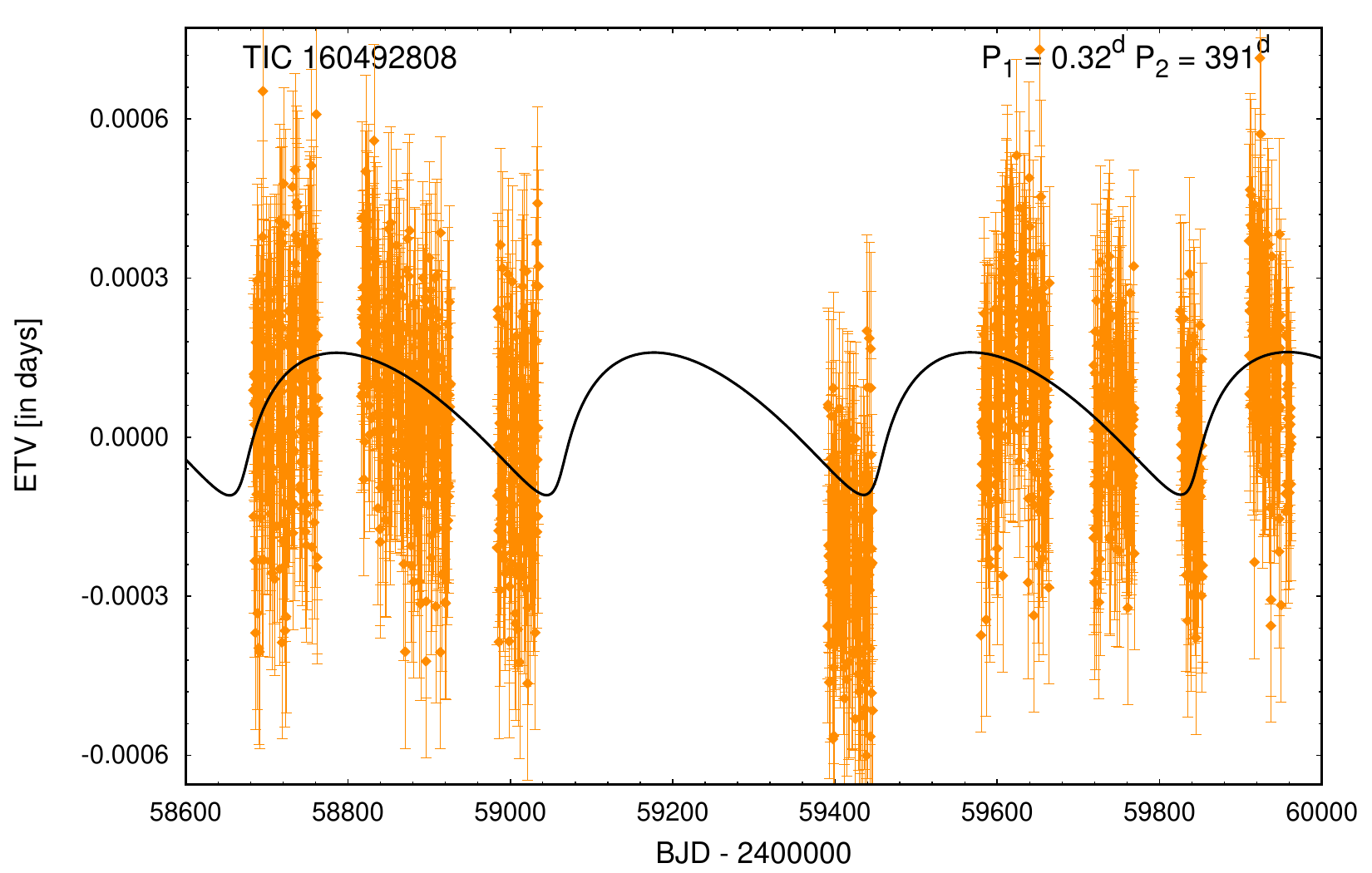}\includegraphics[width=60mm]{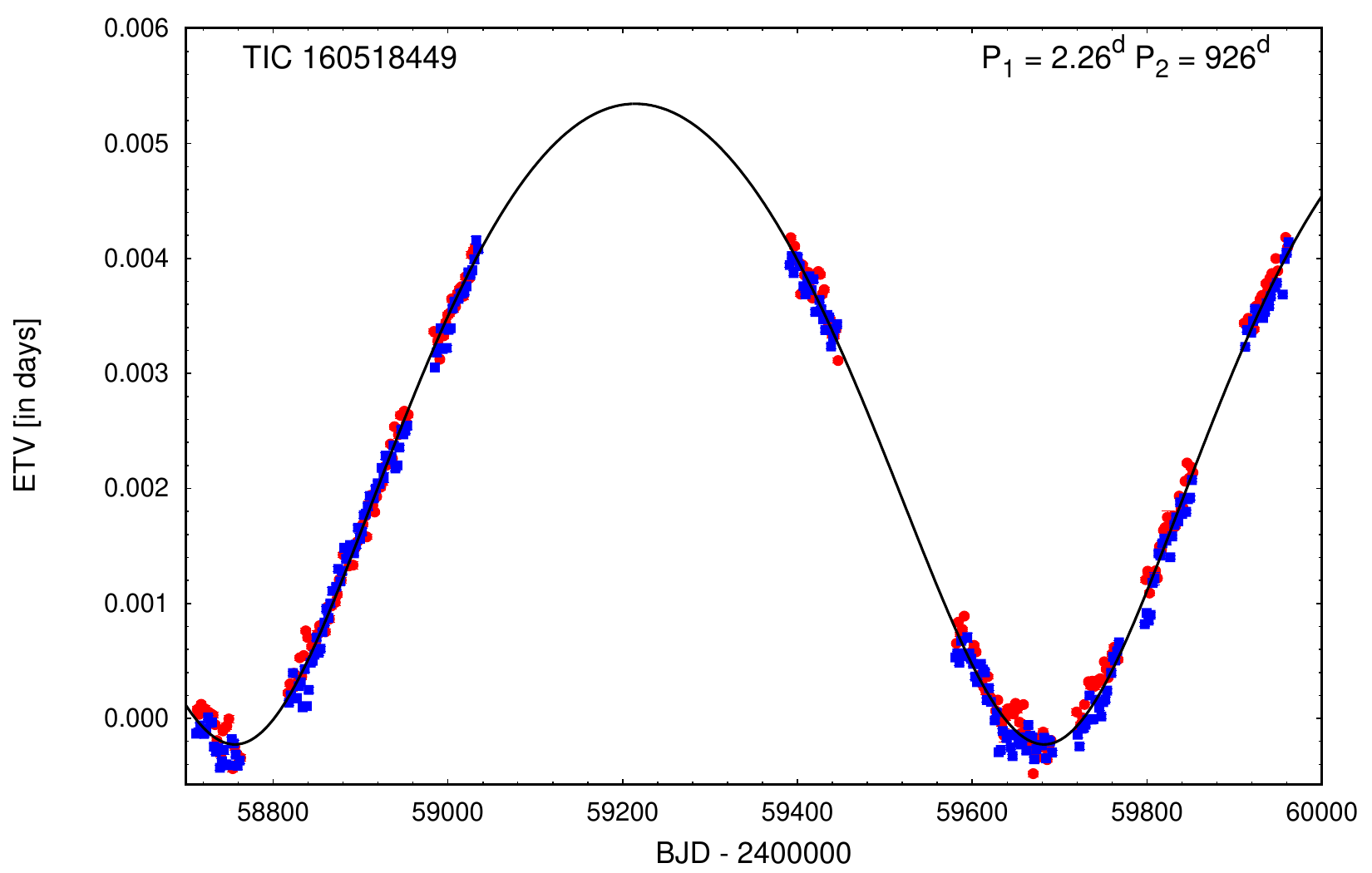}\includegraphics[width=60mm]{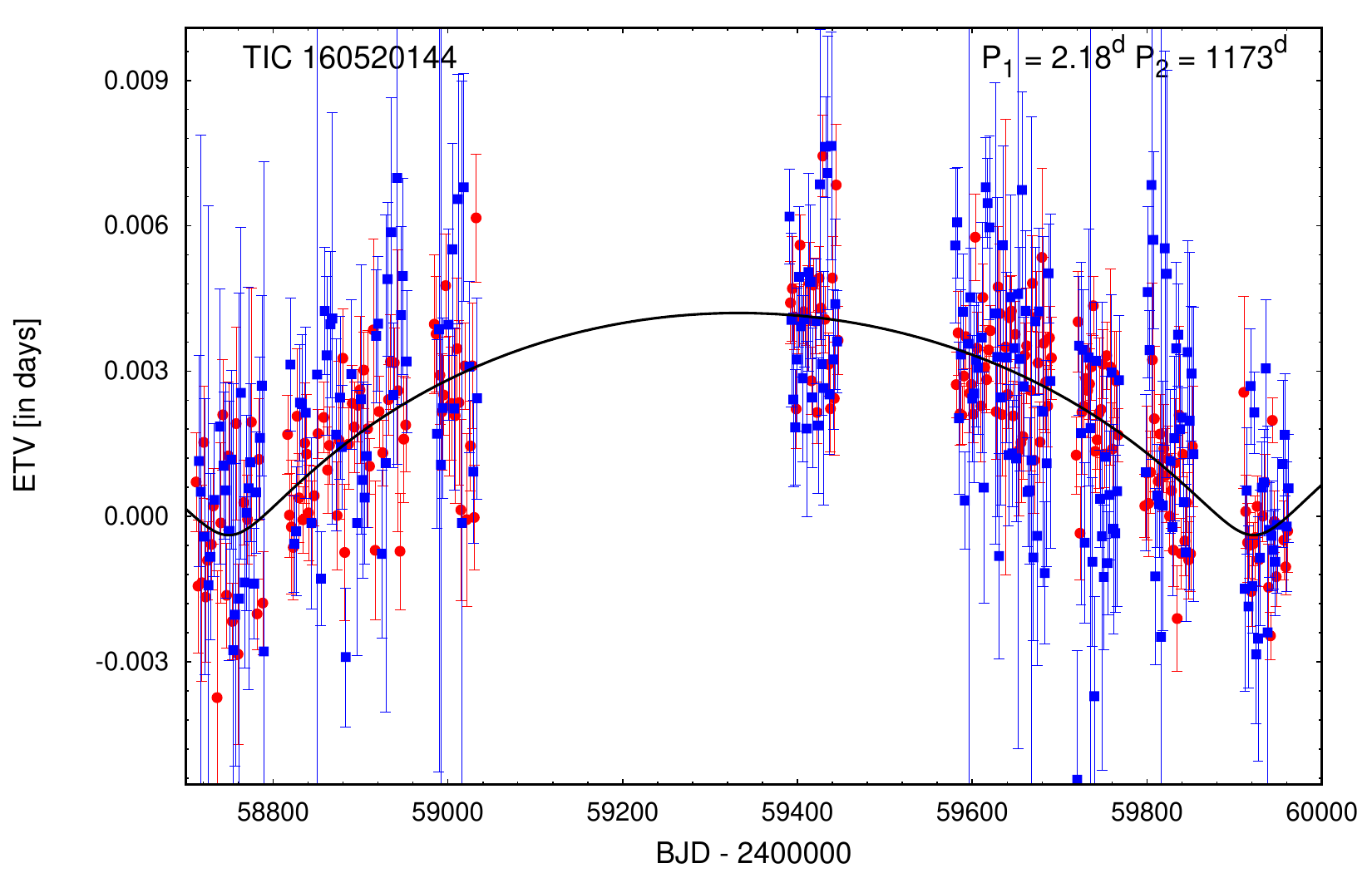}
\includegraphics[width=60mm]{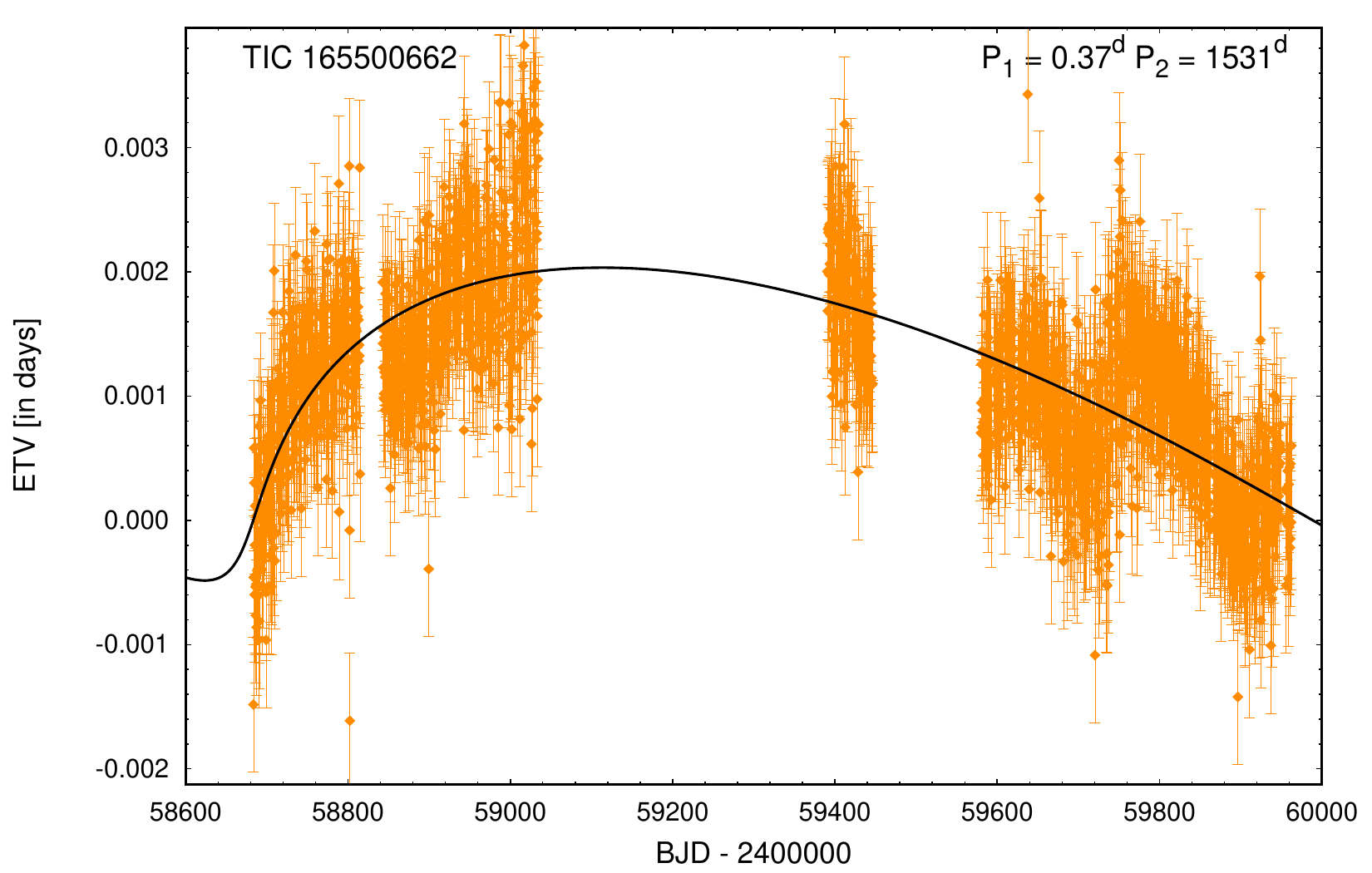}\includegraphics[width=60mm]{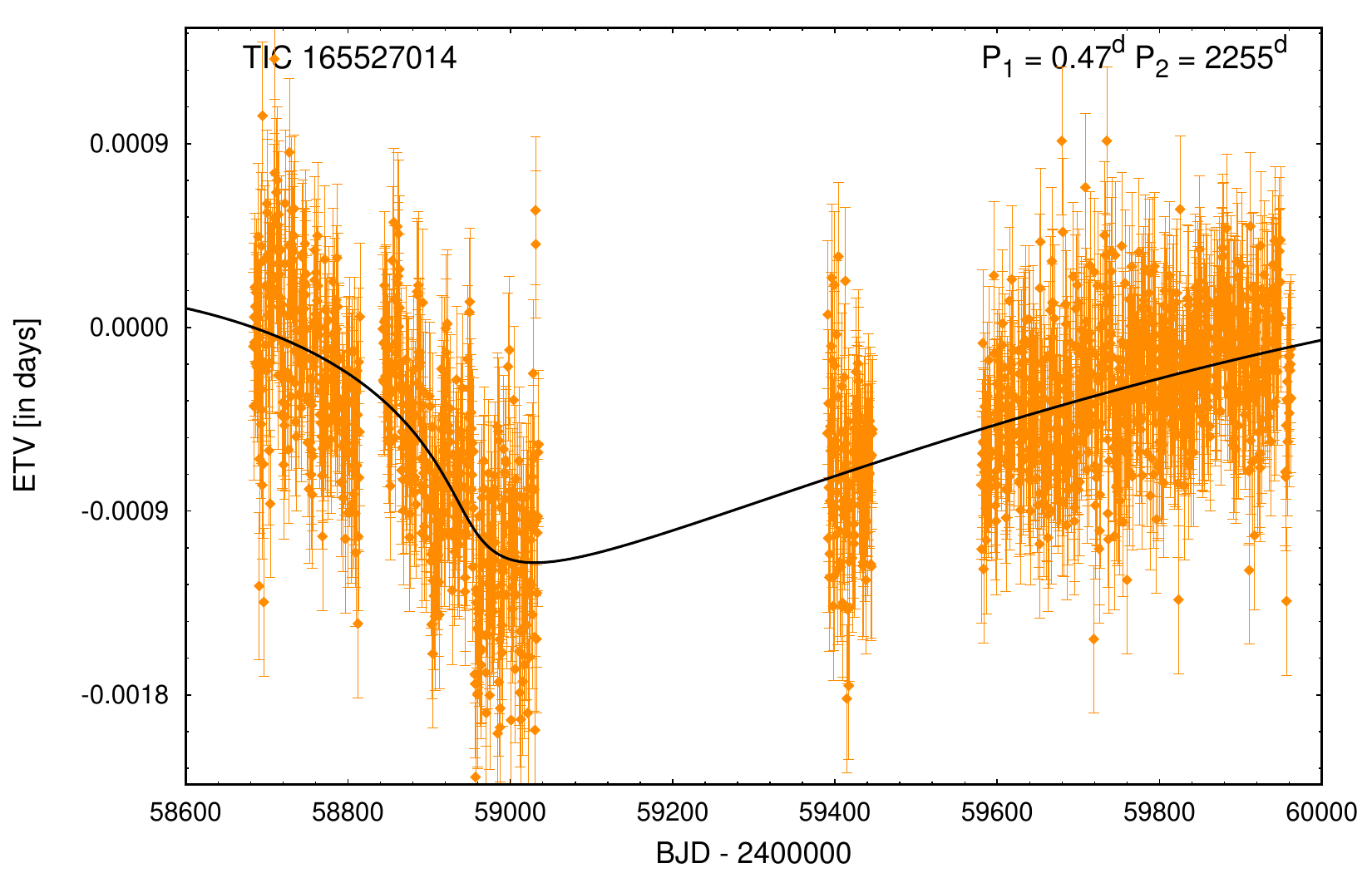}\includegraphics[width=60mm]{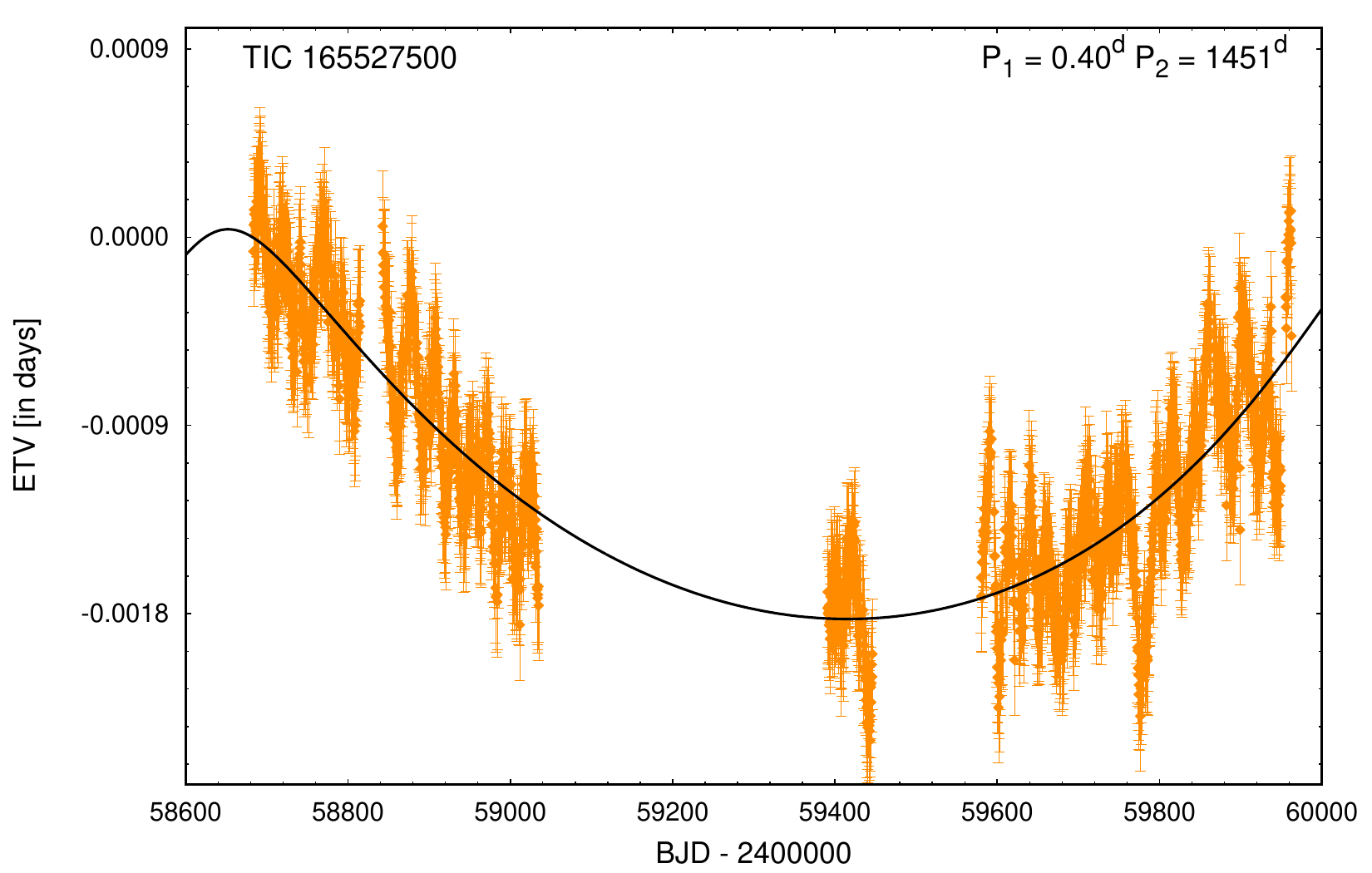}
\includegraphics[width=60mm]{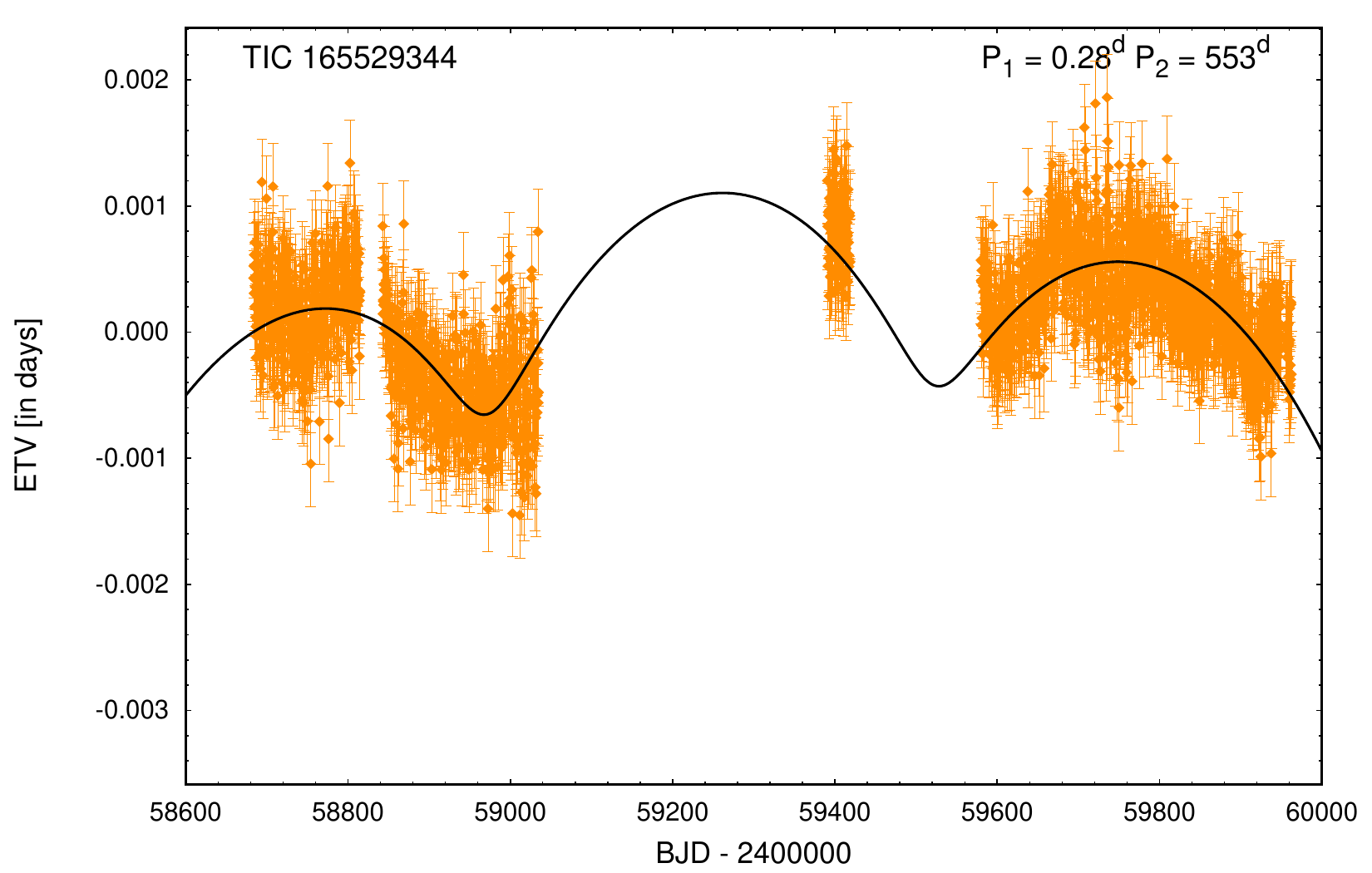}\includegraphics[width=60mm]{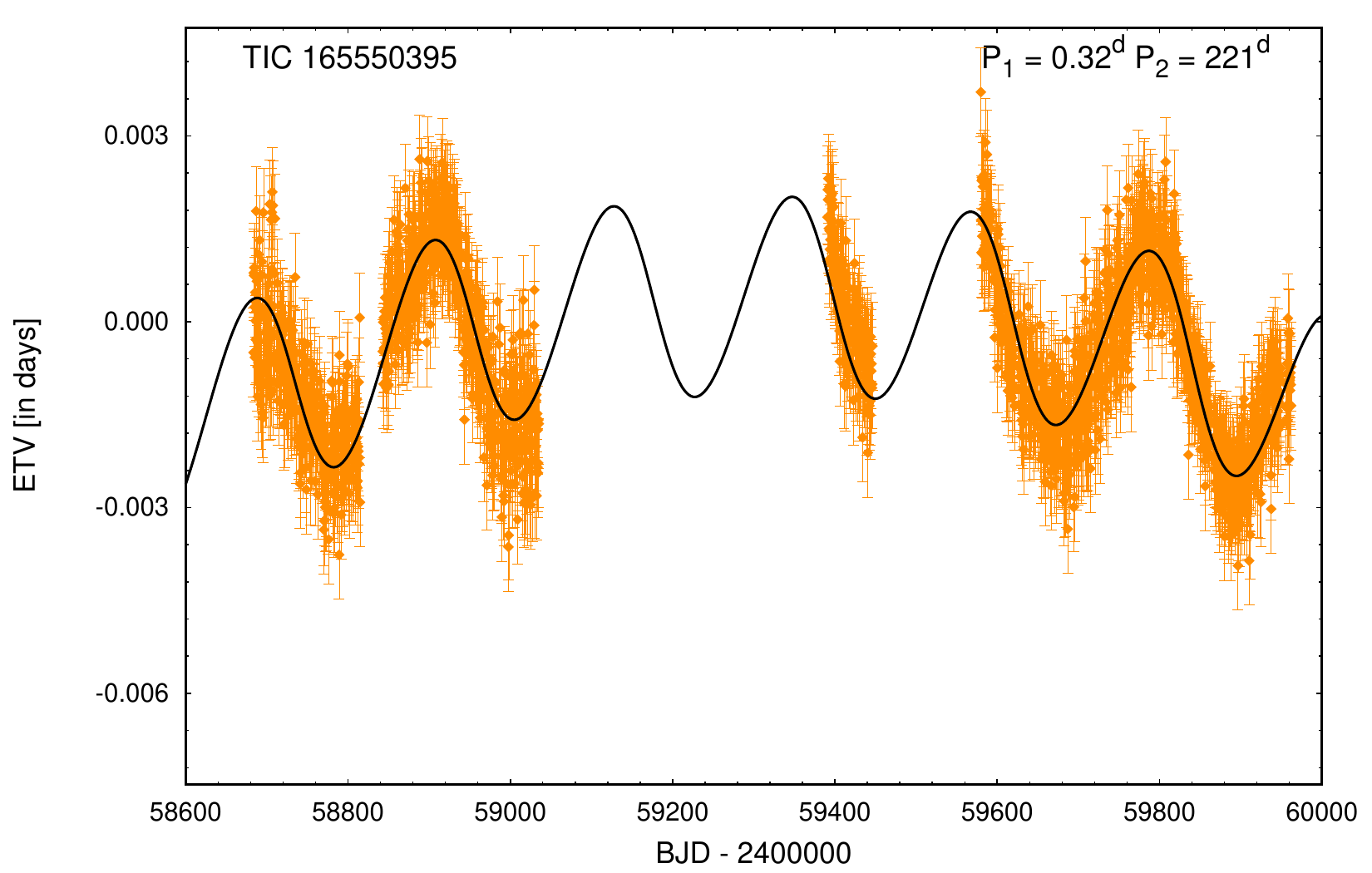}\includegraphics[width=60mm]{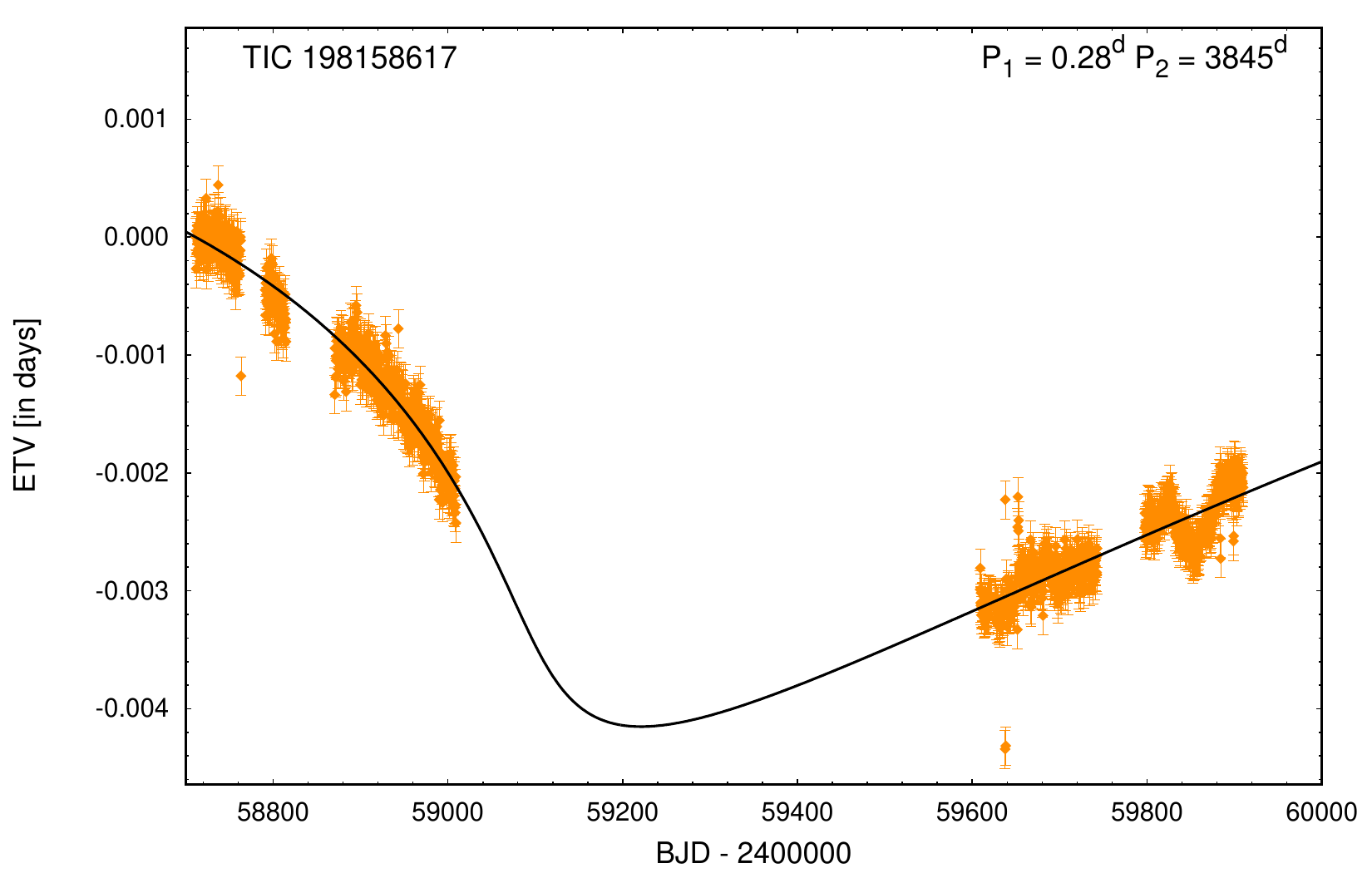}
\caption{ETVs with third body solutions. ETV curves calculated from \textit{TESS} observations of primary and secondary minima, and the average of the two, are denoted by red circles, blue boxes, and orange diamonds, respectively. We display and fit the ETV curves for both the primary and secondary eclipses only when the data quality warrant a joint analysis and the binary is eccentric. If the primary and secondary ETV curves are of comparable quality and the binary eccentricity is nearly zero, we display and fit only the average of the two ETV curves. If the quality of the primary ETV curve is significantly better than that of the secondary curve or, if only primary eclipses are present, we present only the plot and the fit for the primary eclipses. Ground-based minima (taken from the literature, and available only for a few systems) are denoted by upward red triangles (primary) and downward blue triangles (secondary); their estimated uncertainties are also indicated.  Pure LTTE solutions are plotted with black lines, while combined dynamical and LTTE solutions are drawn with grey lines. (Note, the use of quadratic or cubic terms is not indicated; for these and other details, see Table~\ref{Tab:Orbelem_LTTE1}--\ref{Tab:Orbelemdyn2}}
\label{Fig:ETVs}
\end{figure*}

\addtocounter{figure}{-1}

\begin{figure*}
\includegraphics[width=60mm]{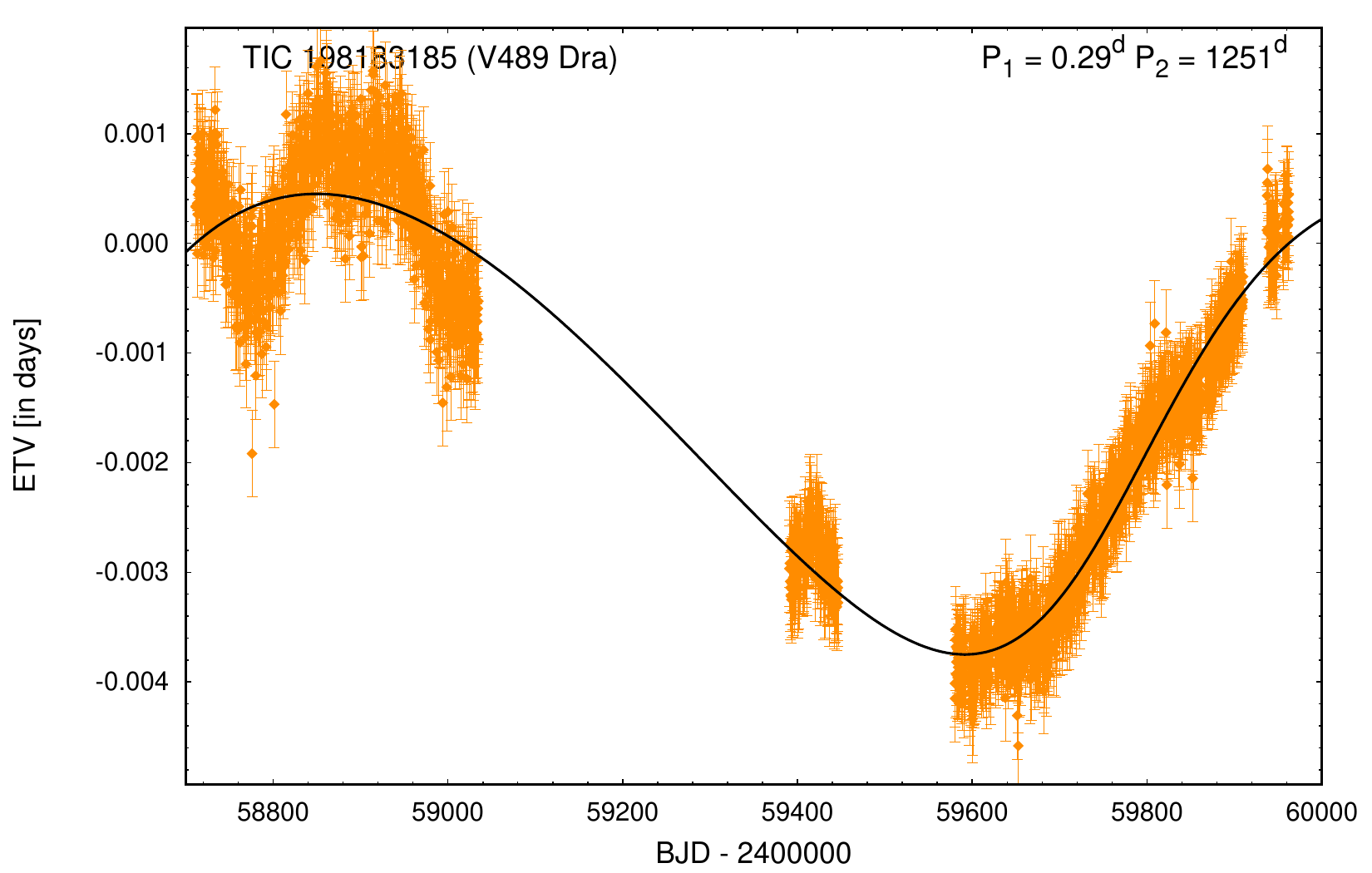}\includegraphics[width=60mm]{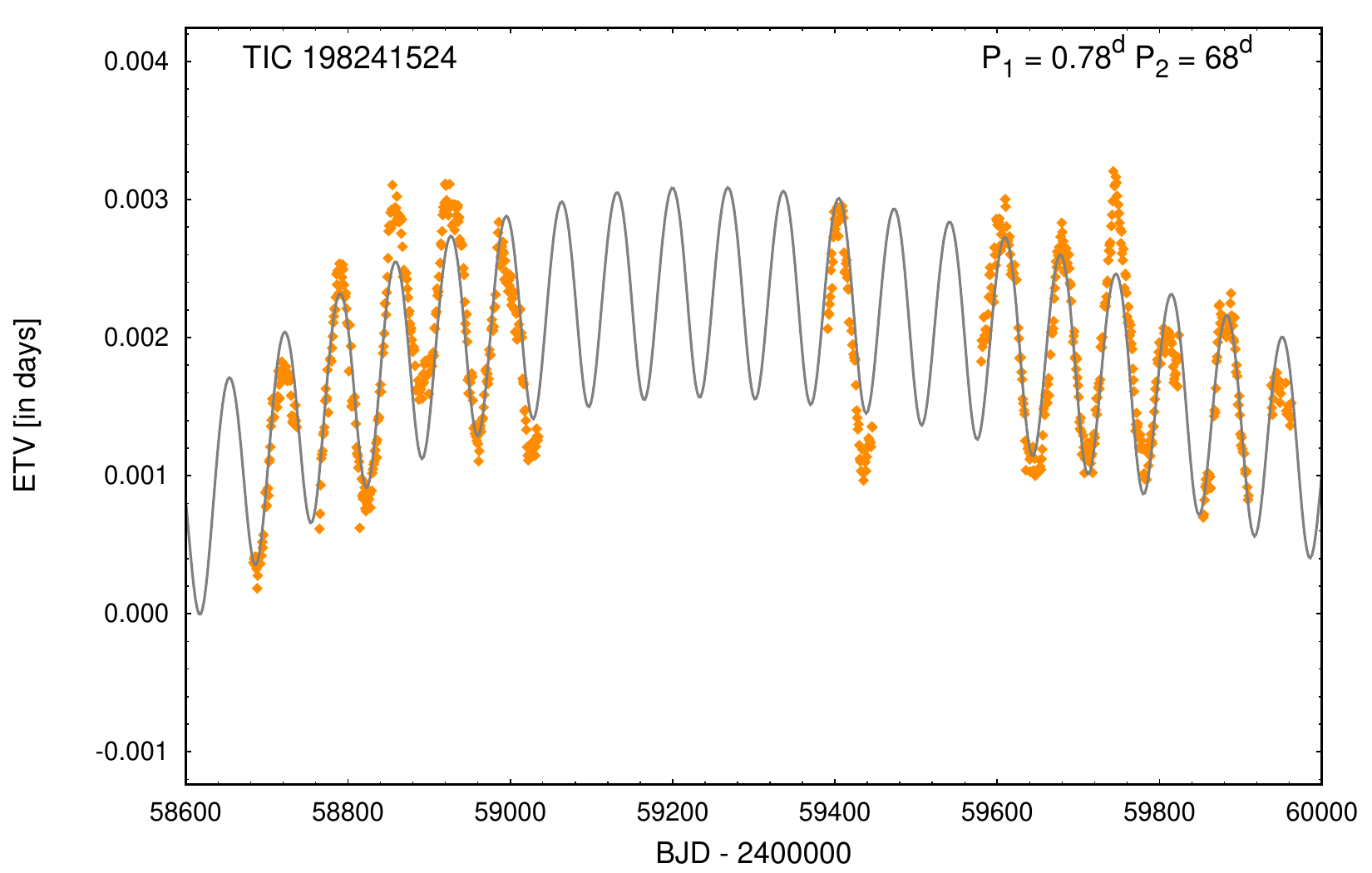}\includegraphics[width=60mm]{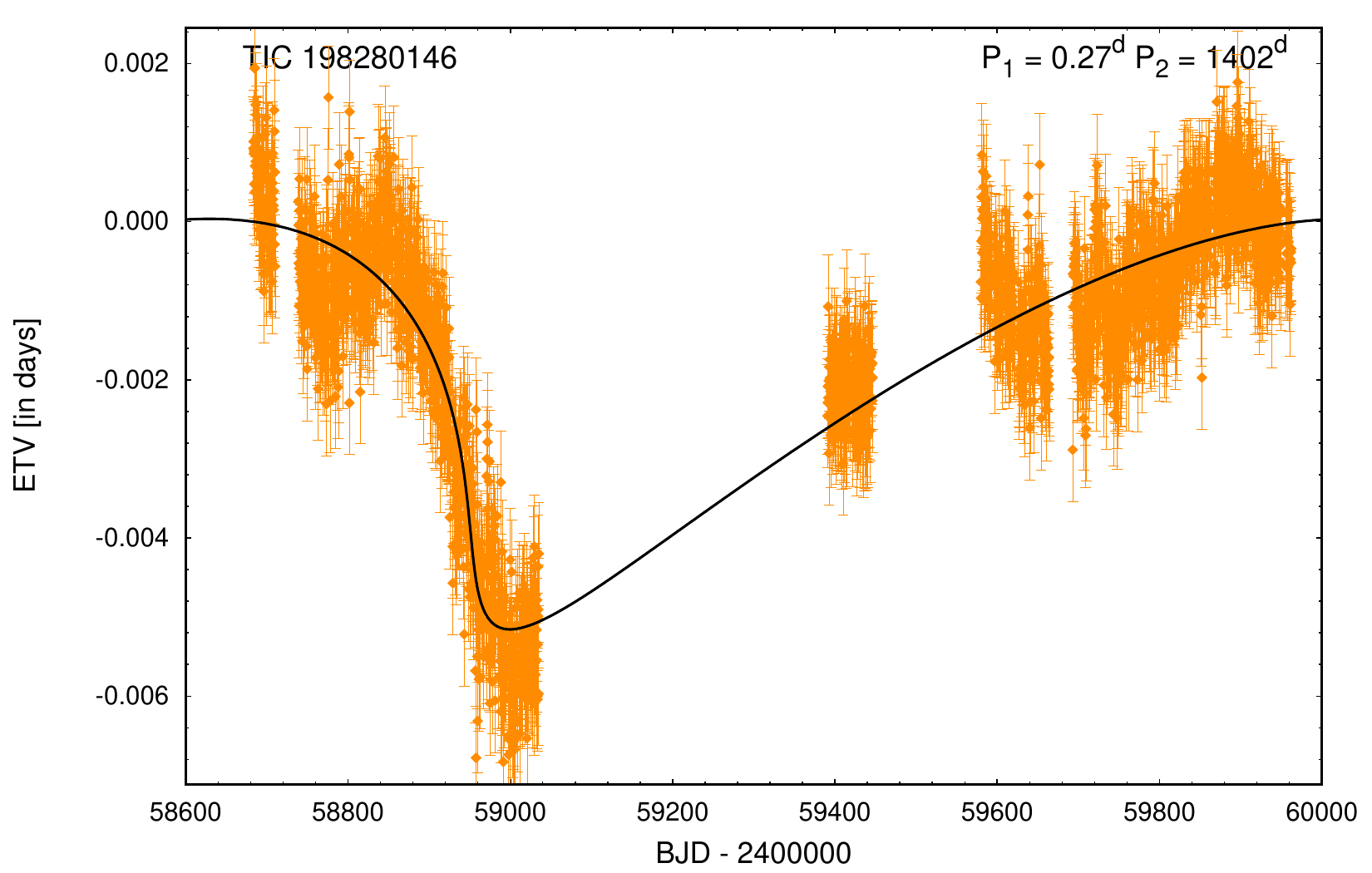}
\includegraphics[width=60mm]{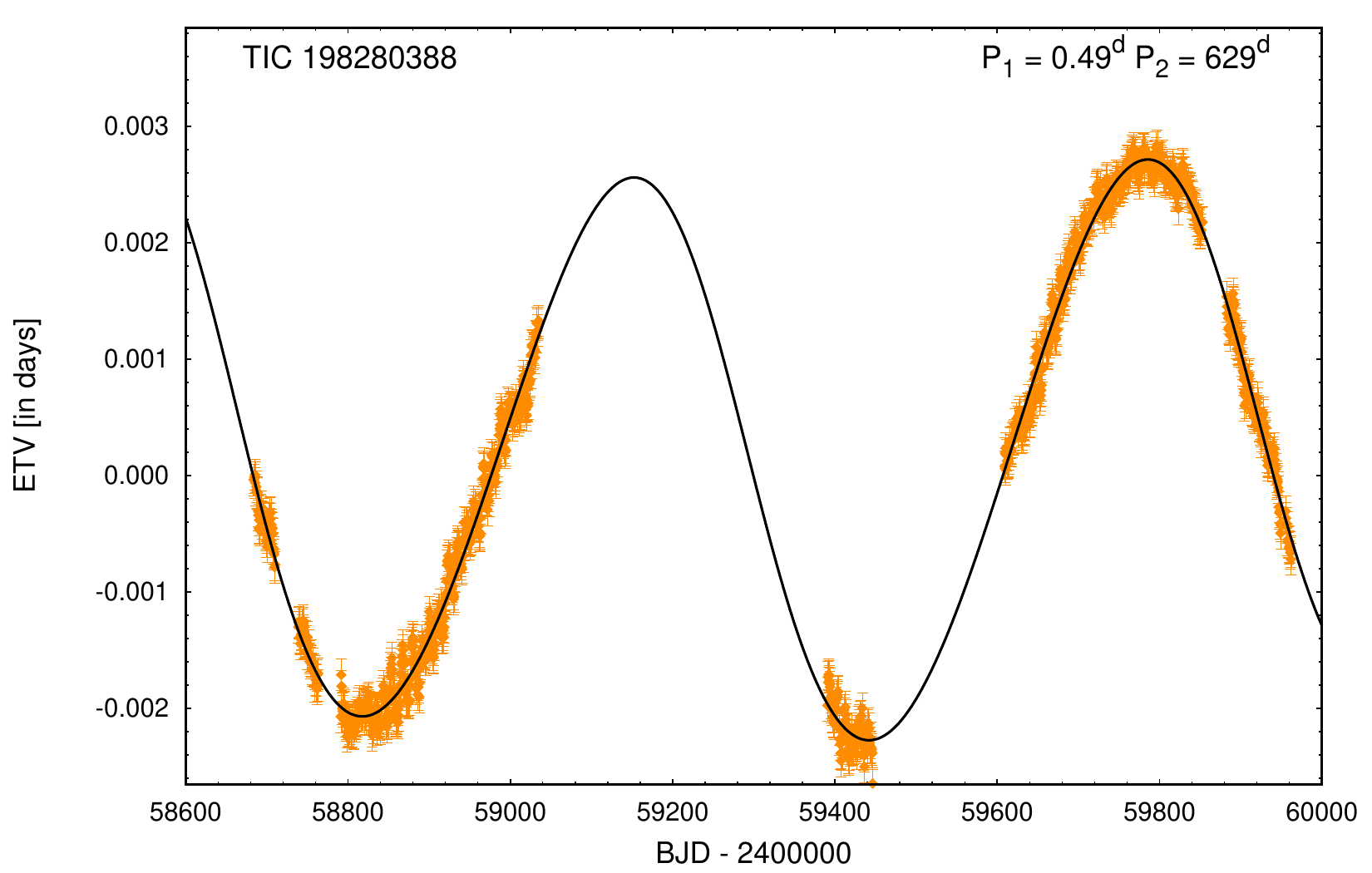}\includegraphics[width=60mm]{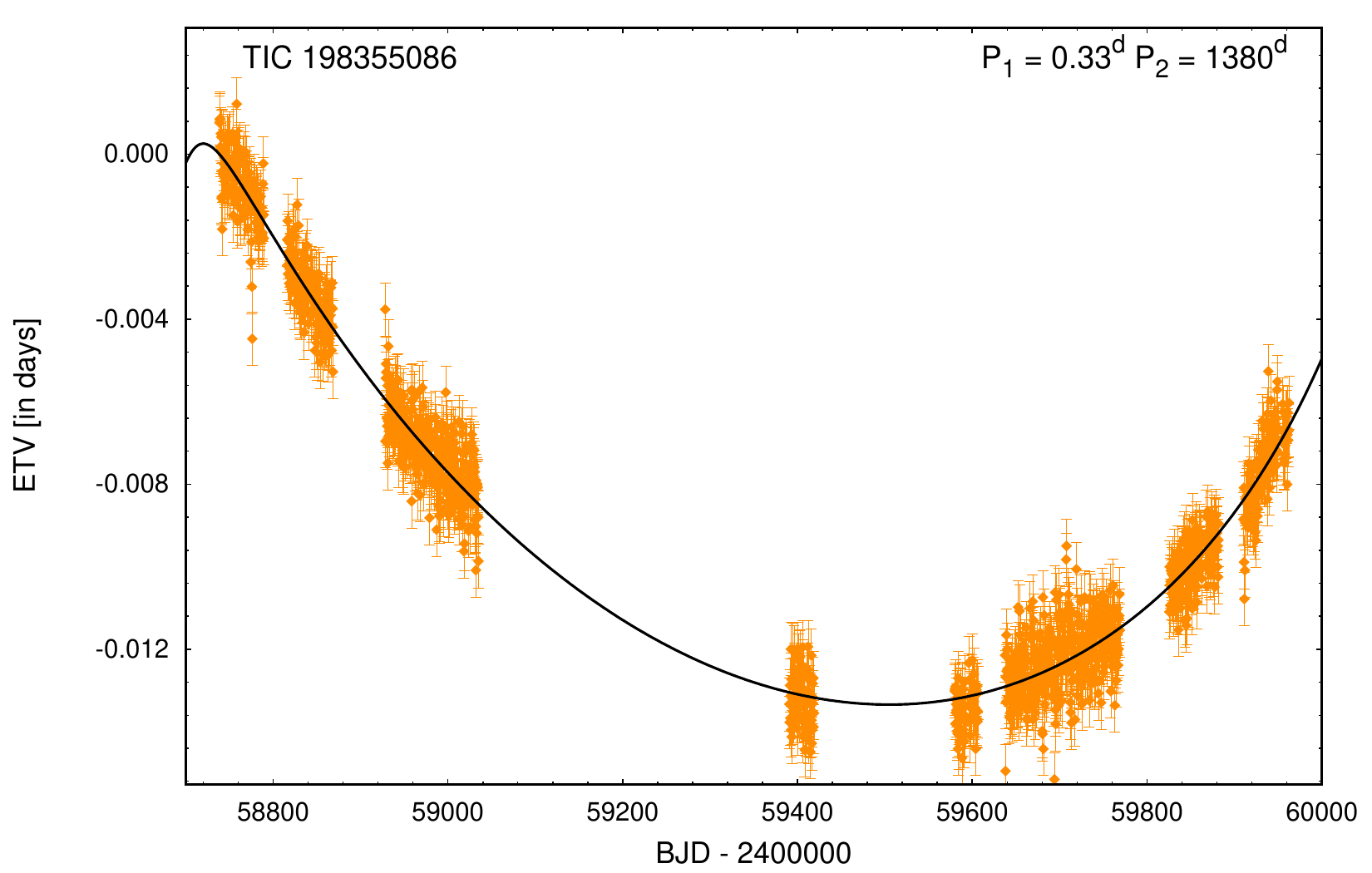}\includegraphics[width=60mm]{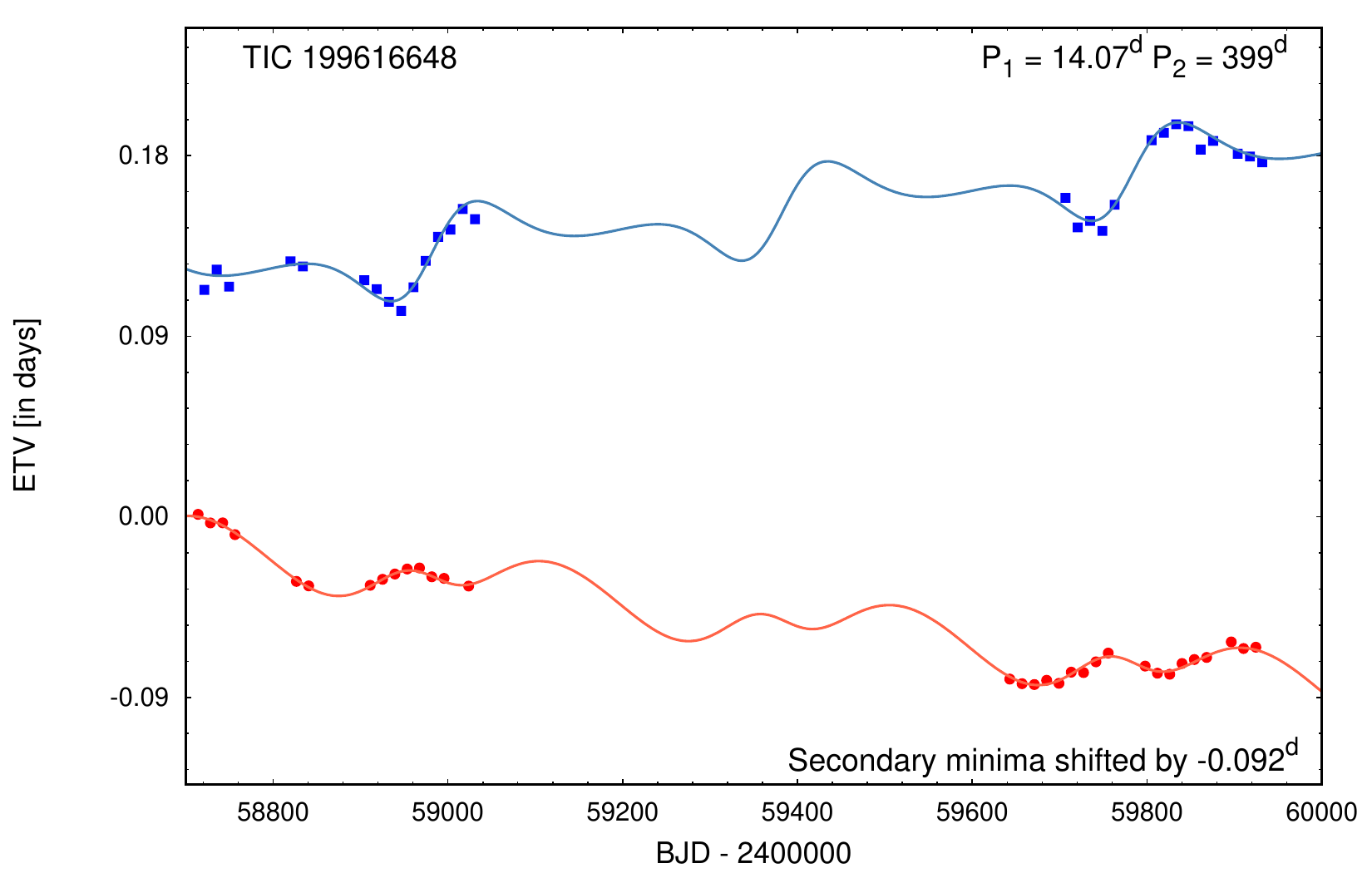}
\includegraphics[width=60mm]{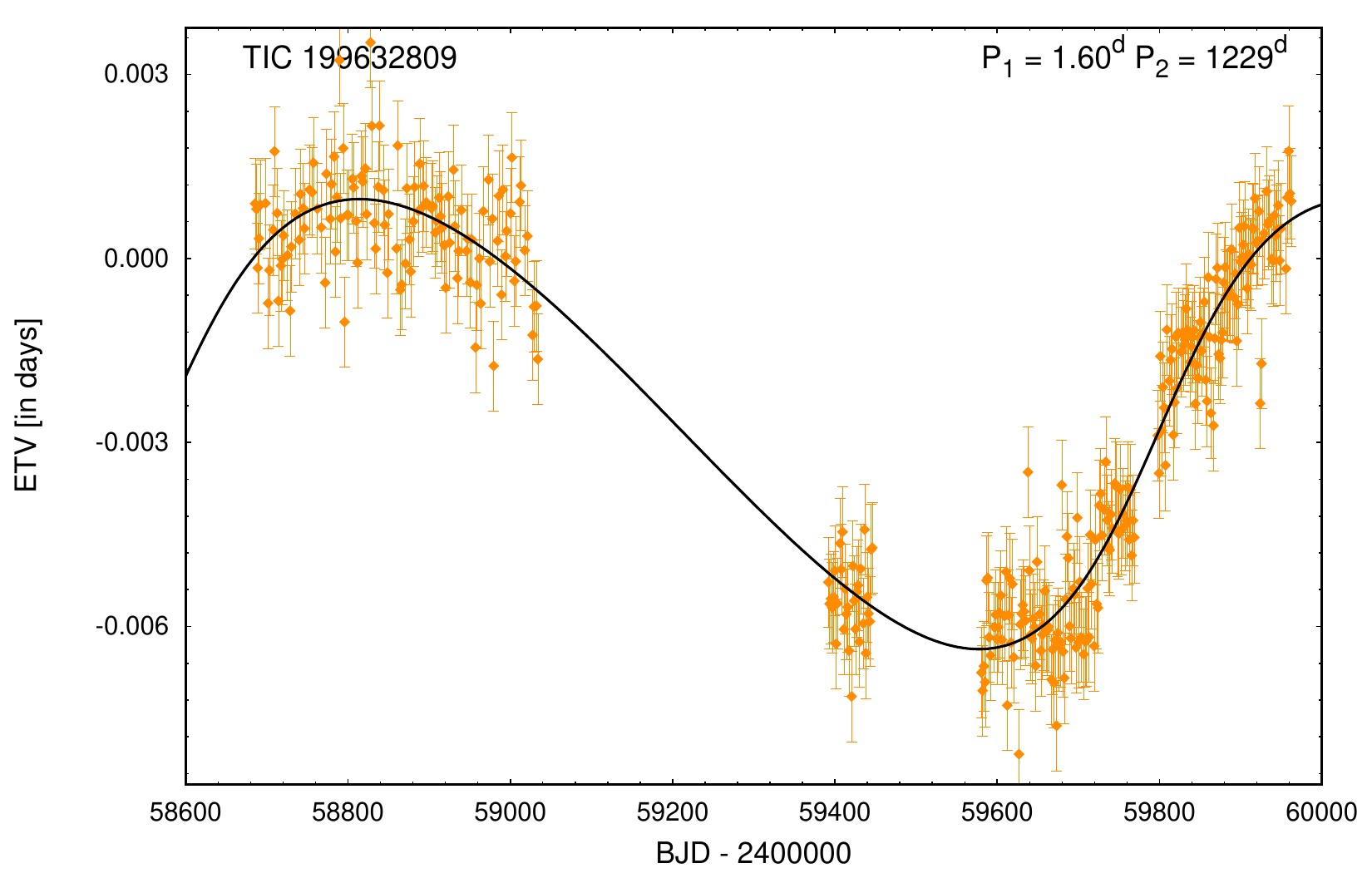}\includegraphics[width=60mm]{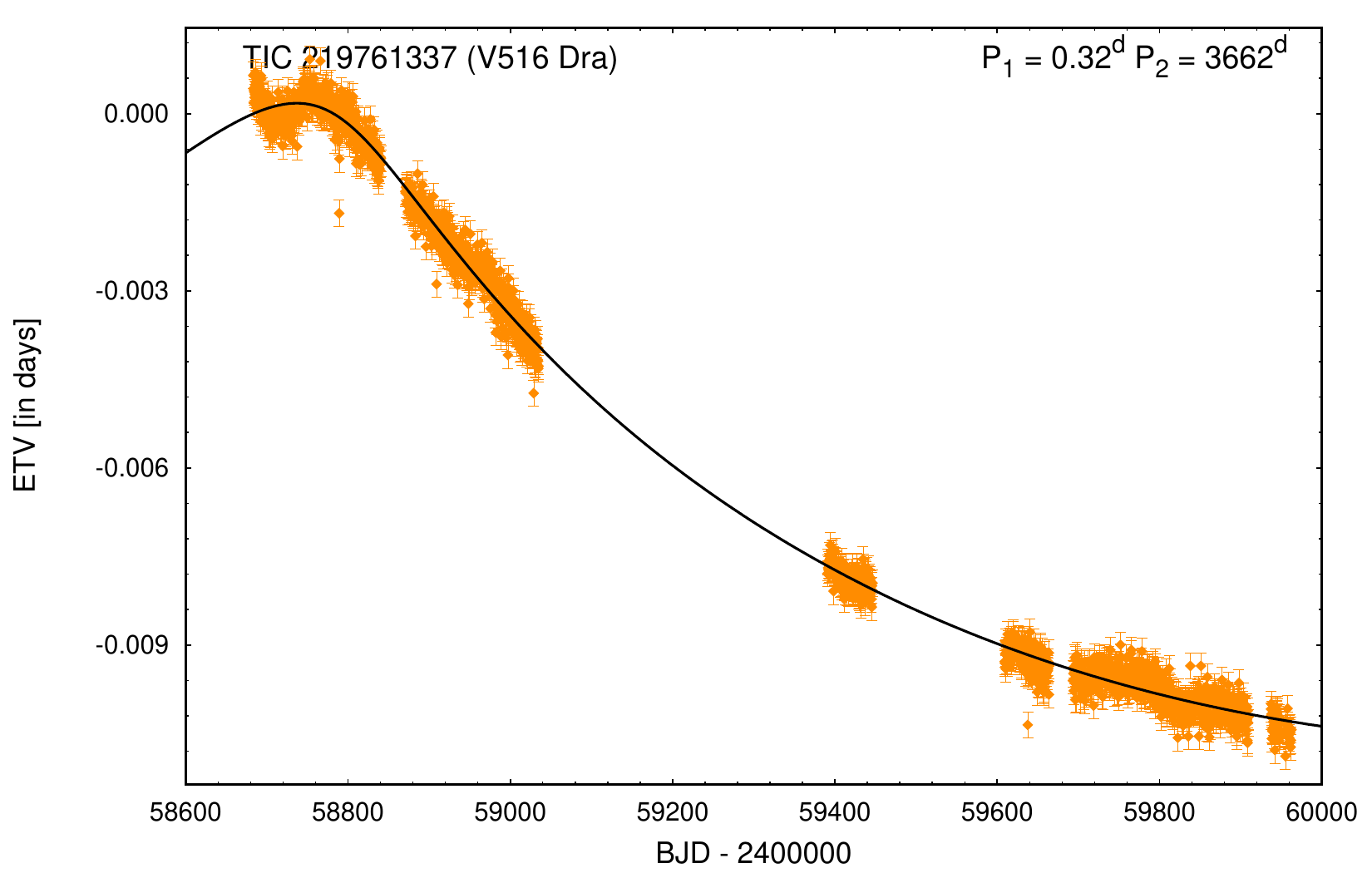}\includegraphics[width=60mm]{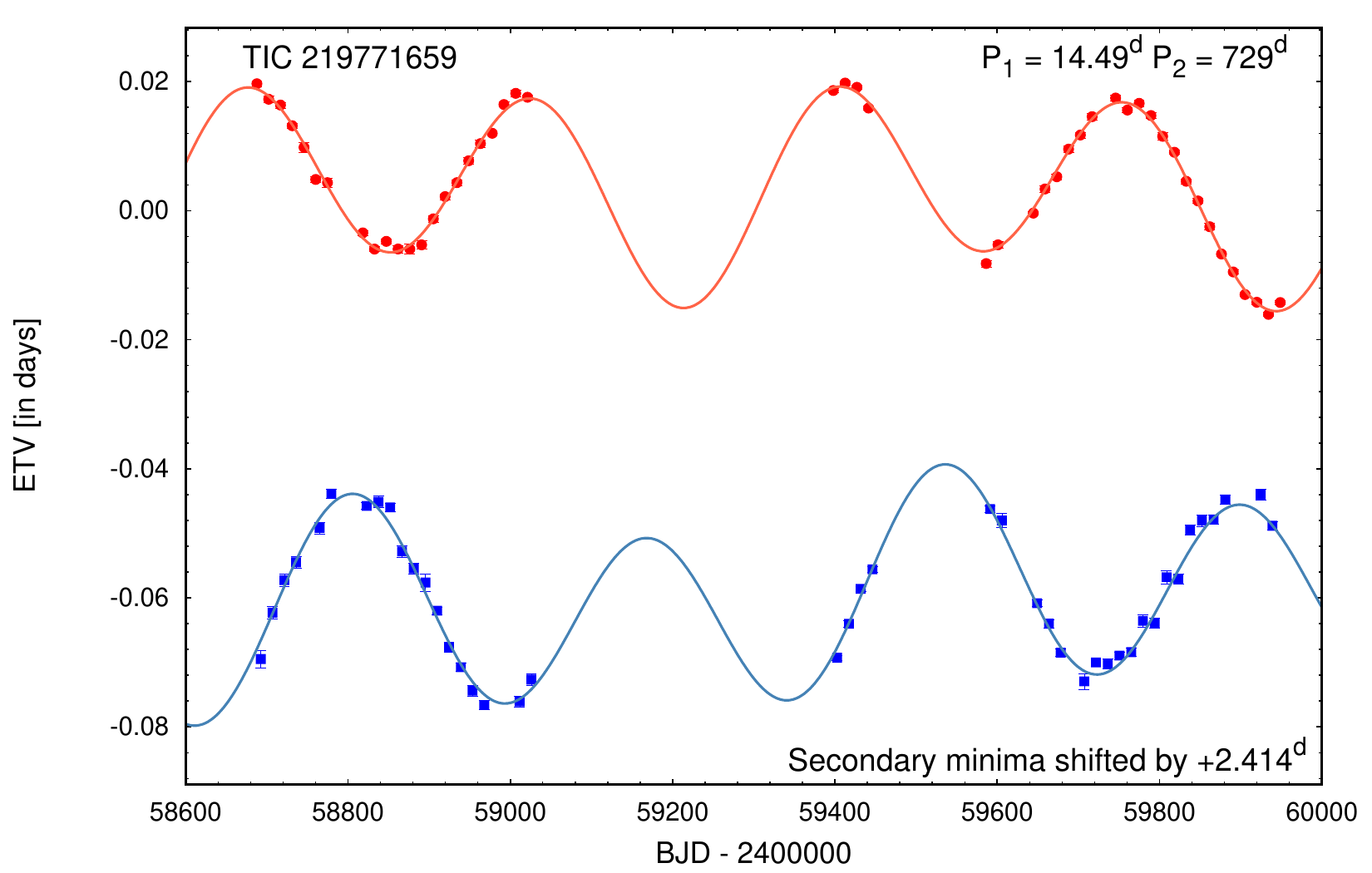}
\includegraphics[width=60mm]{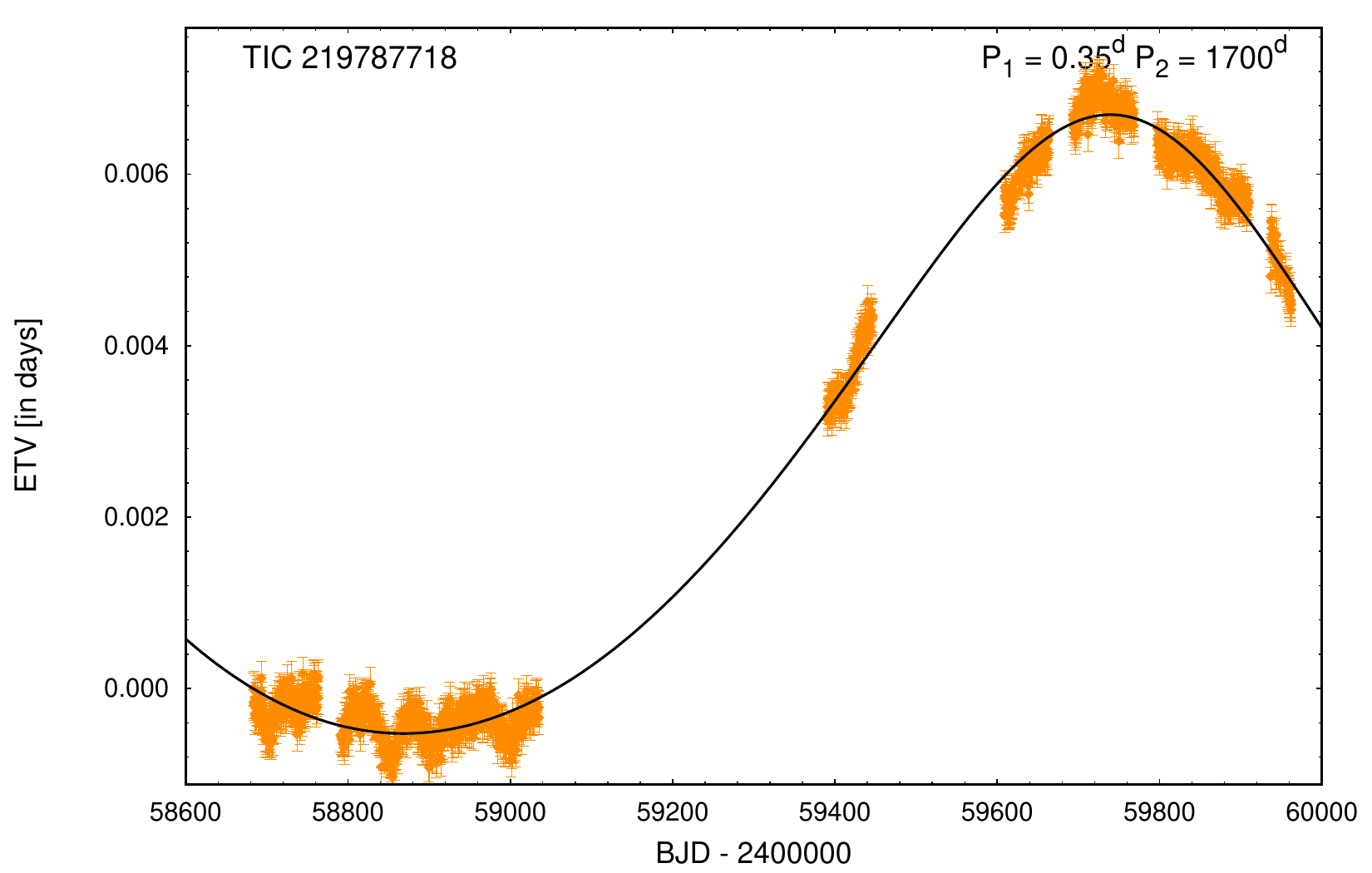}\includegraphics[width=60mm]{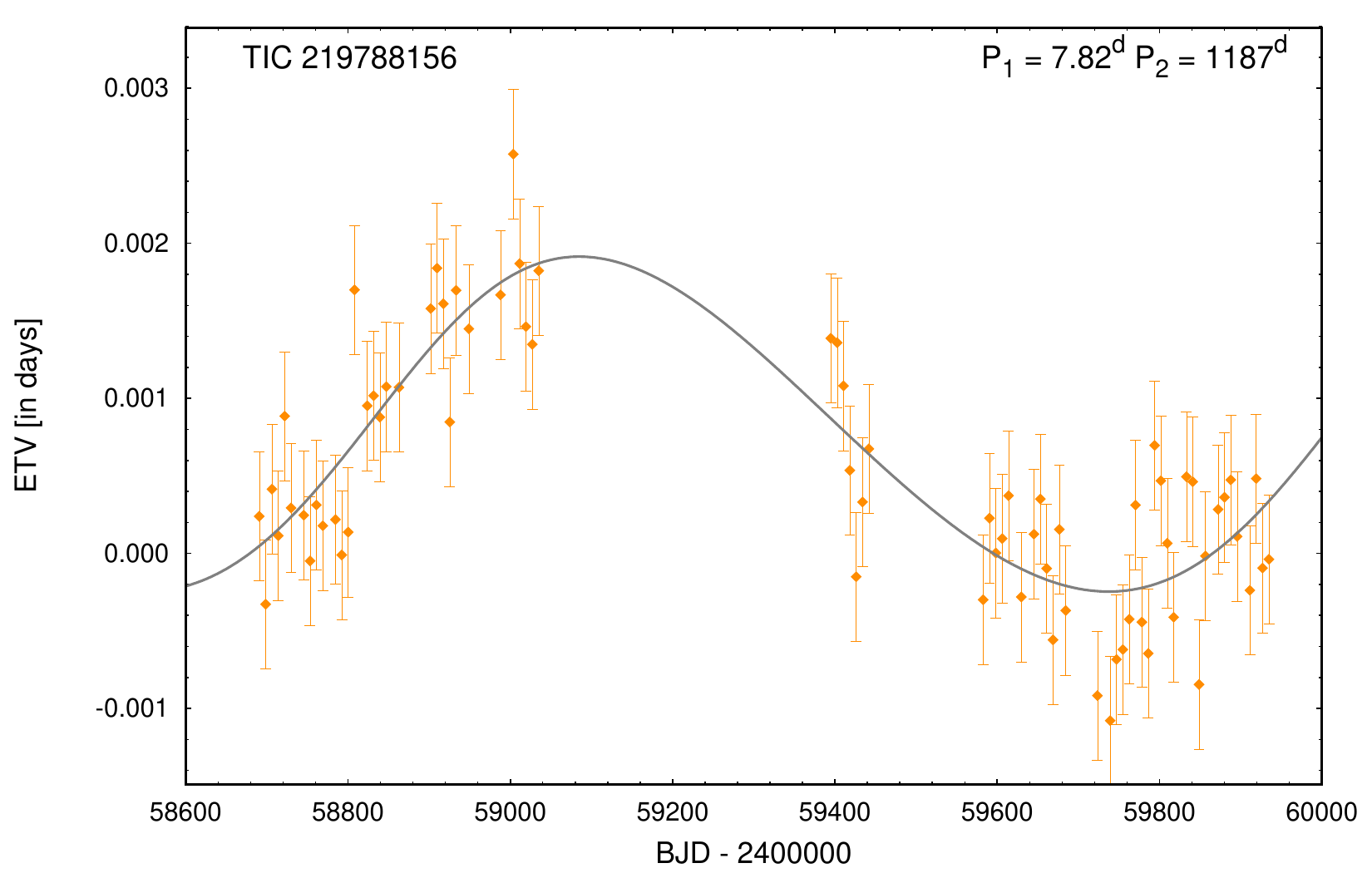}\includegraphics[width=60mm]{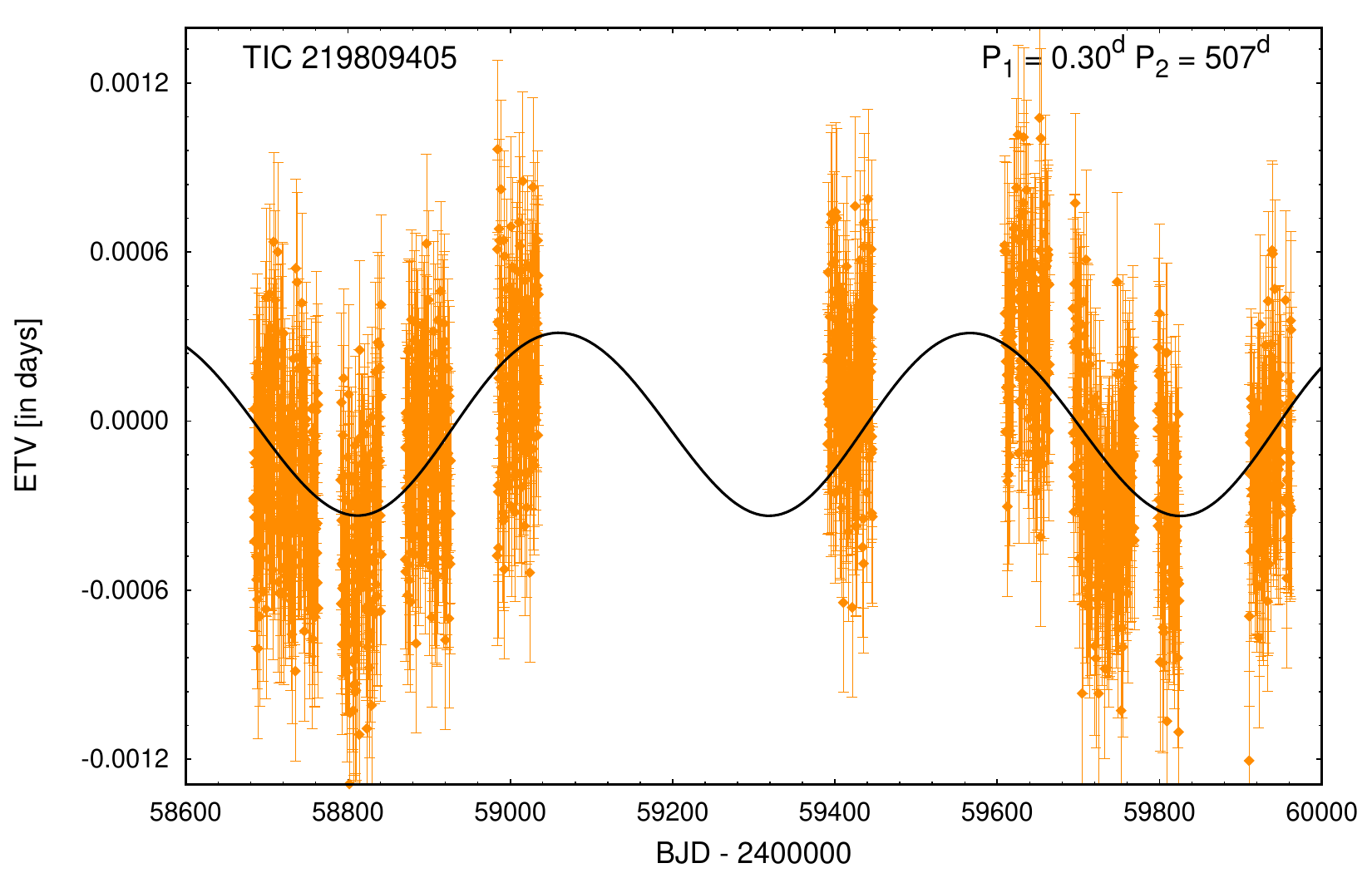}
\includegraphics[width=60mm]{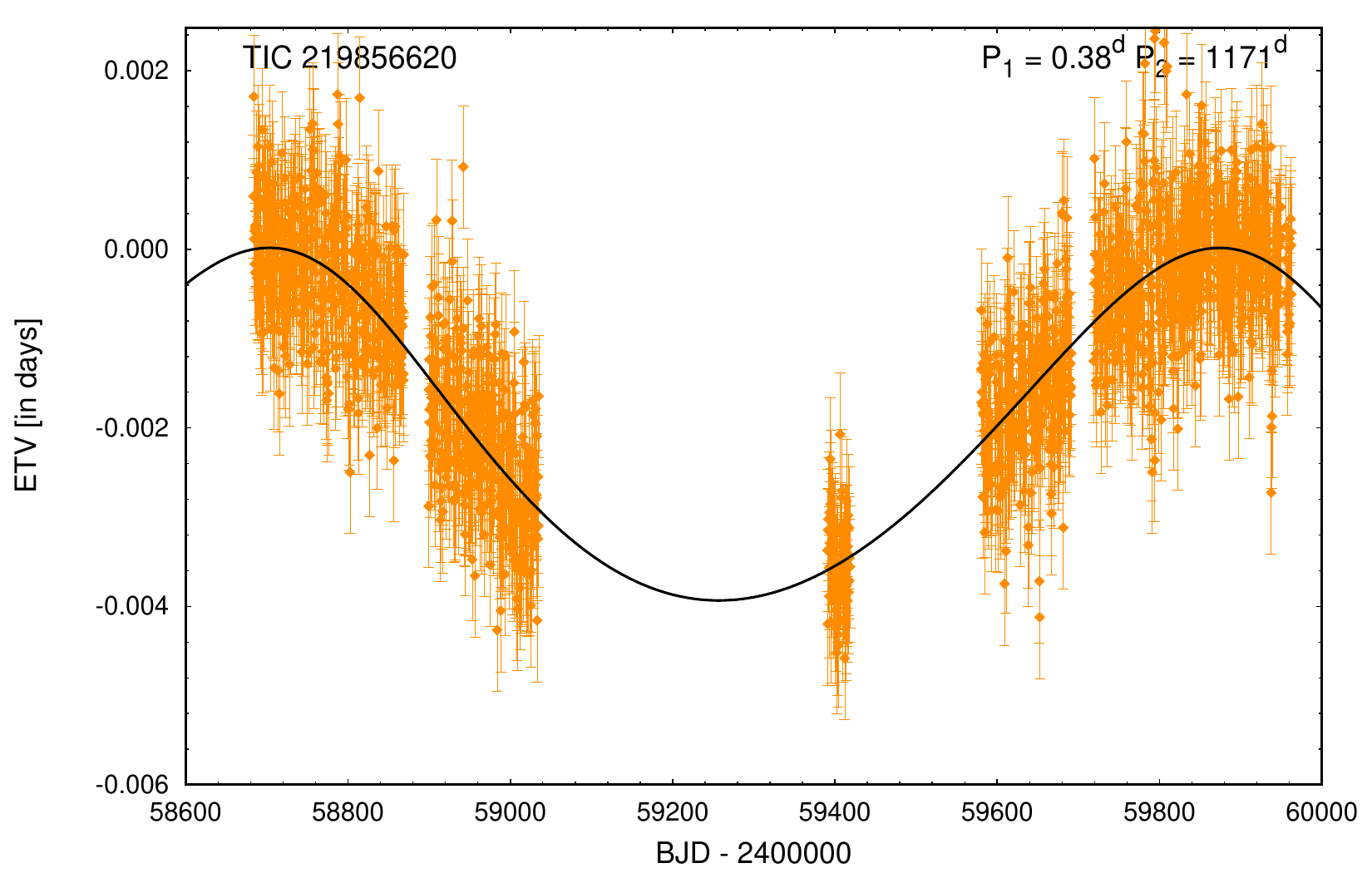}\includegraphics[width=60mm]{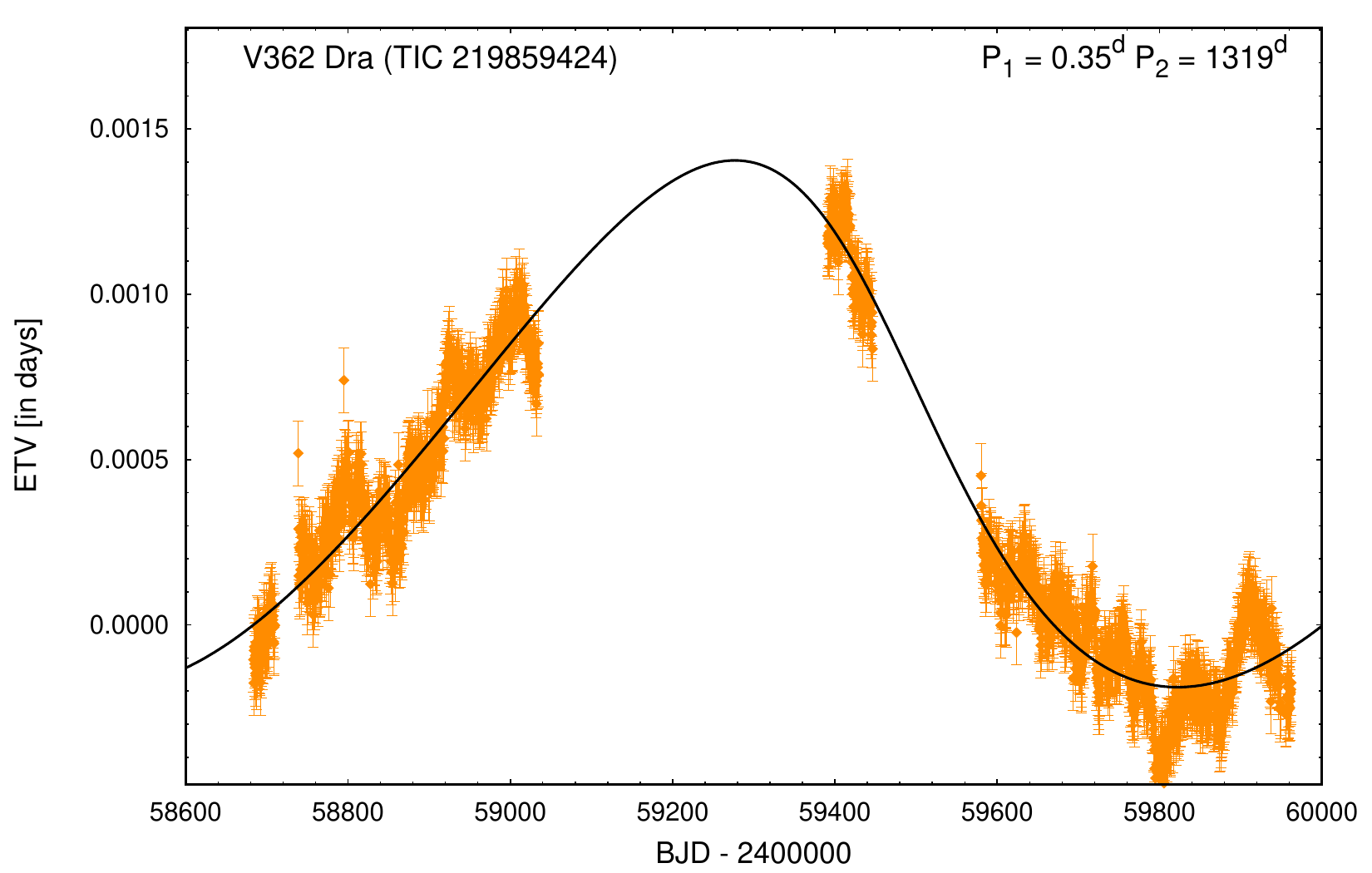}\includegraphics[width=60mm]{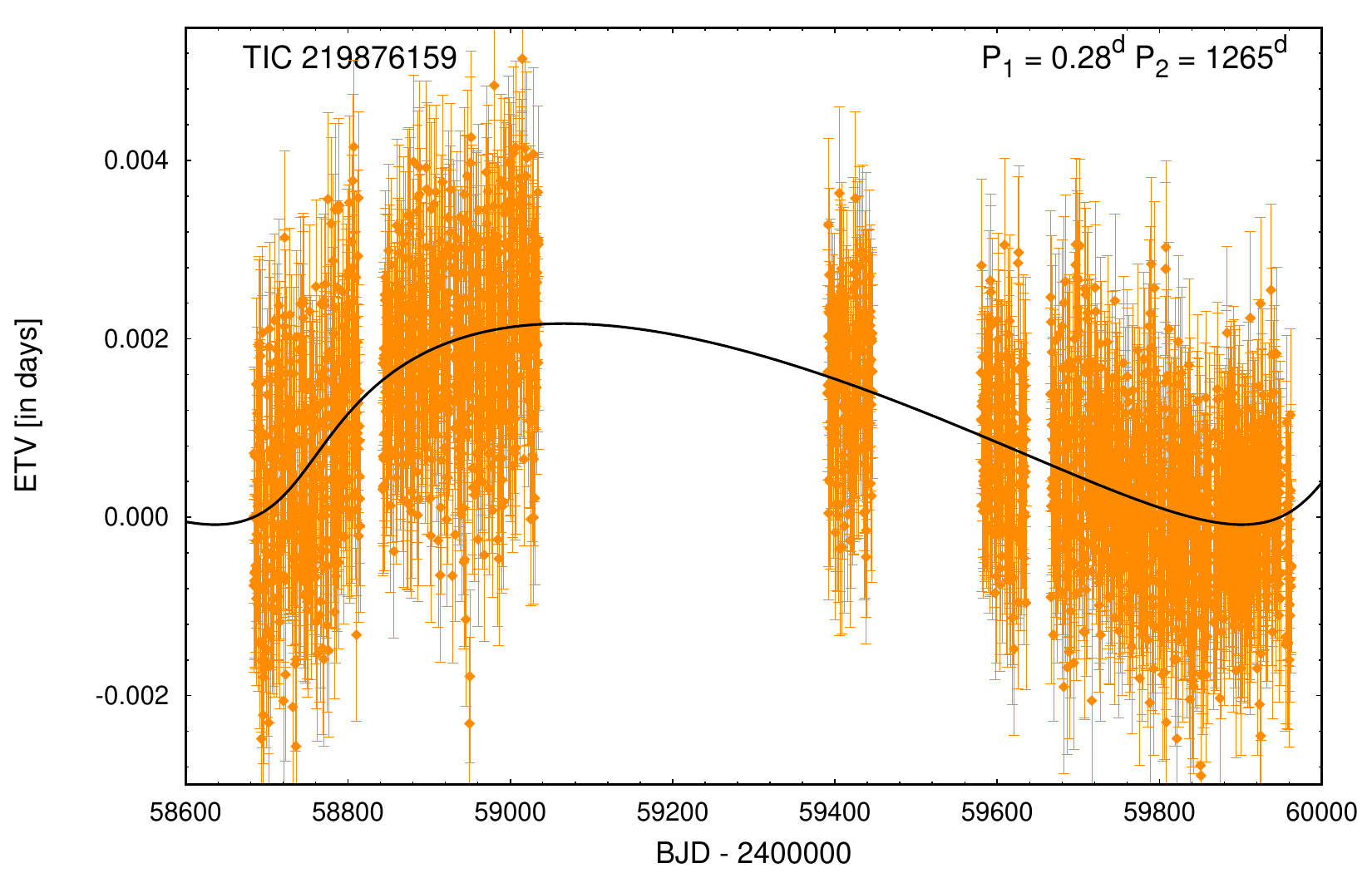}
\includegraphics[width=60mm]{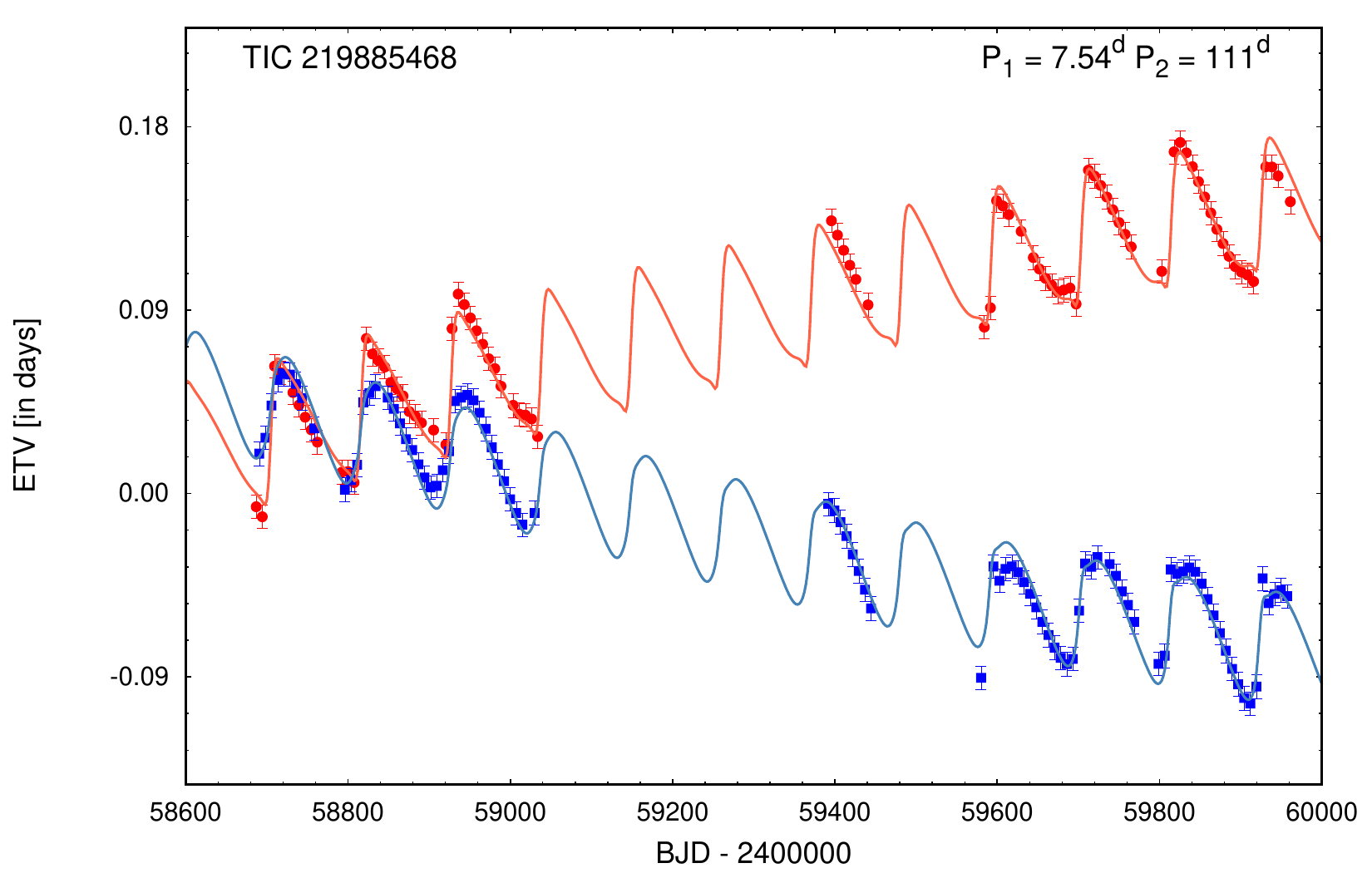}\includegraphics[width=60mm]{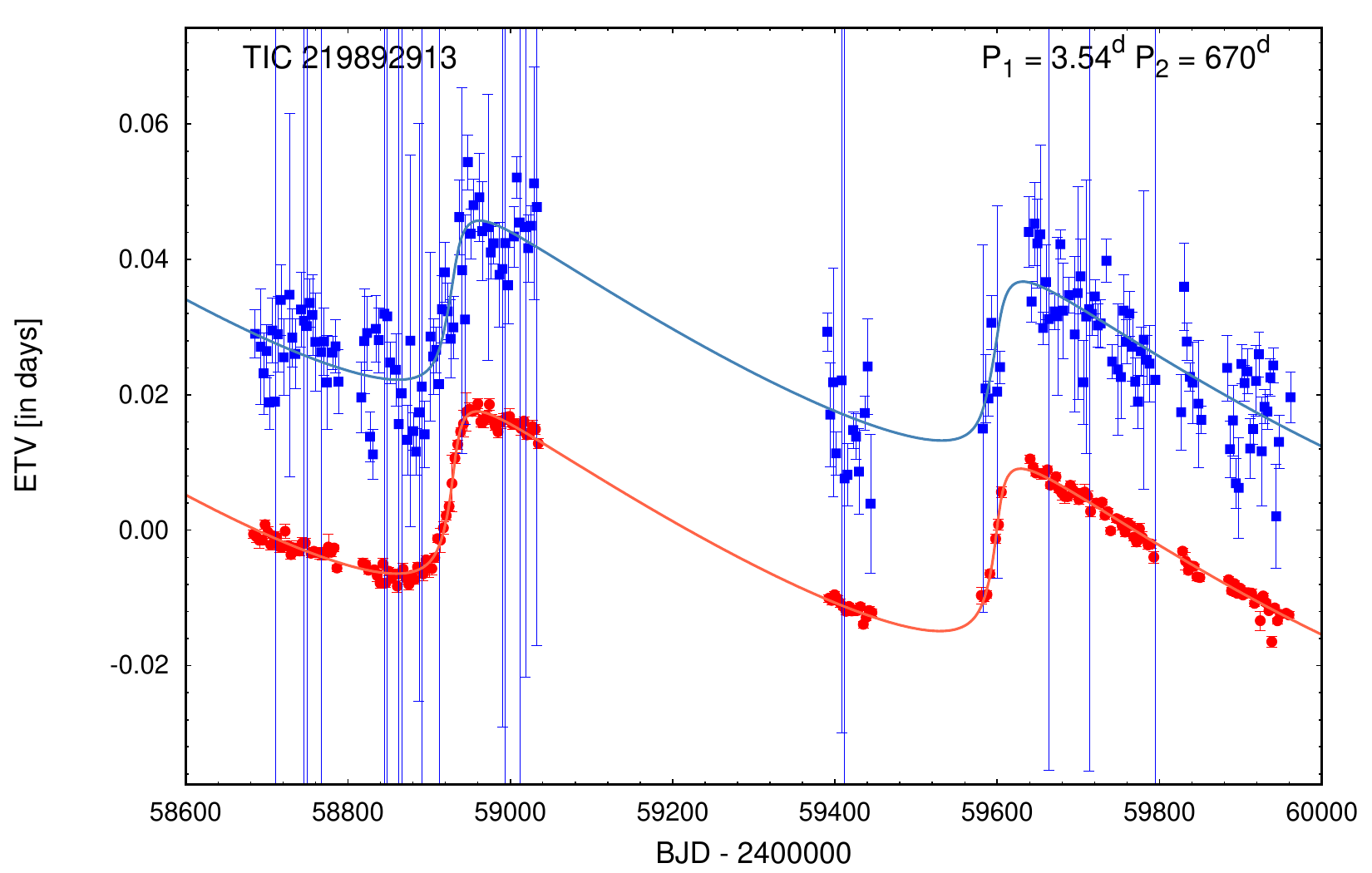}\includegraphics[width=60mm]{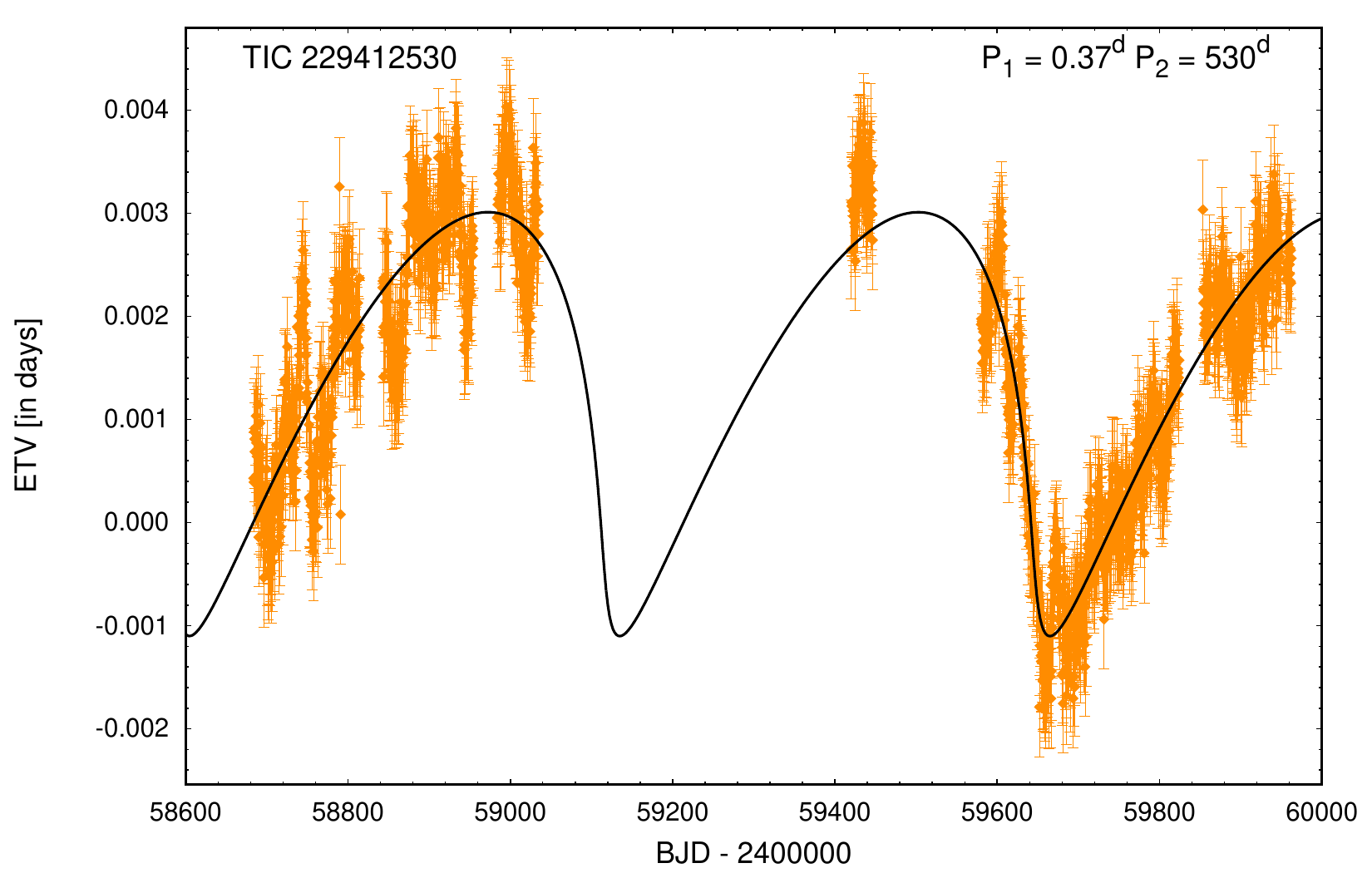}
\caption{(continued)}
\end{figure*}

\addtocounter{figure}{-1}

\begin{figure*}
\includegraphics[width=60mm]{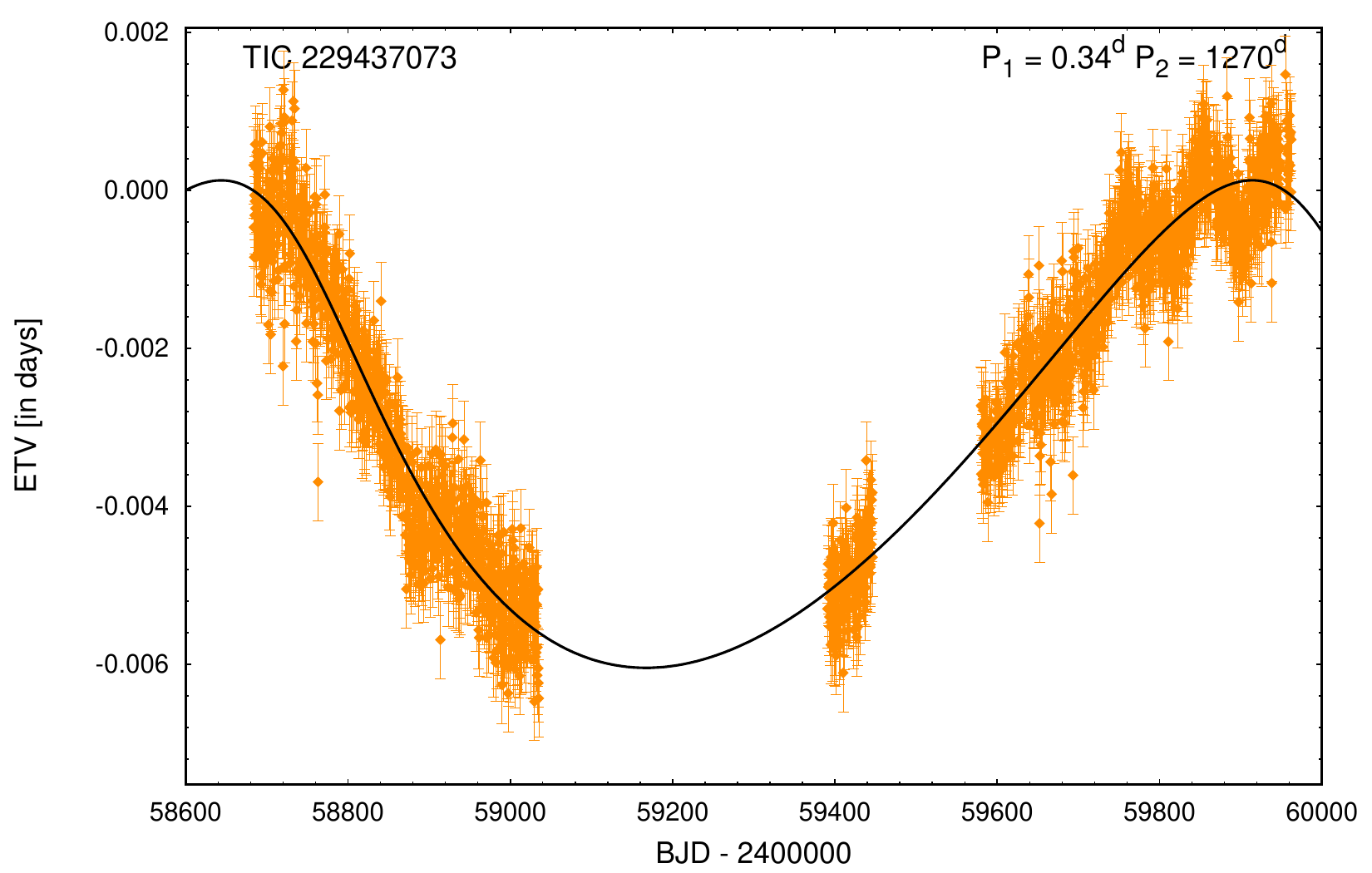}\includegraphics[width=60mm]{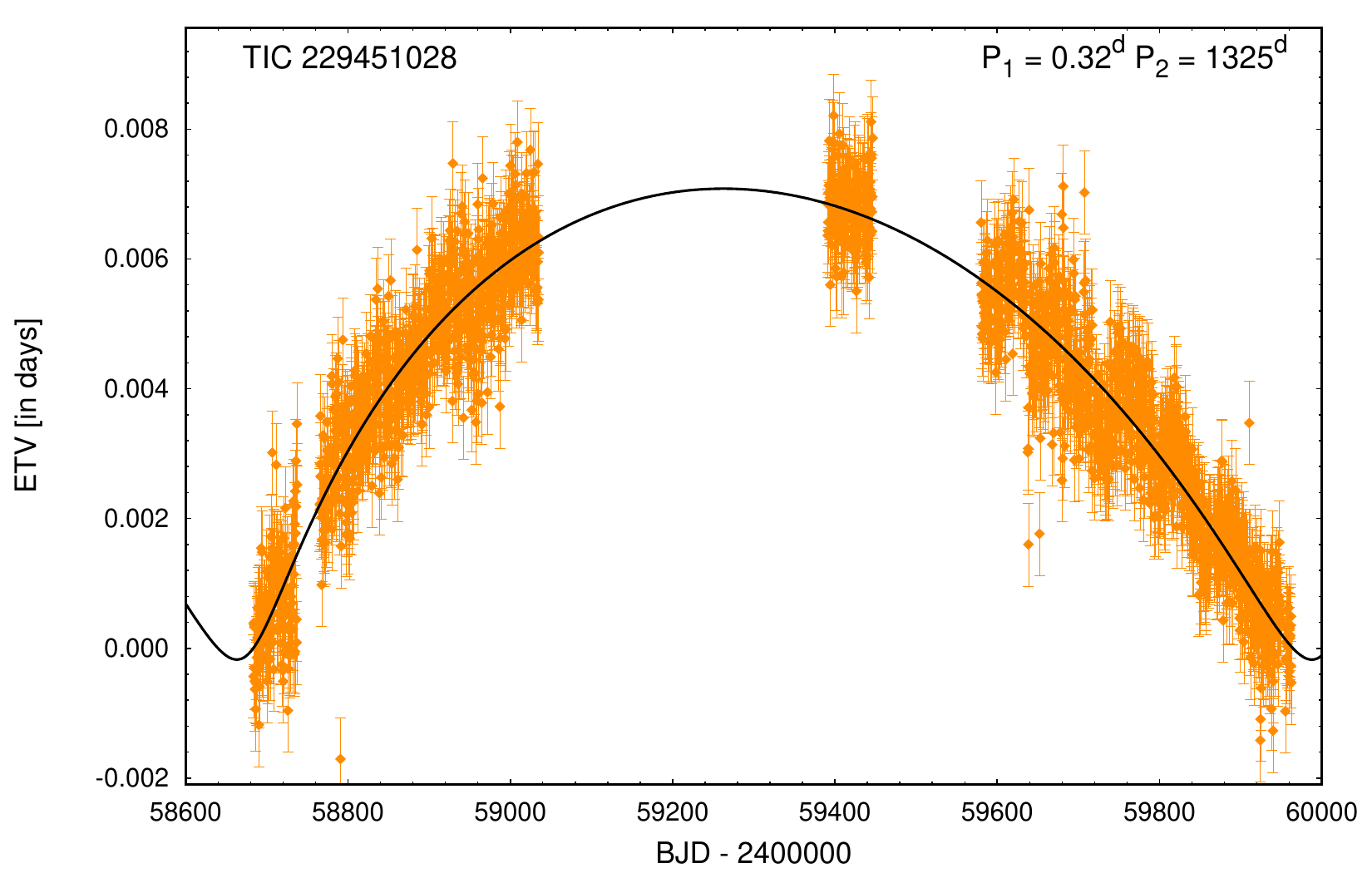}\includegraphics[width=60mm]{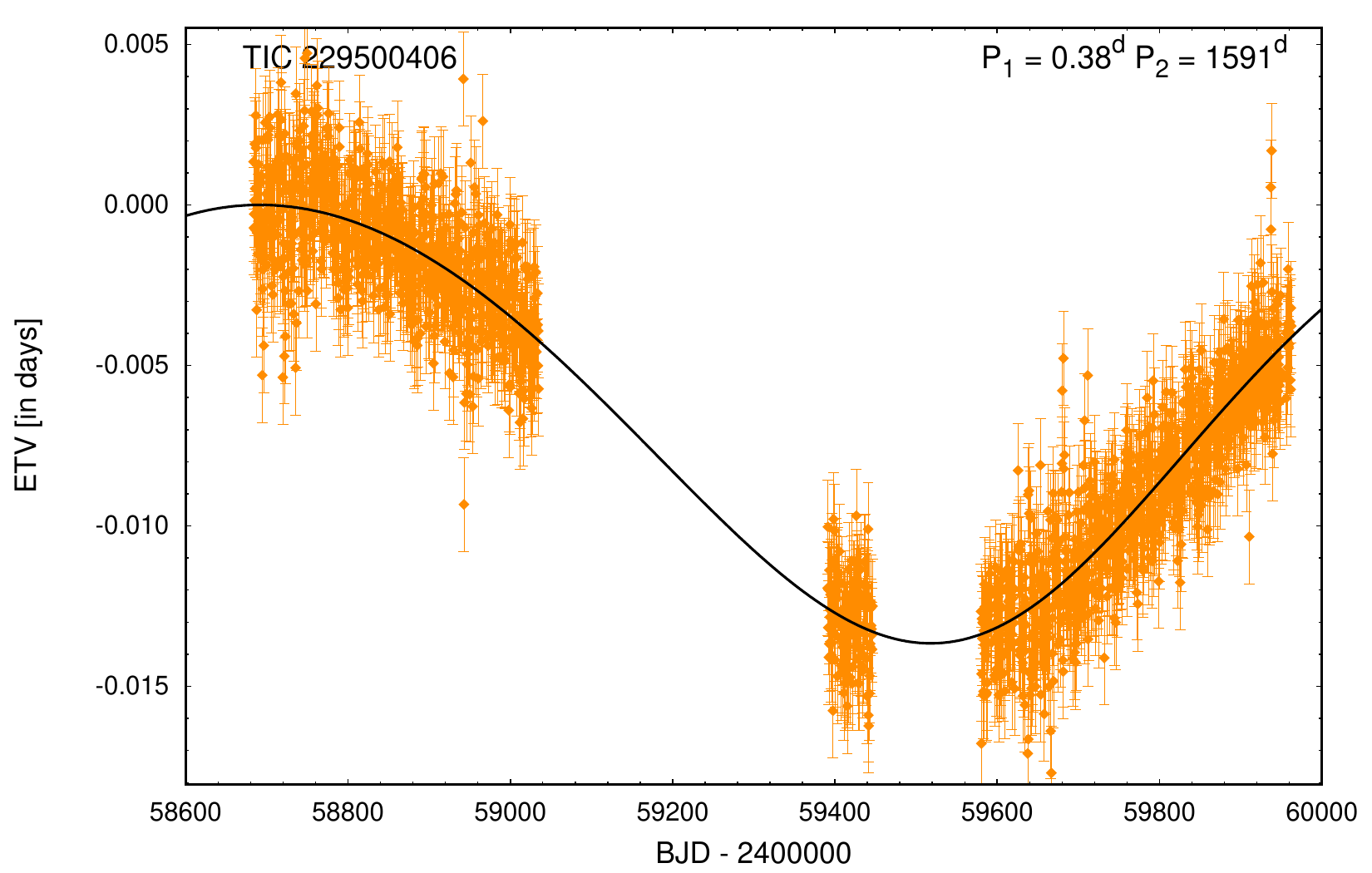}
\includegraphics[width=60mm]{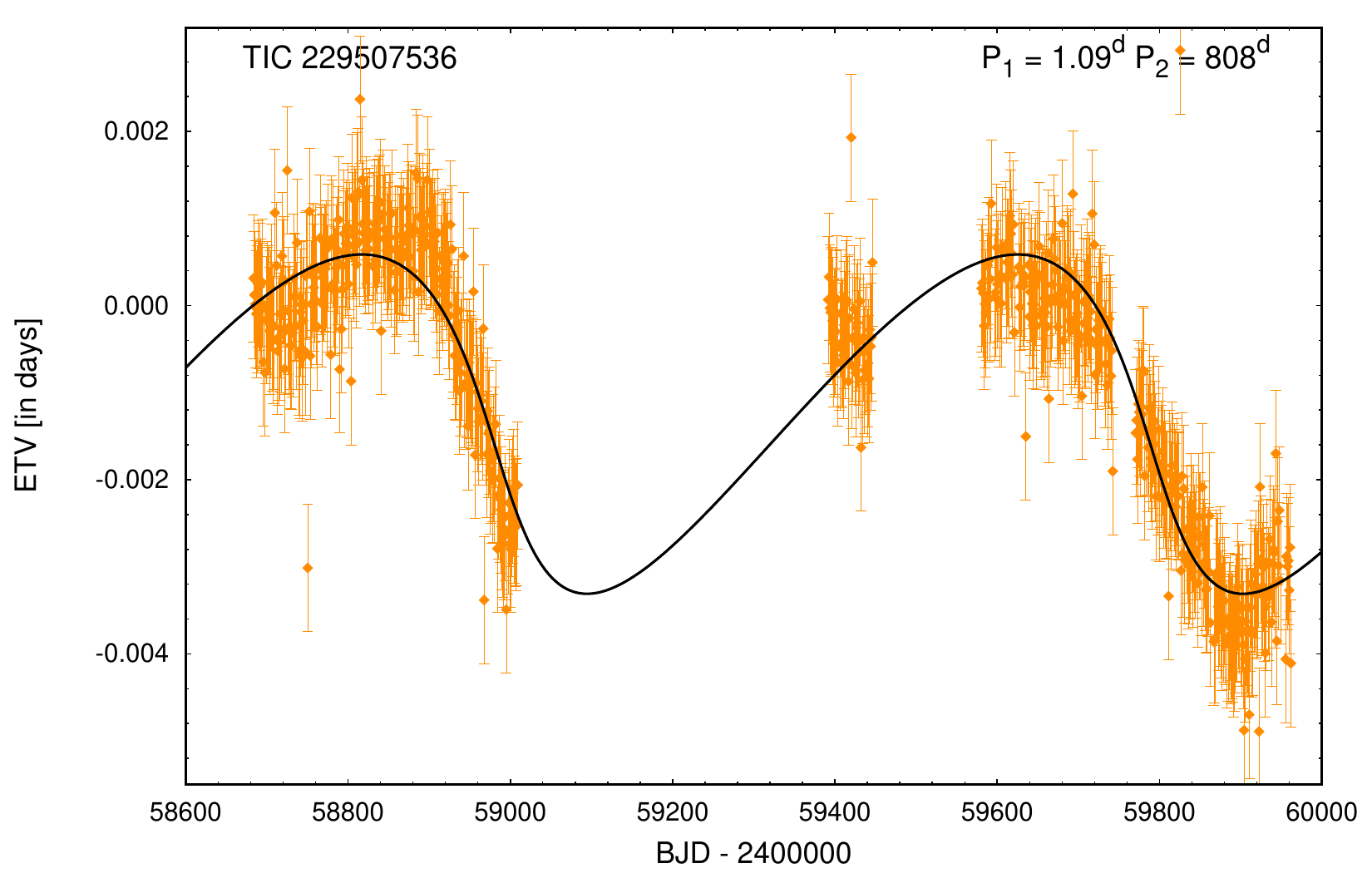}\includegraphics[width=60mm]{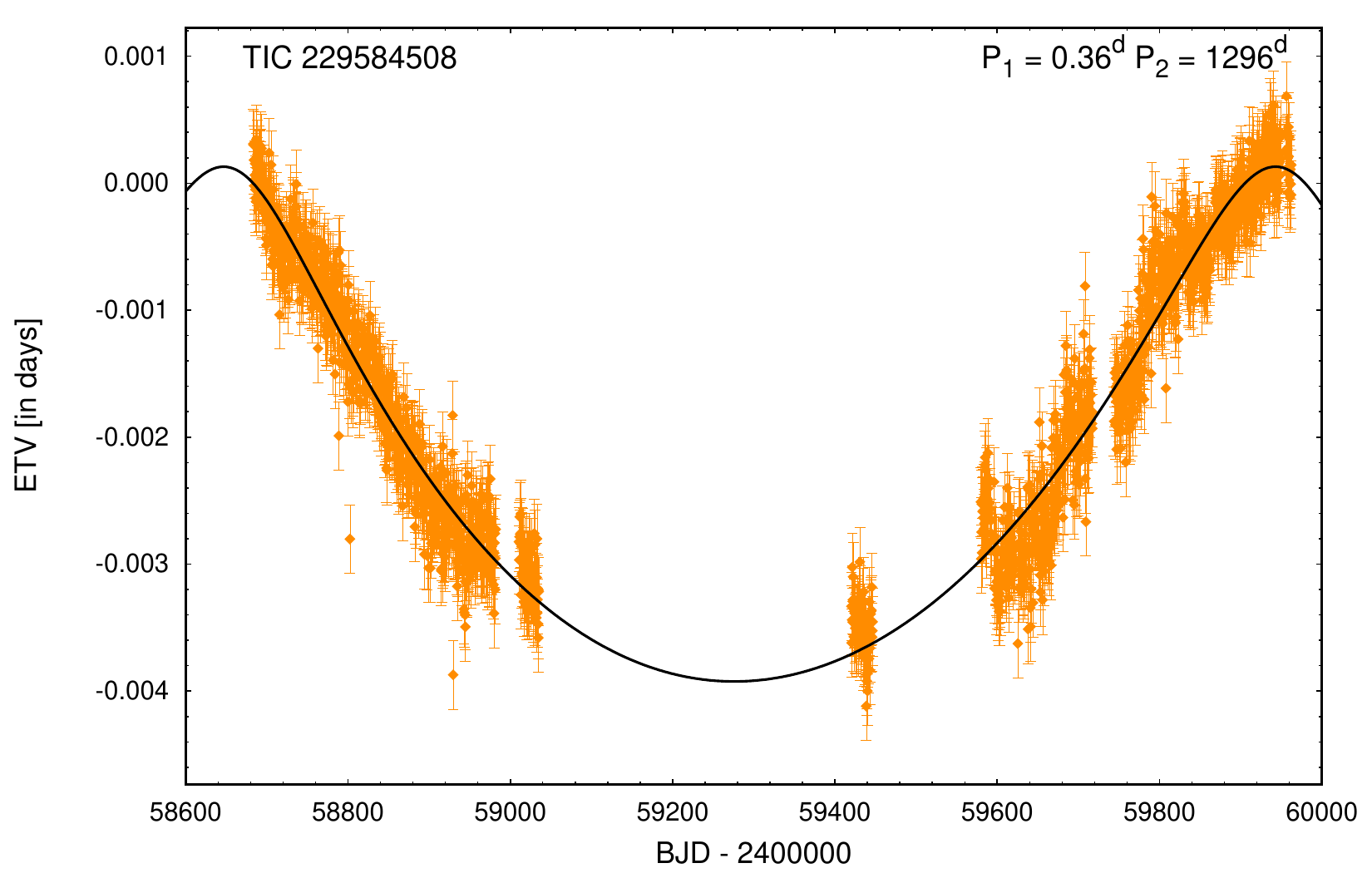}\includegraphics[width=60mm]{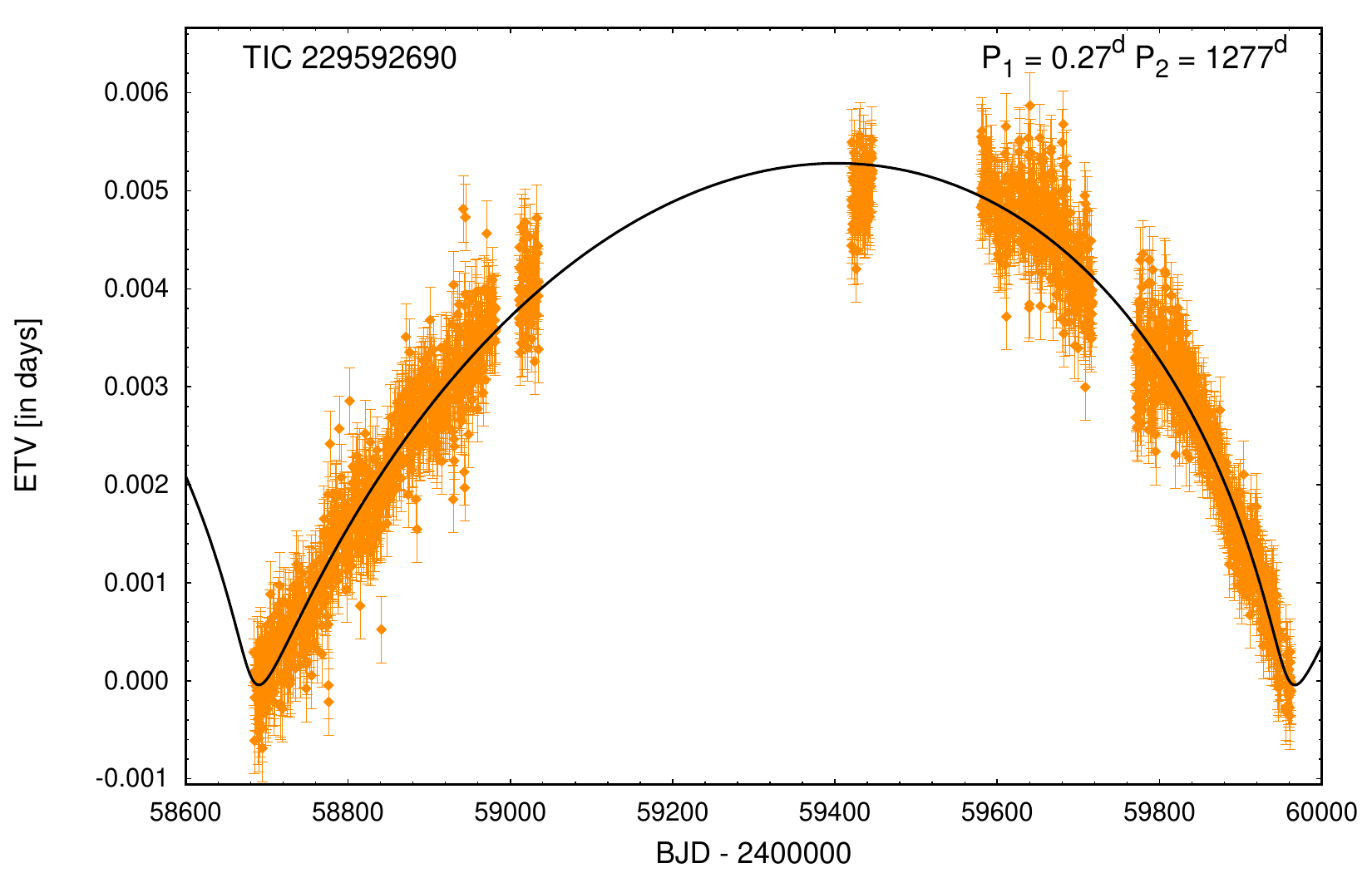}
\includegraphics[width=60mm]{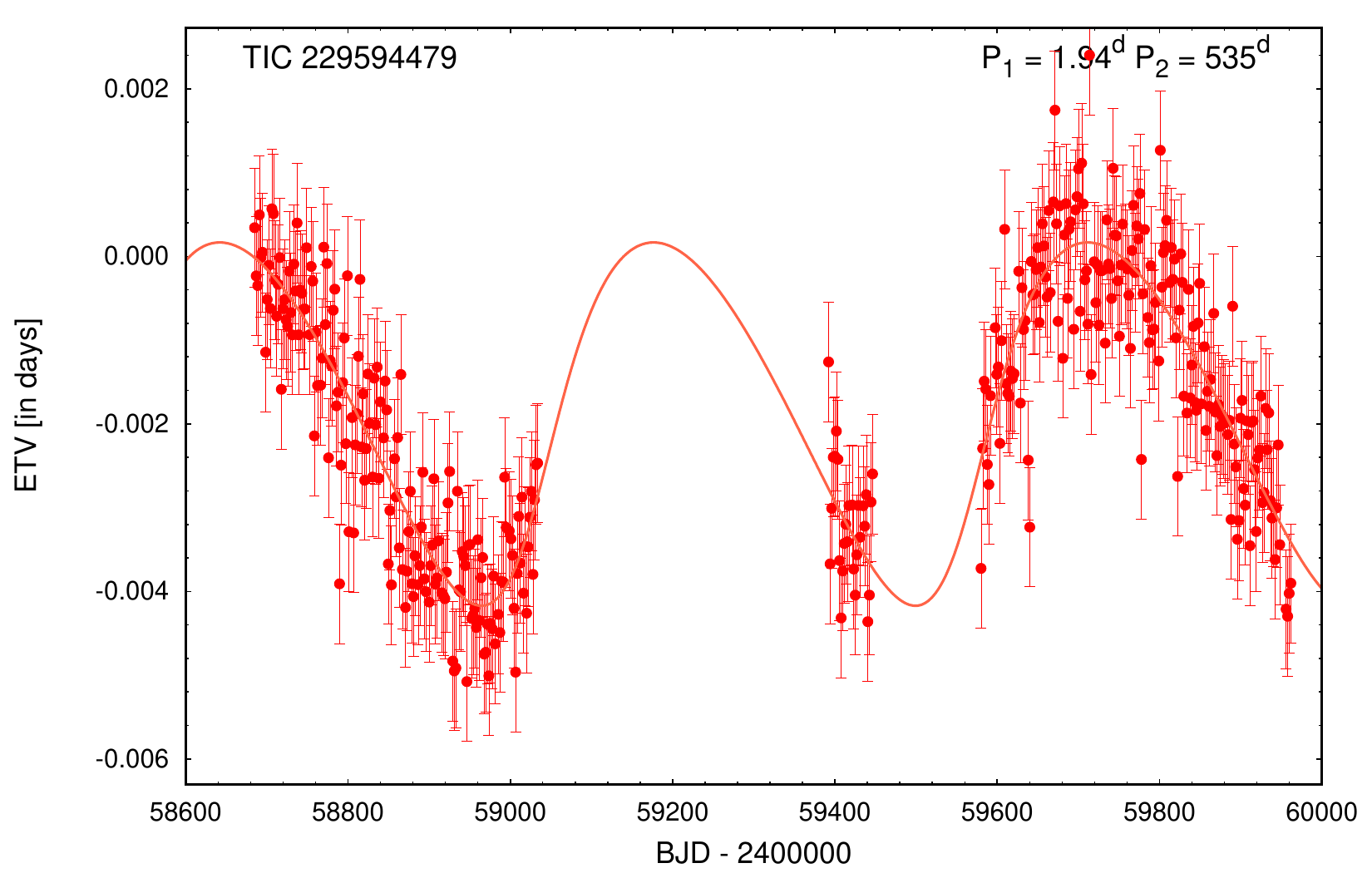}\includegraphics[width=60mm]{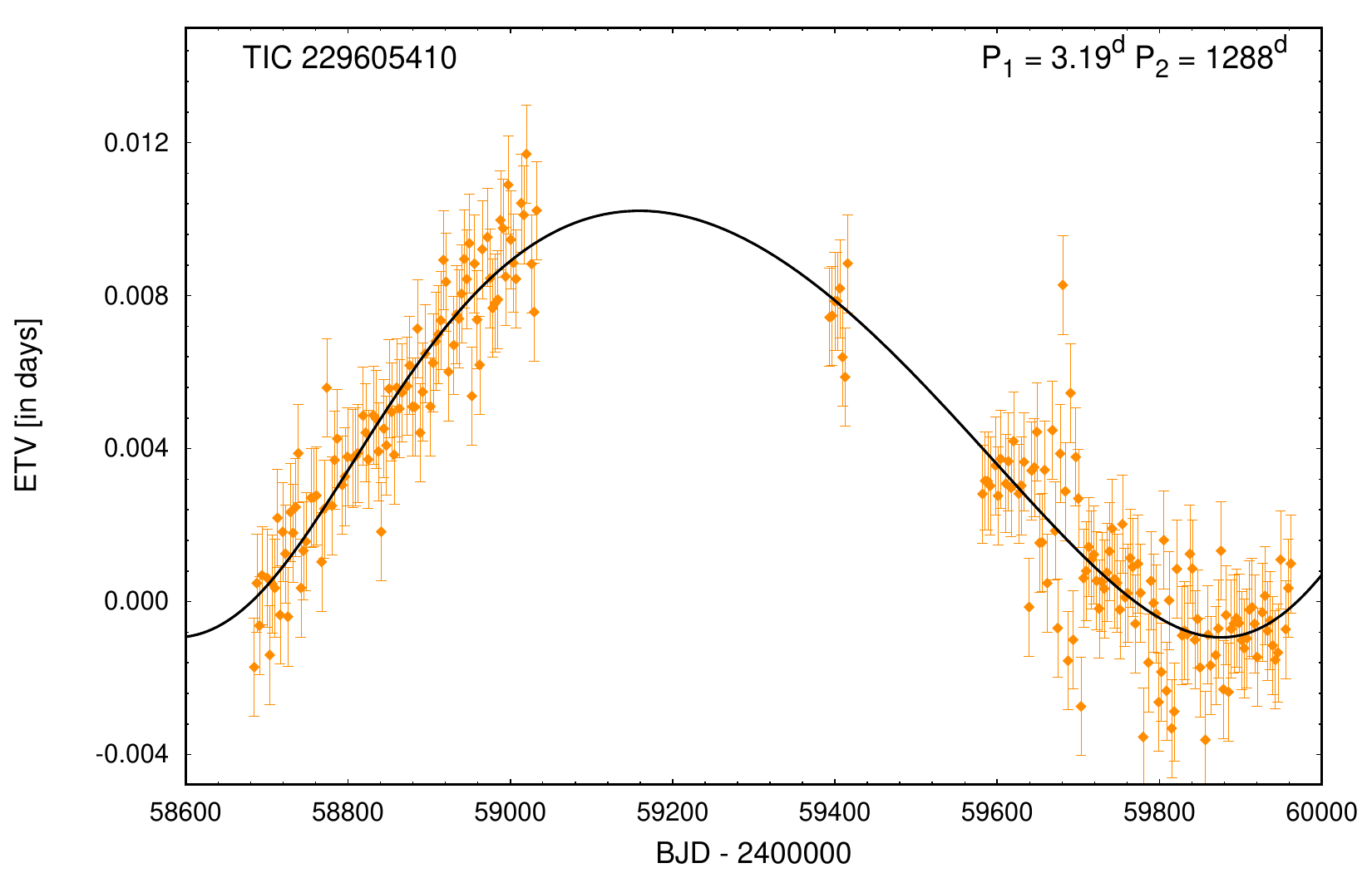}\includegraphics[width=60mm]{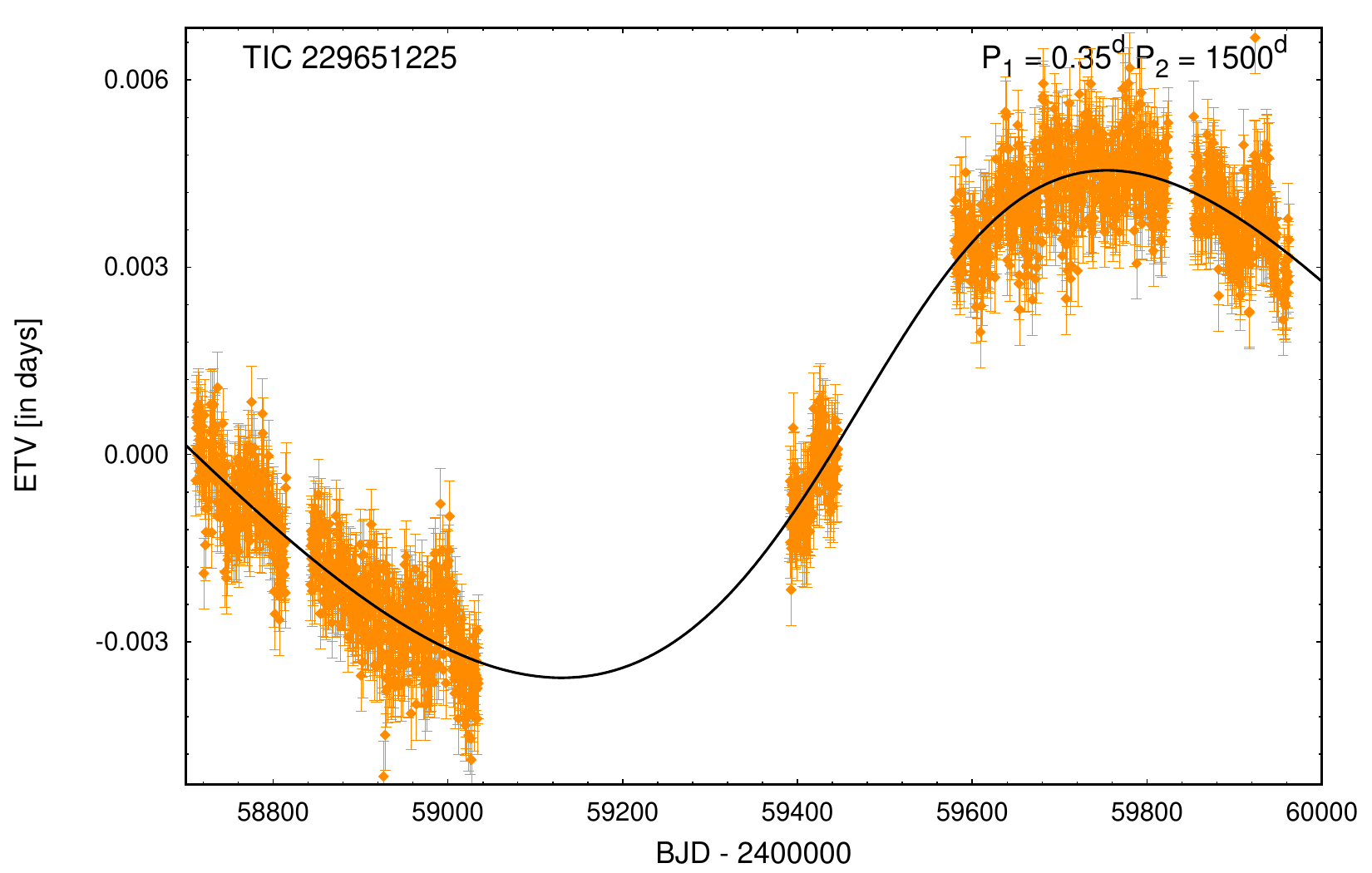}
\includegraphics[width=60mm]{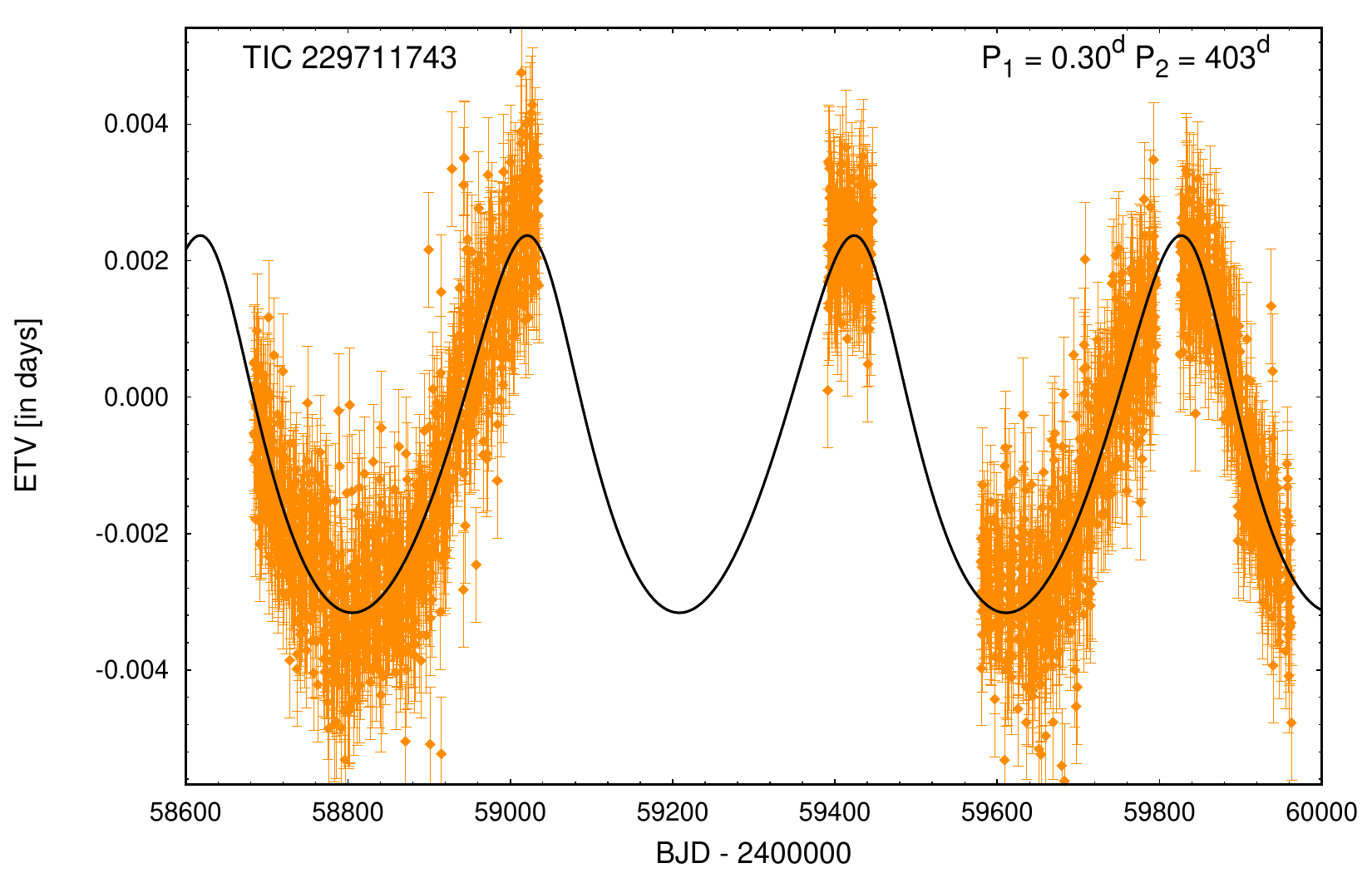}\includegraphics[width=60mm]{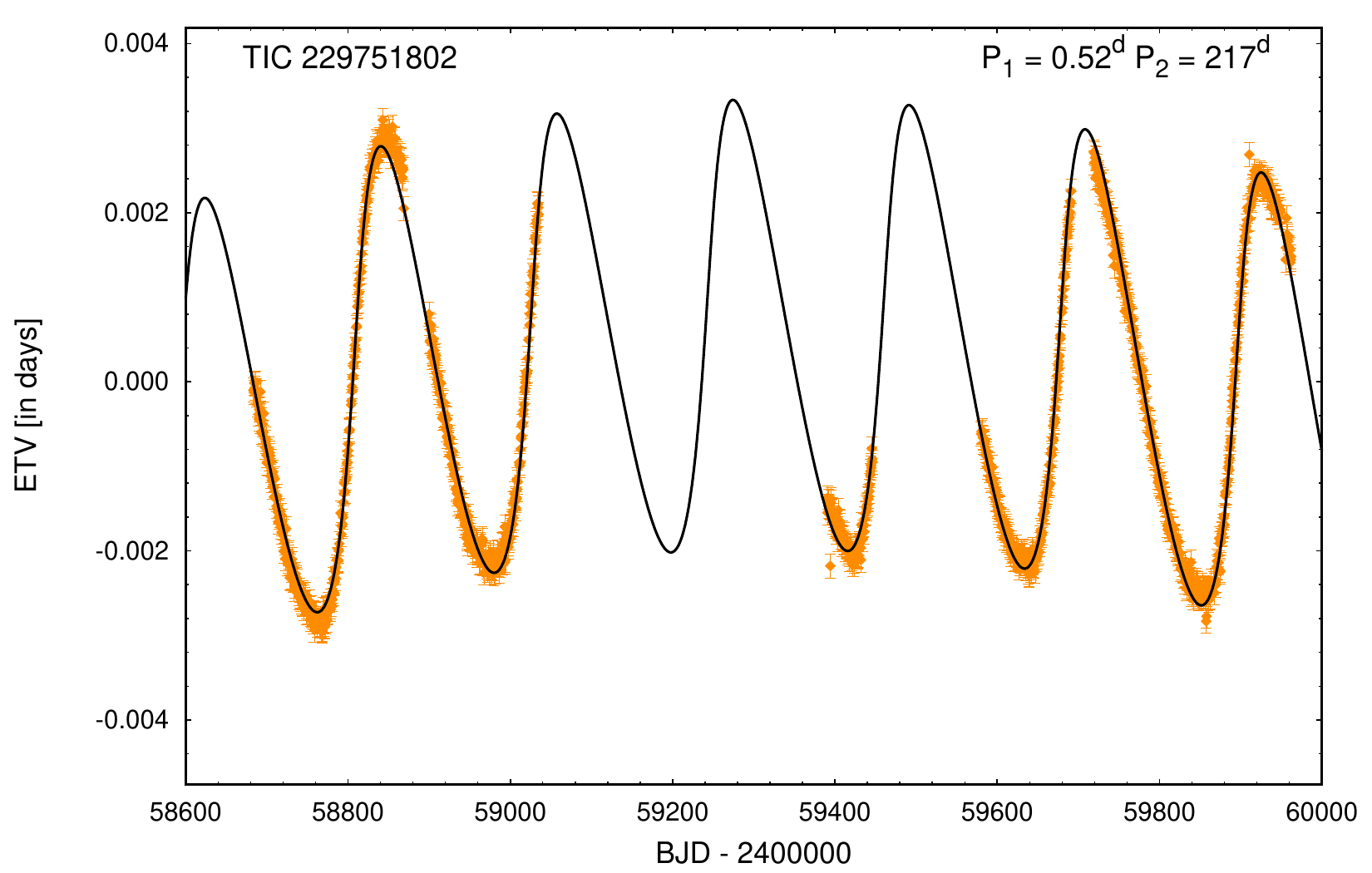}\includegraphics[width=60mm]{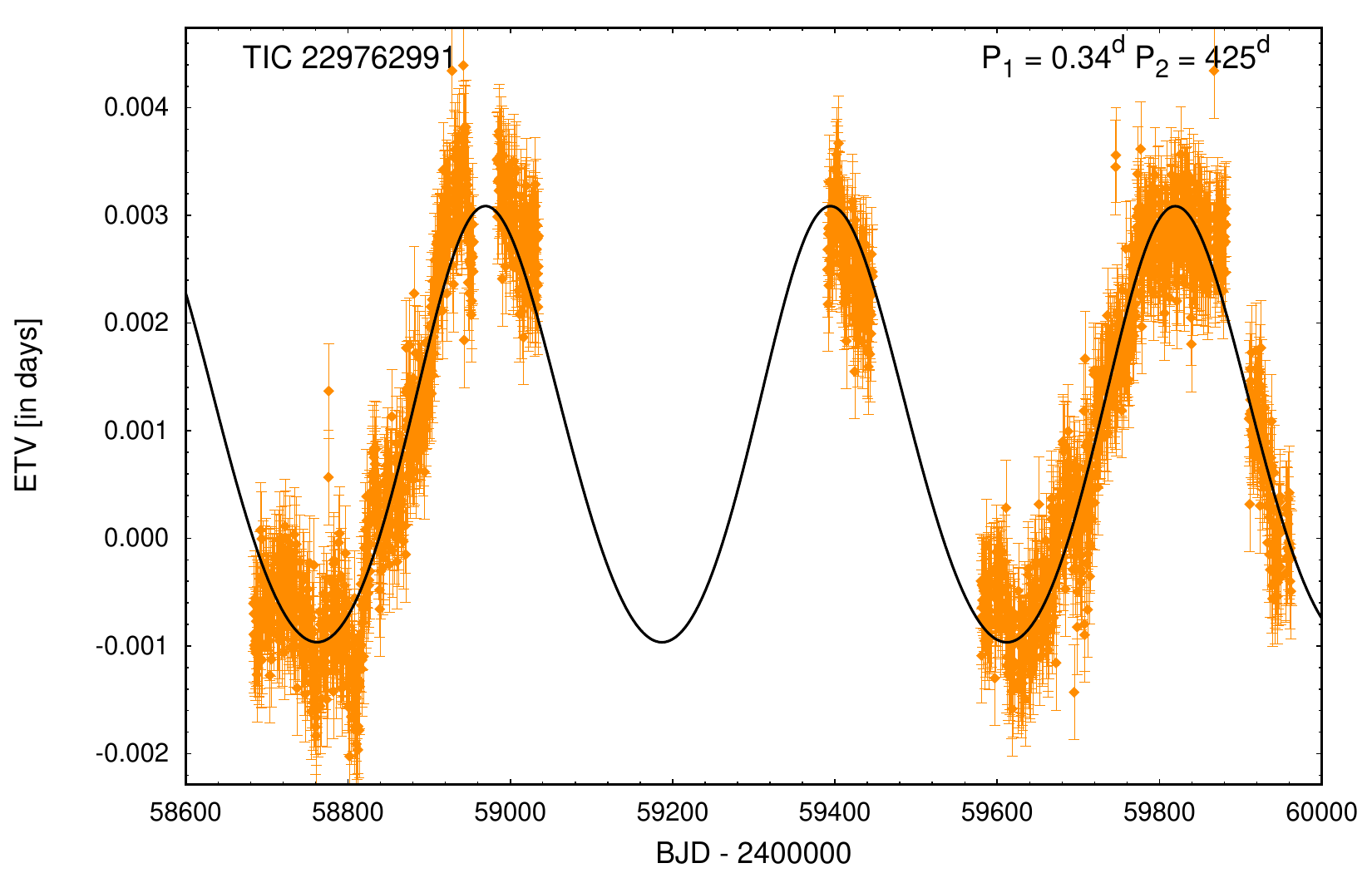}
\includegraphics[width=60mm]{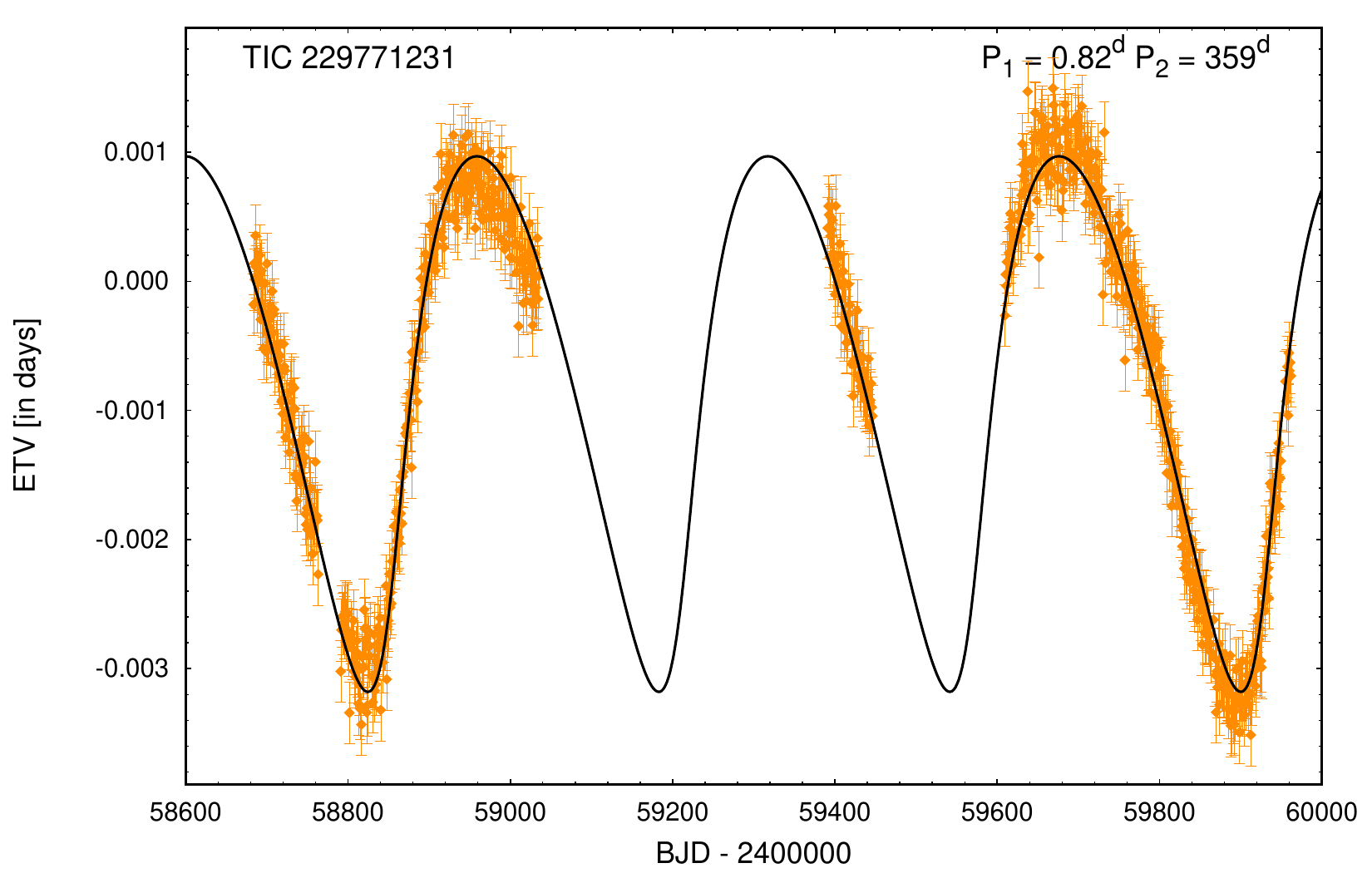}\includegraphics[width=60mm]{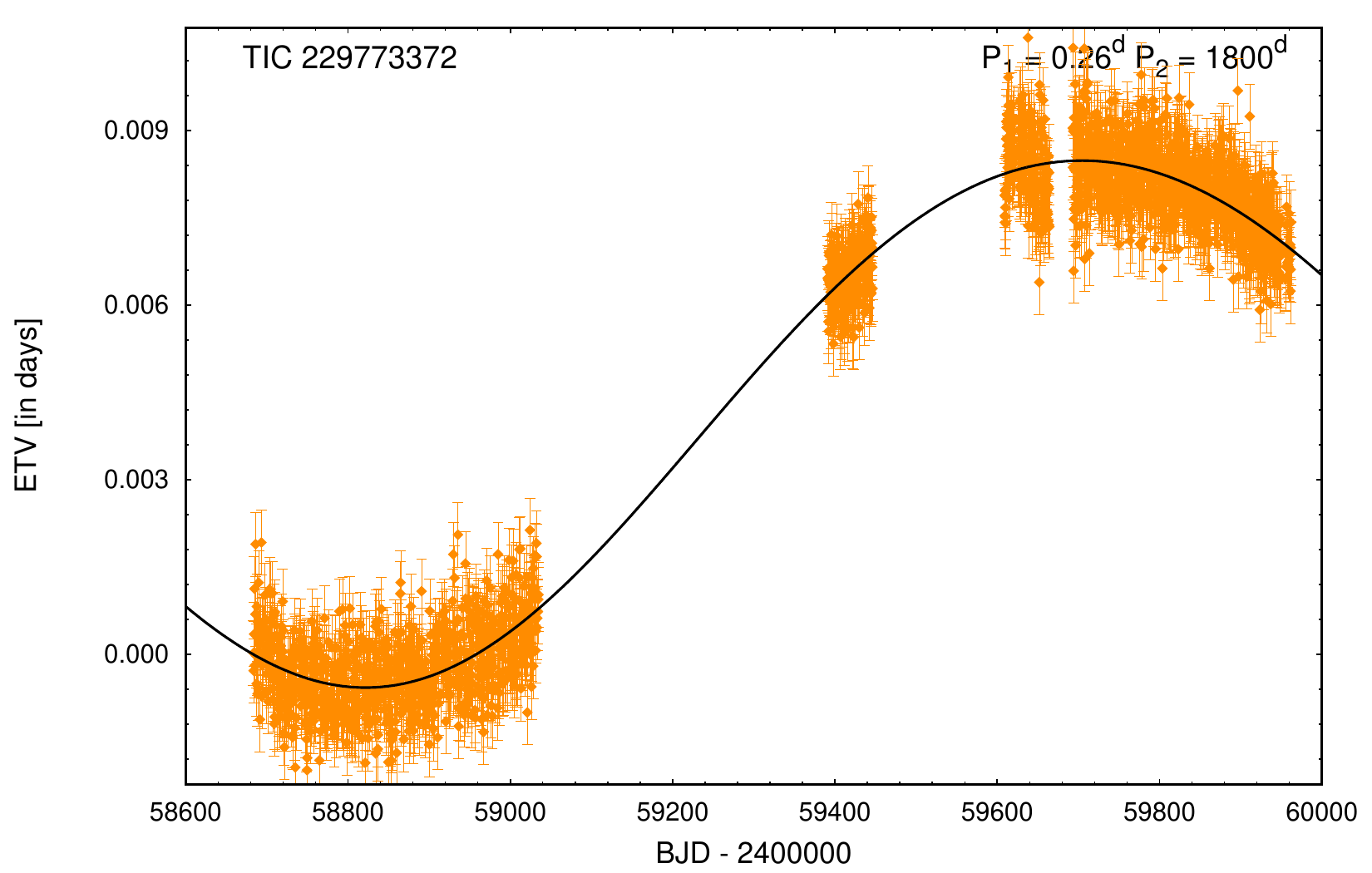}\includegraphics[width=60mm]{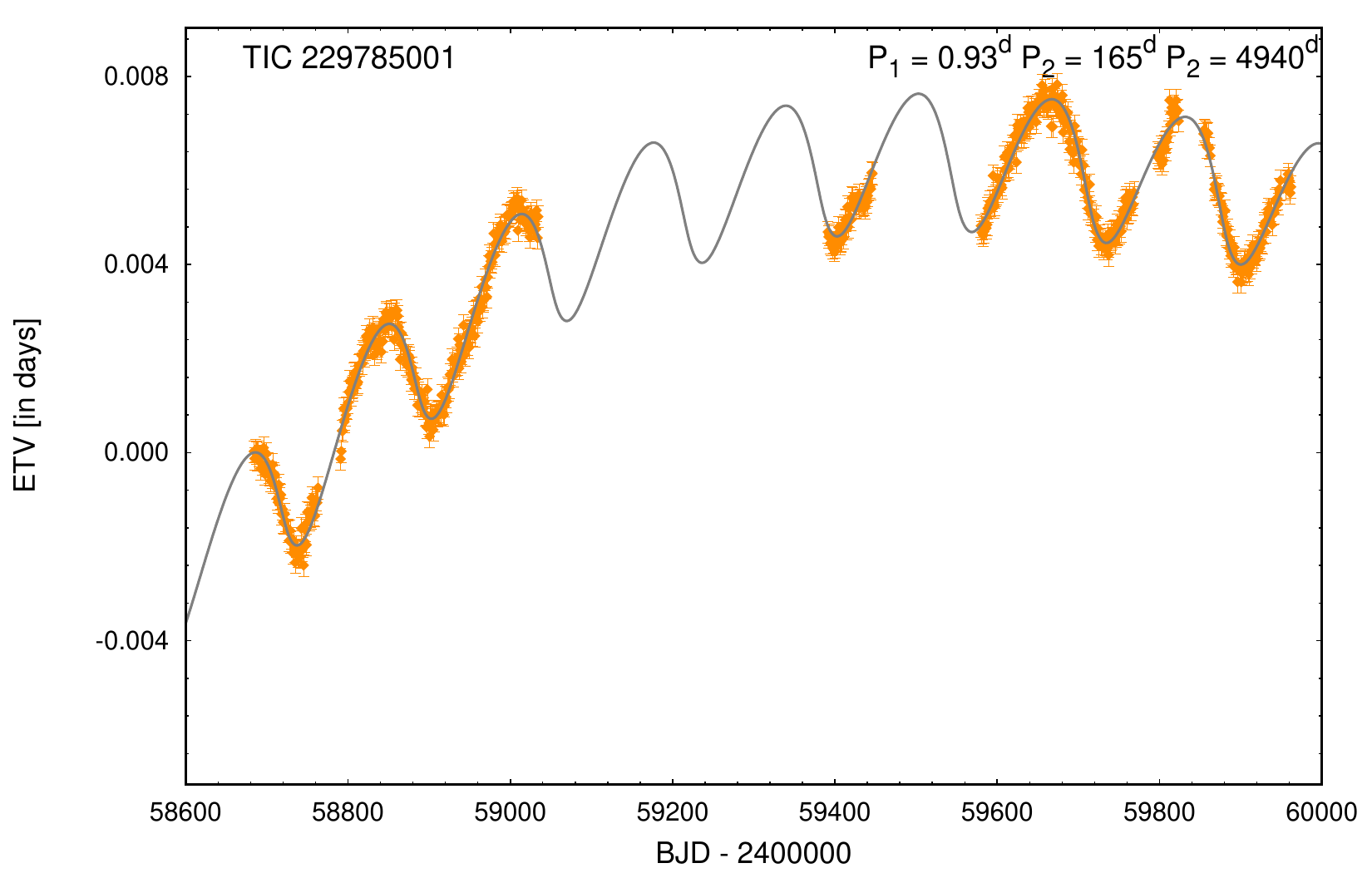}
\includegraphics[width=60mm]{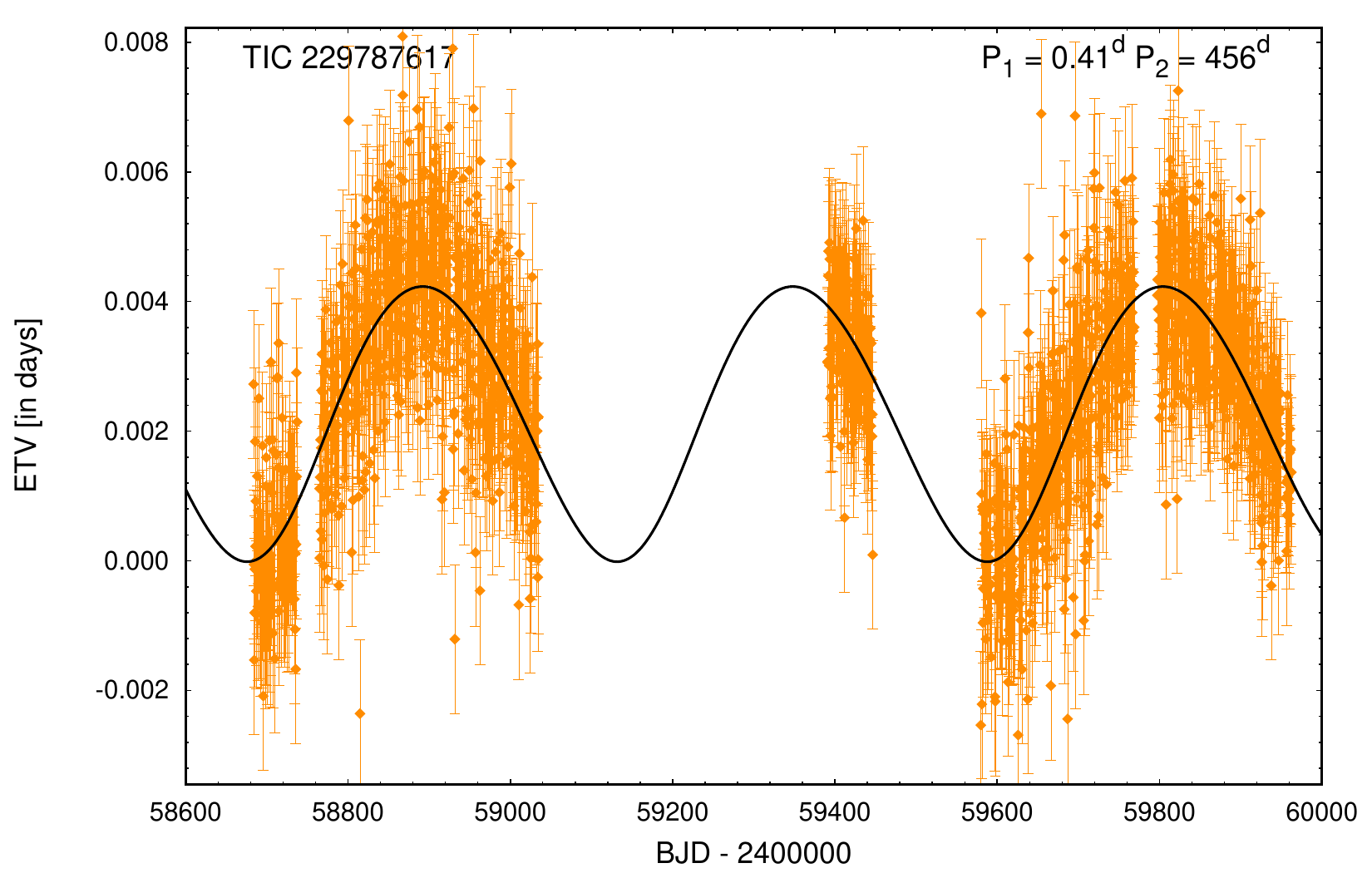}\includegraphics[width=60mm]{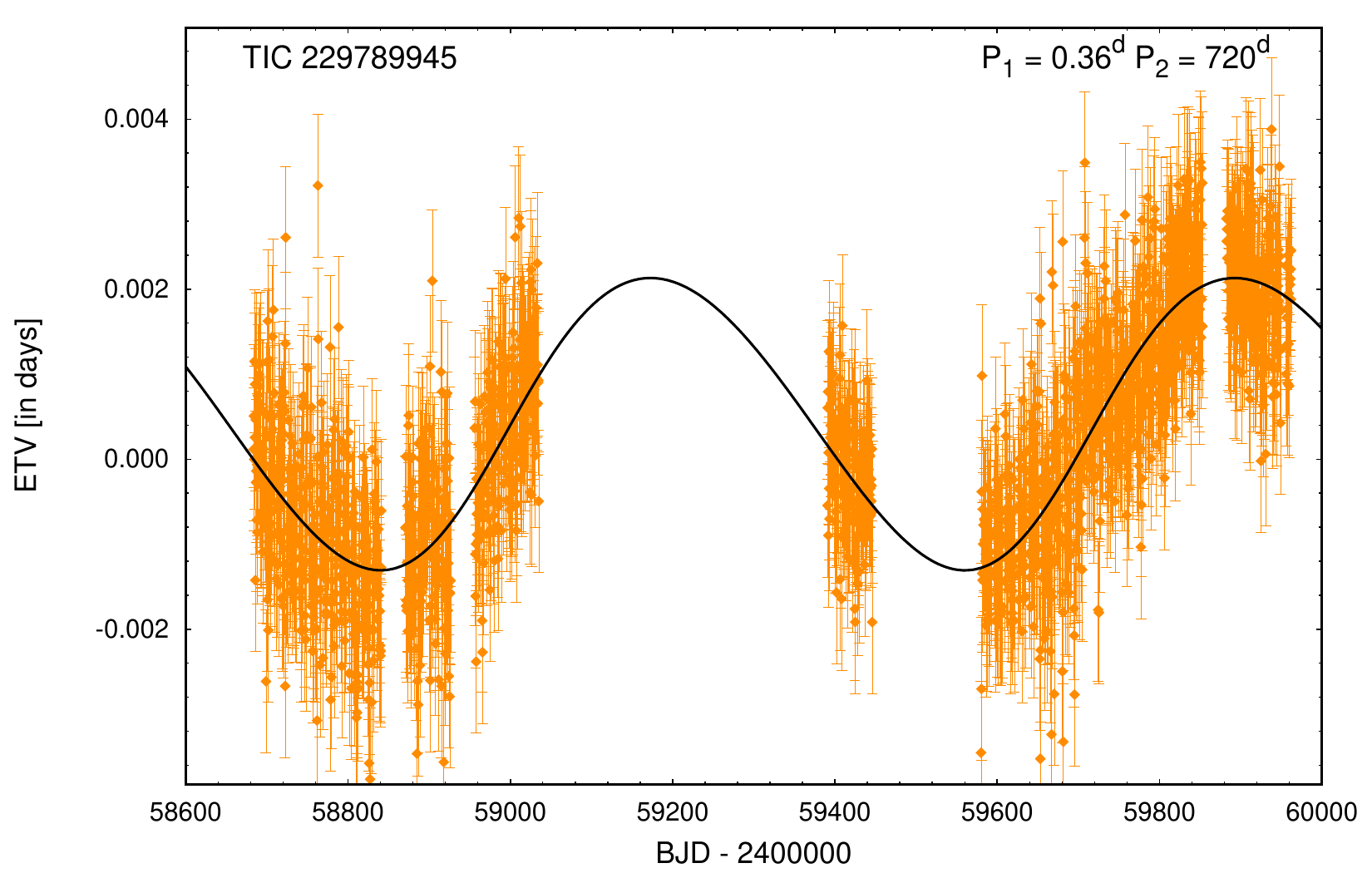}\includegraphics[width=60mm]{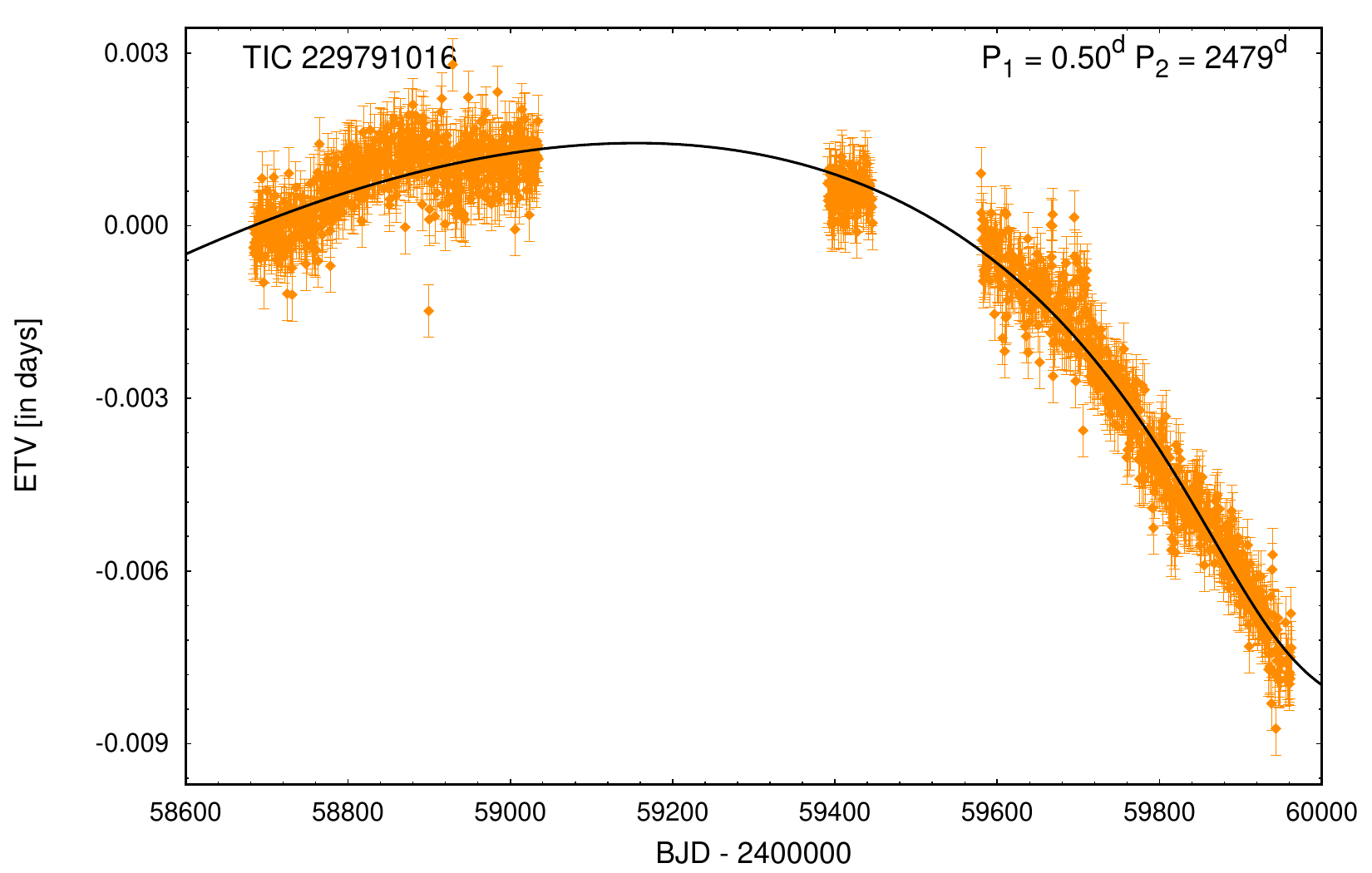}
\caption{(continued)}
\end{figure*}

\addtocounter{figure}{-1}

\begin{figure*}
\includegraphics[width=60mm]{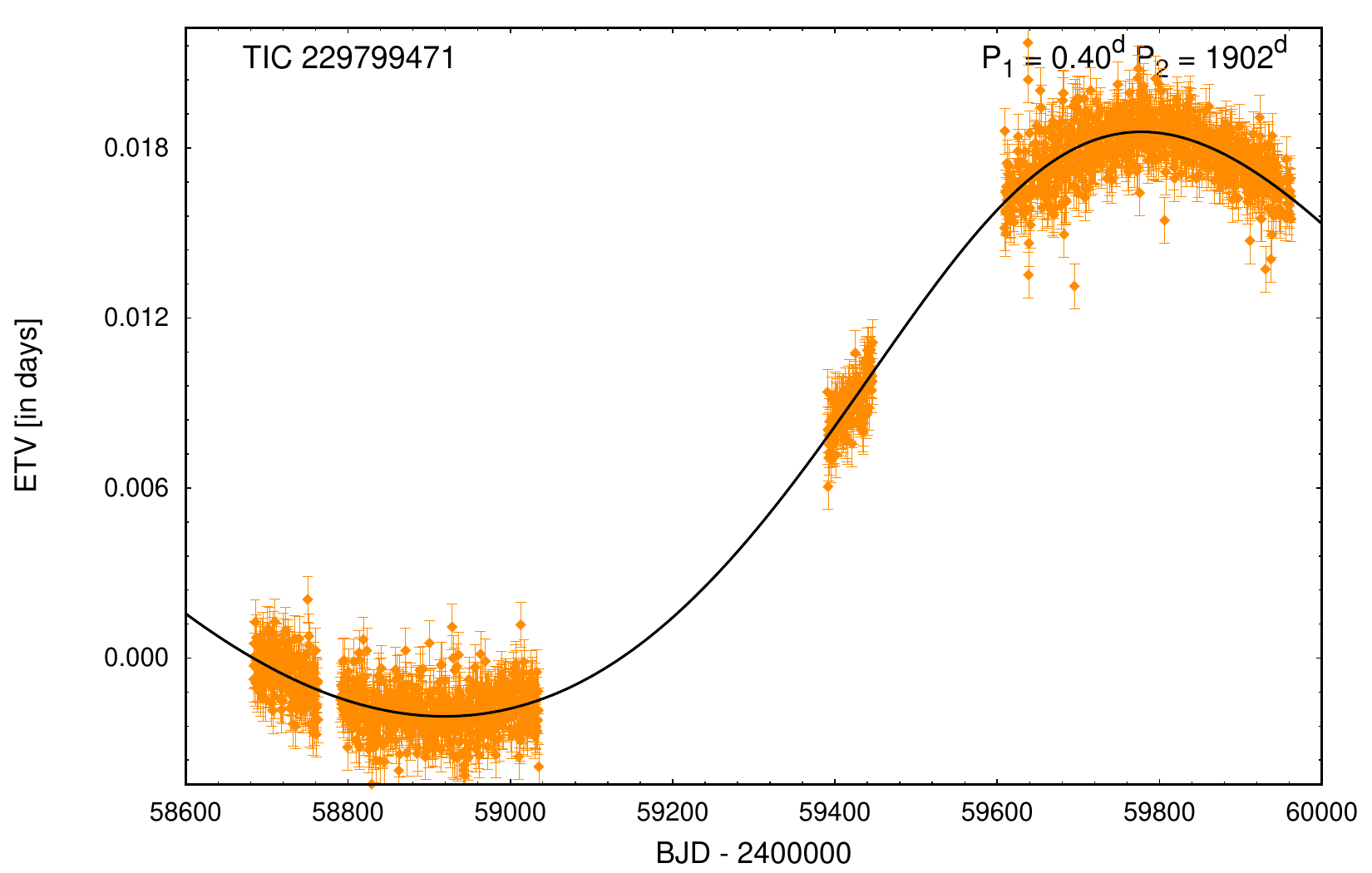}\includegraphics[width=60mm]{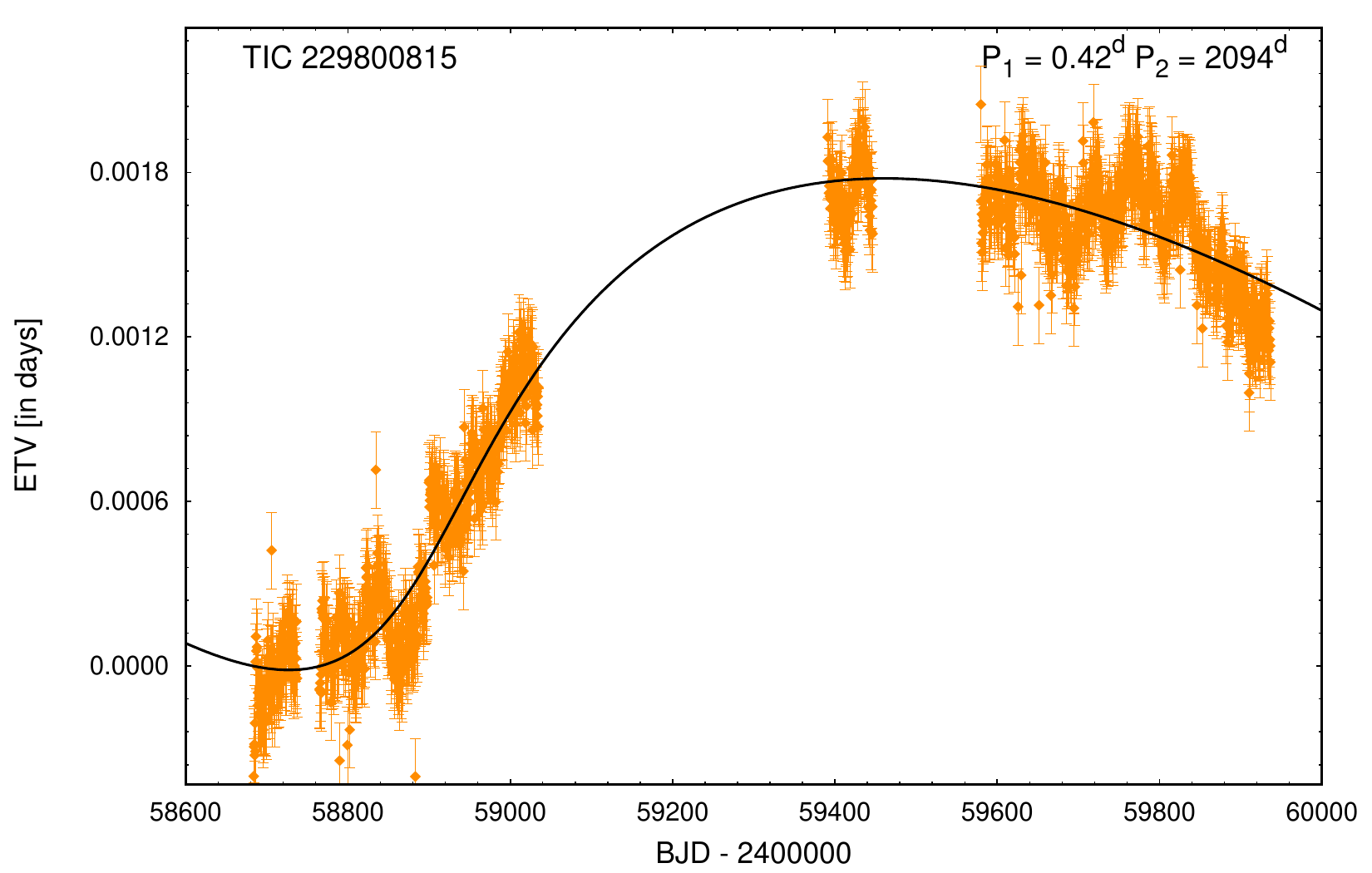}\includegraphics[width=60mm]{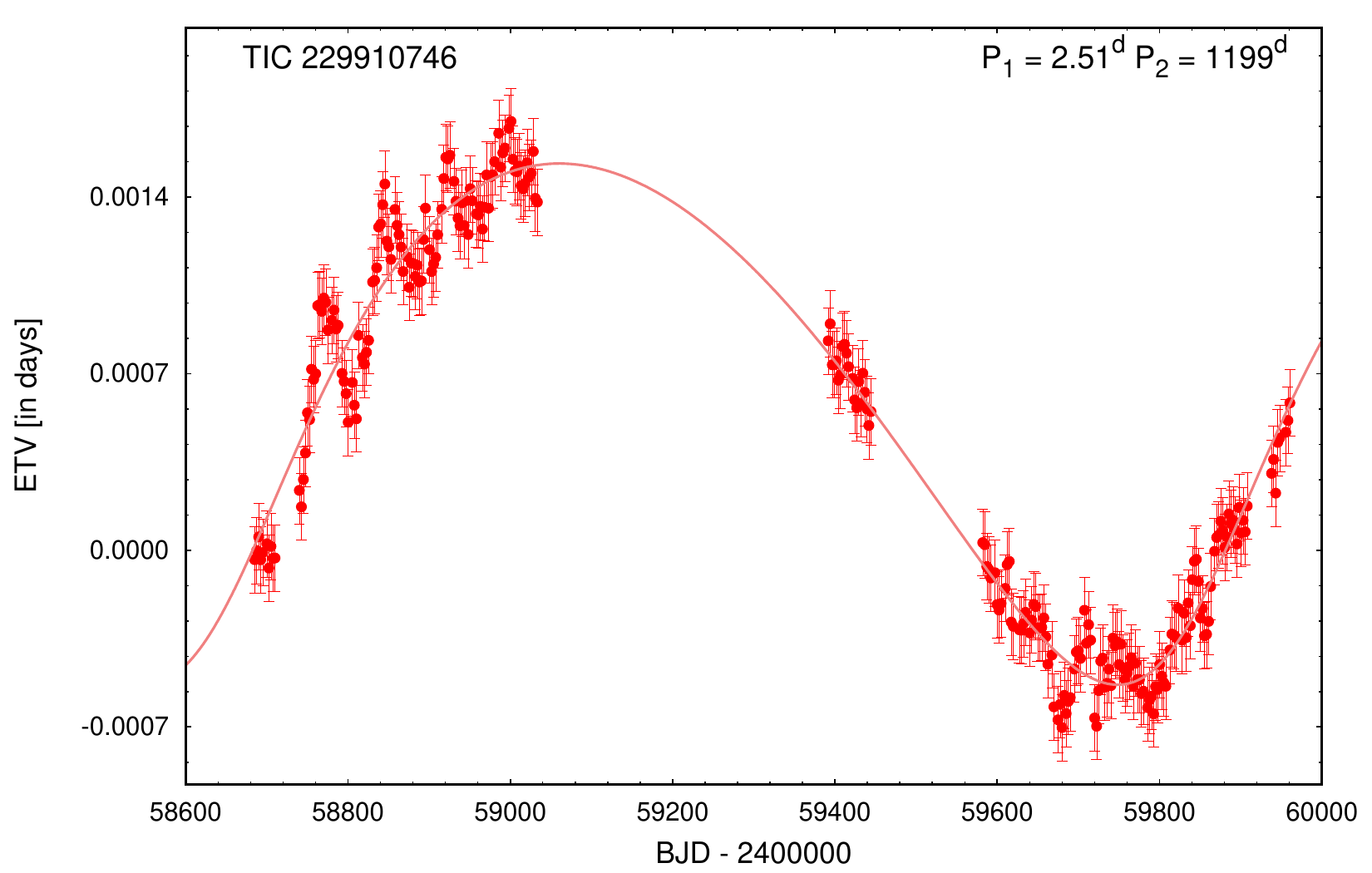}
\includegraphics[width=60mm]{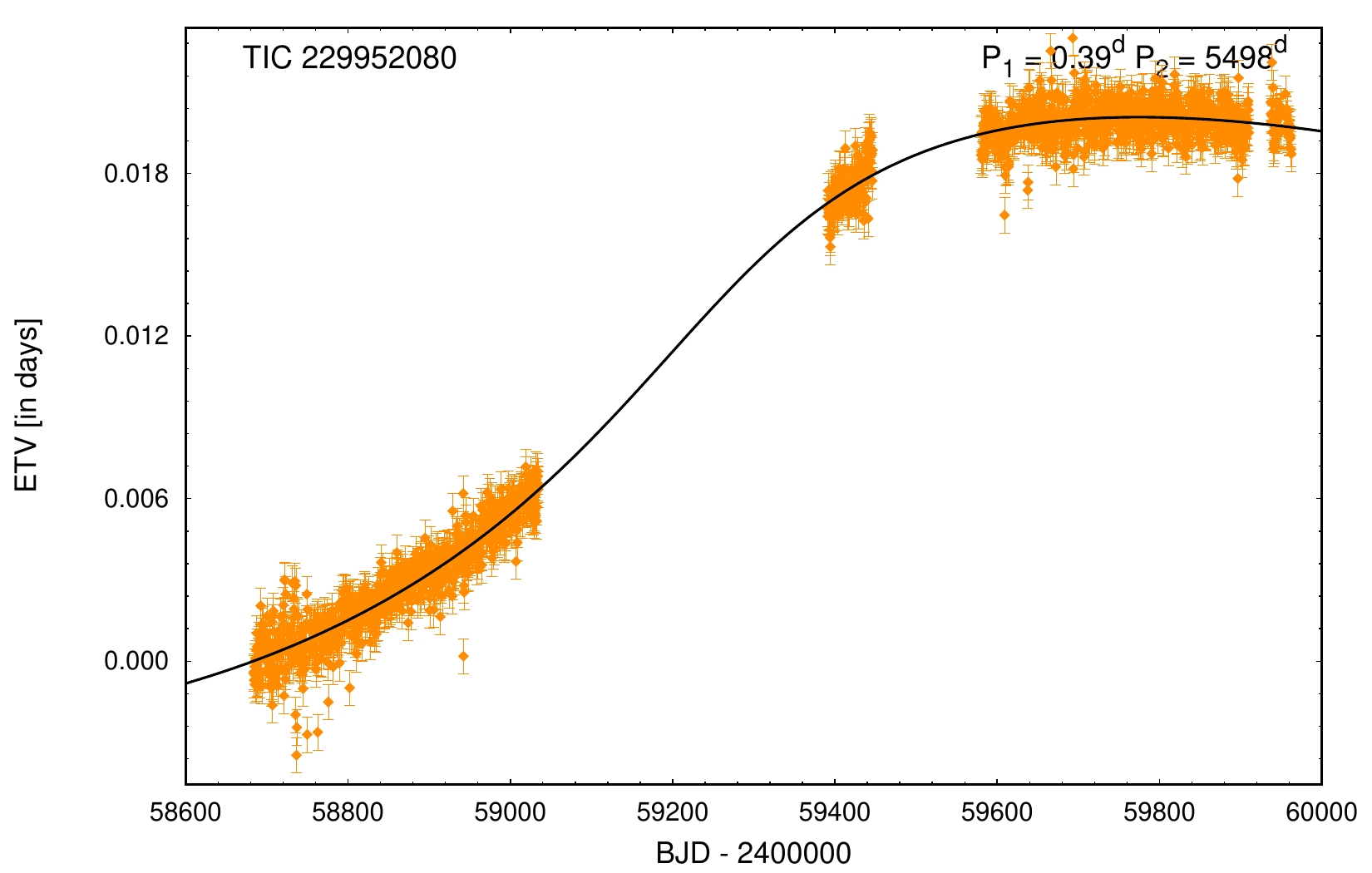}\includegraphics[width=60mm]{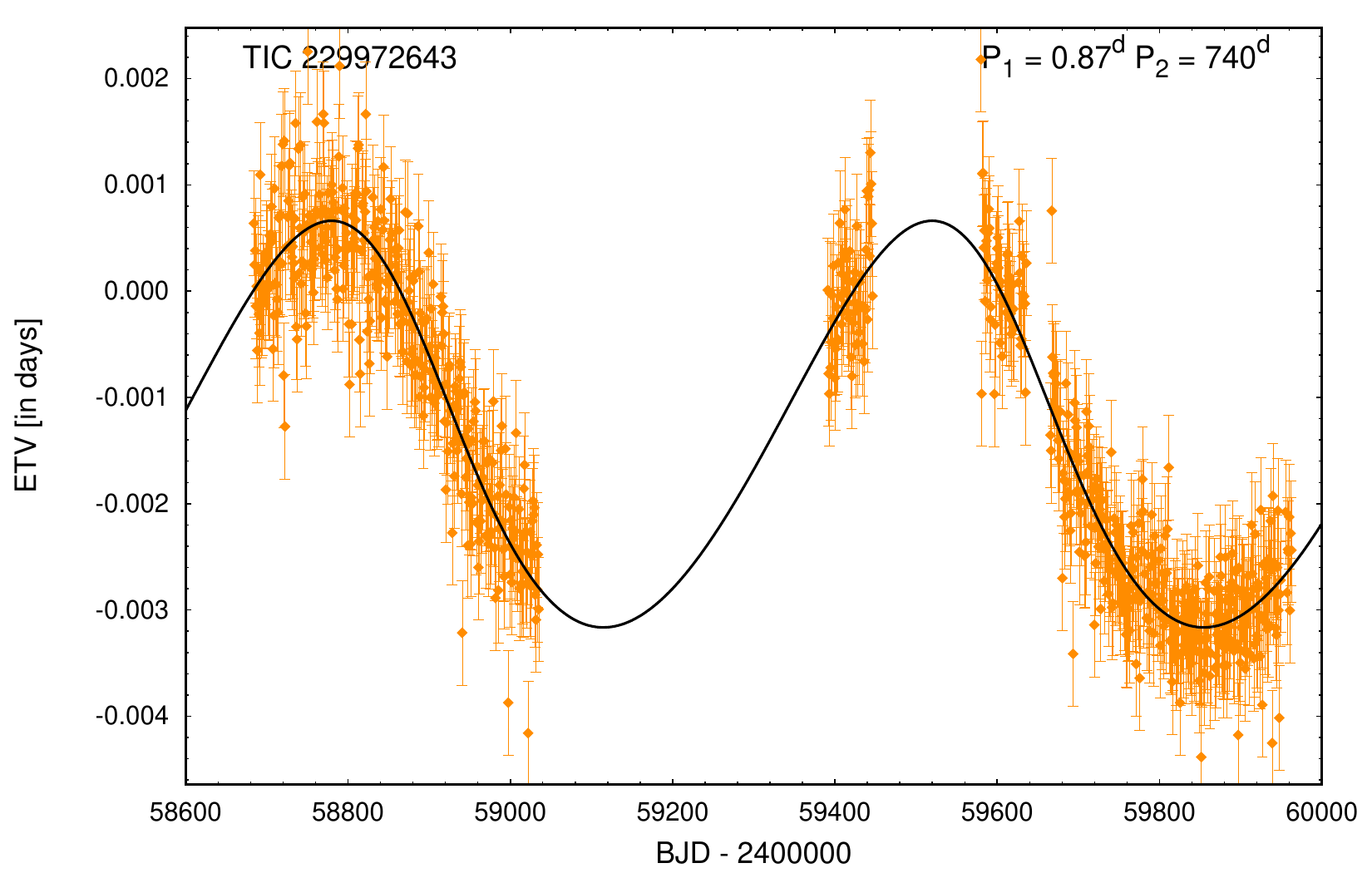}\includegraphics[width=60mm]{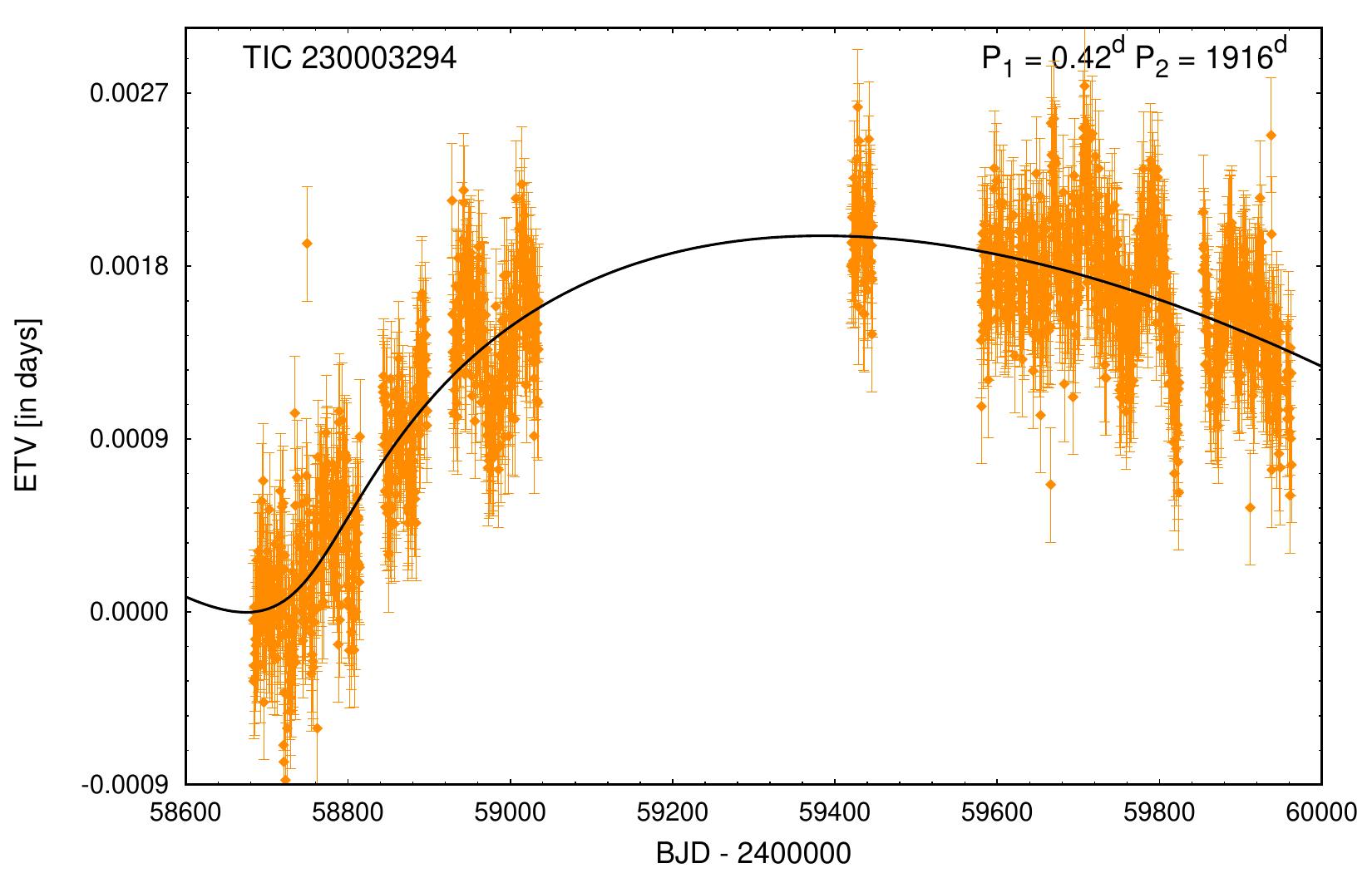}
\includegraphics[width=60mm]{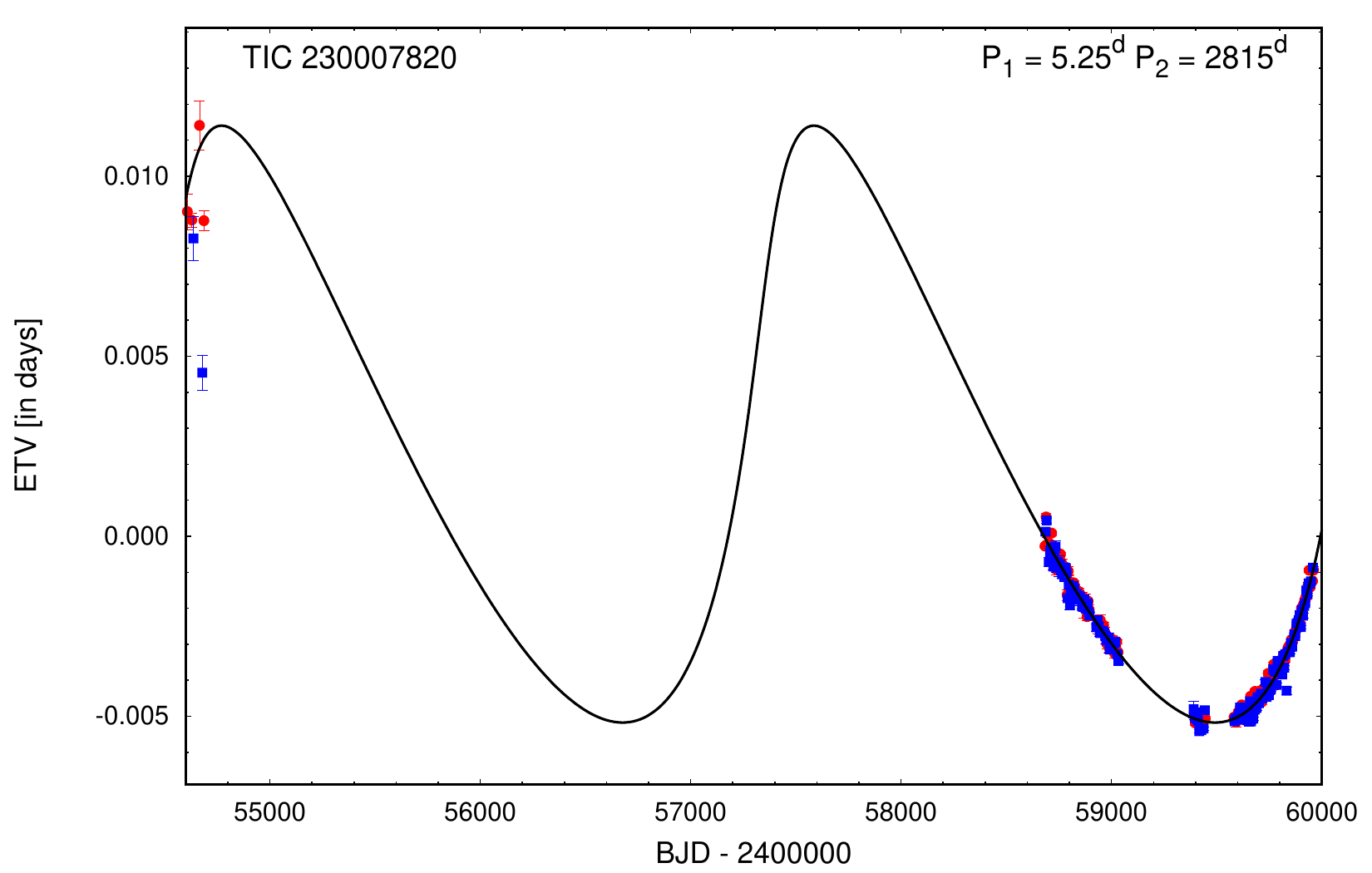}\includegraphics[width=60mm]{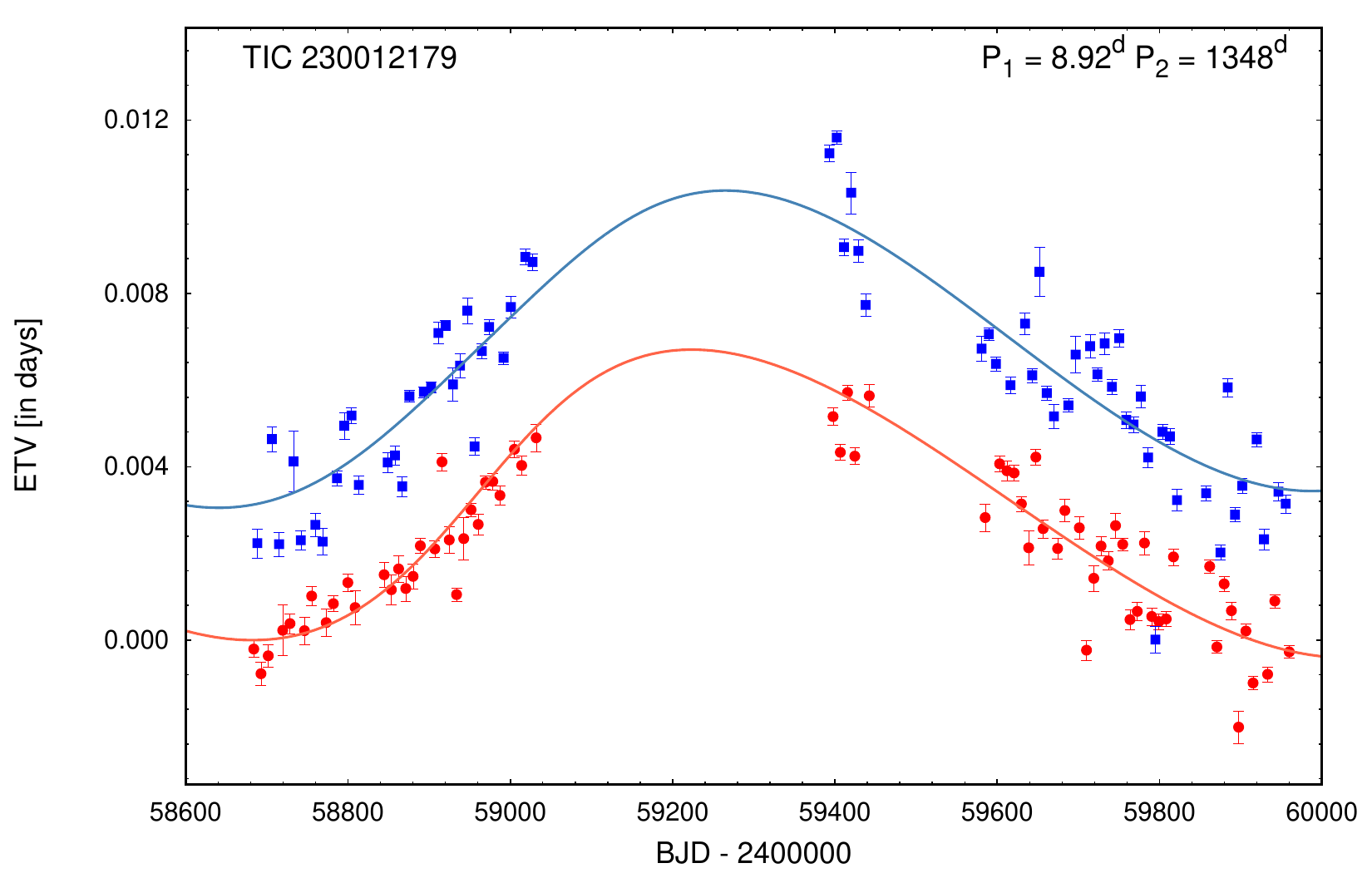}\includegraphics[width=60mm]{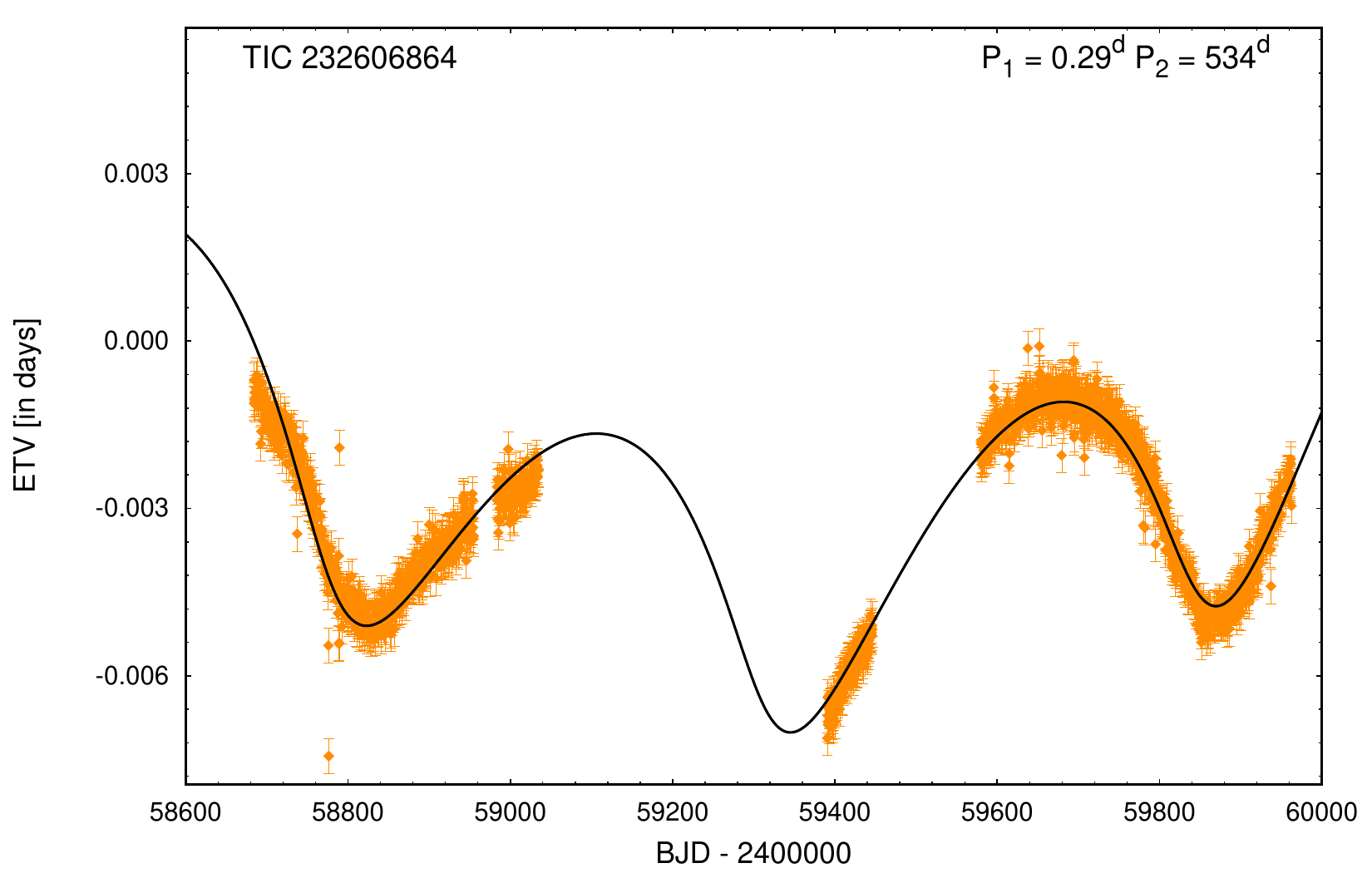}
\includegraphics[width=60mm]{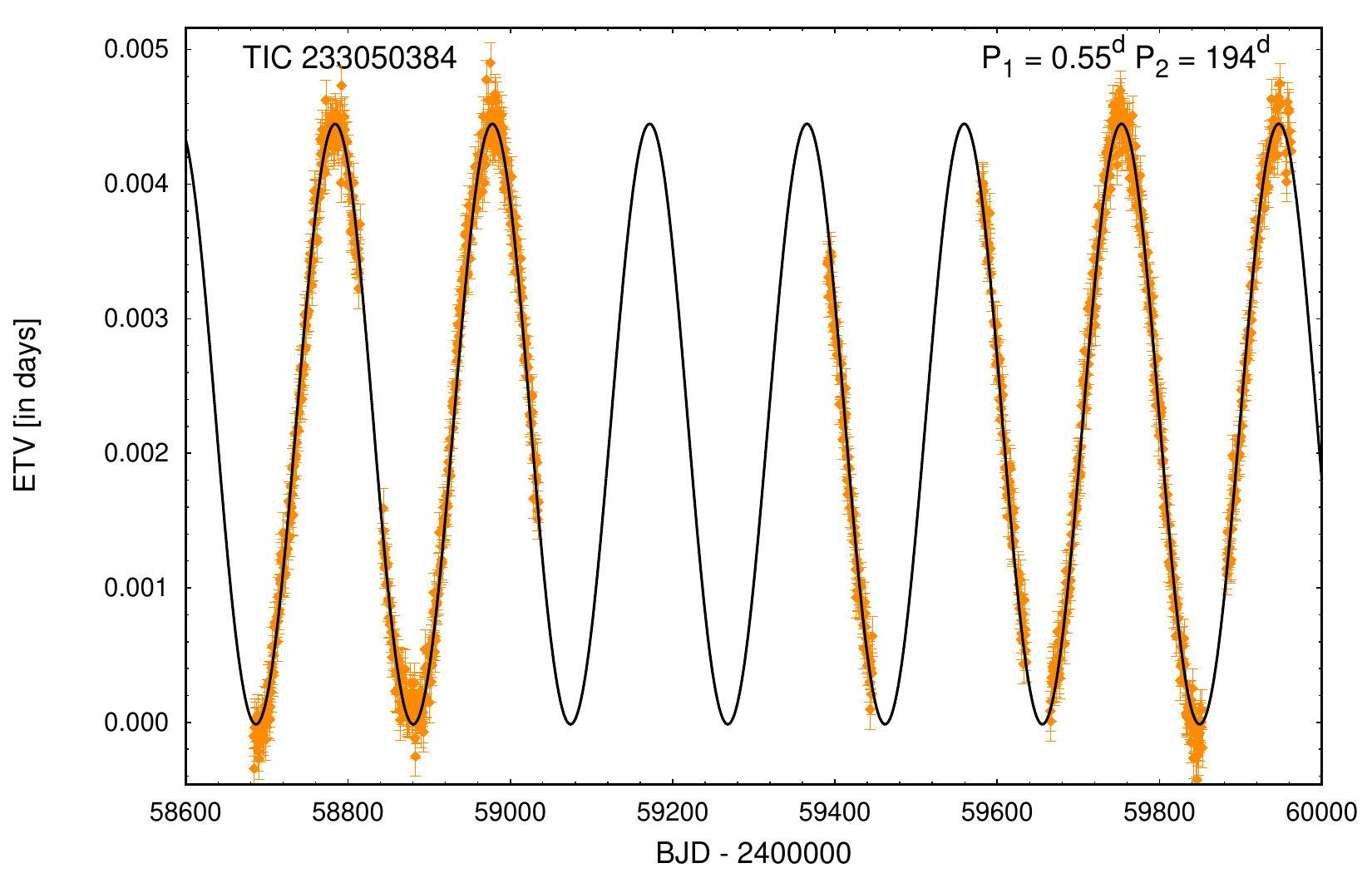}\includegraphics[width=60mm]{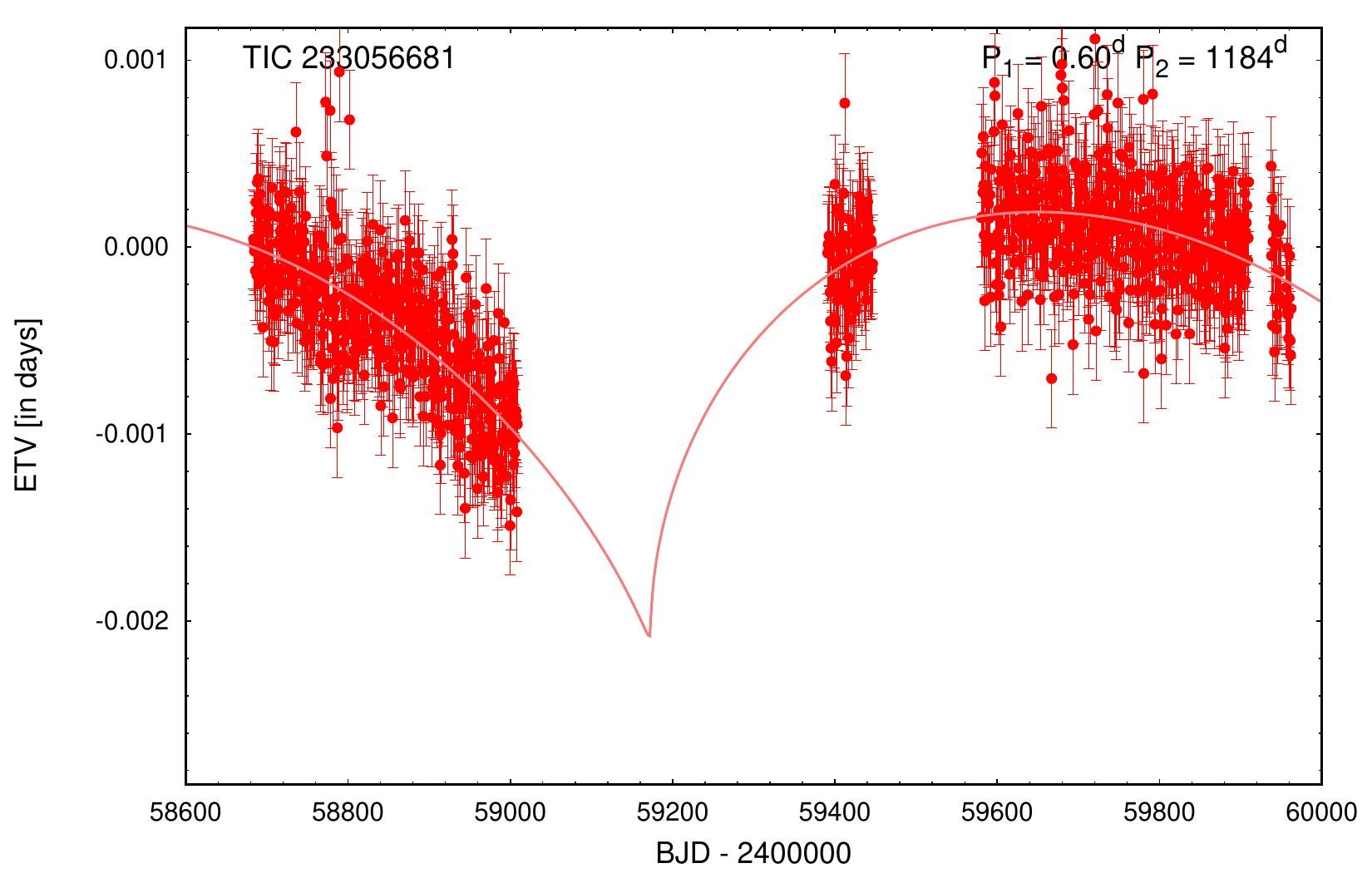}\includegraphics[width=60mm]{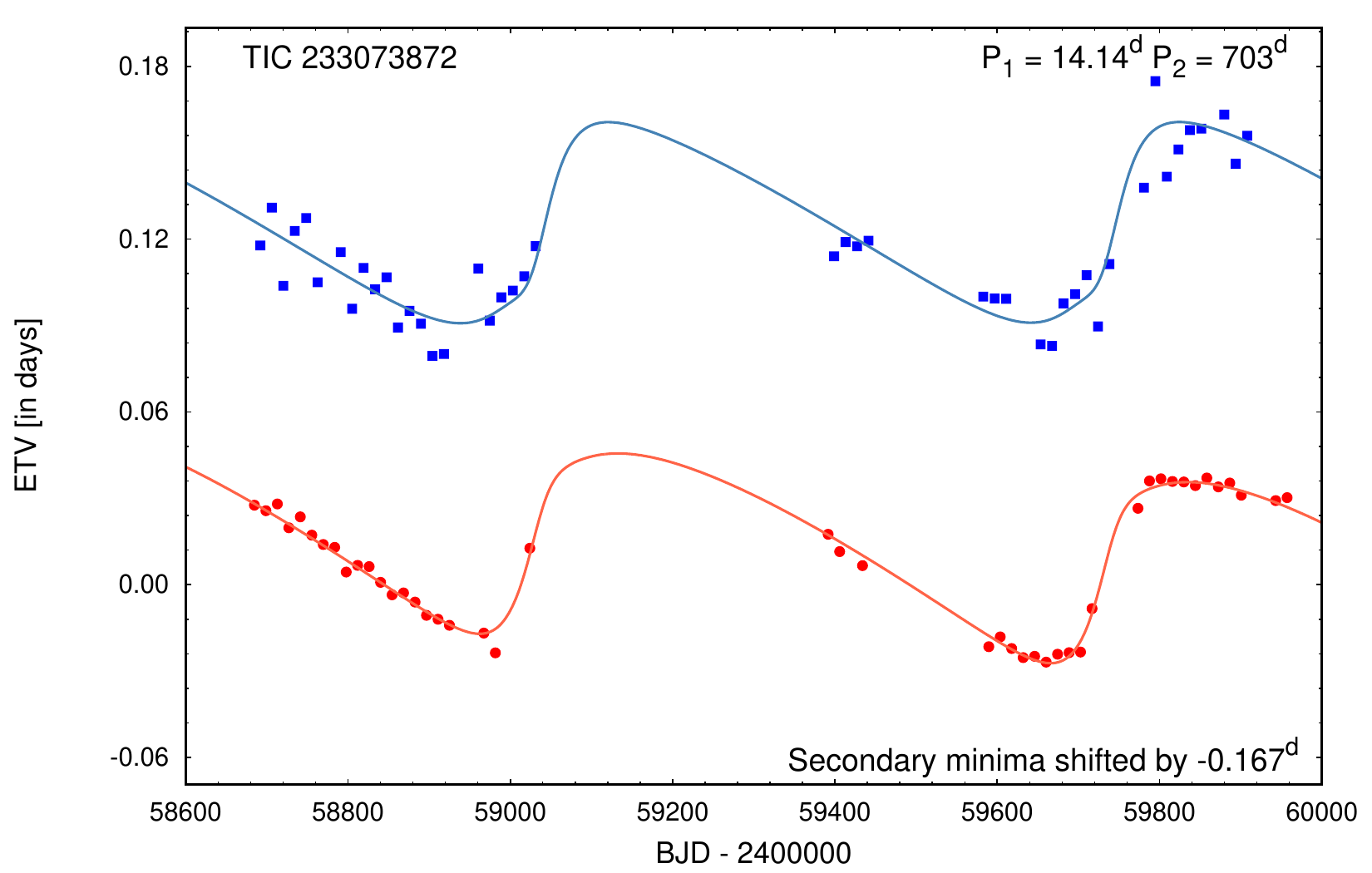}
\includegraphics[width=60mm]{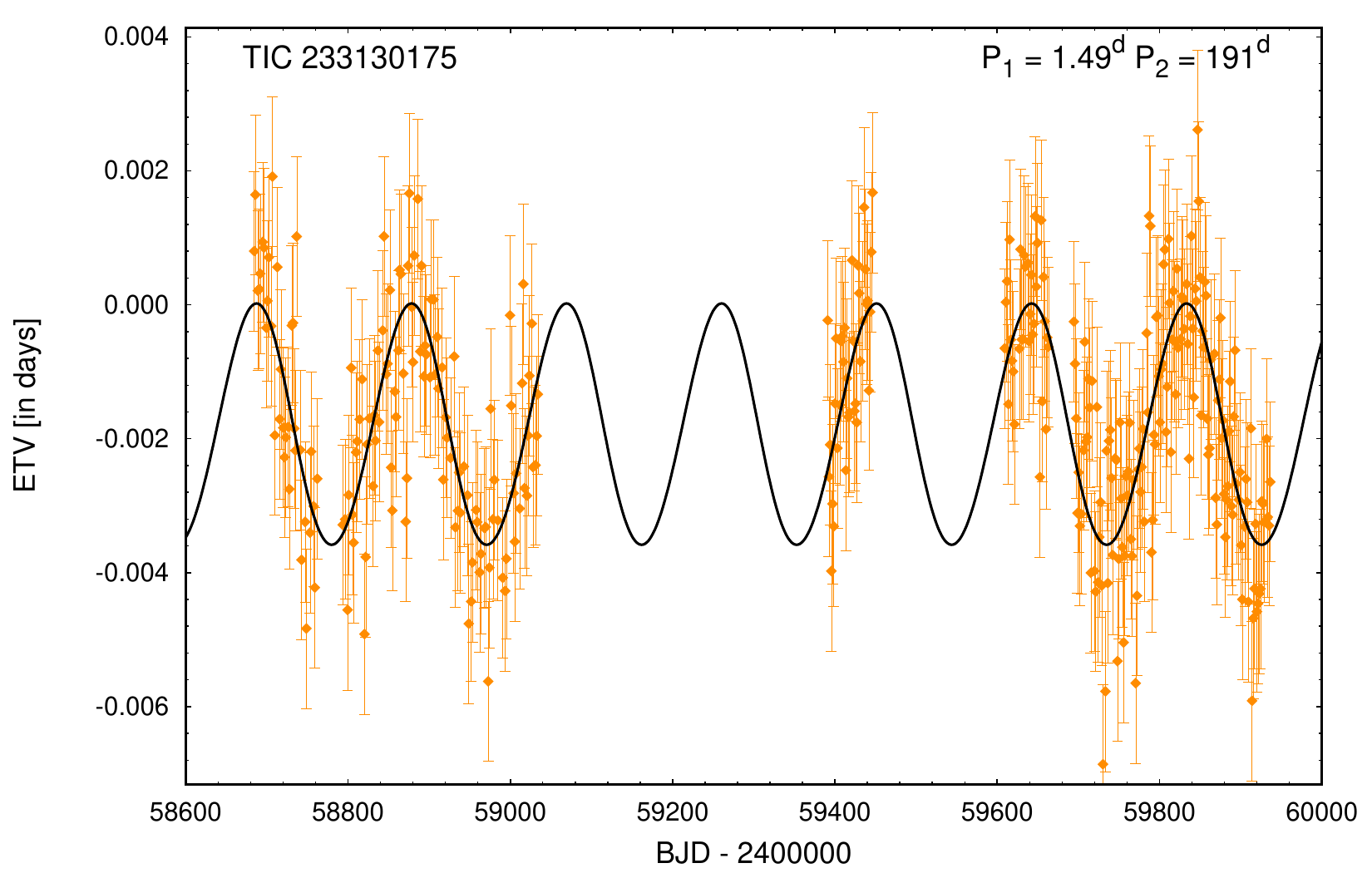}\includegraphics[width=60mm]{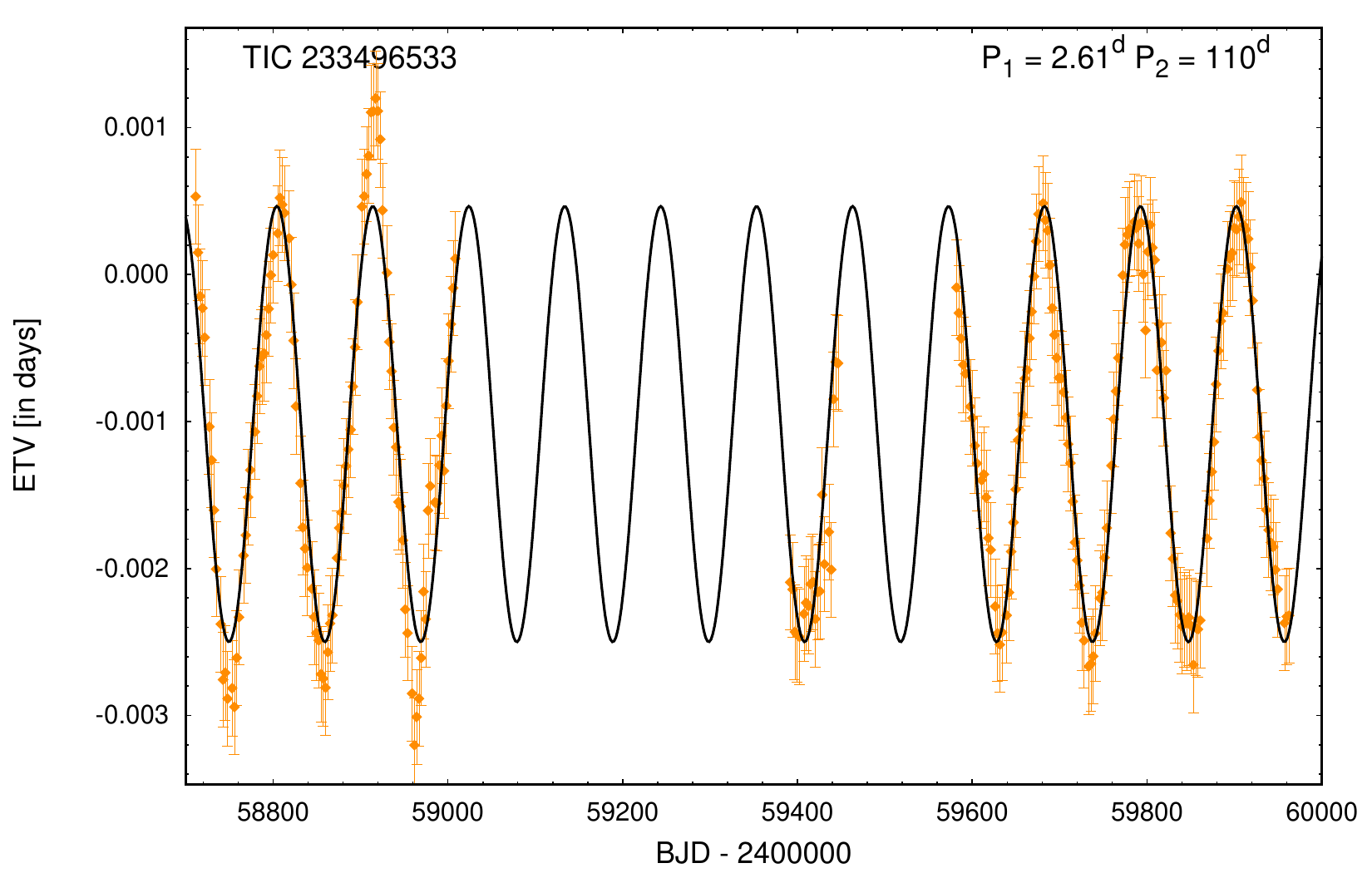}\includegraphics[width=60mm]{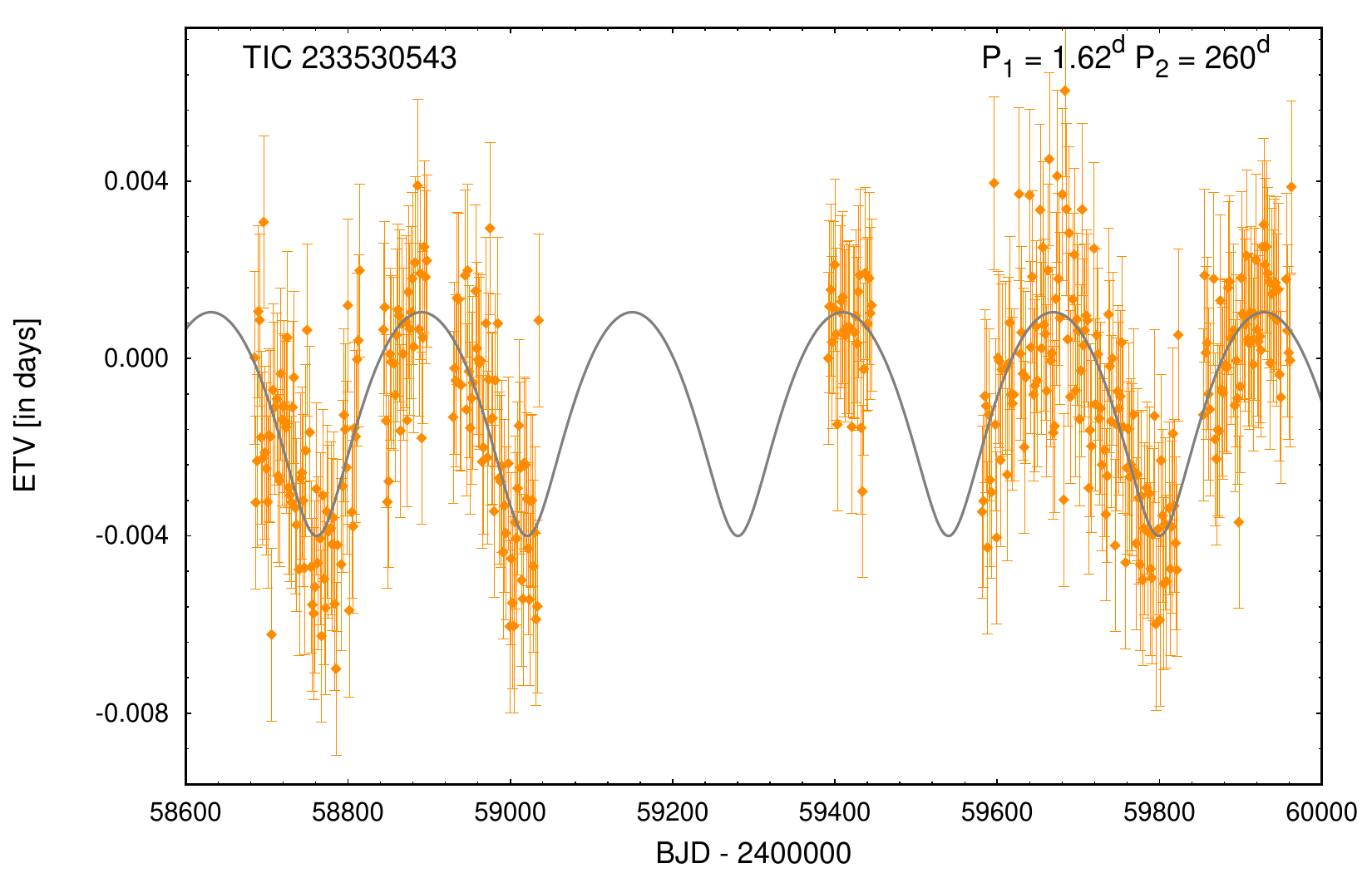}
\includegraphics[width=60mm]{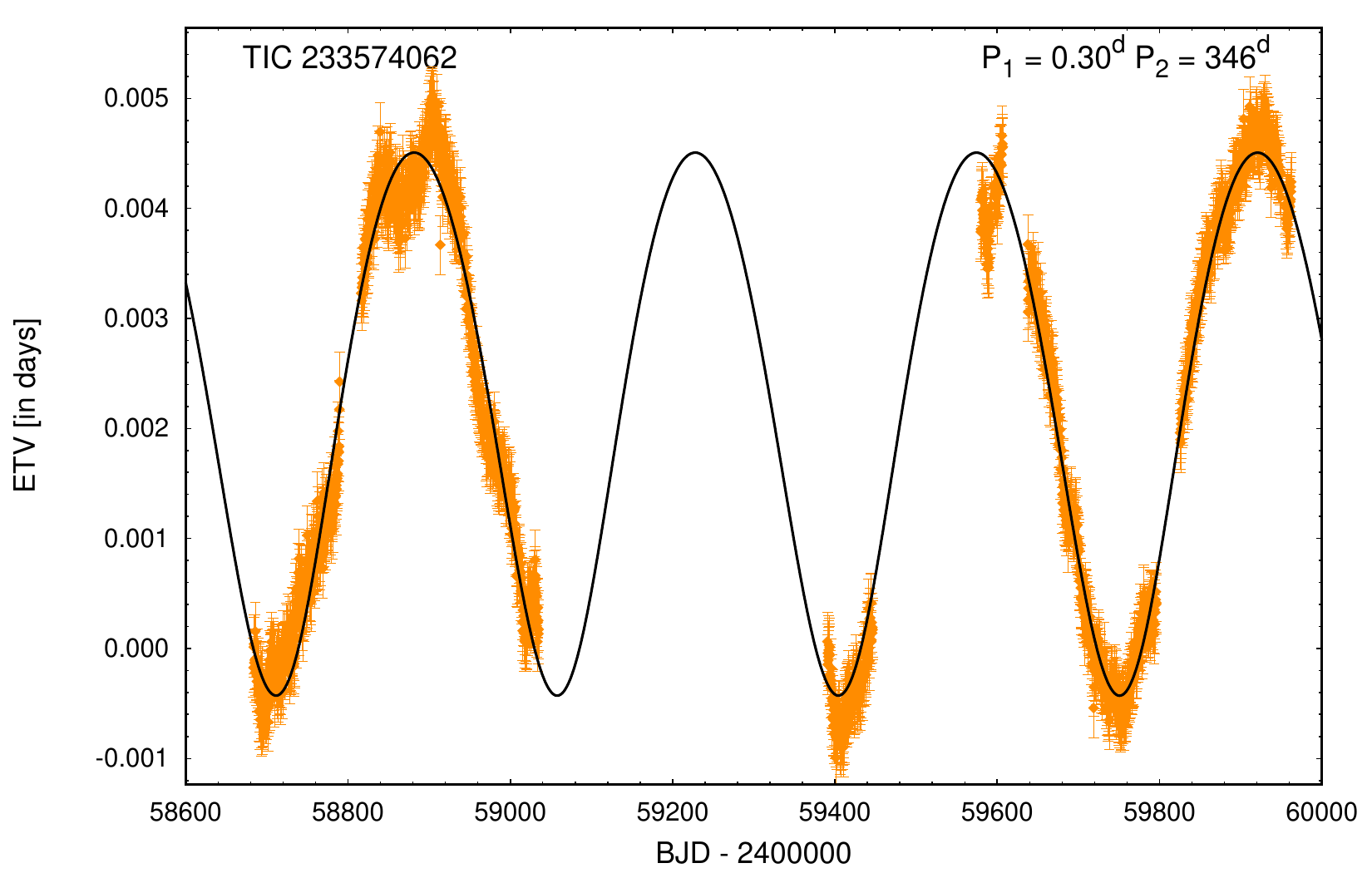}\includegraphics[width=60mm]{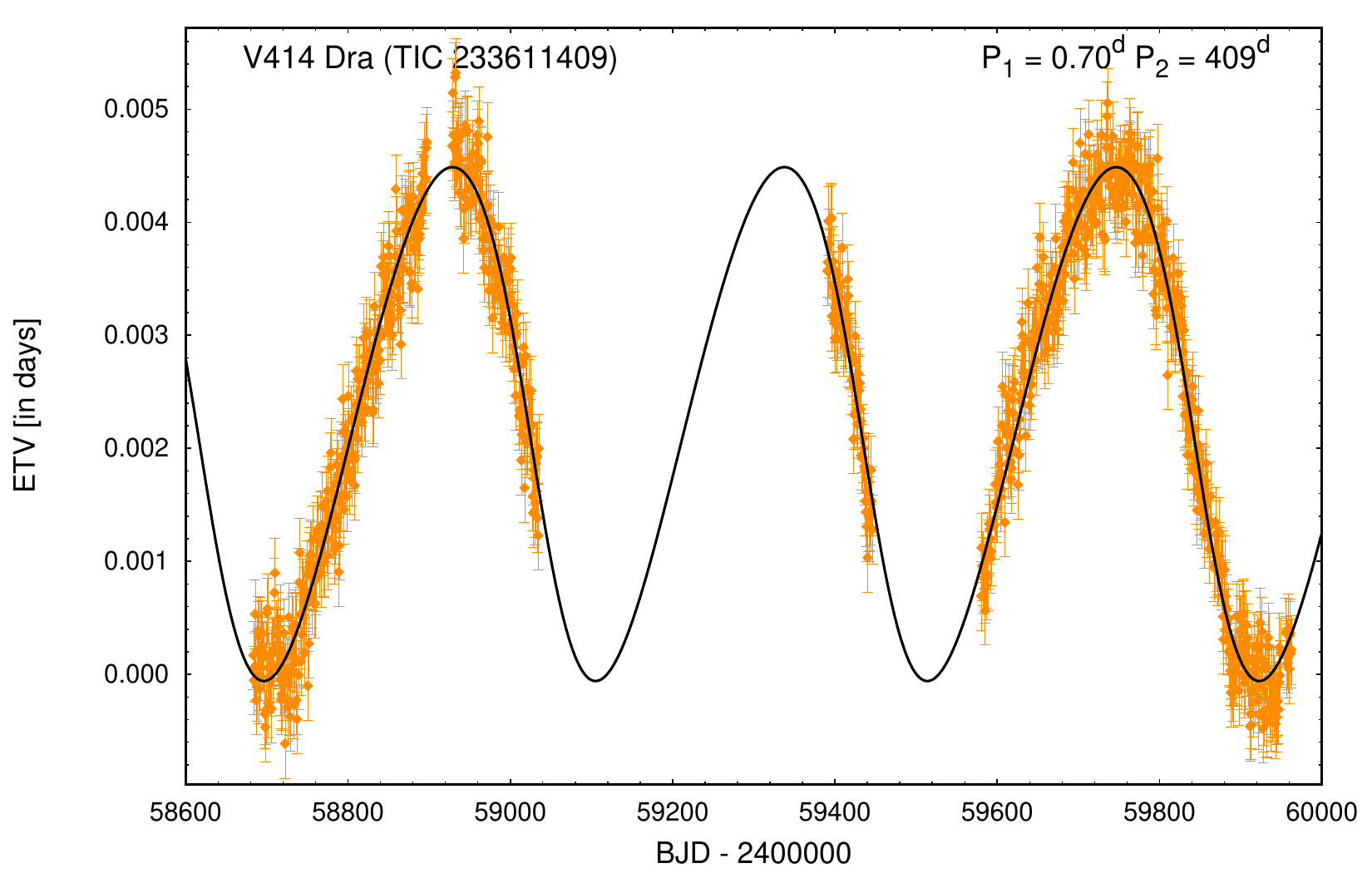}\includegraphics[width=60mm]{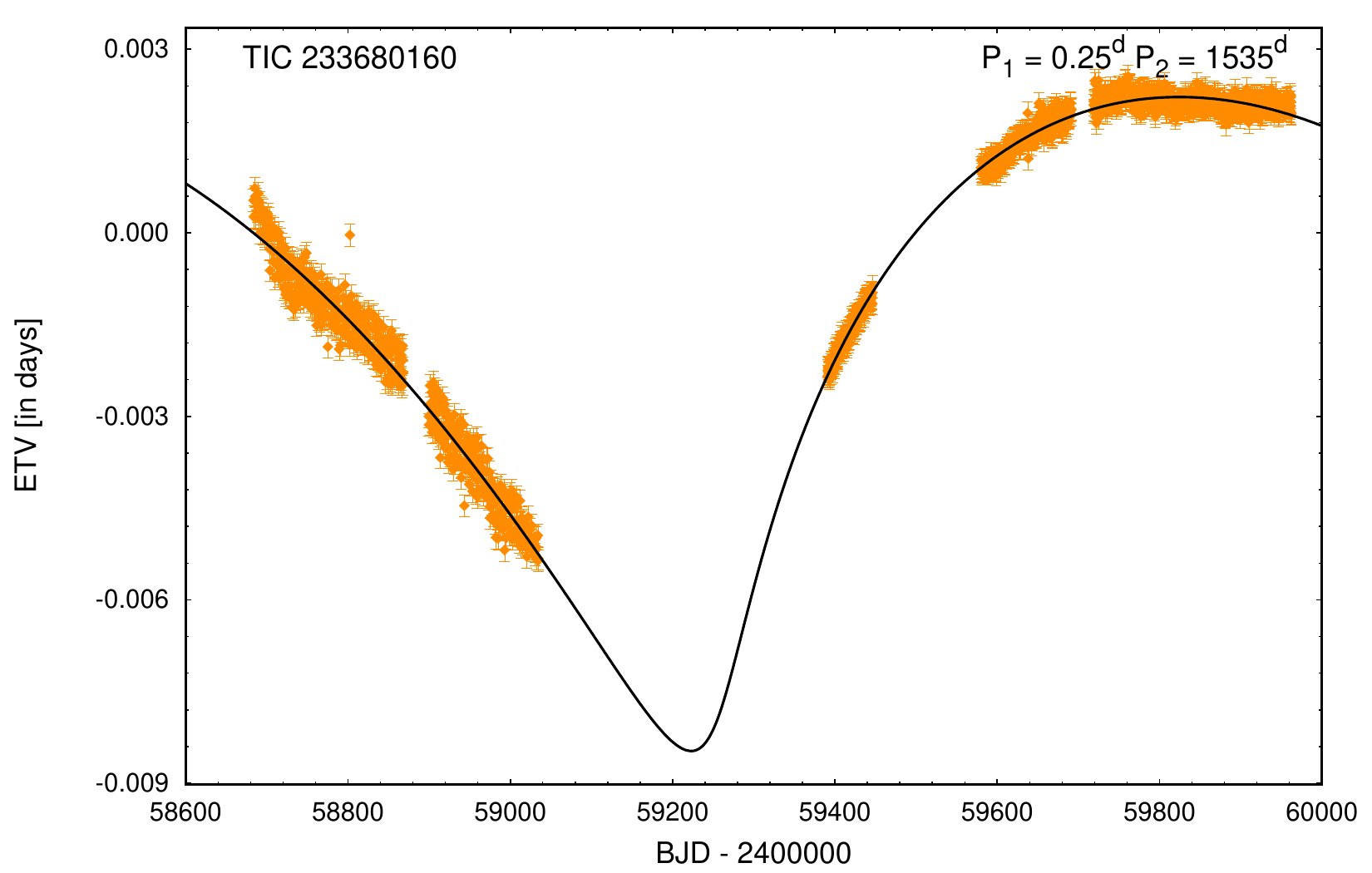}
\caption{(continued)}
\end{figure*}

\addtocounter{figure}{-1}

\begin{figure*}
\includegraphics[width=60mm]{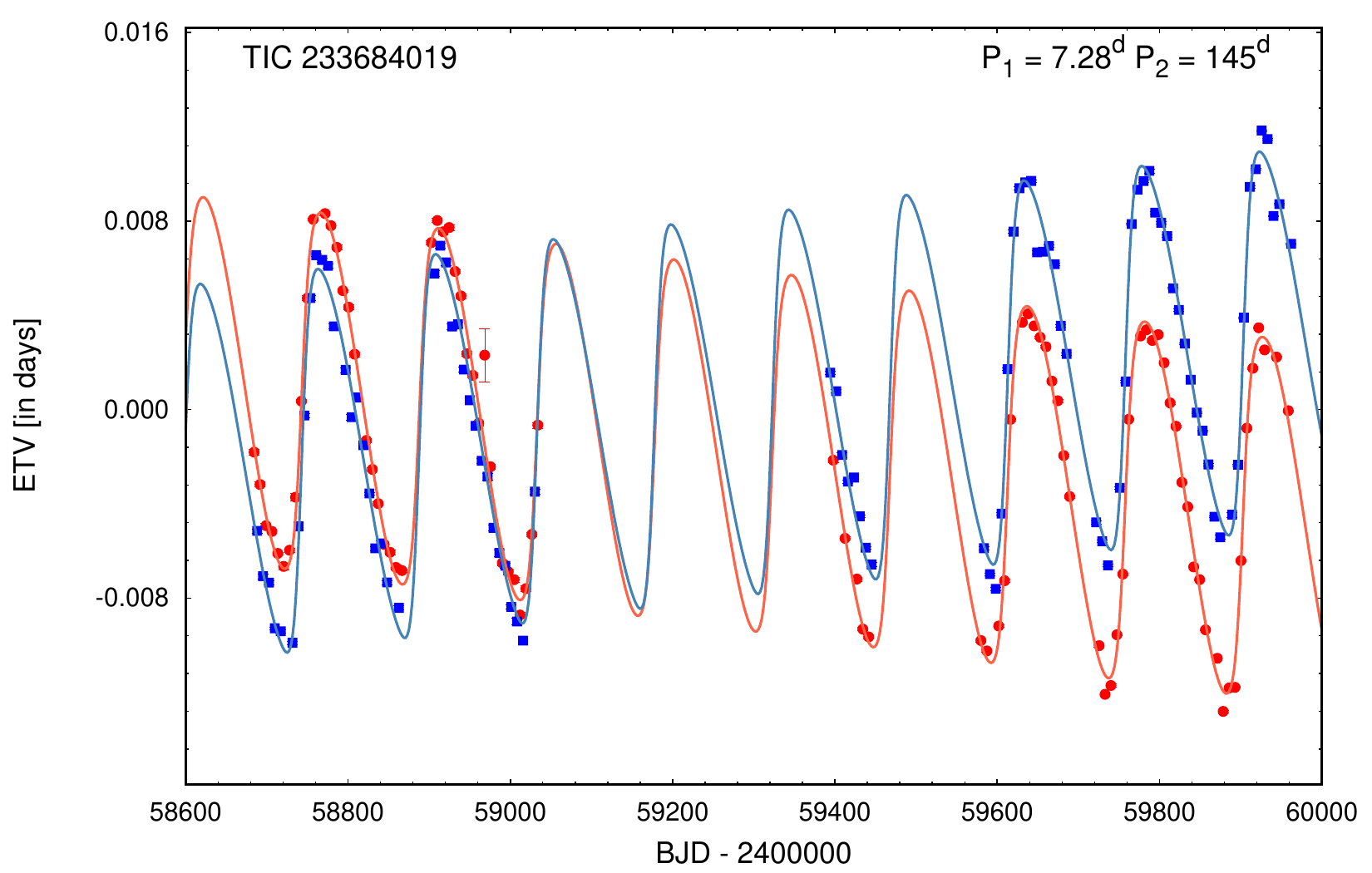}\includegraphics[width=60mm]{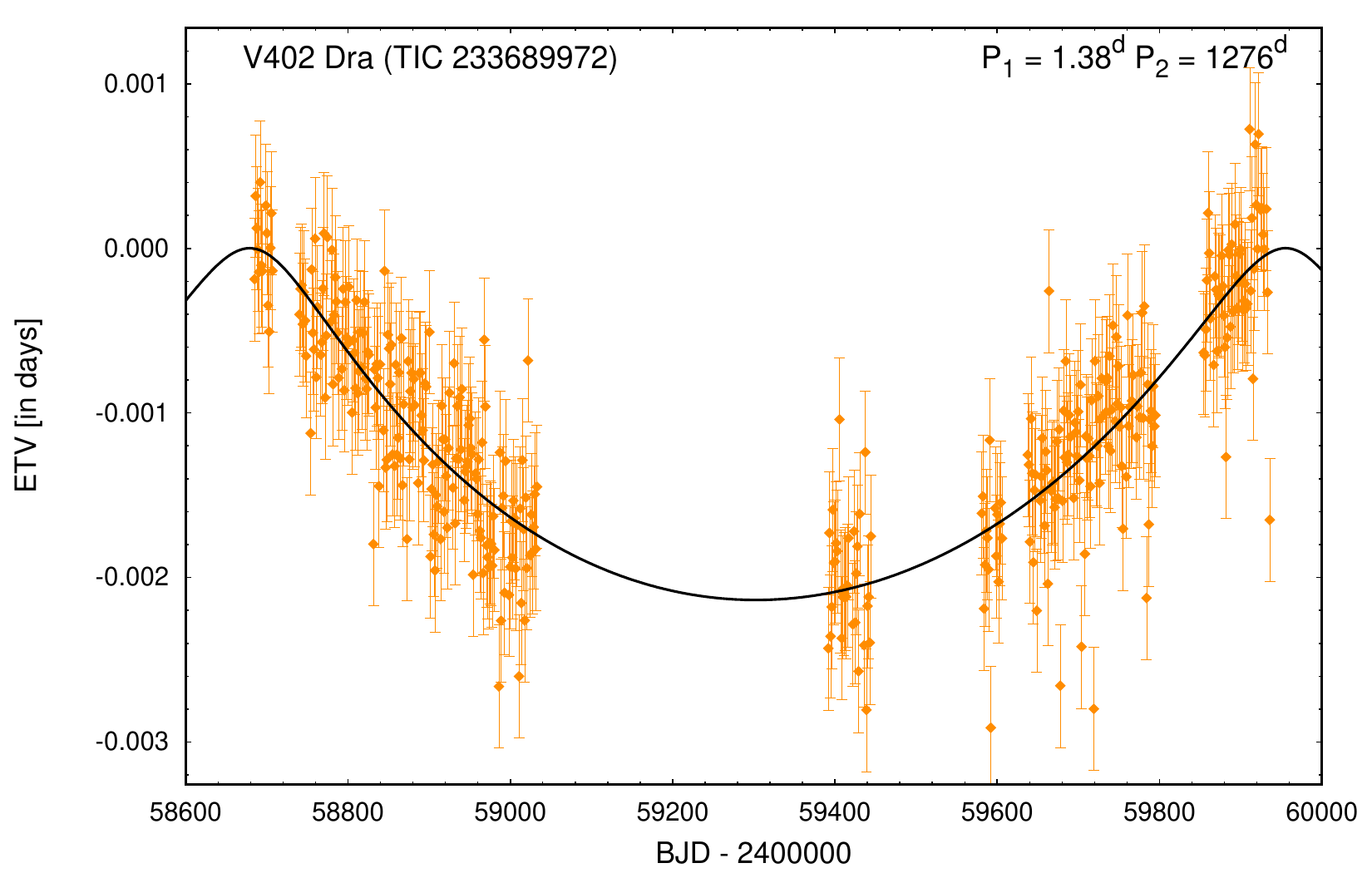}\includegraphics[width=60mm]{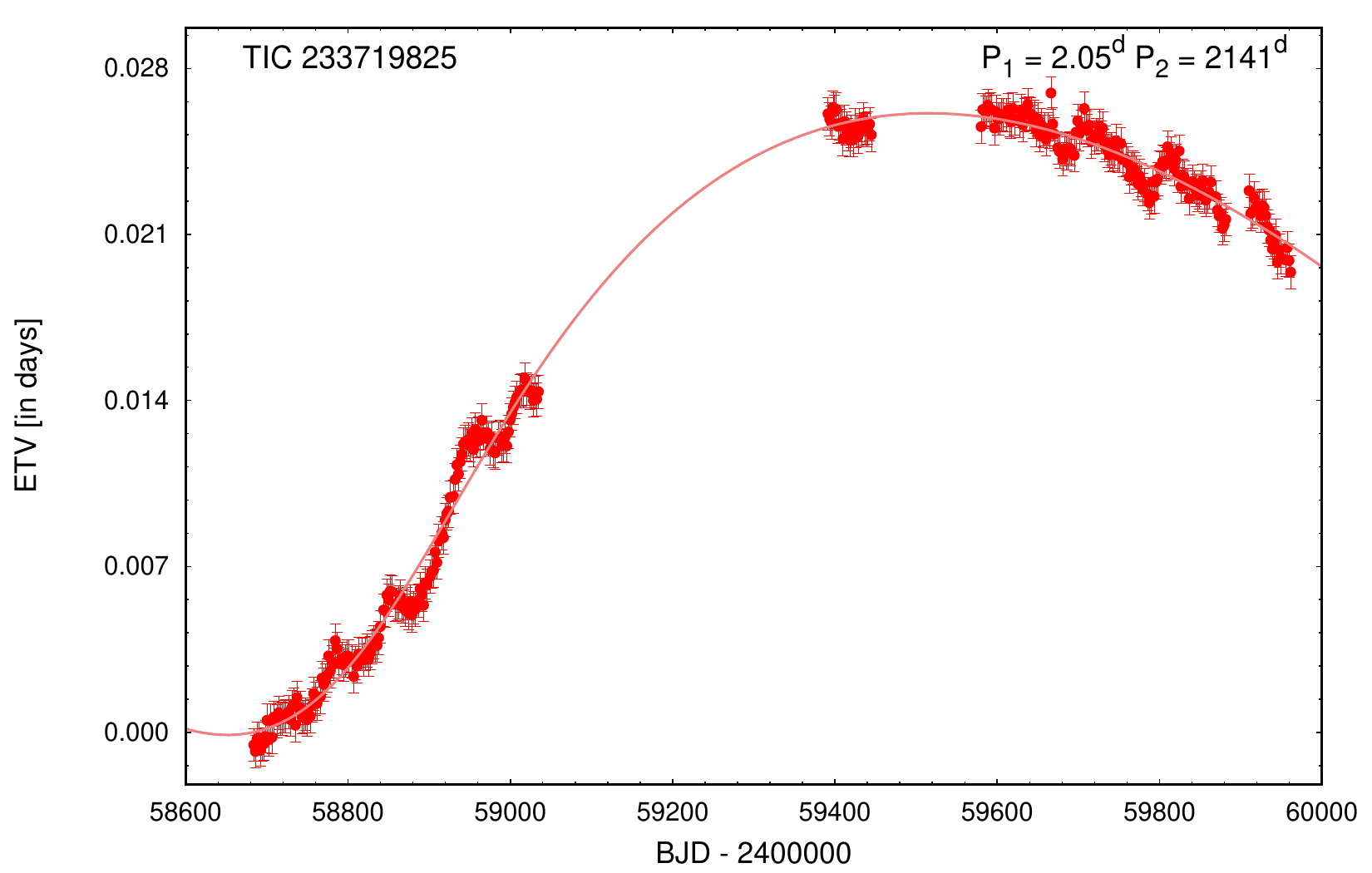}
\includegraphics[width=60mm]{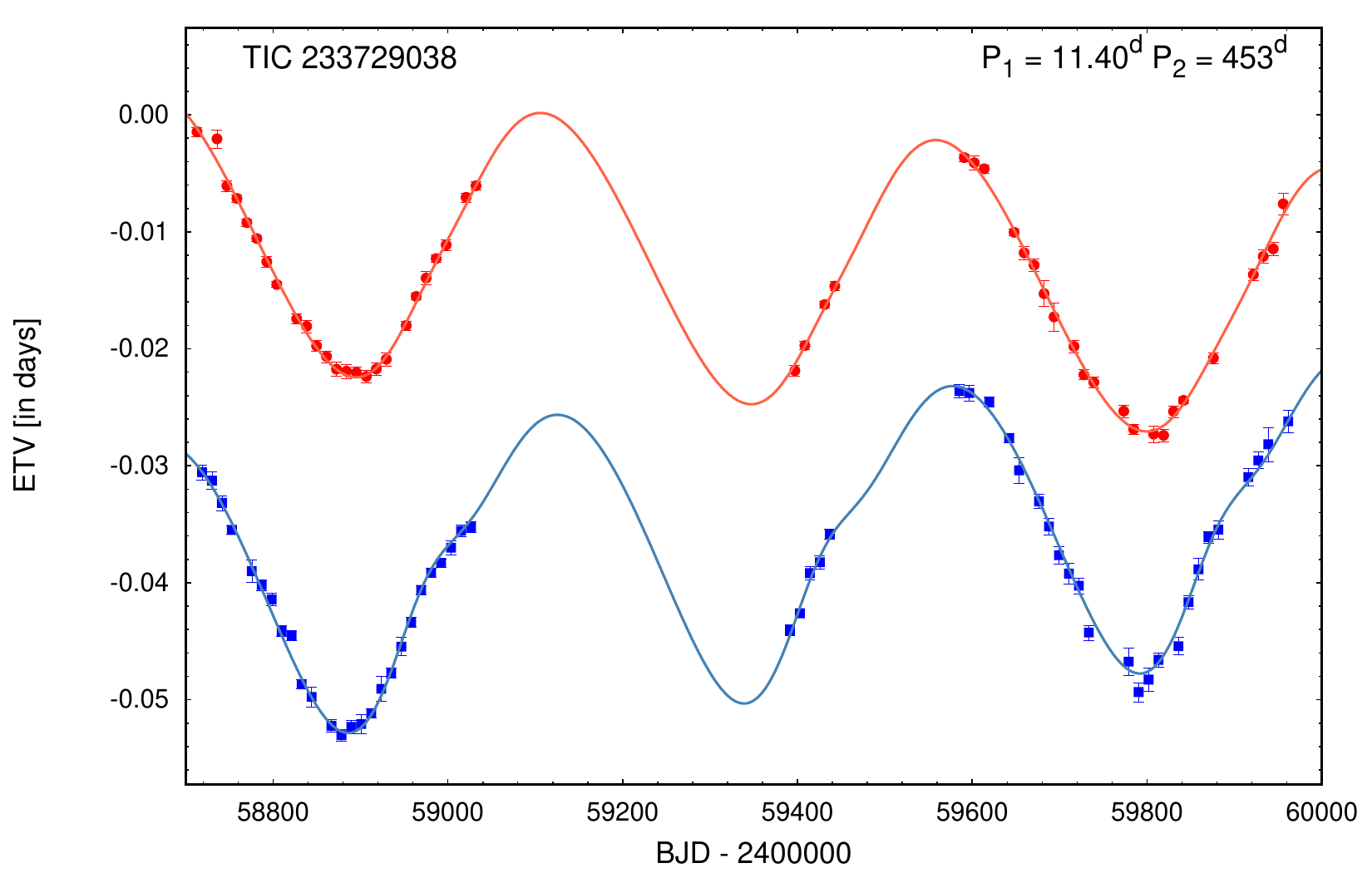}\includegraphics[width=60mm]{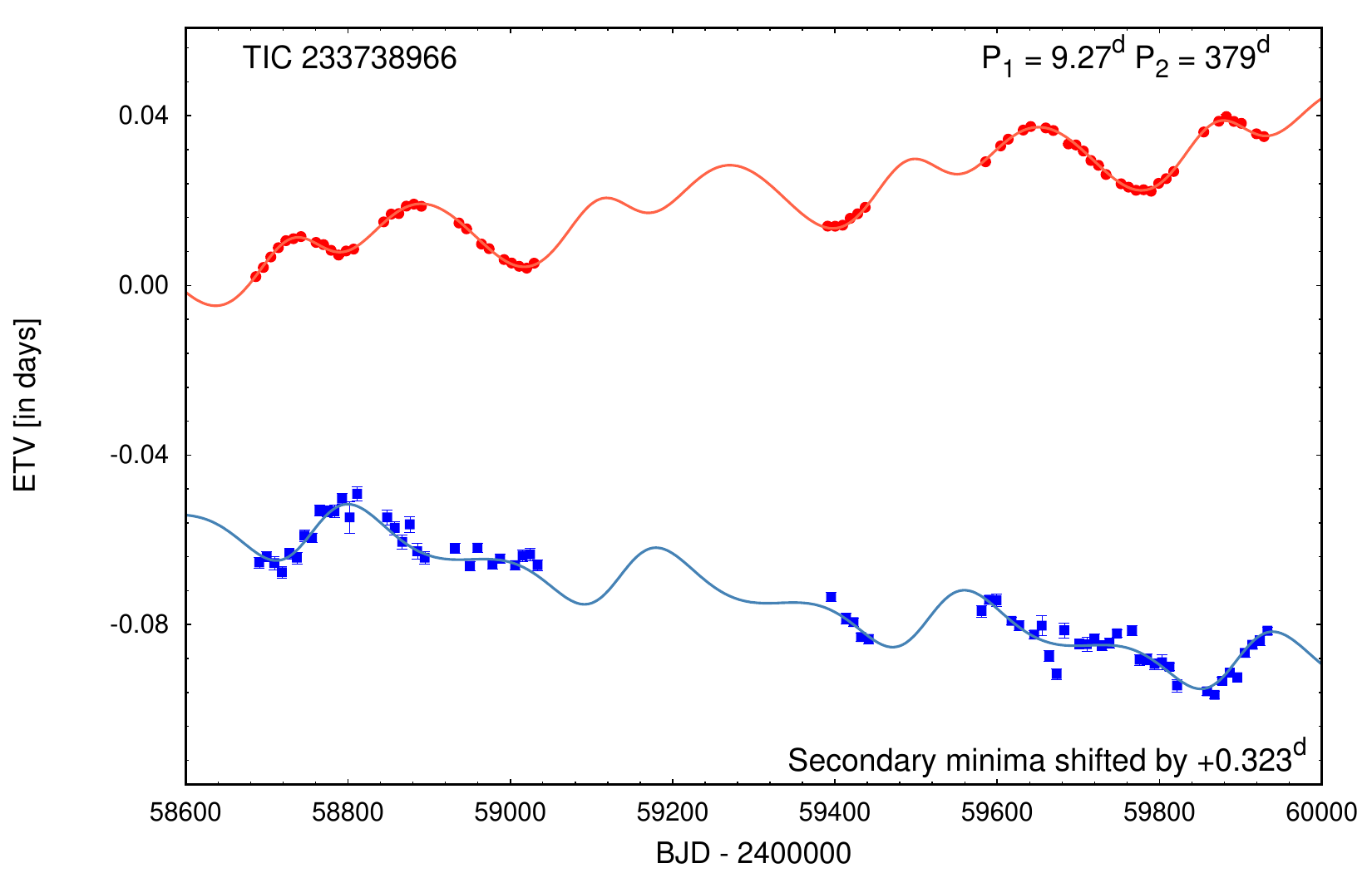}\includegraphics[width=60mm]{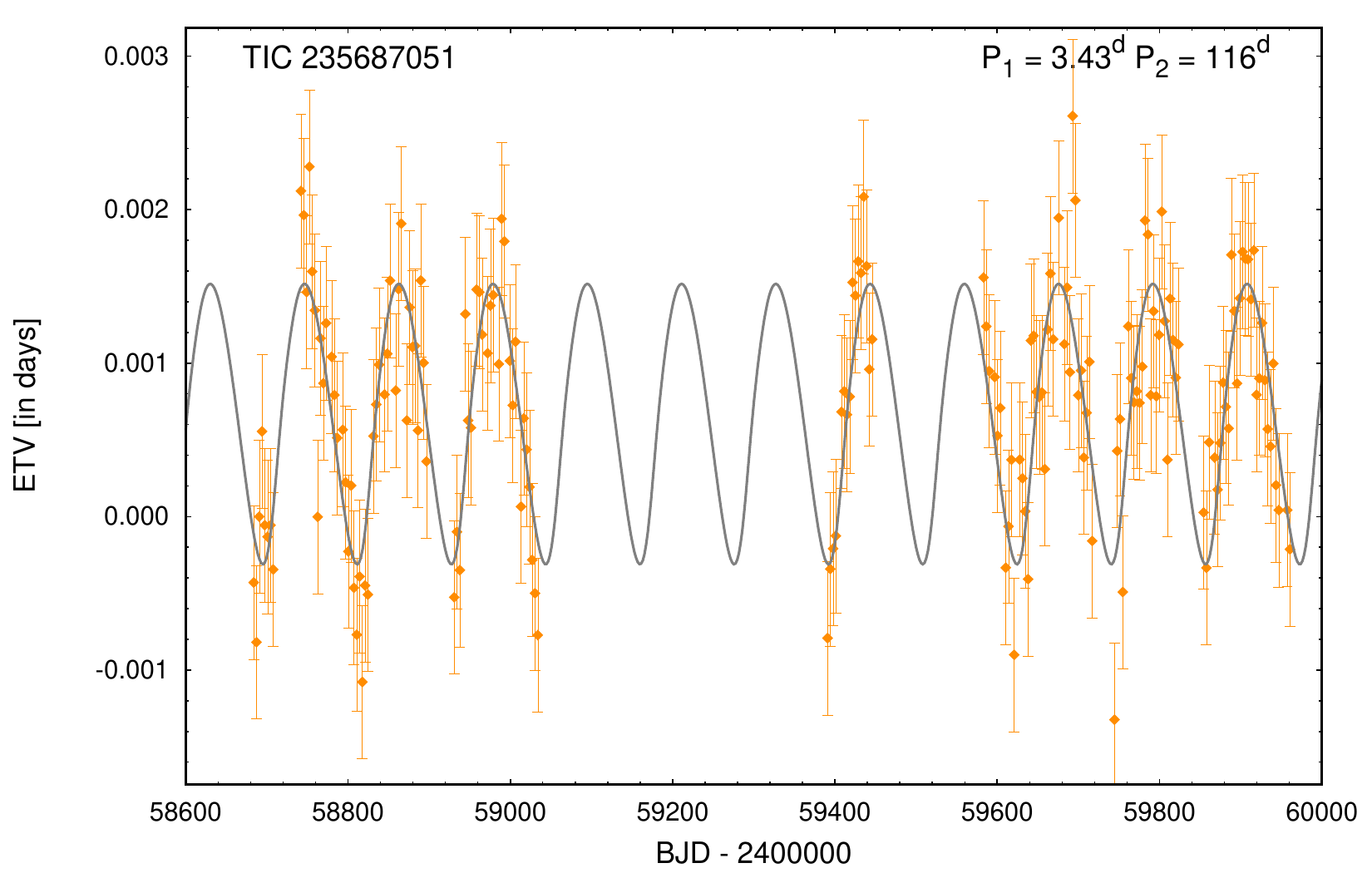}
\includegraphics[width=60mm]{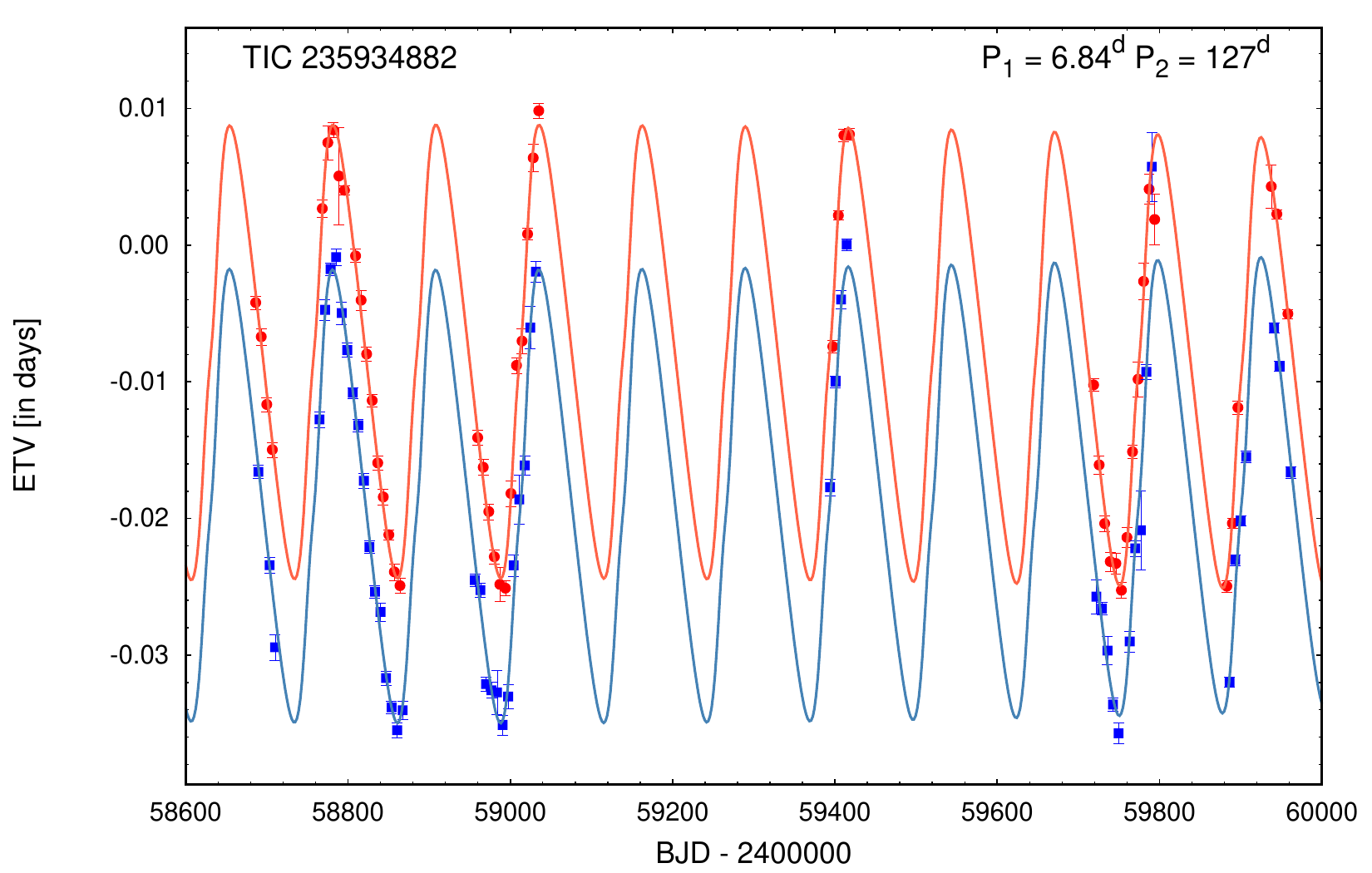}\includegraphics[width=60mm]{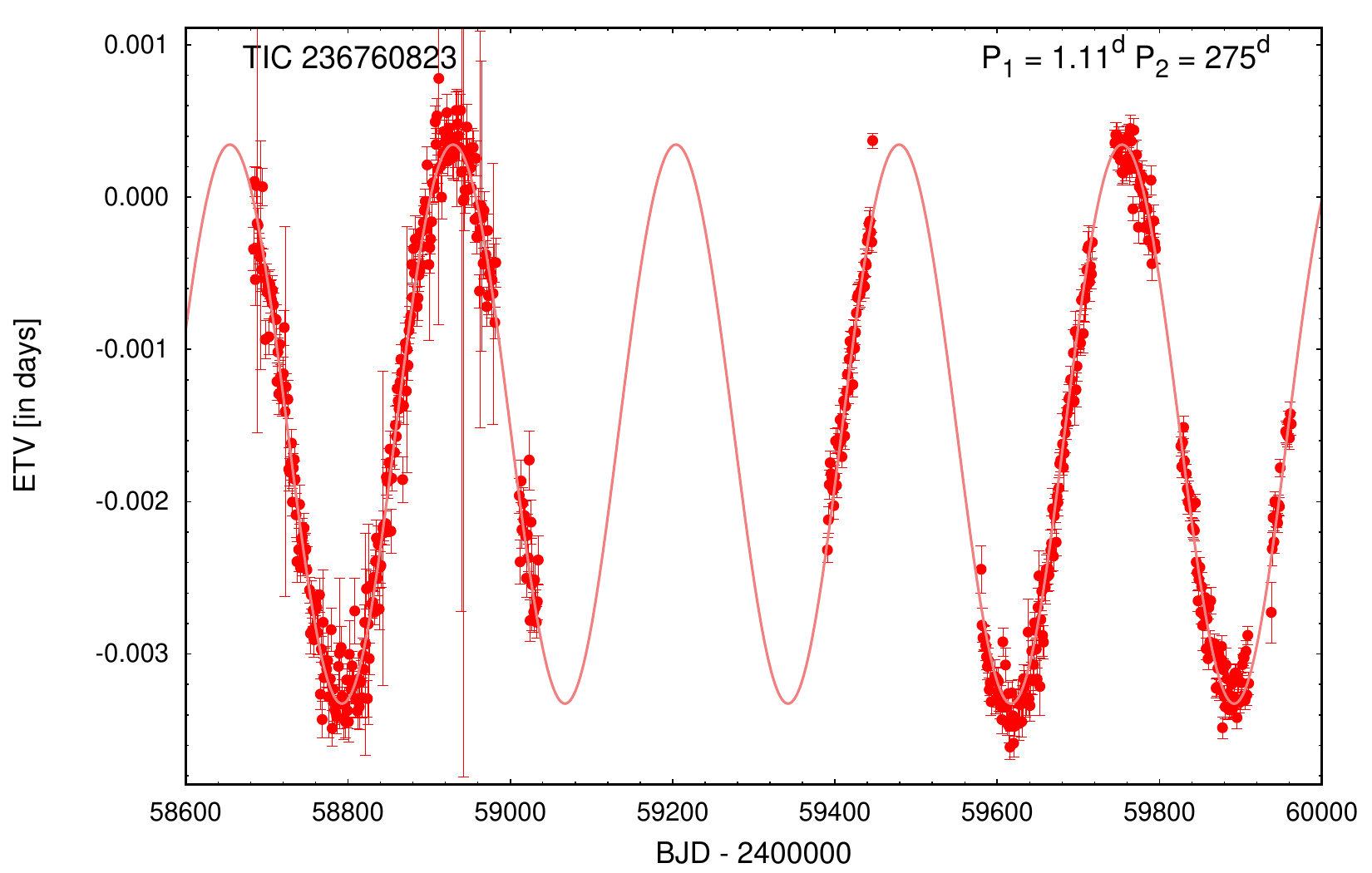}\includegraphics[width=60mm]{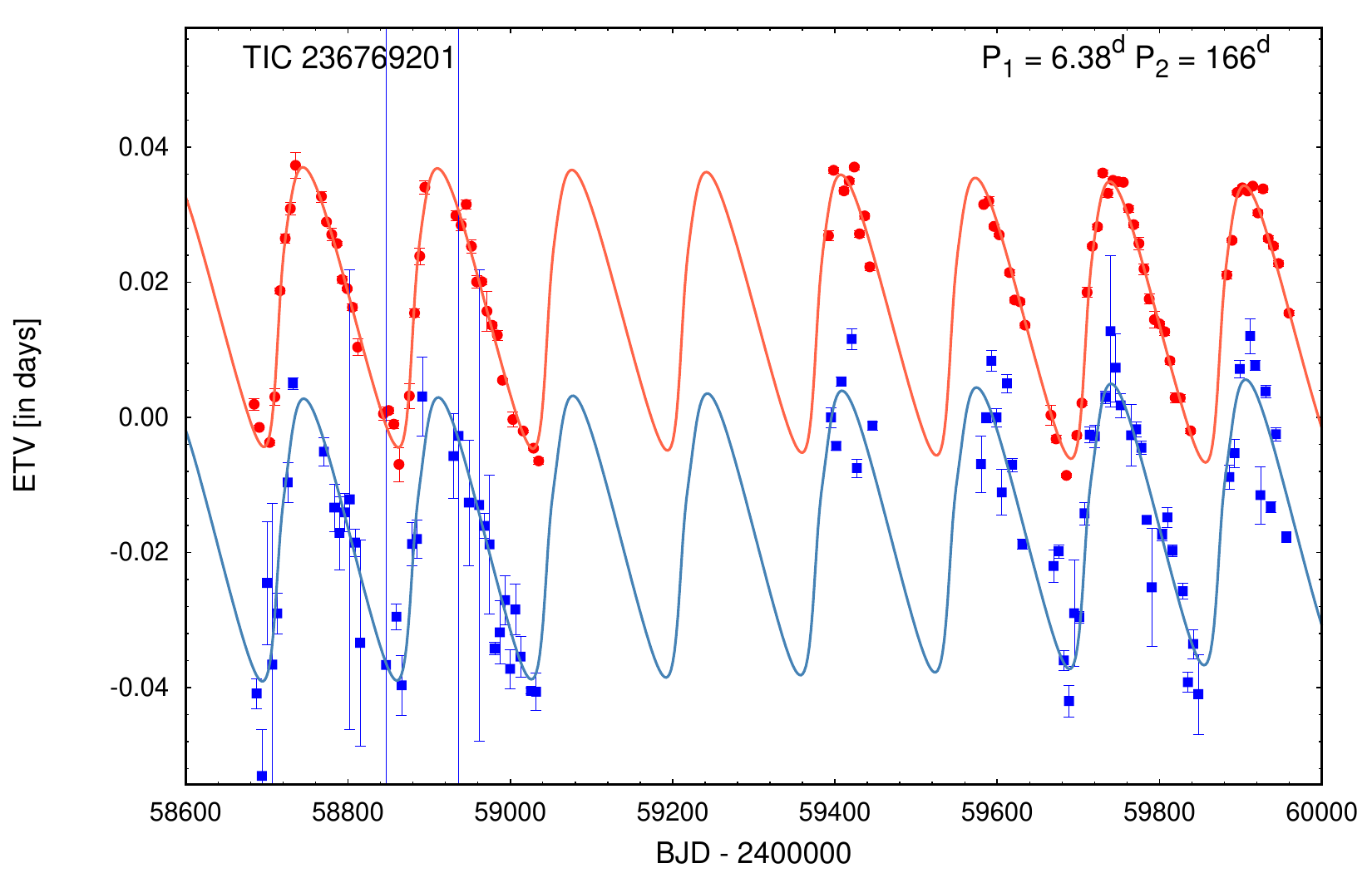}
\includegraphics[width=60mm]{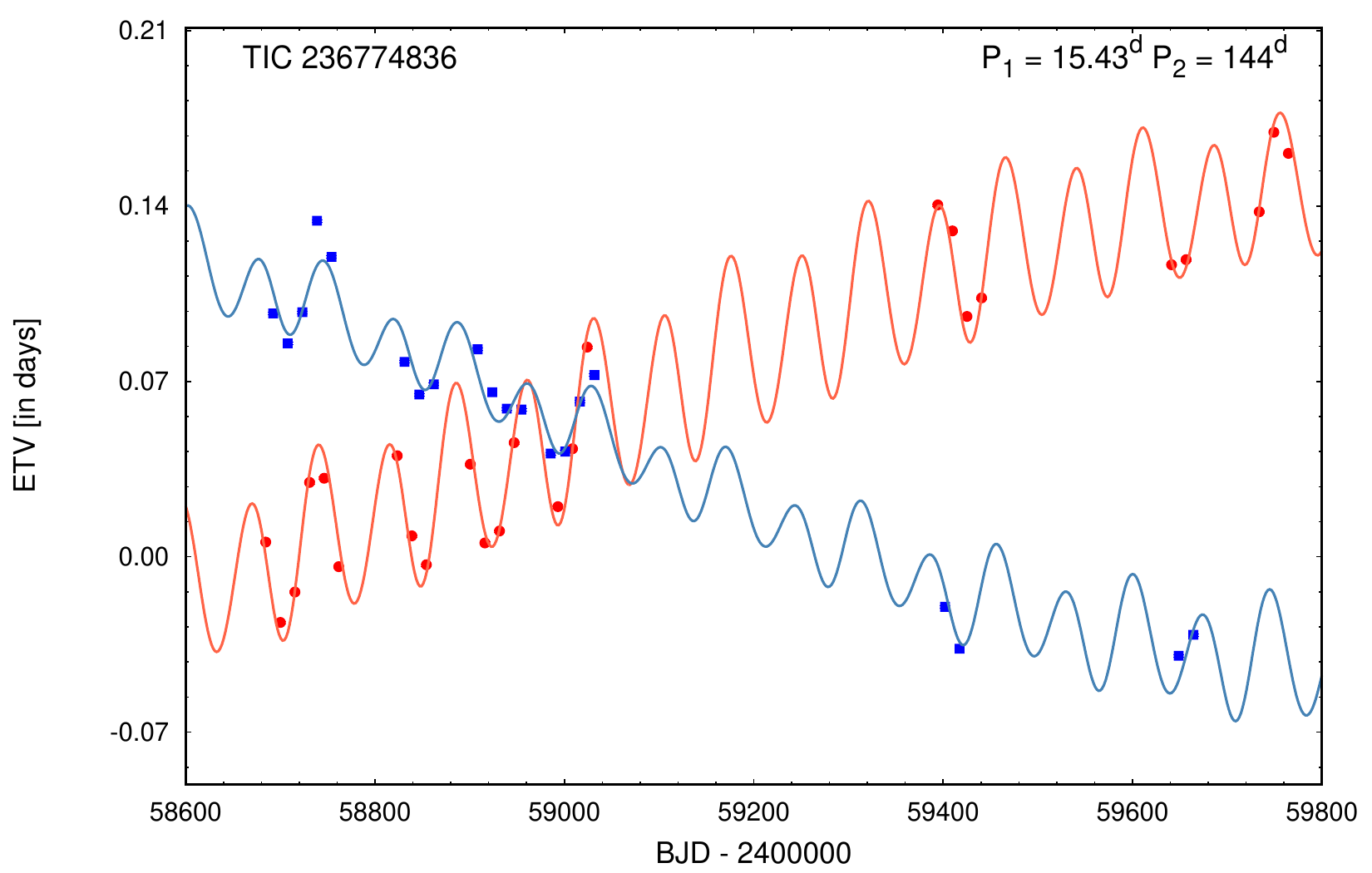}\includegraphics[width=60mm]{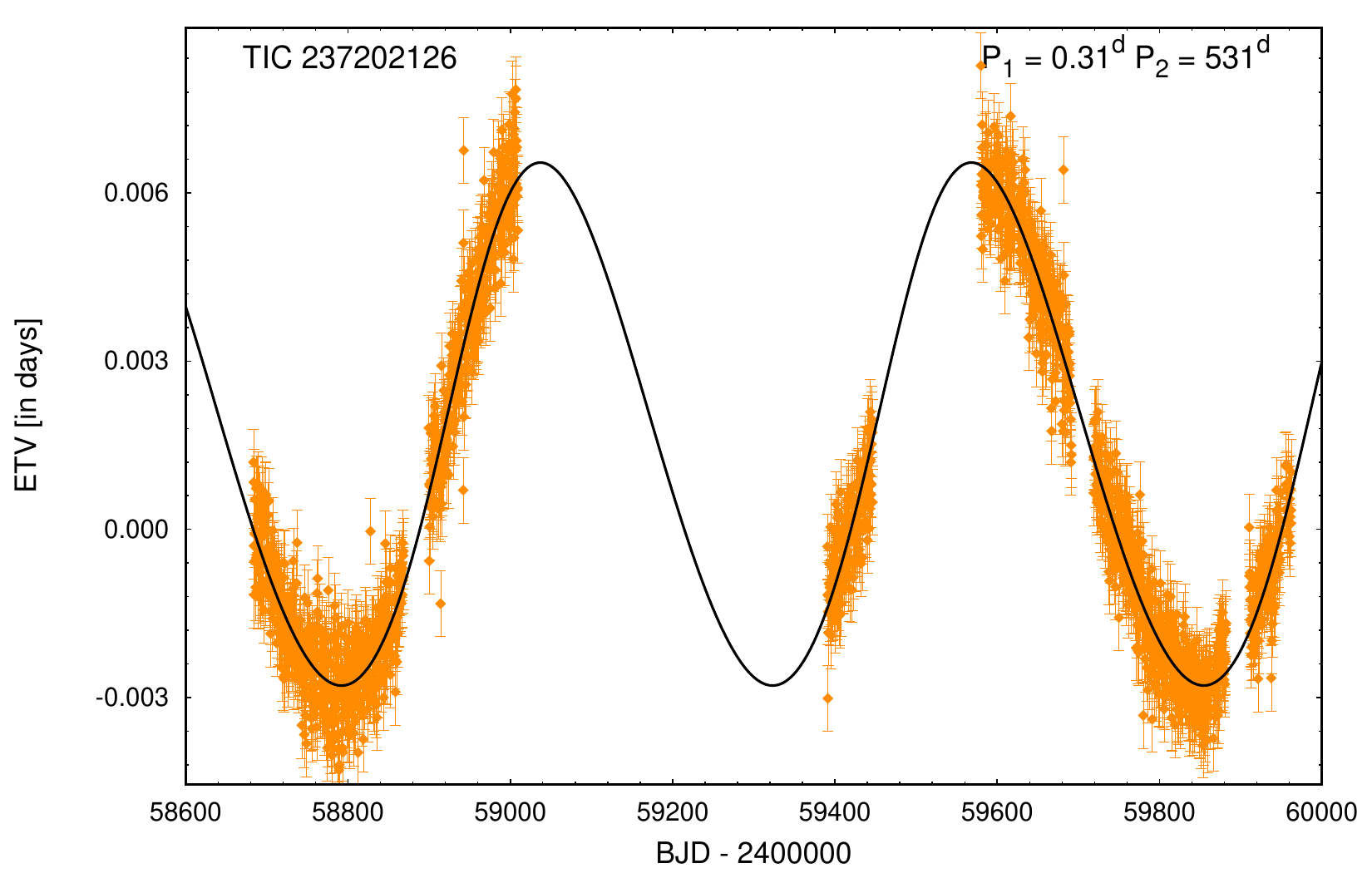}\includegraphics[width=60mm]{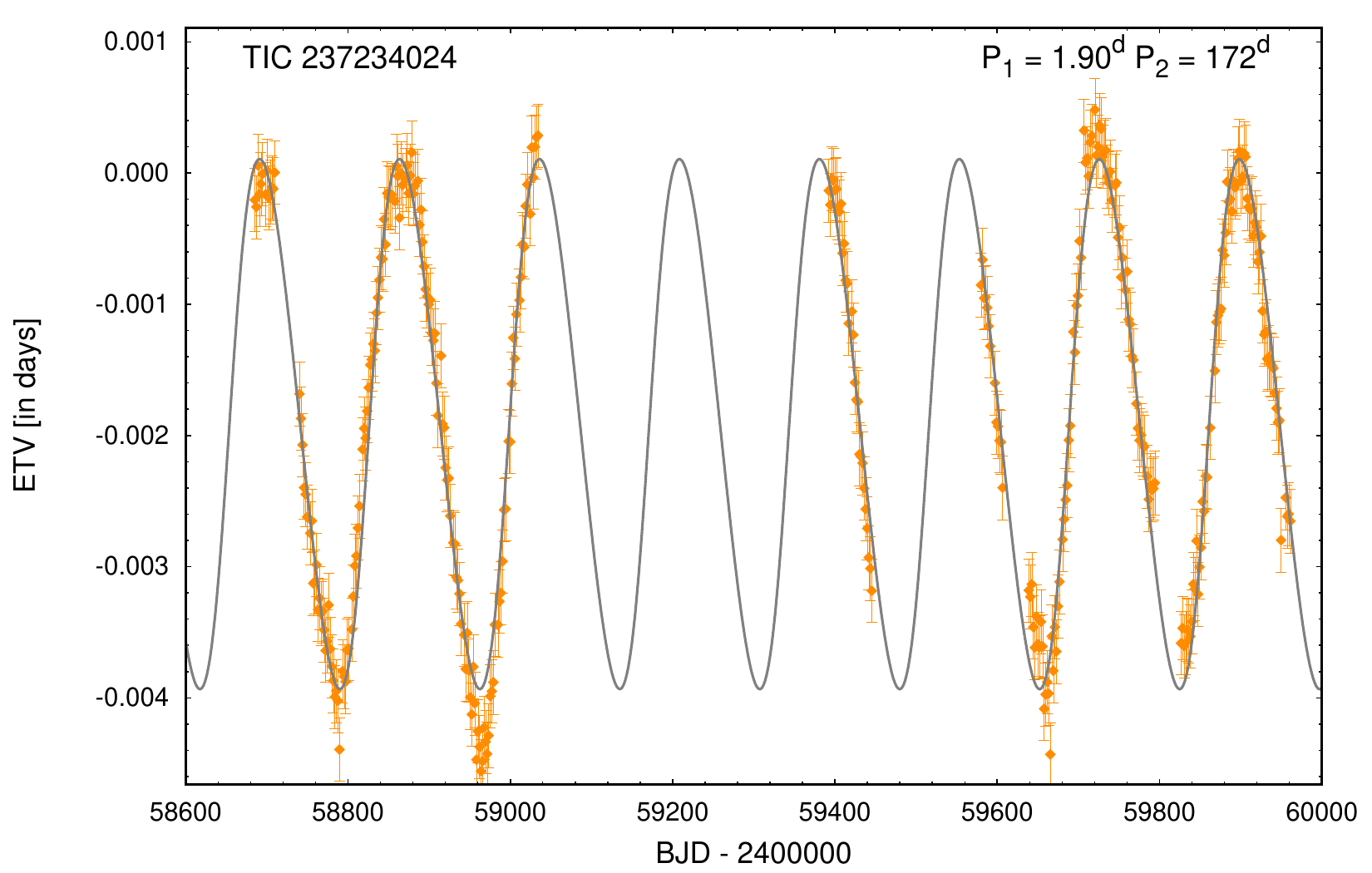}
\includegraphics[width=60mm]{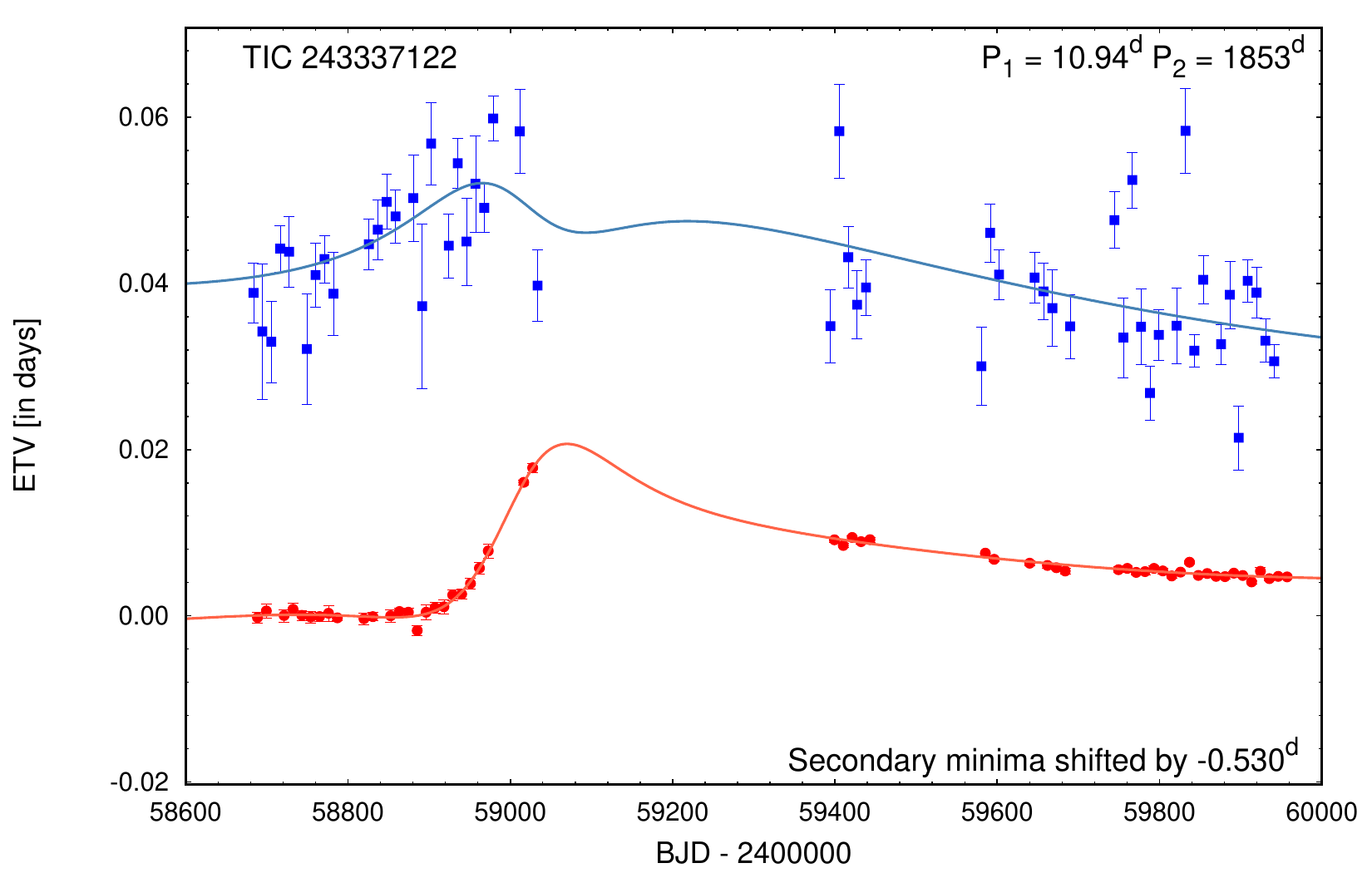}\includegraphics[width=60mm]{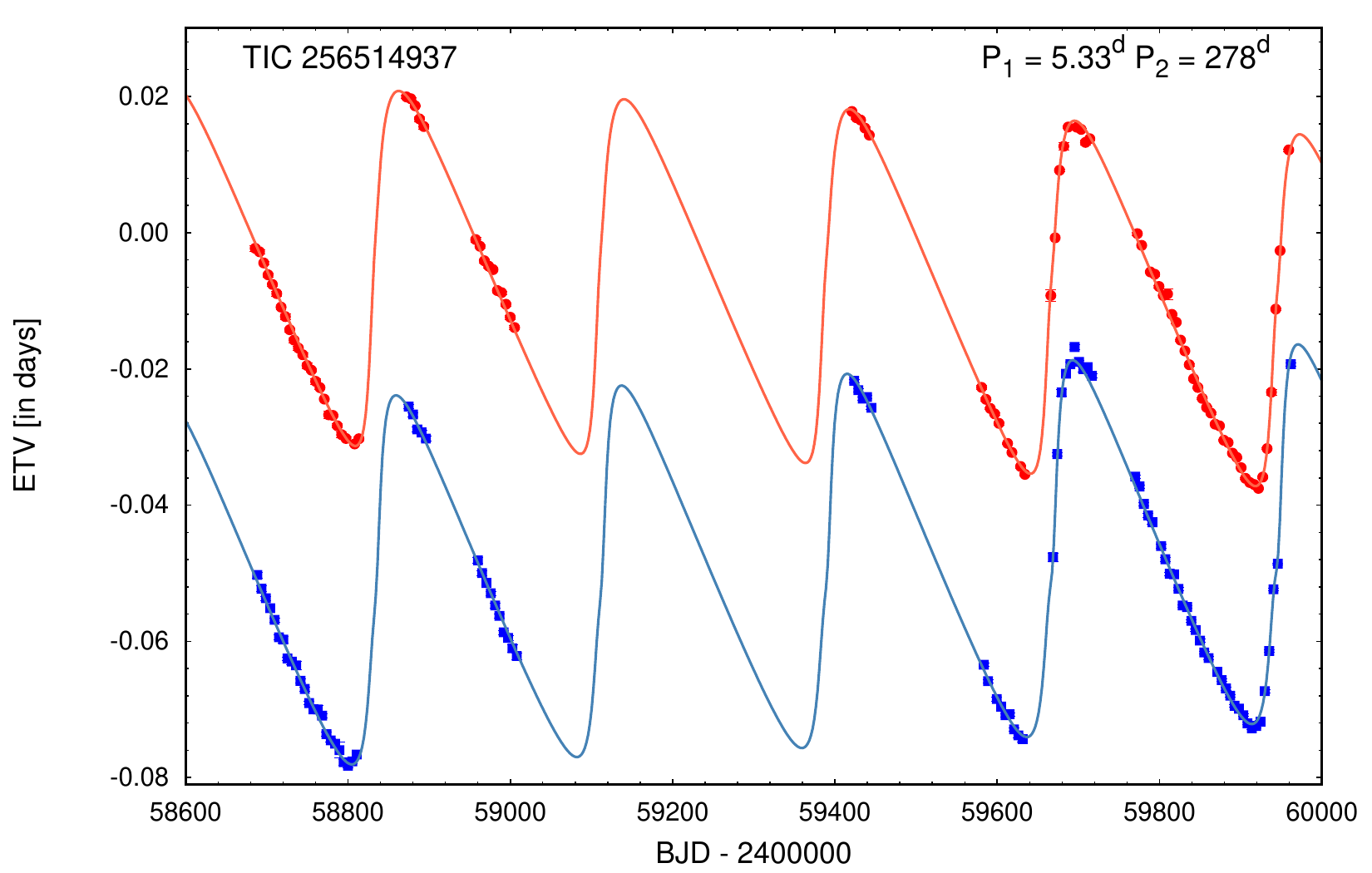}\includegraphics[width=60mm]{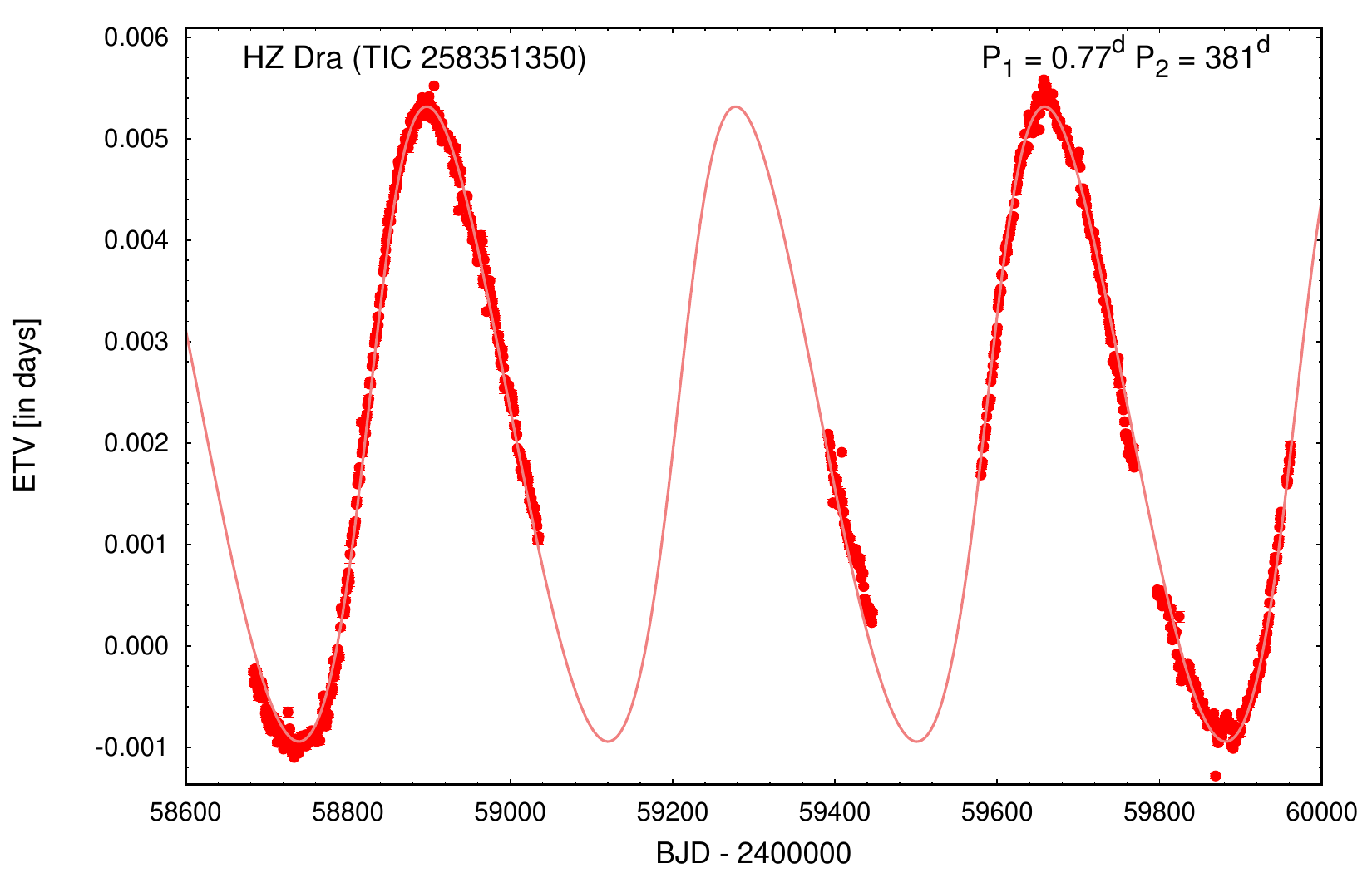}
\includegraphics[width=60mm]{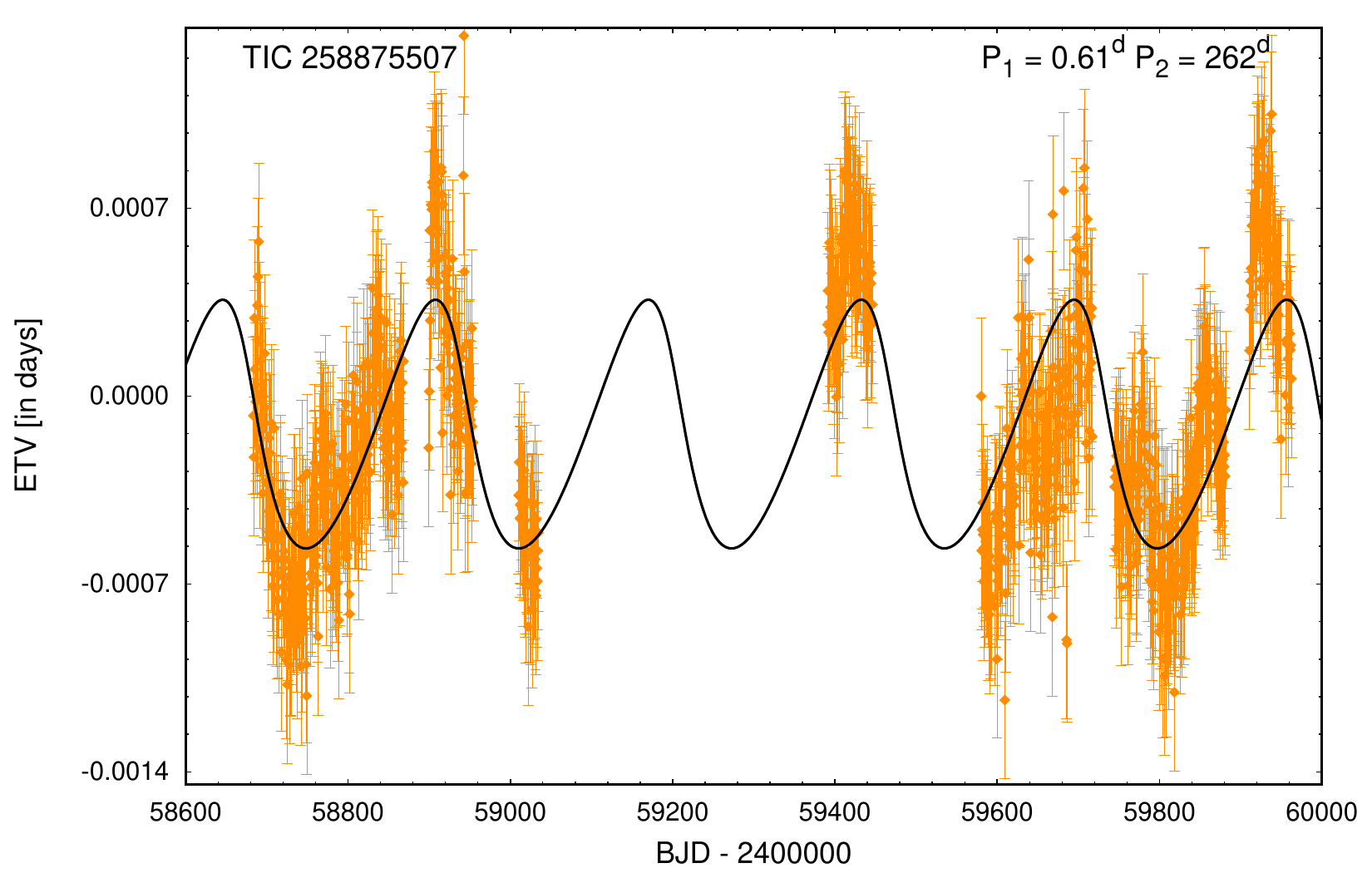}\includegraphics[width=60mm]{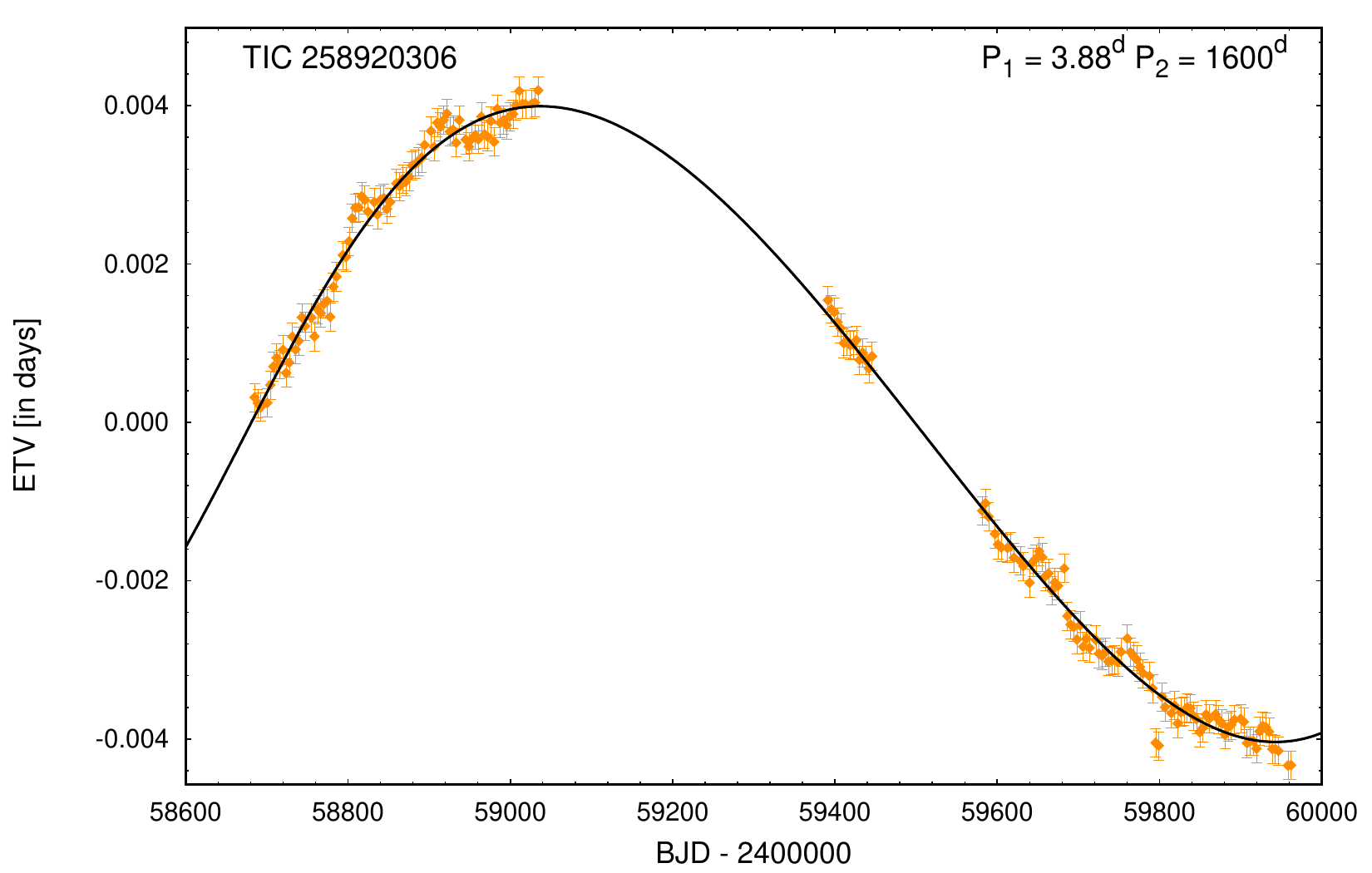}\includegraphics[width=60mm]{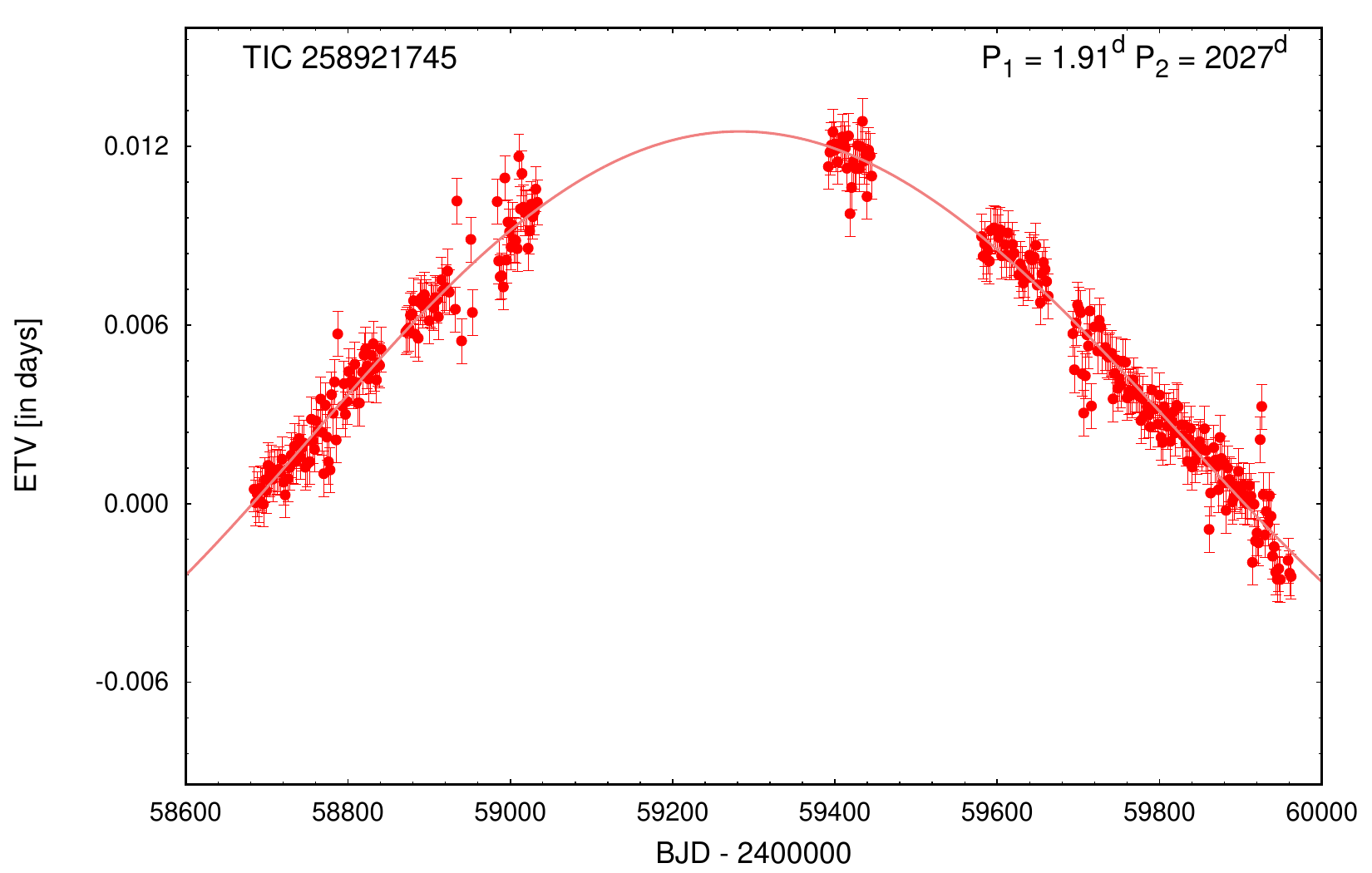}
\caption{(continued)}
\end{figure*}

\addtocounter{figure}{-1}

\begin{figure*}
\includegraphics[width=60mm]{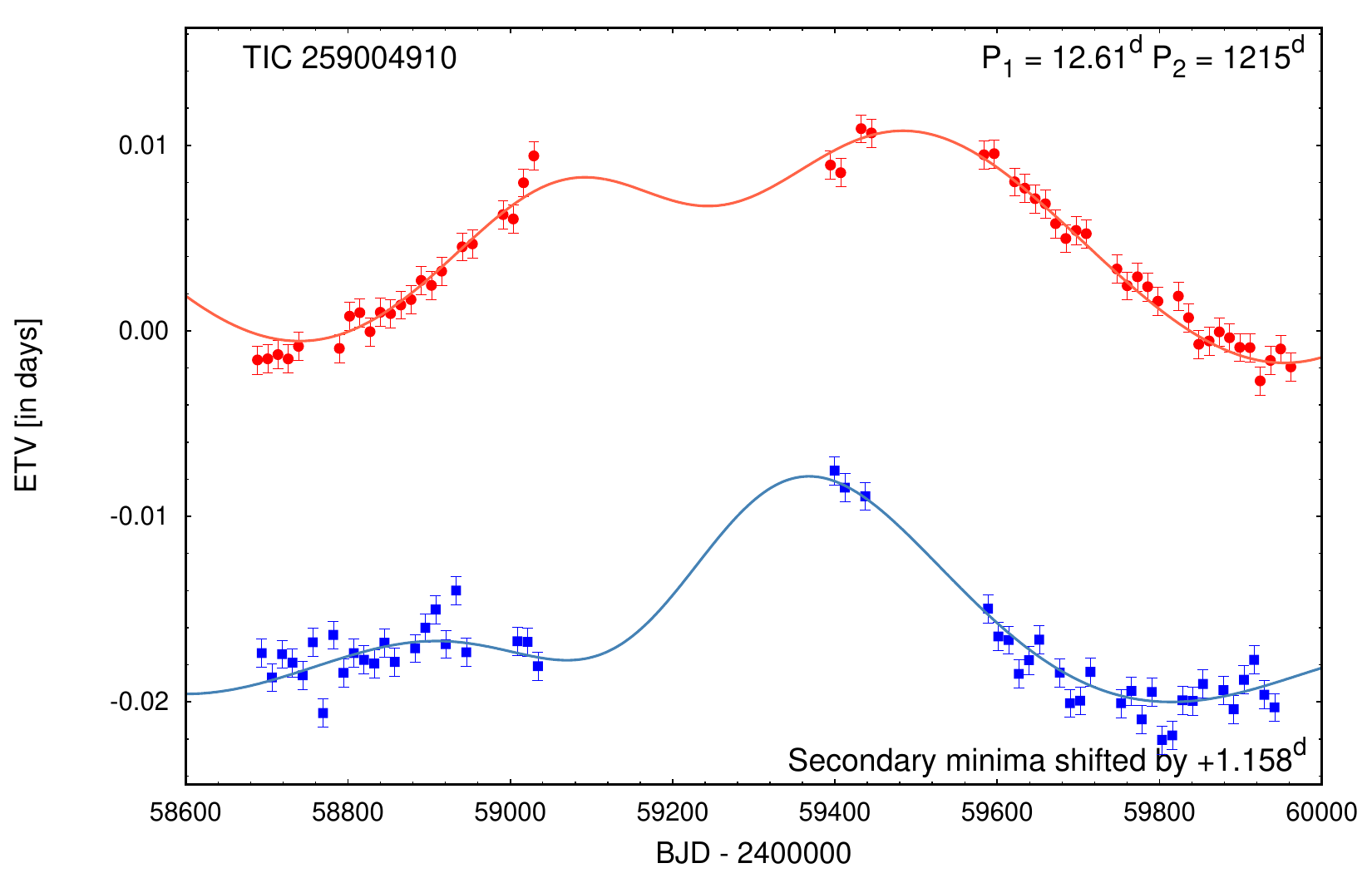}\includegraphics[width=60mm]{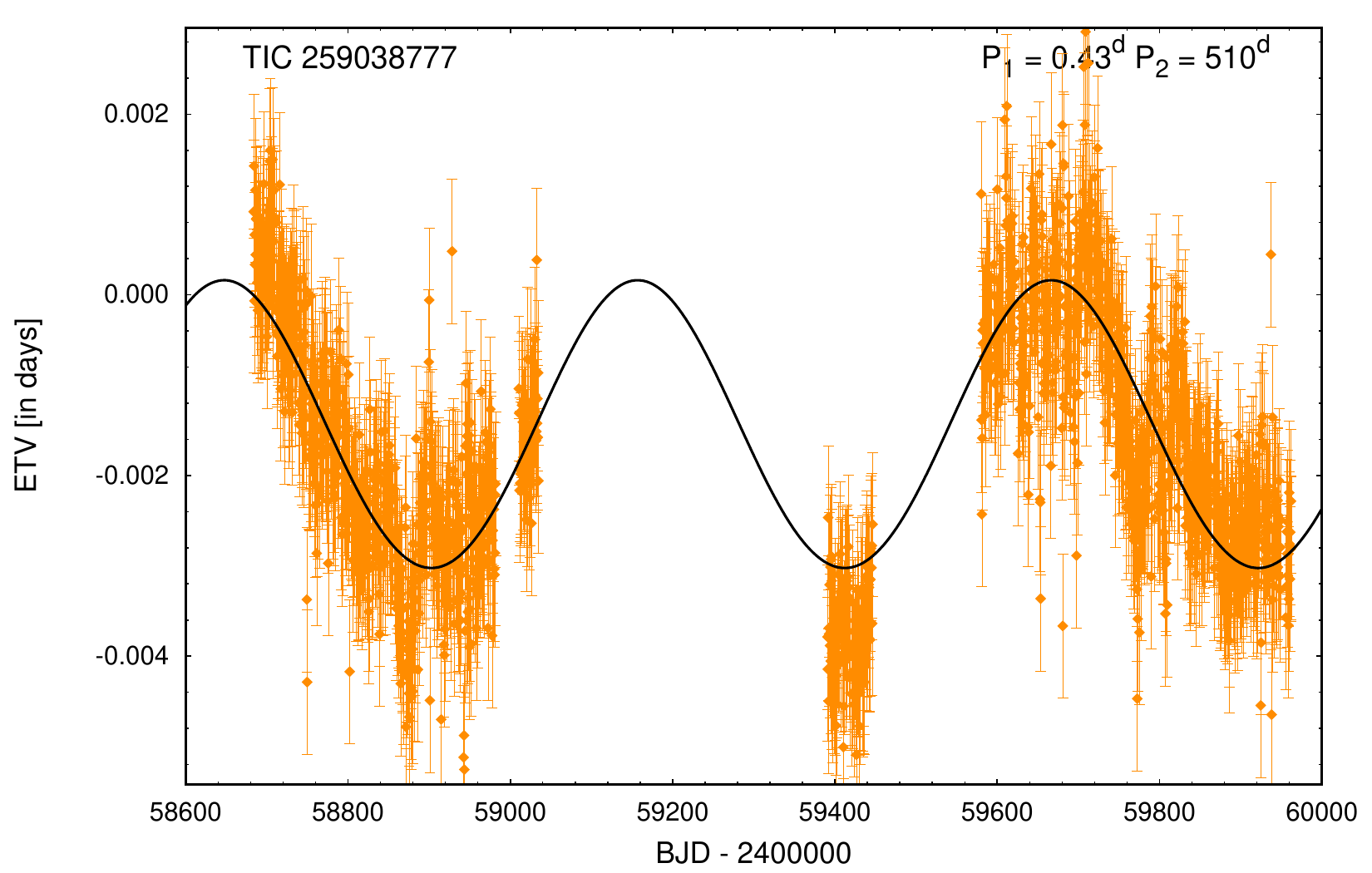}\includegraphics[width=60mm]{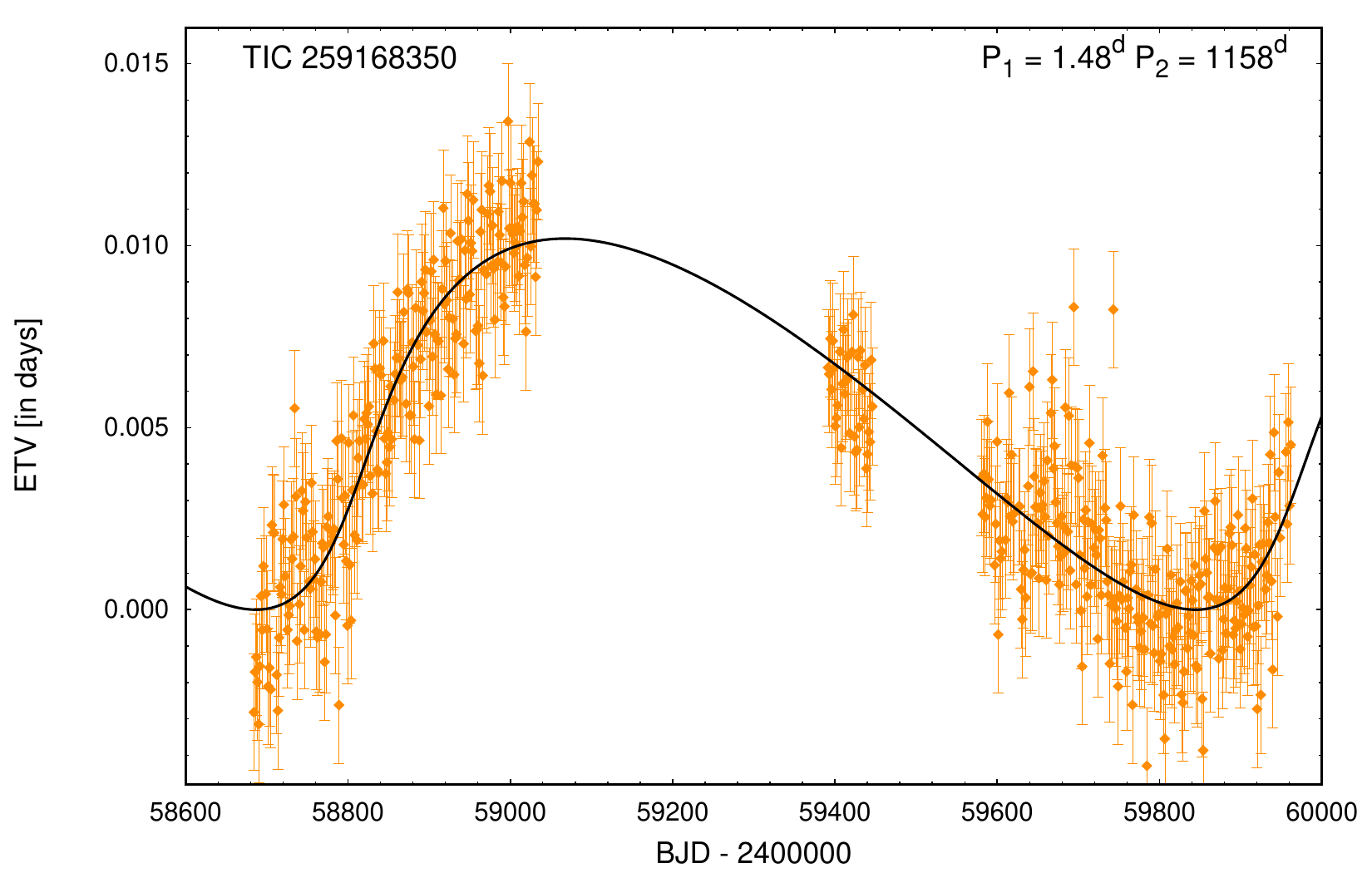}
\includegraphics[width=60mm]{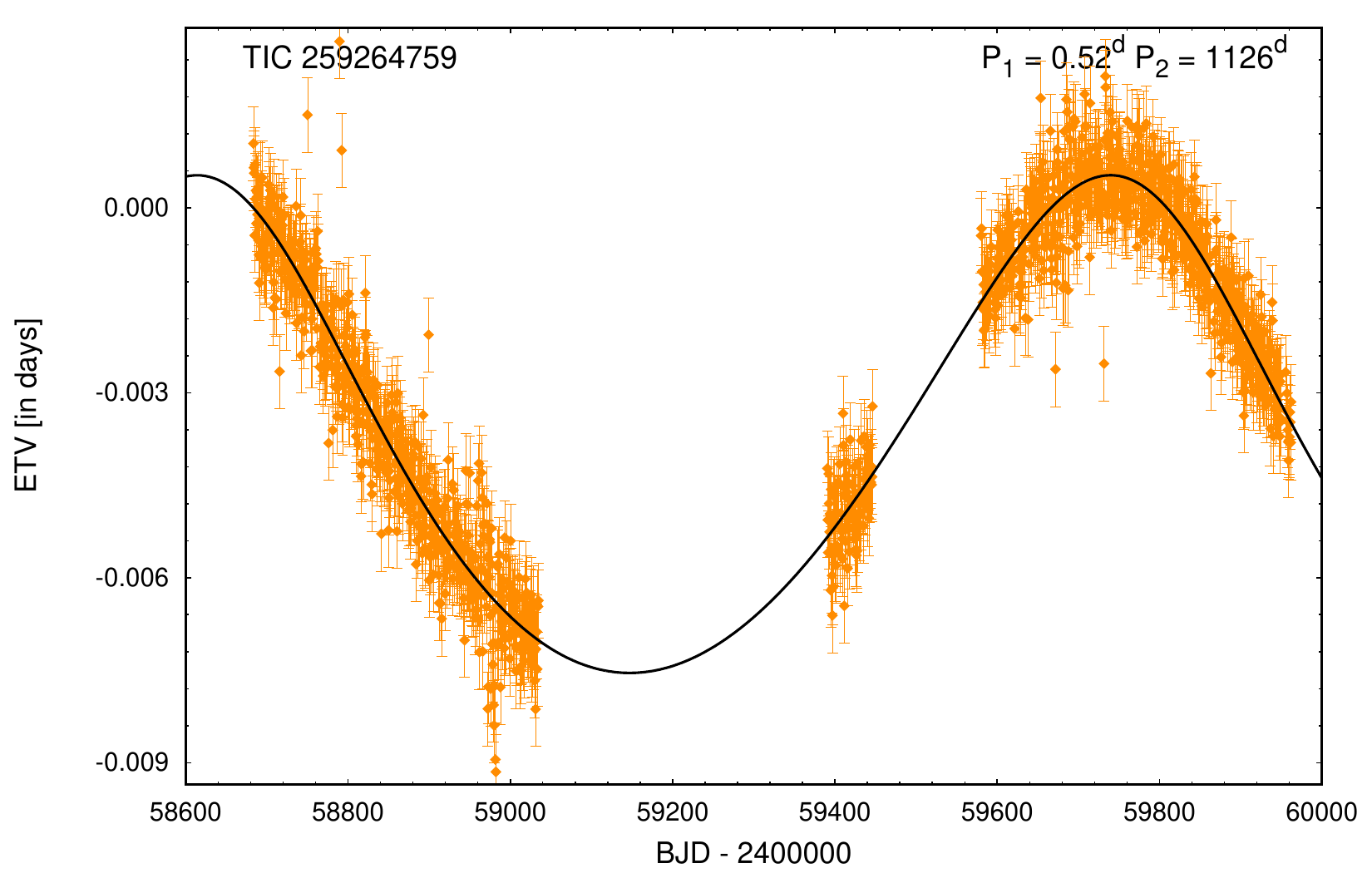}\includegraphics[width=60mm]{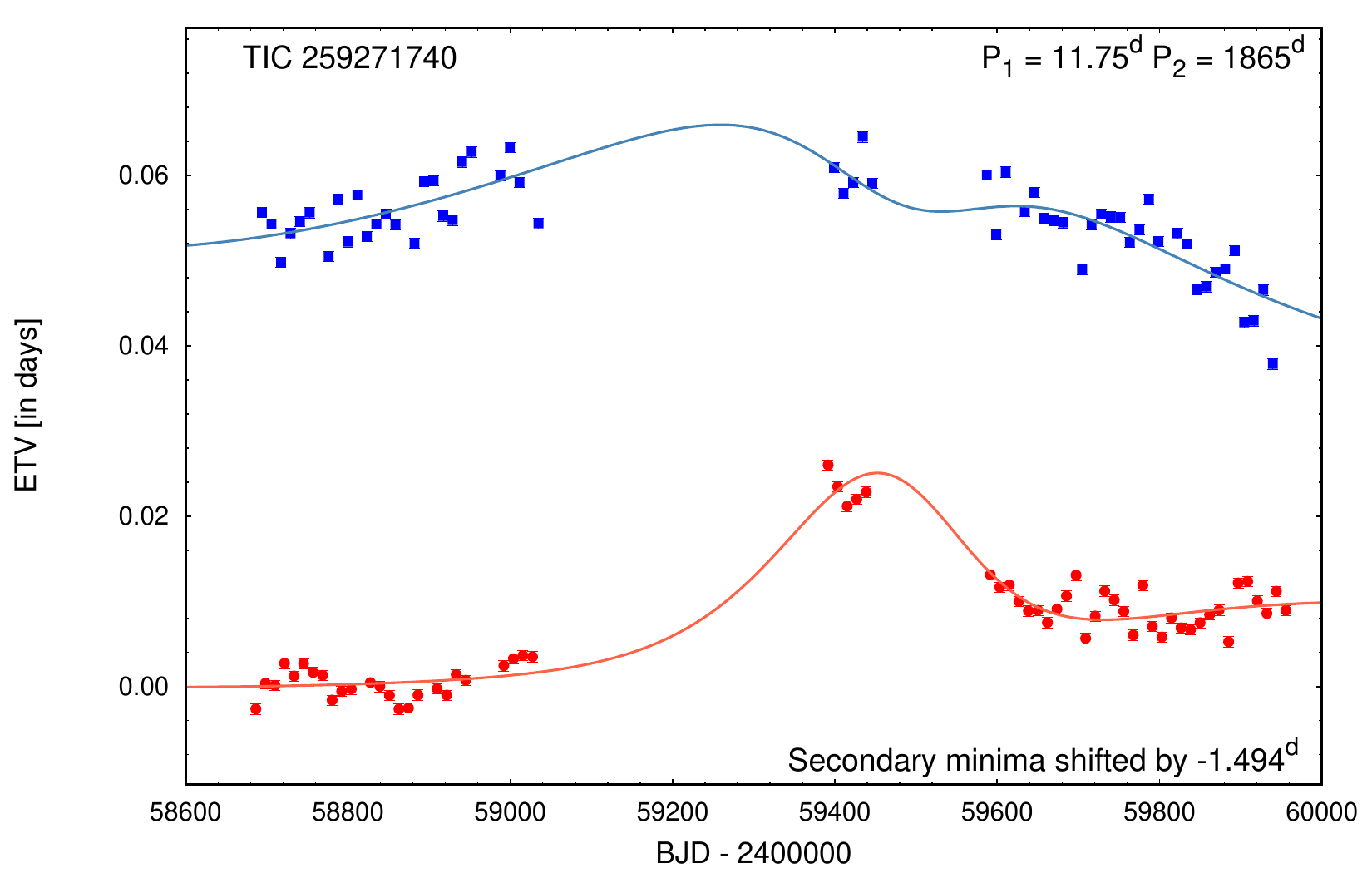}\includegraphics[width=60mm]{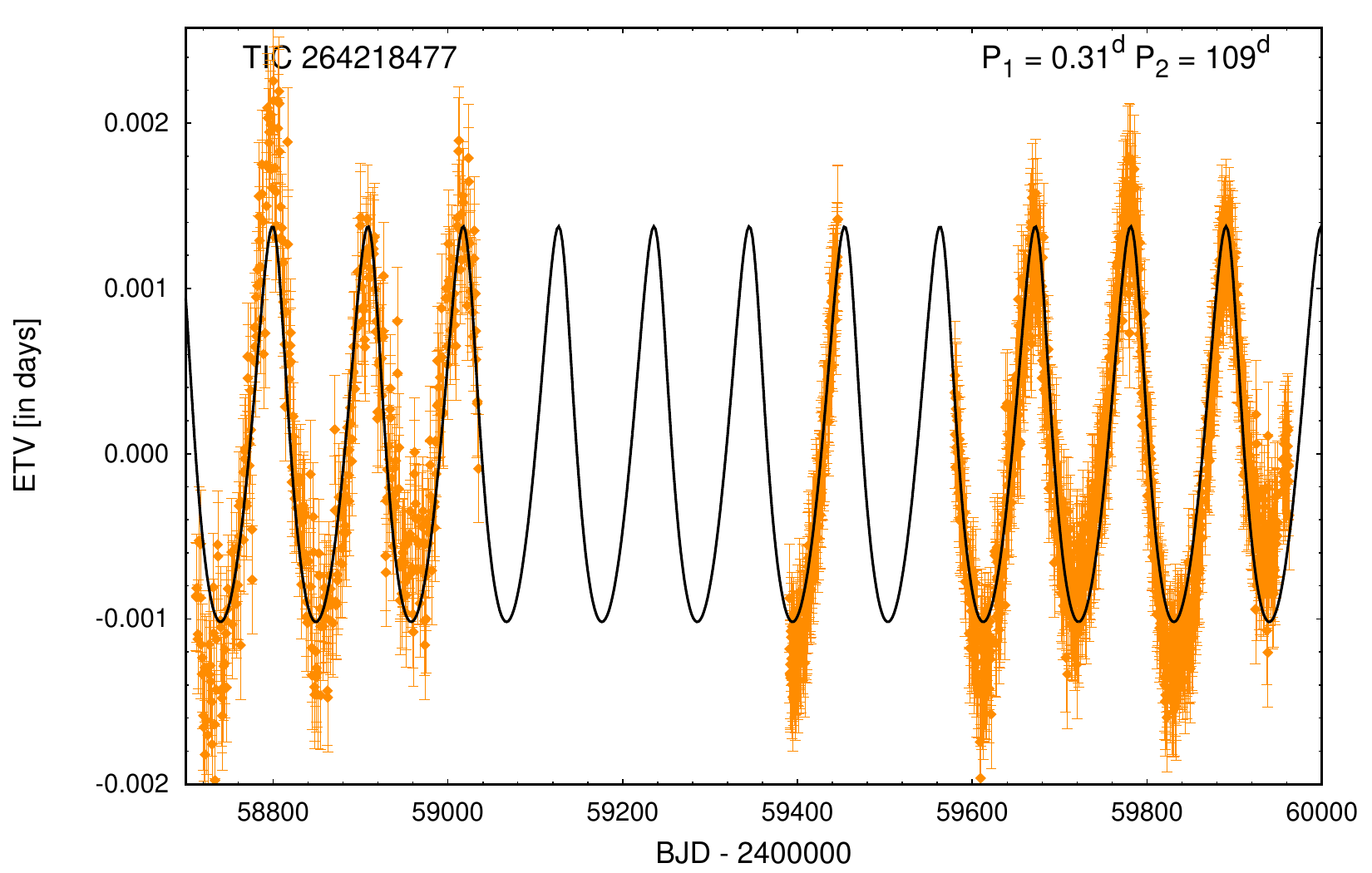}
\includegraphics[width=60mm]{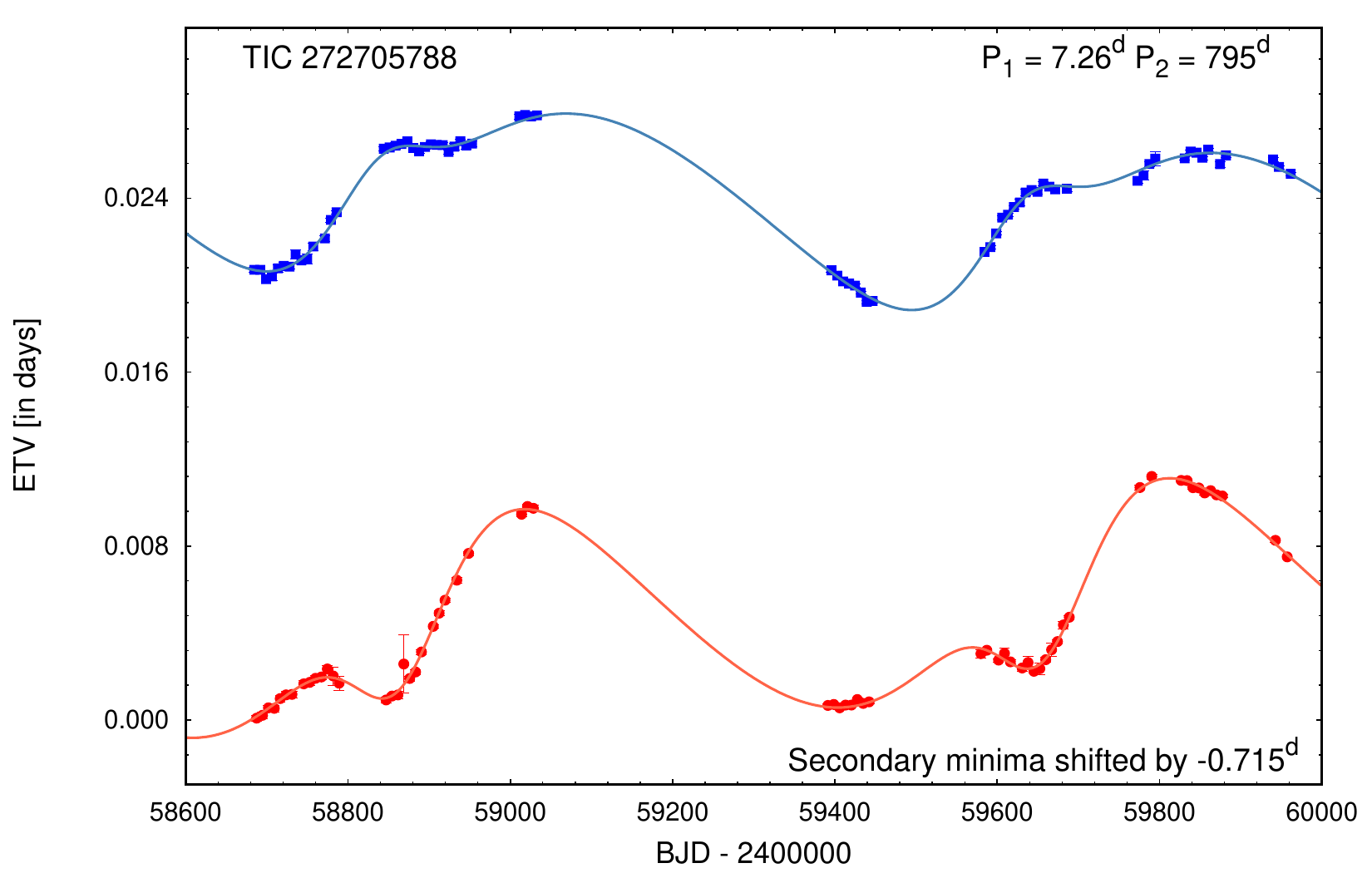}\includegraphics[width=60mm]{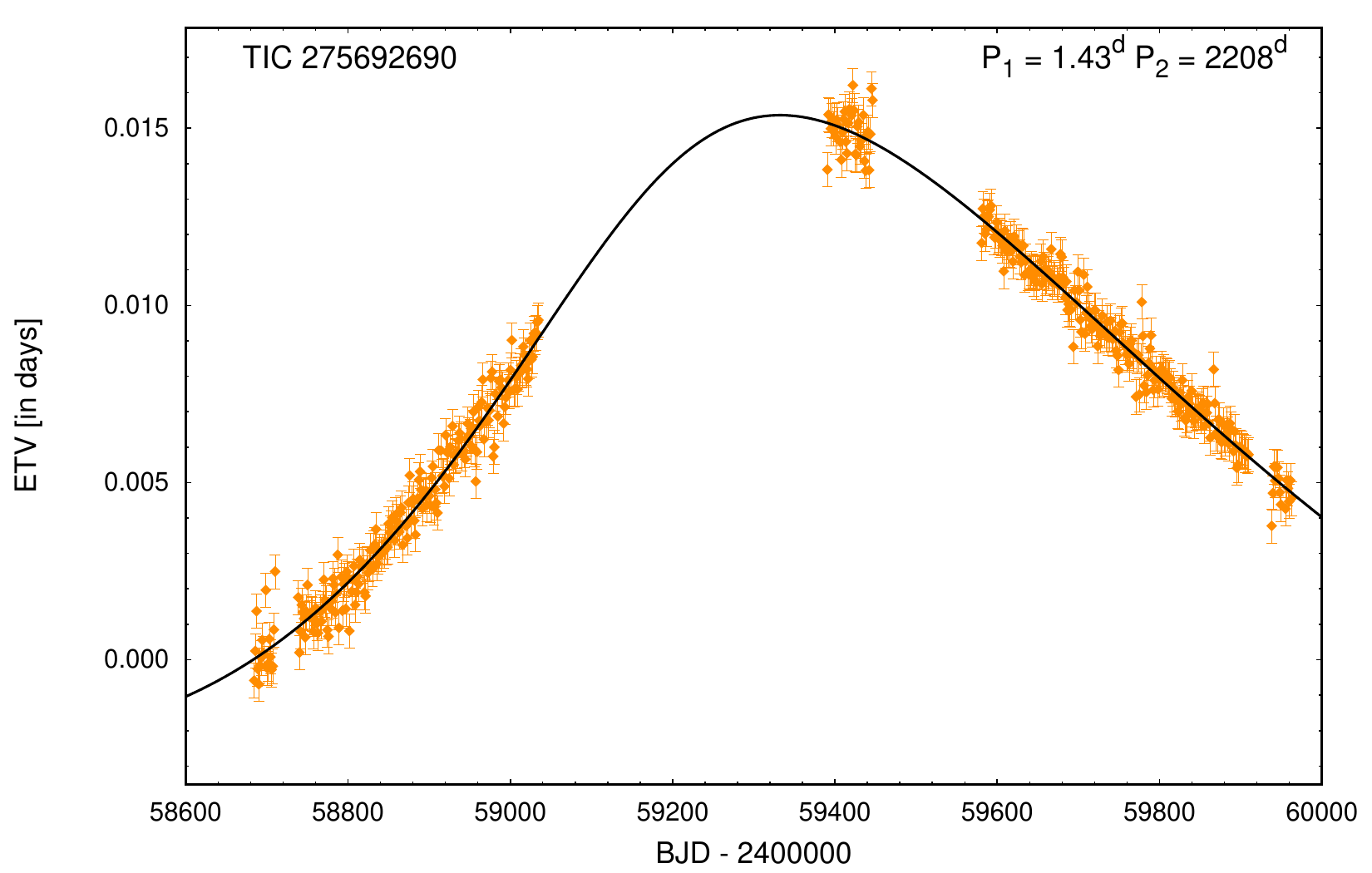}\includegraphics[width=60mm]{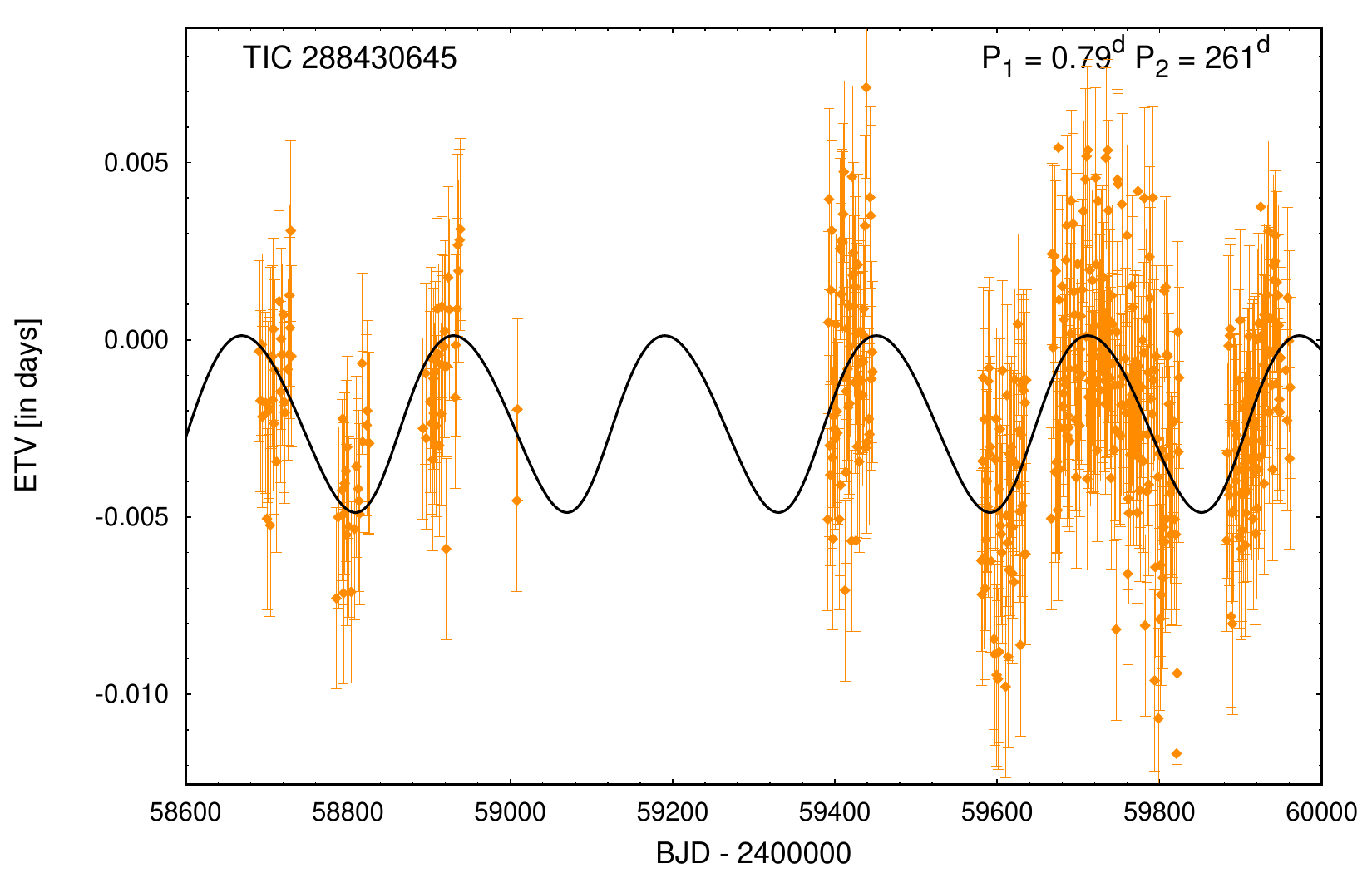}
\includegraphics[width=60mm]{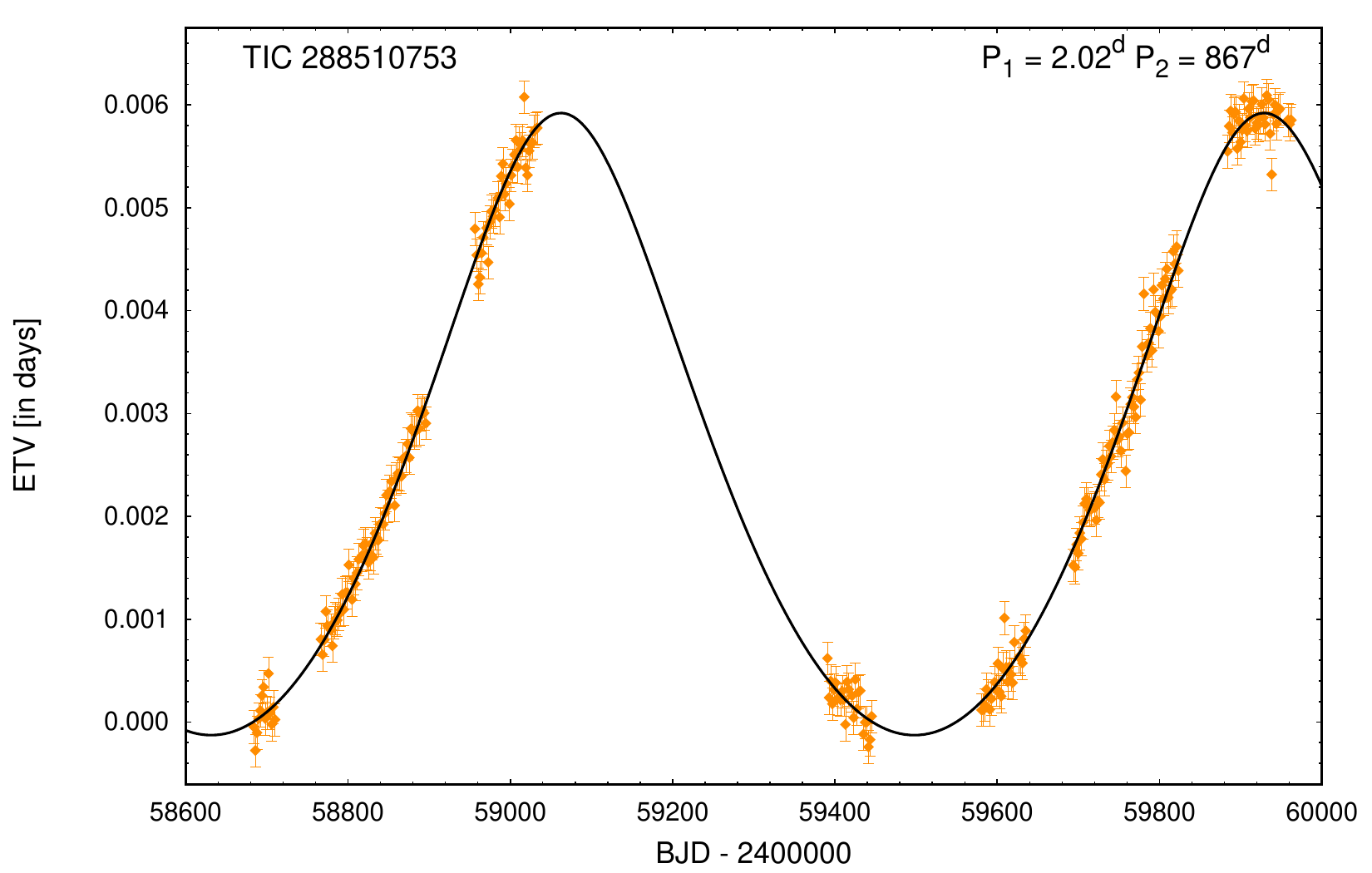}\includegraphics[width=60mm]{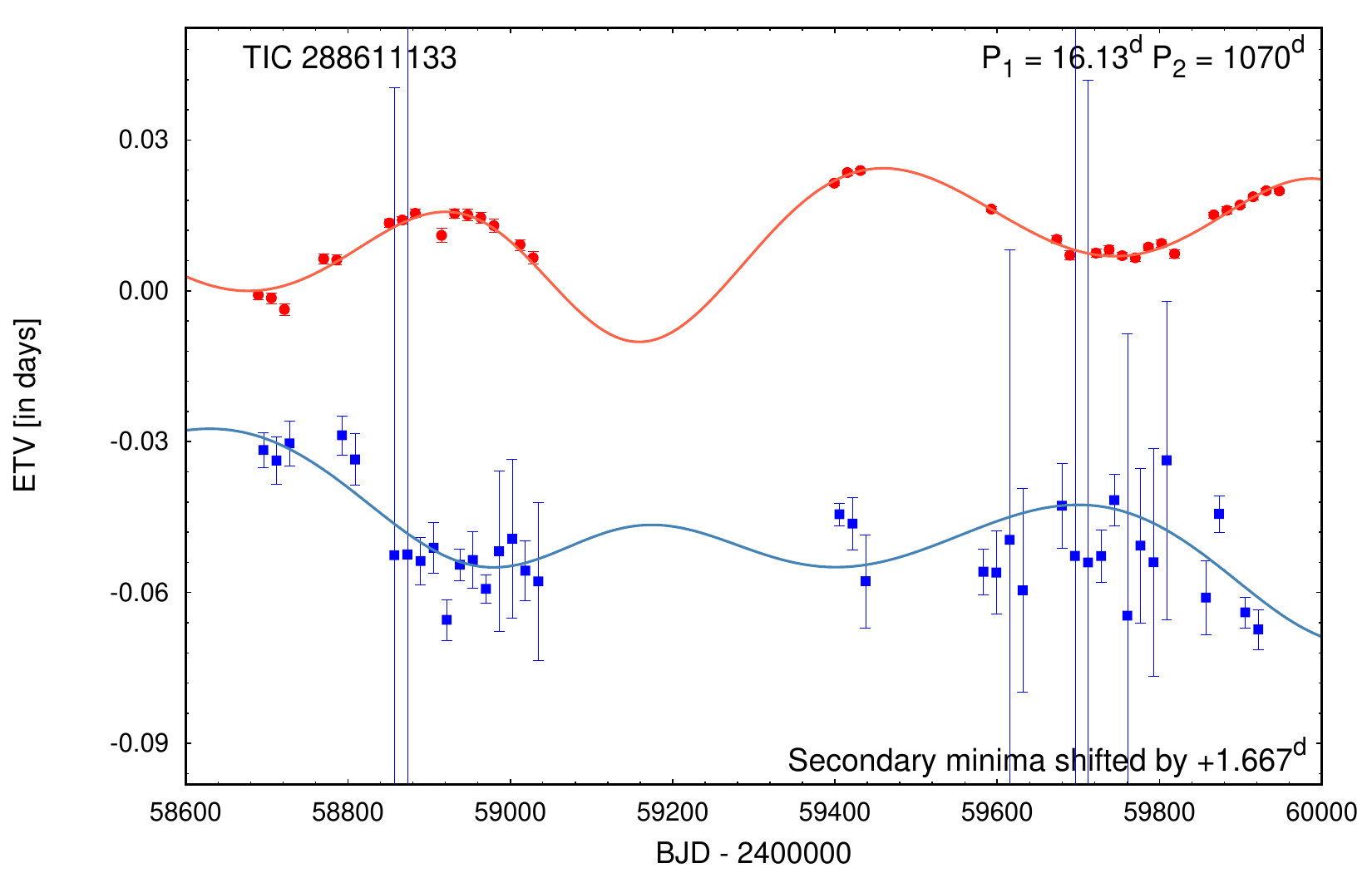}\includegraphics[width=60mm]{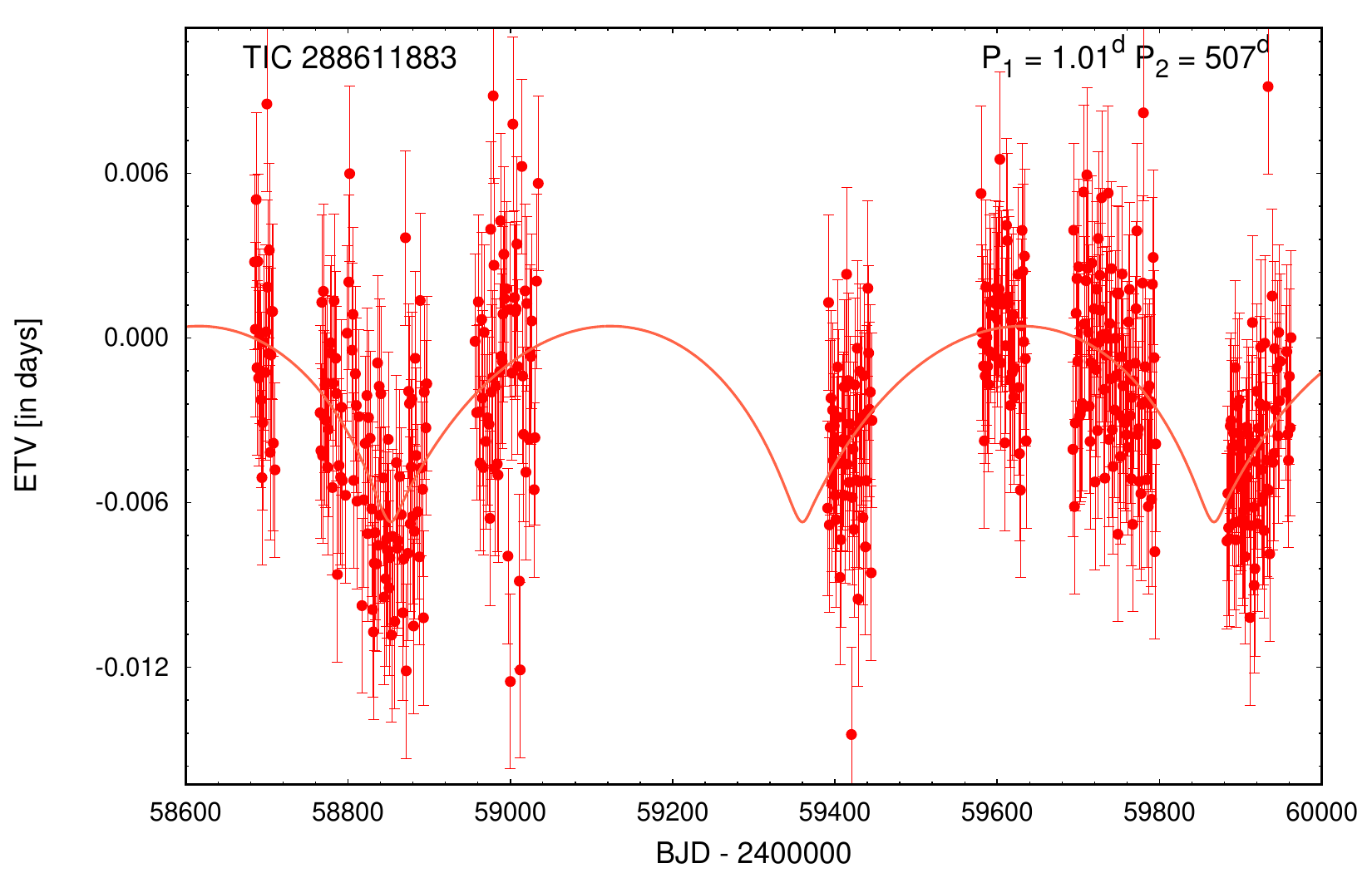}
\includegraphics[width=60mm]{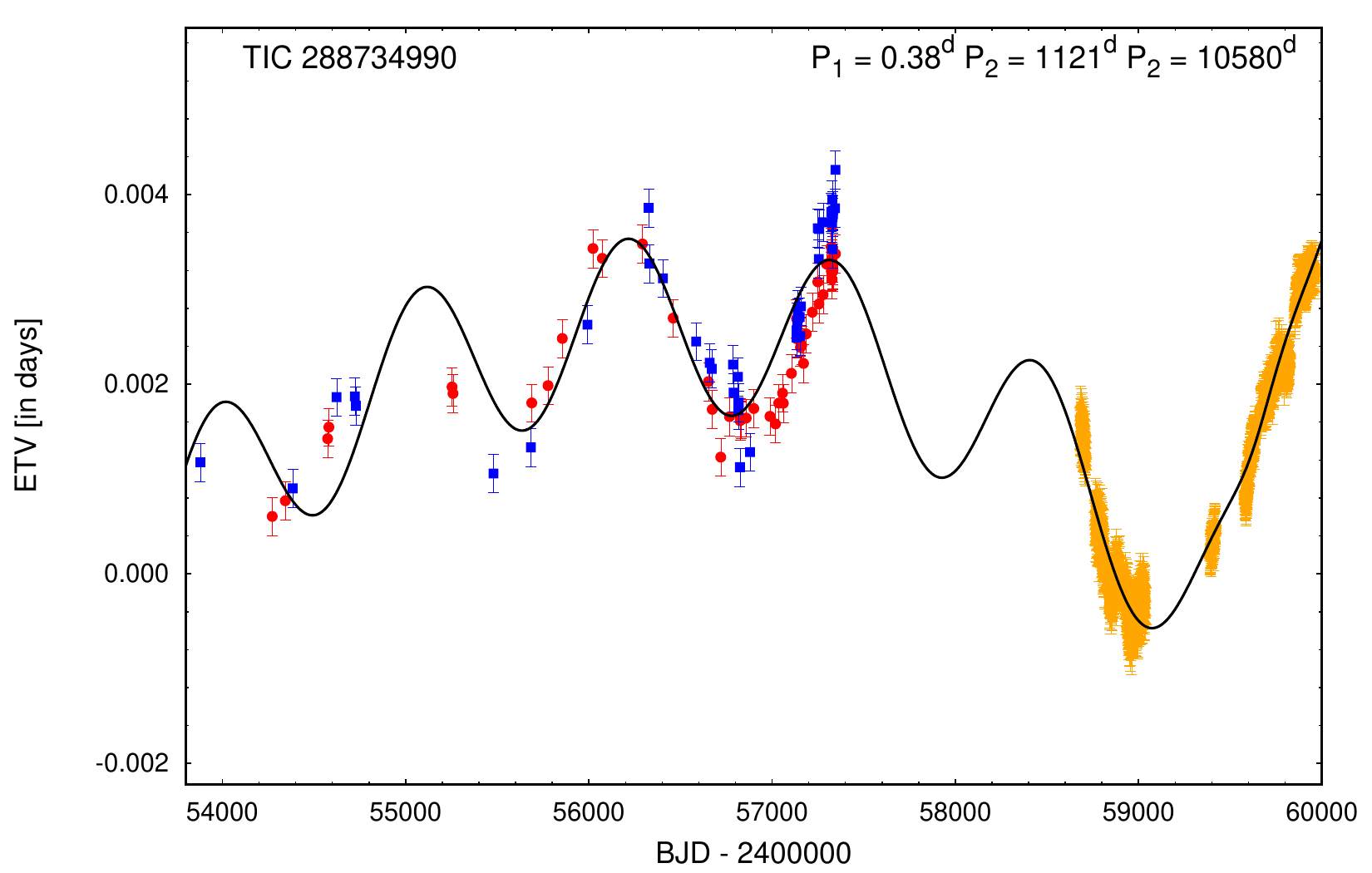}\includegraphics[width=60mm]{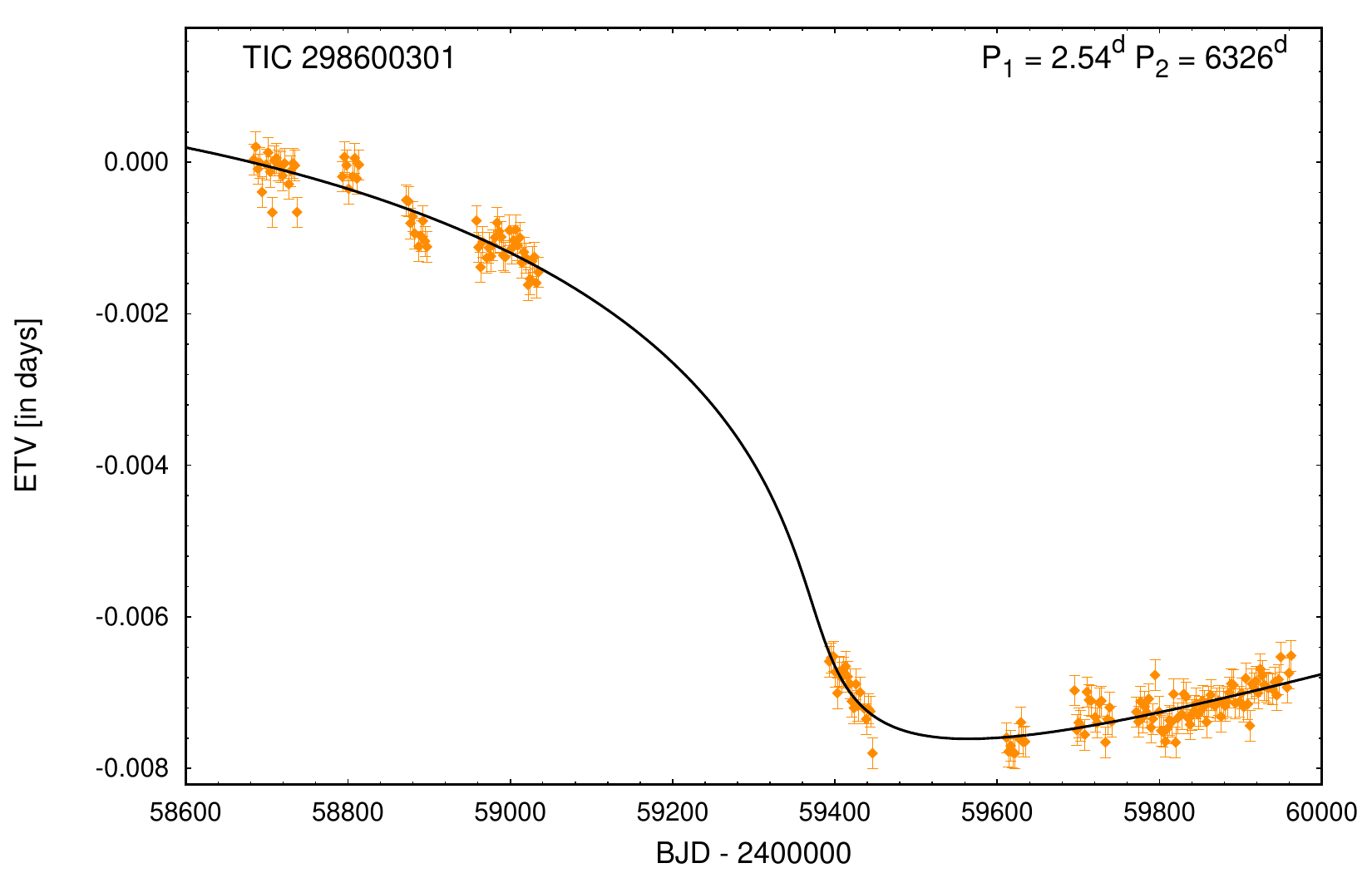}\includegraphics[width=60mm]{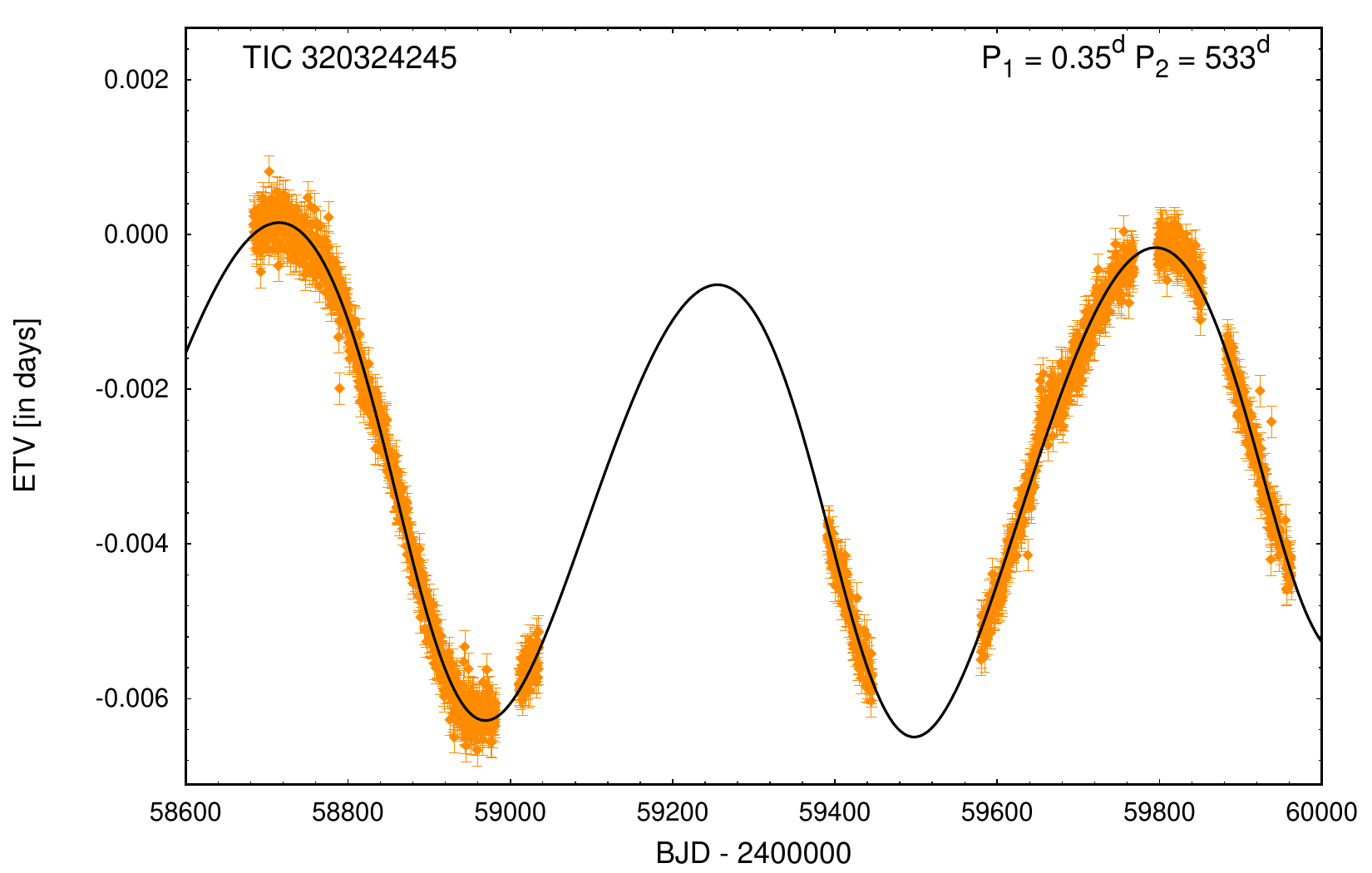}
\includegraphics[width=60mm]{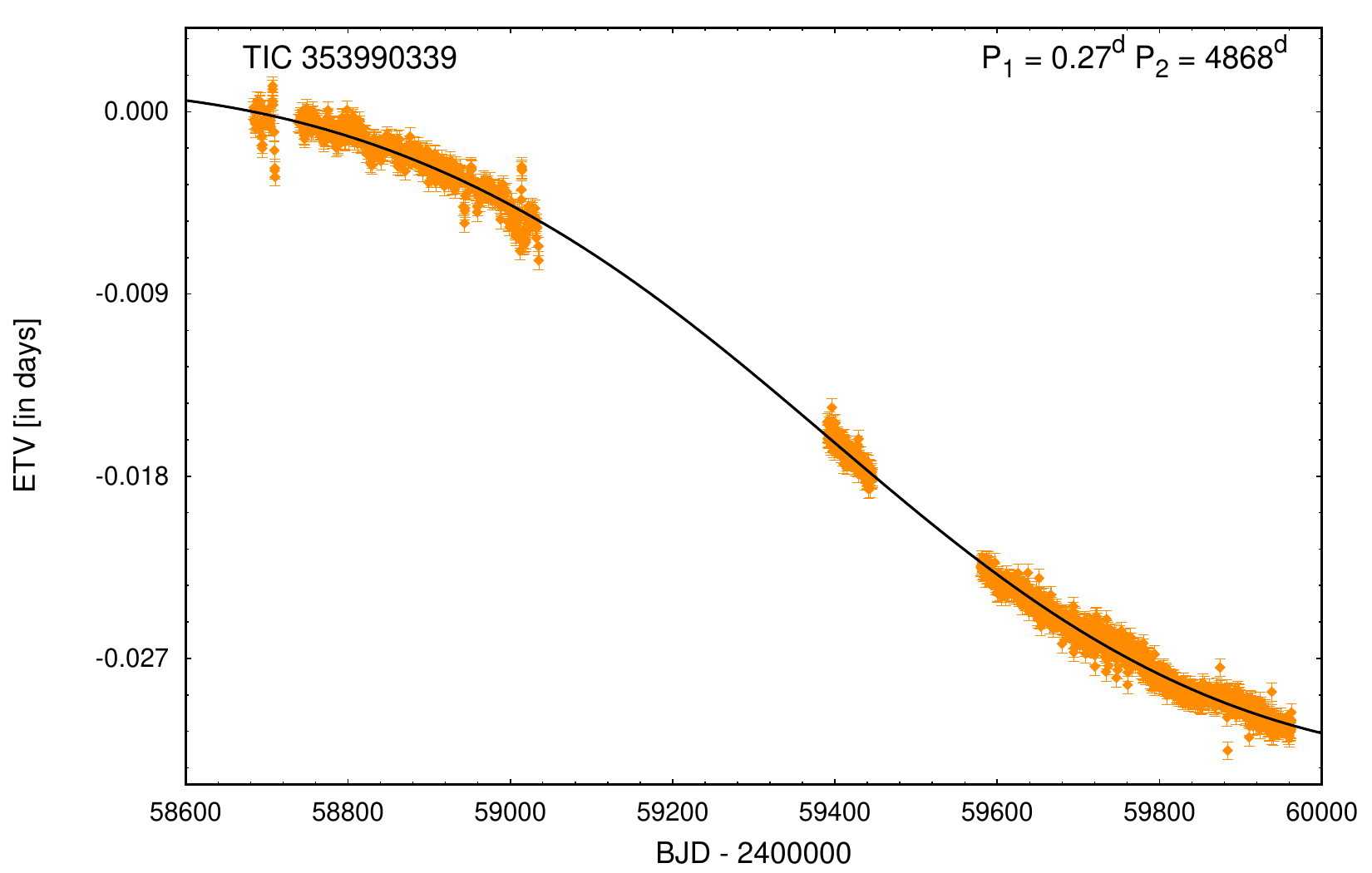}\includegraphics[width=60mm]{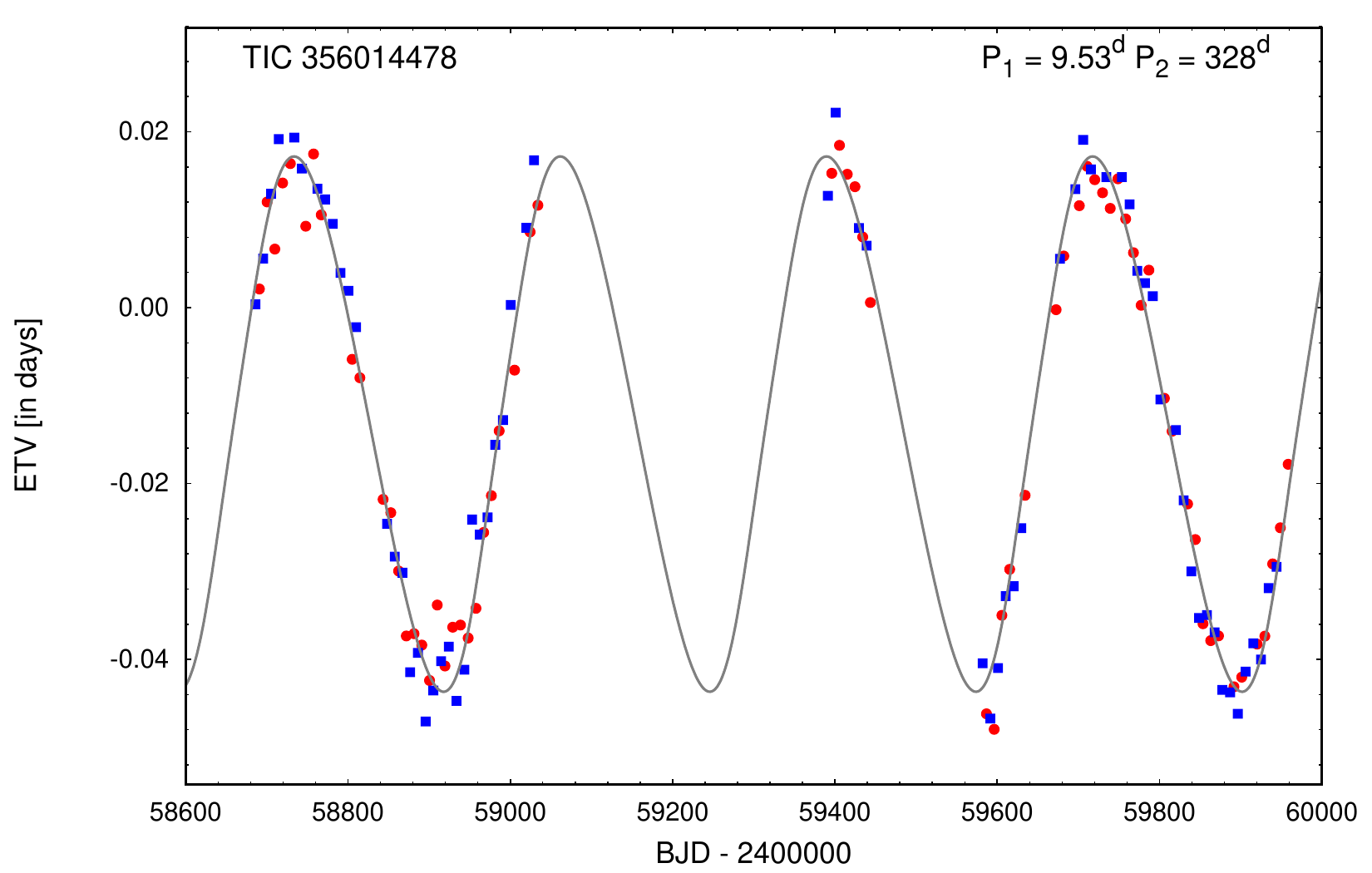}\includegraphics[width=60mm]{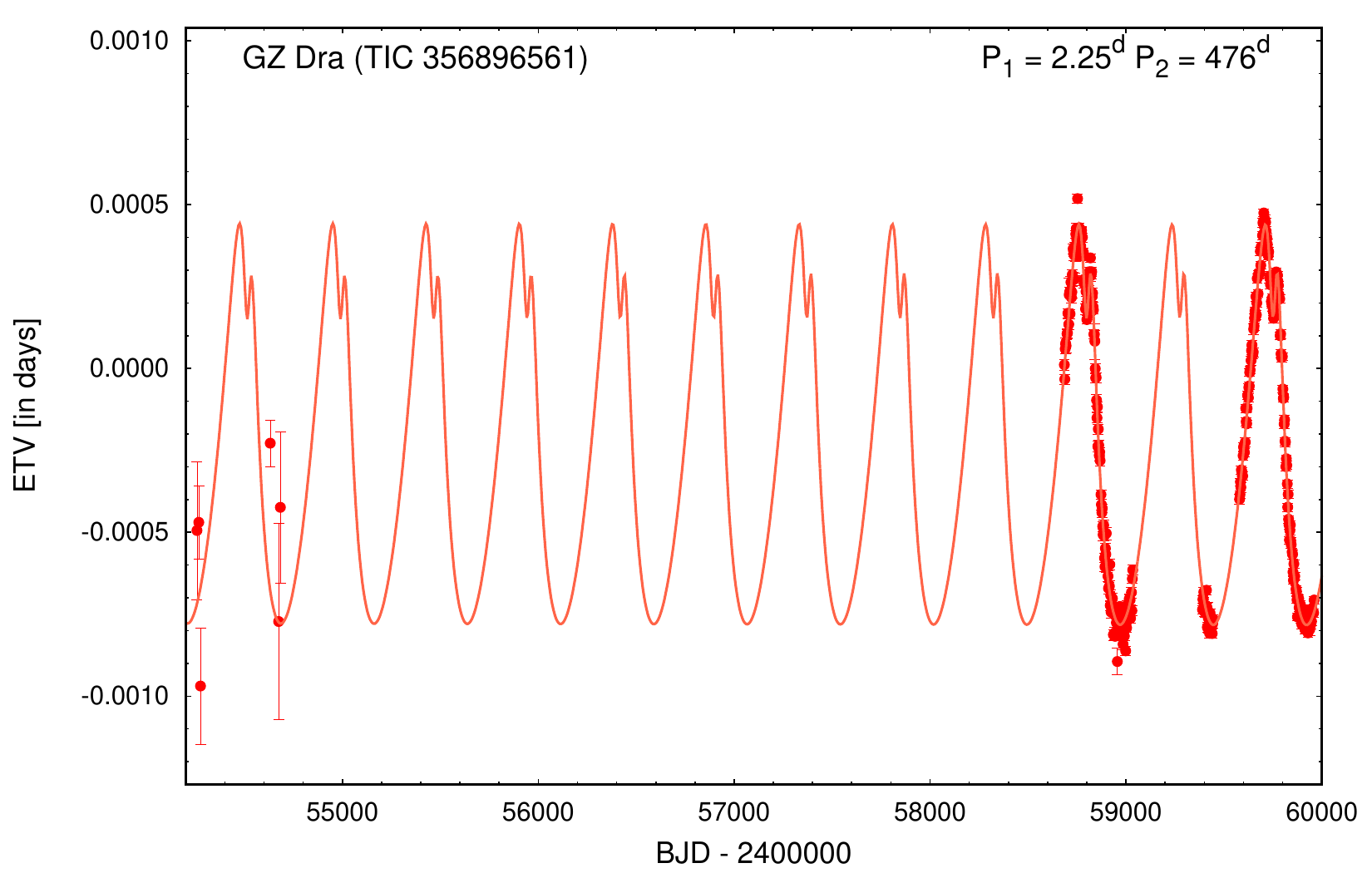}
\caption{(continued)}
\end{figure*}

\addtocounter{figure}{-1}

\begin{figure*}
\includegraphics[width=60mm]{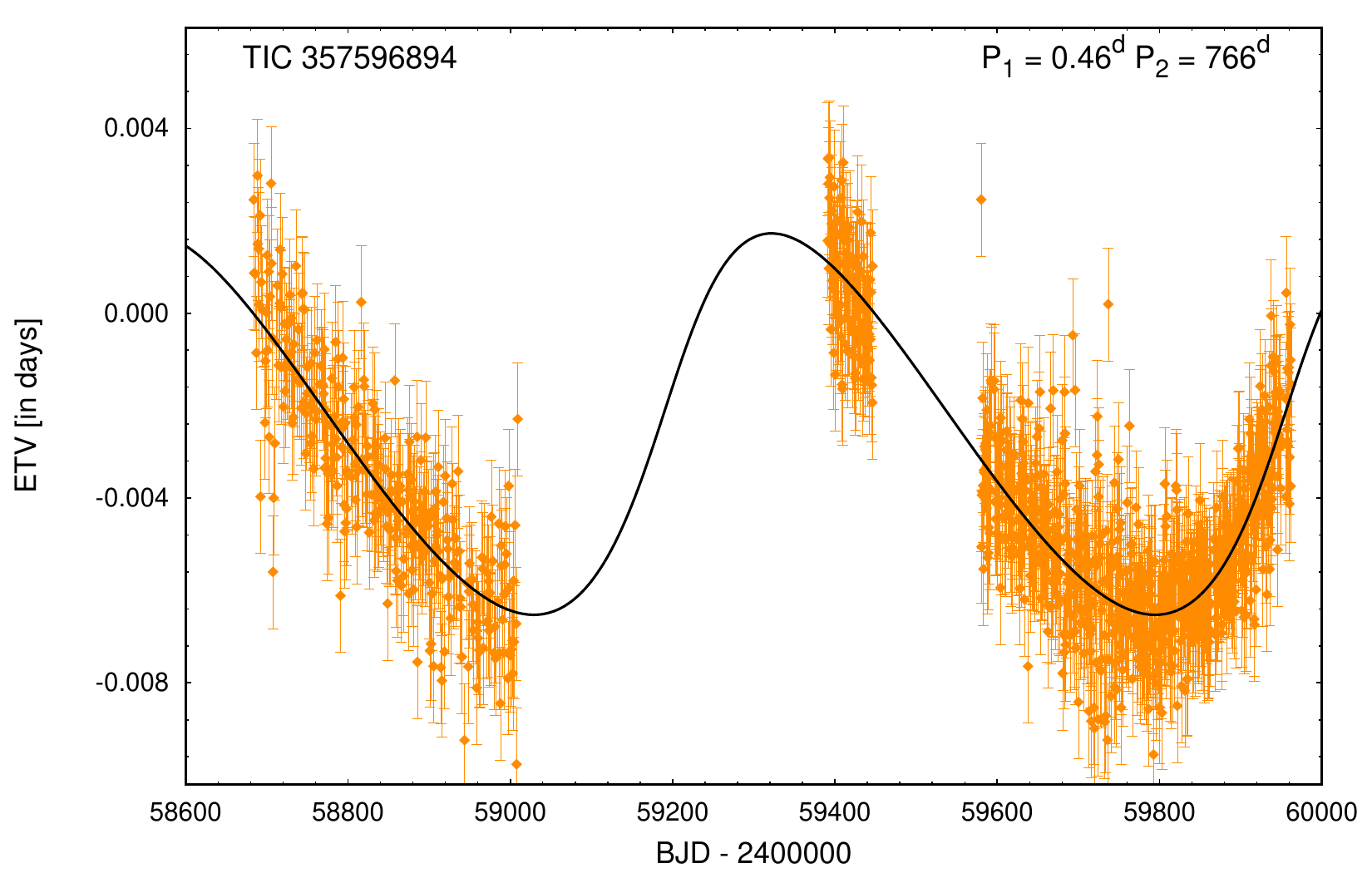}\includegraphics[width=60mm]{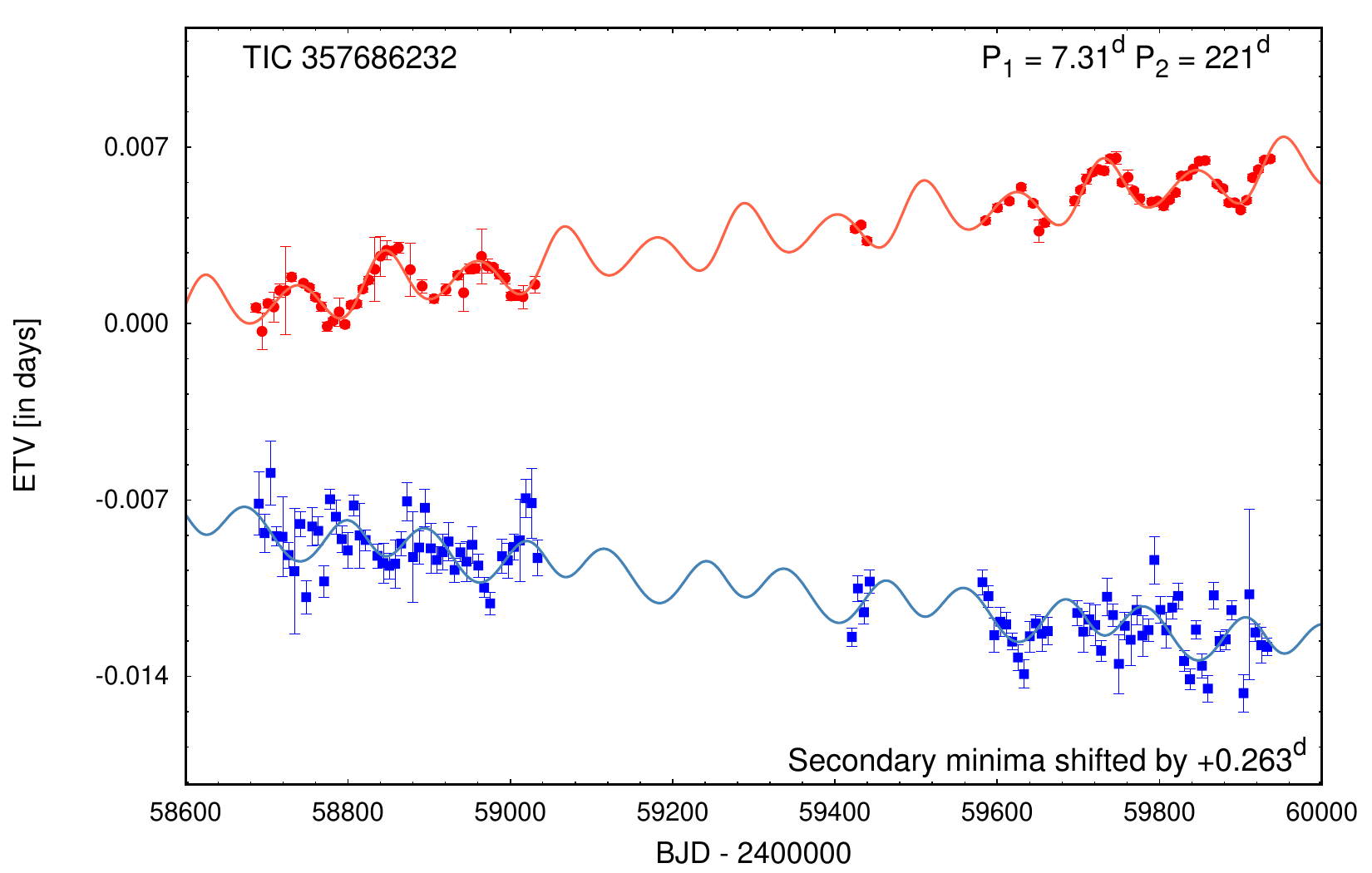}\includegraphics[width=60mm]{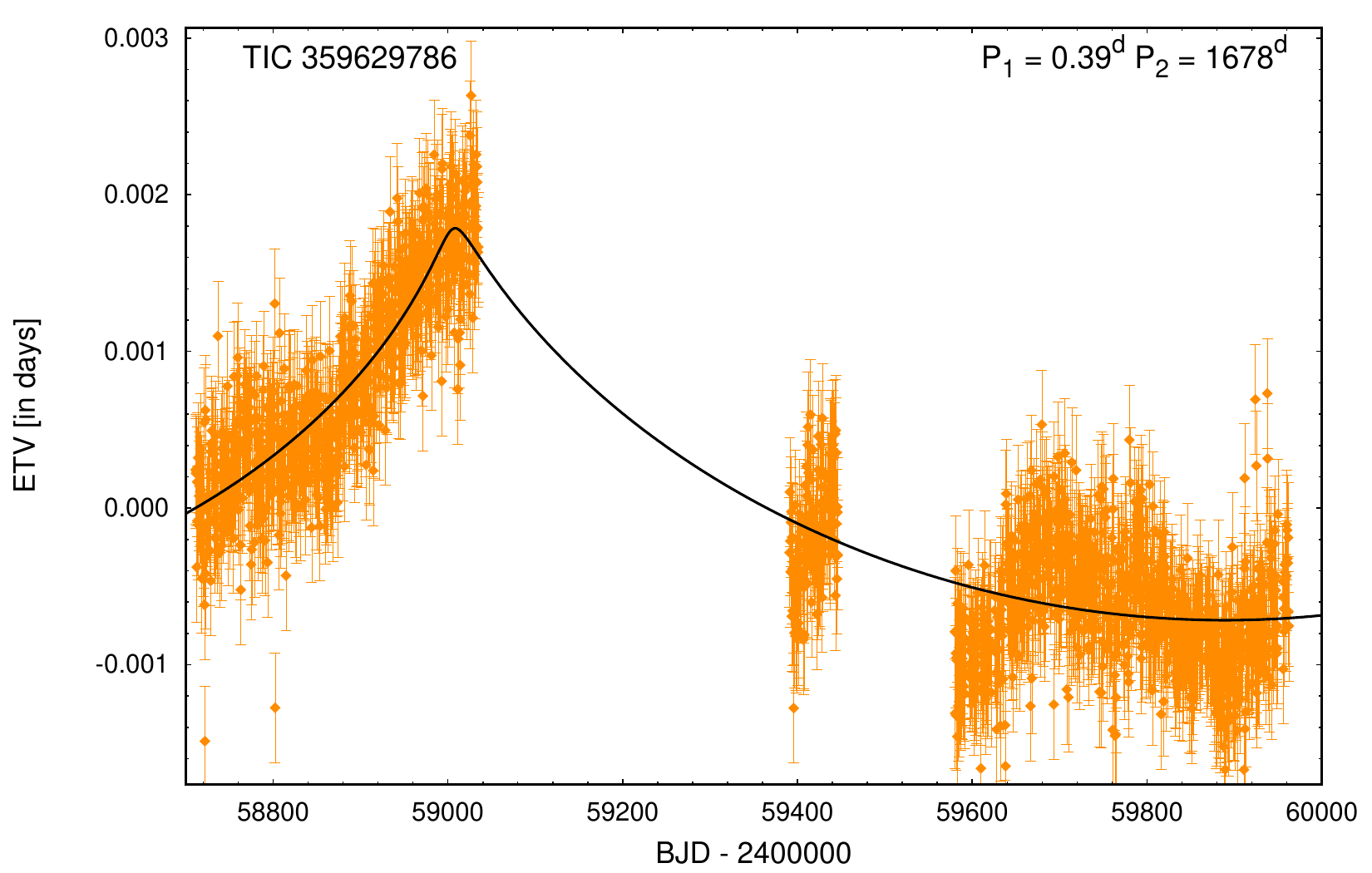}
\includegraphics[width=60mm]{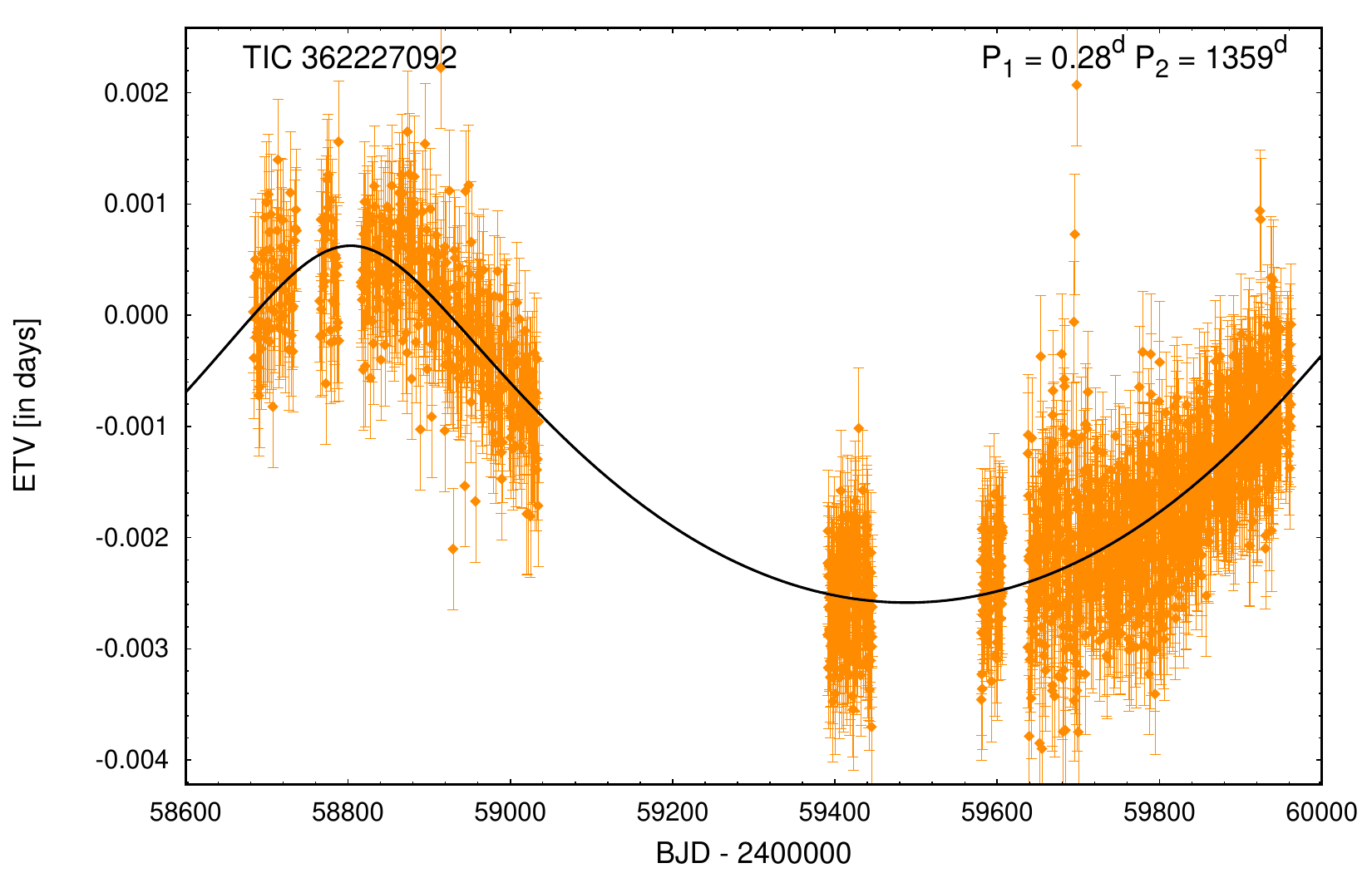}\includegraphics[width=60mm]{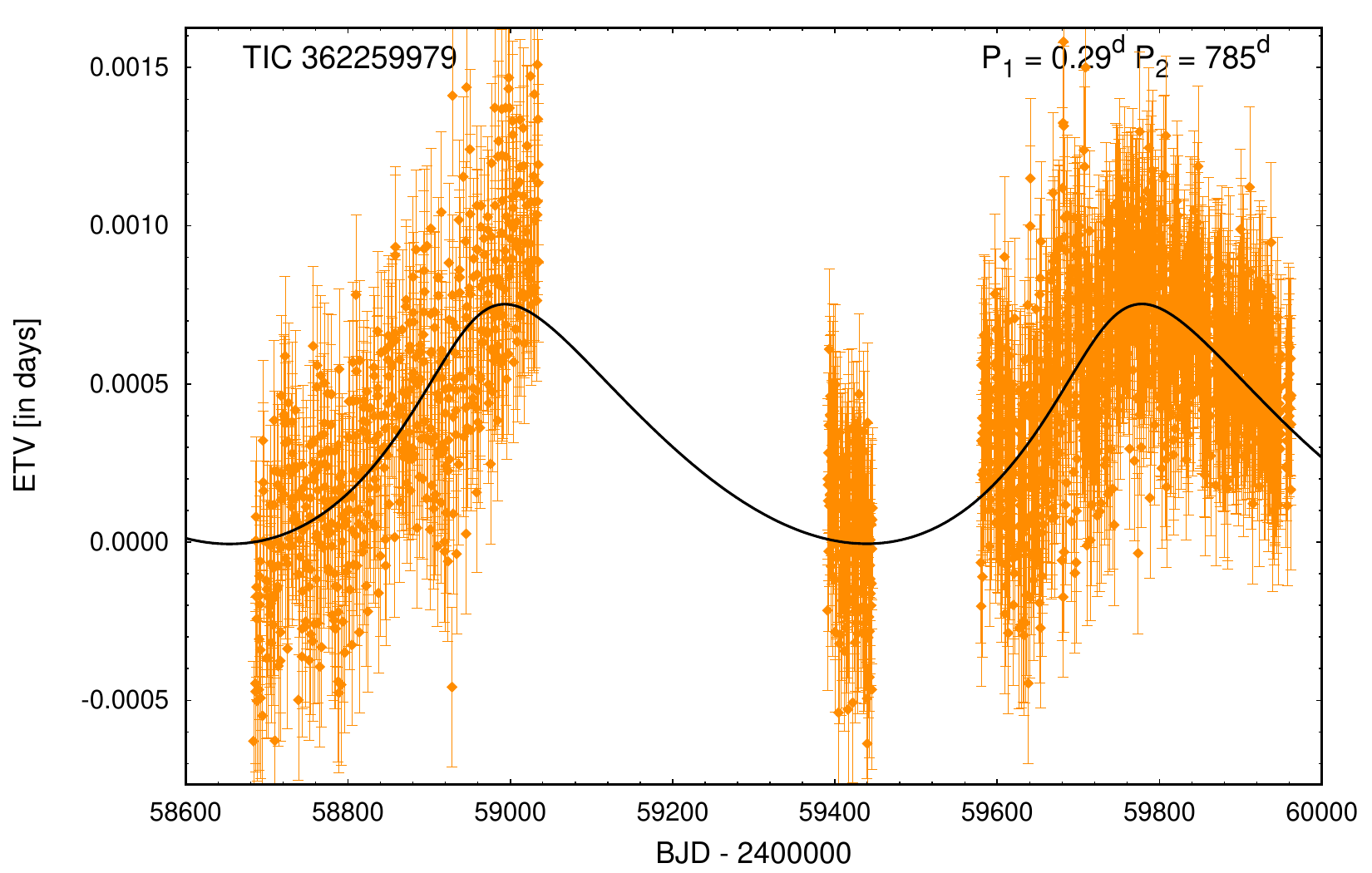}\includegraphics[width=60mm]{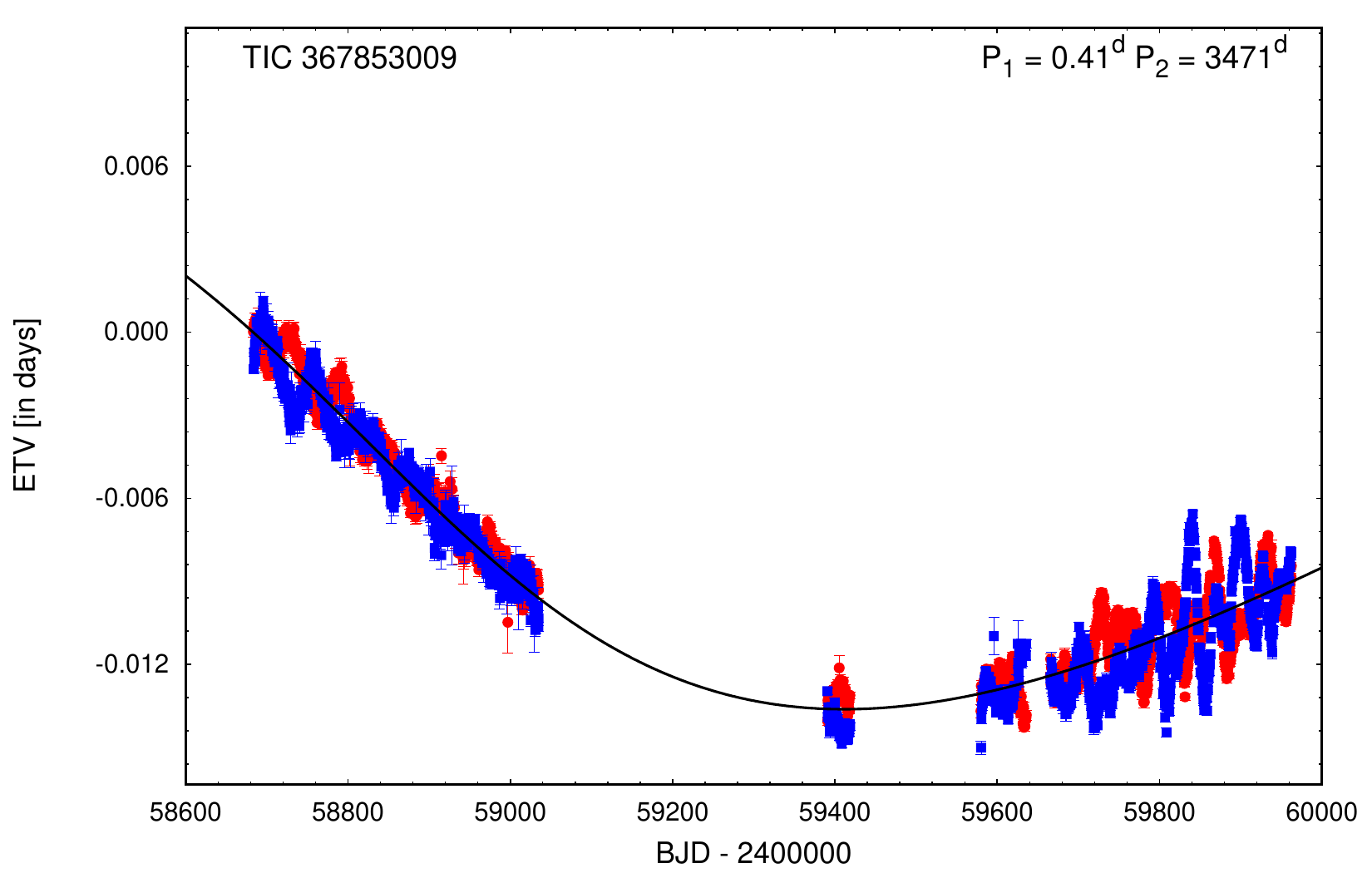}
\includegraphics[width=60mm]{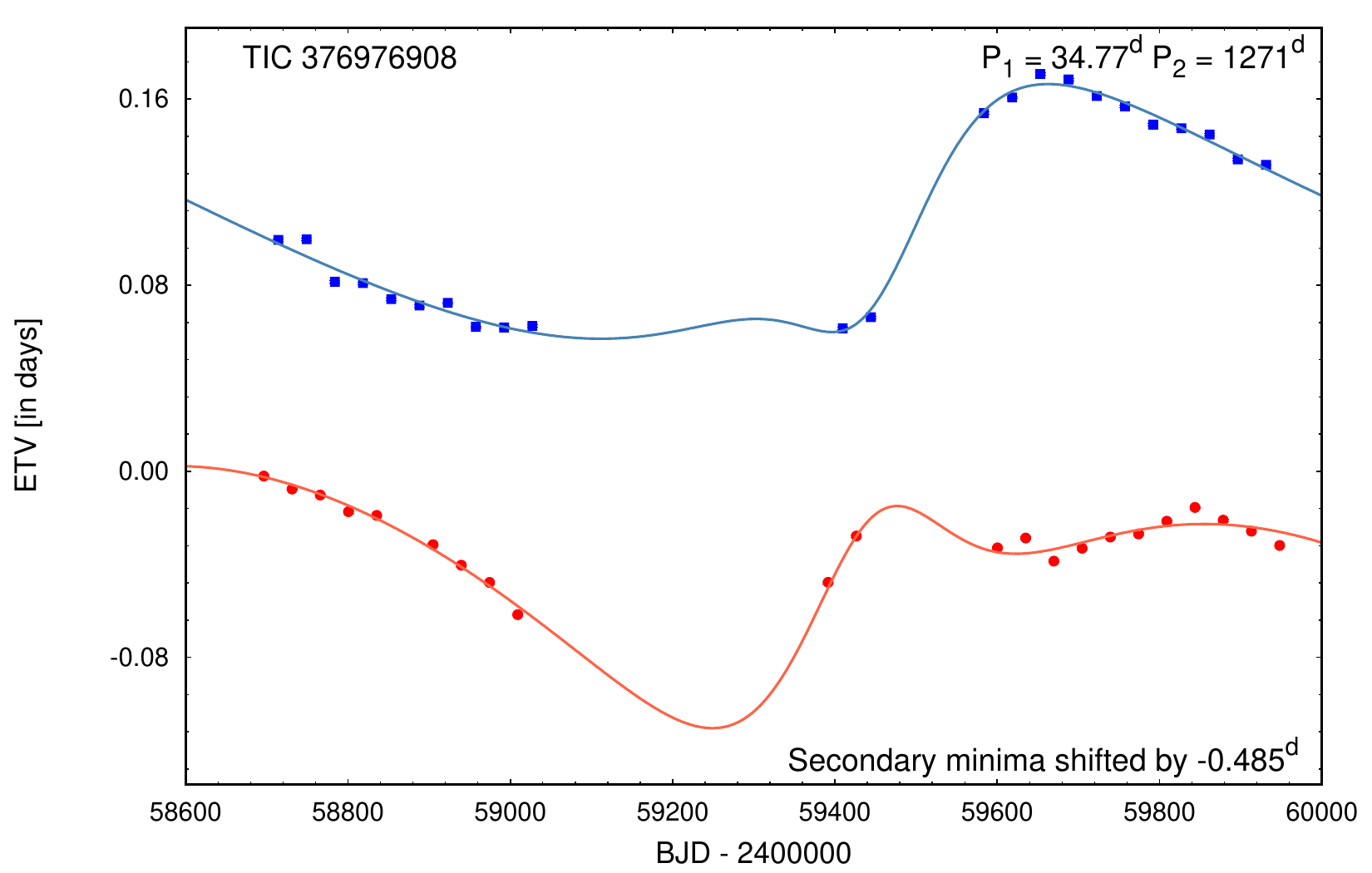}\includegraphics[width=60mm]{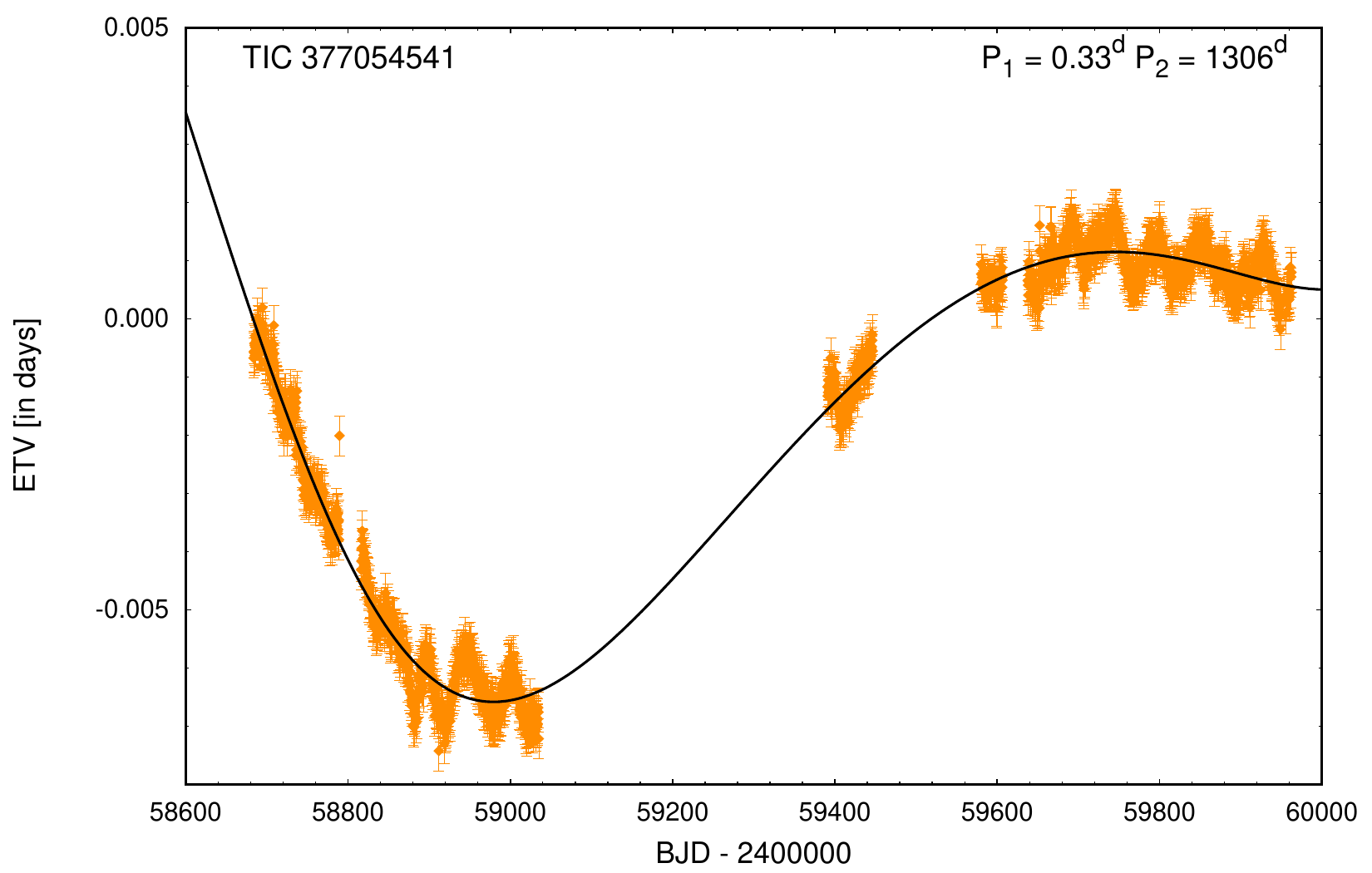}\includegraphics[width=60mm]{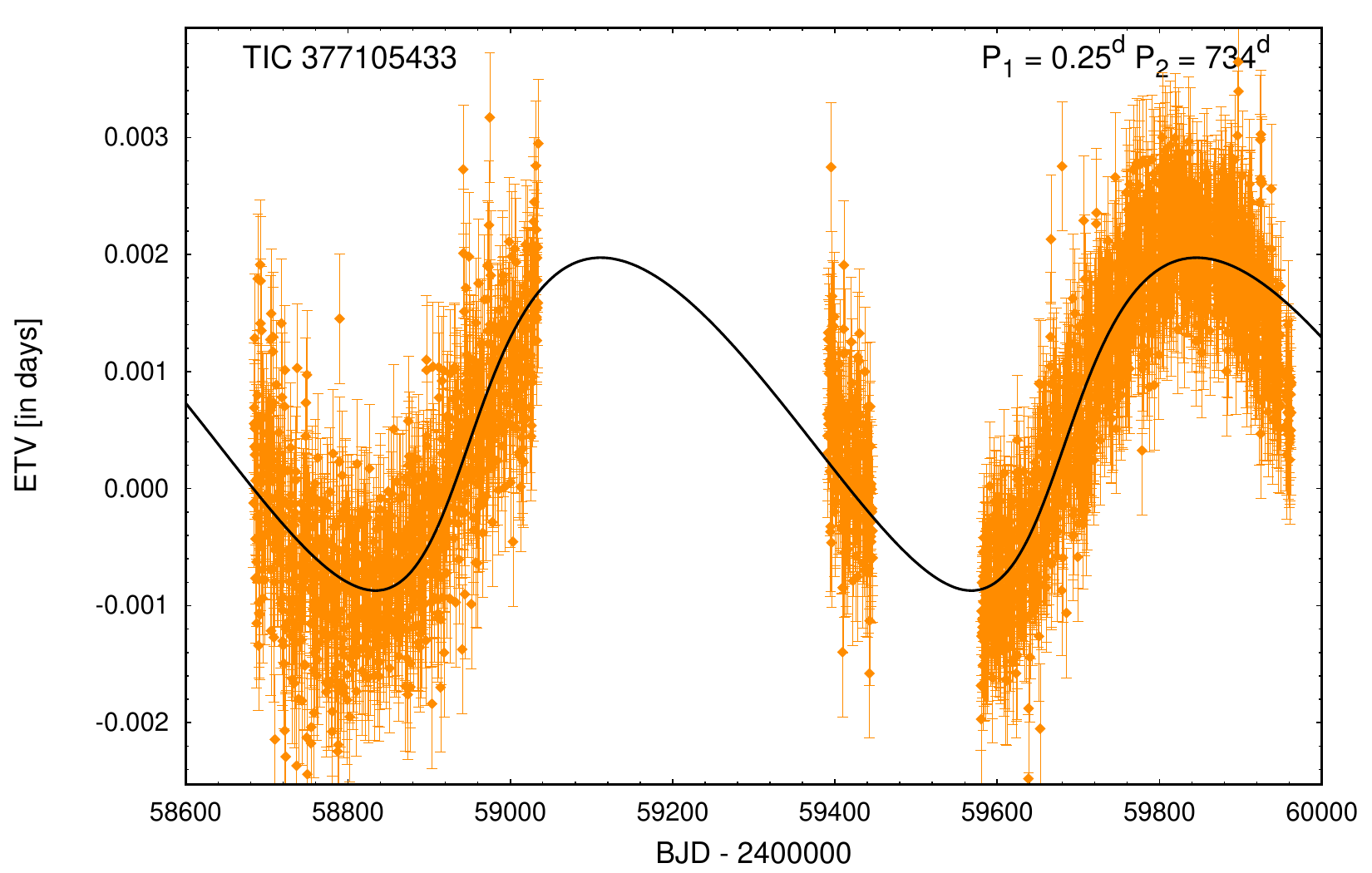}
\includegraphics[width=60mm]{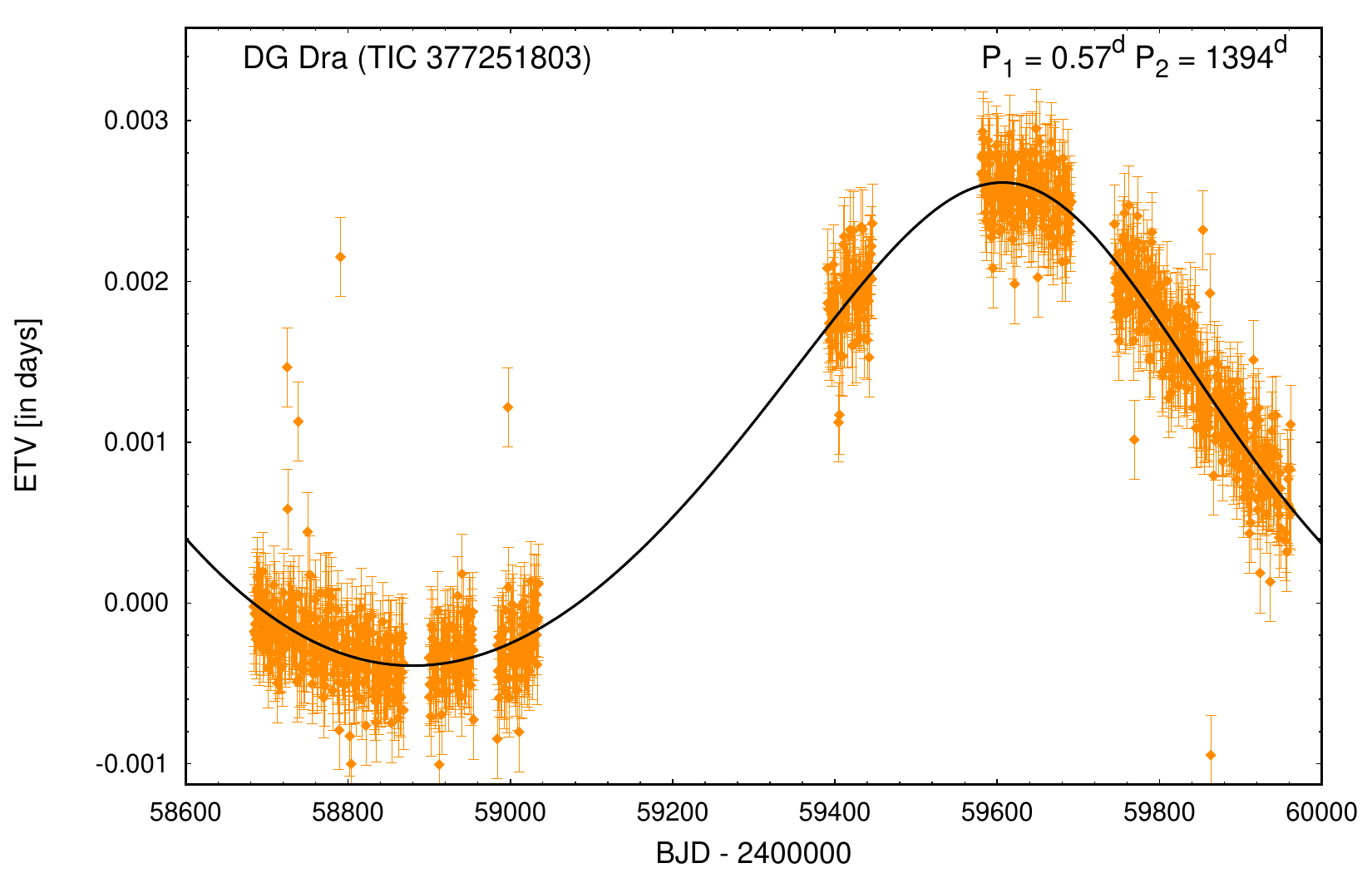}\includegraphics[width=60mm]{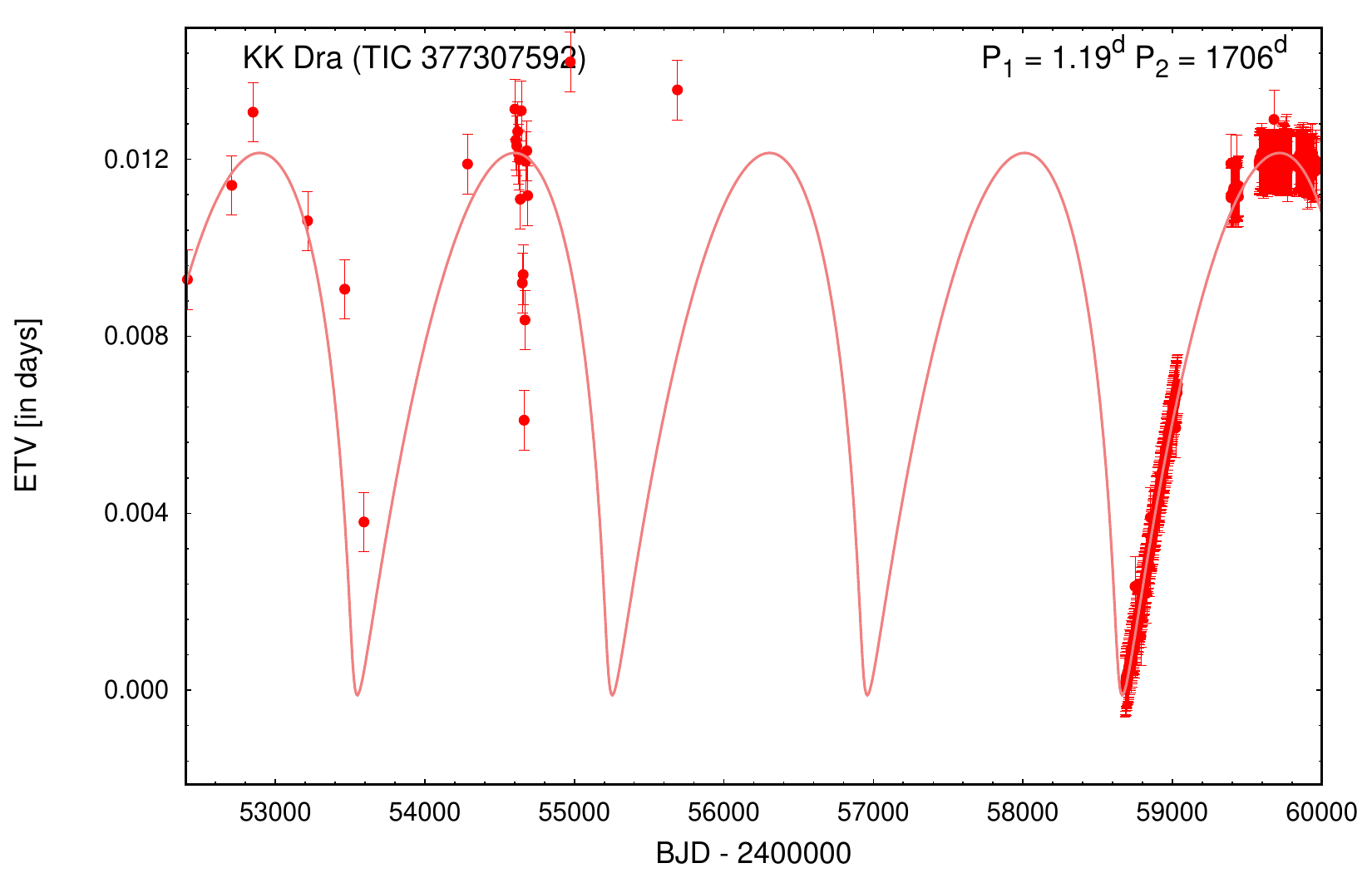}\includegraphics[width=60mm]{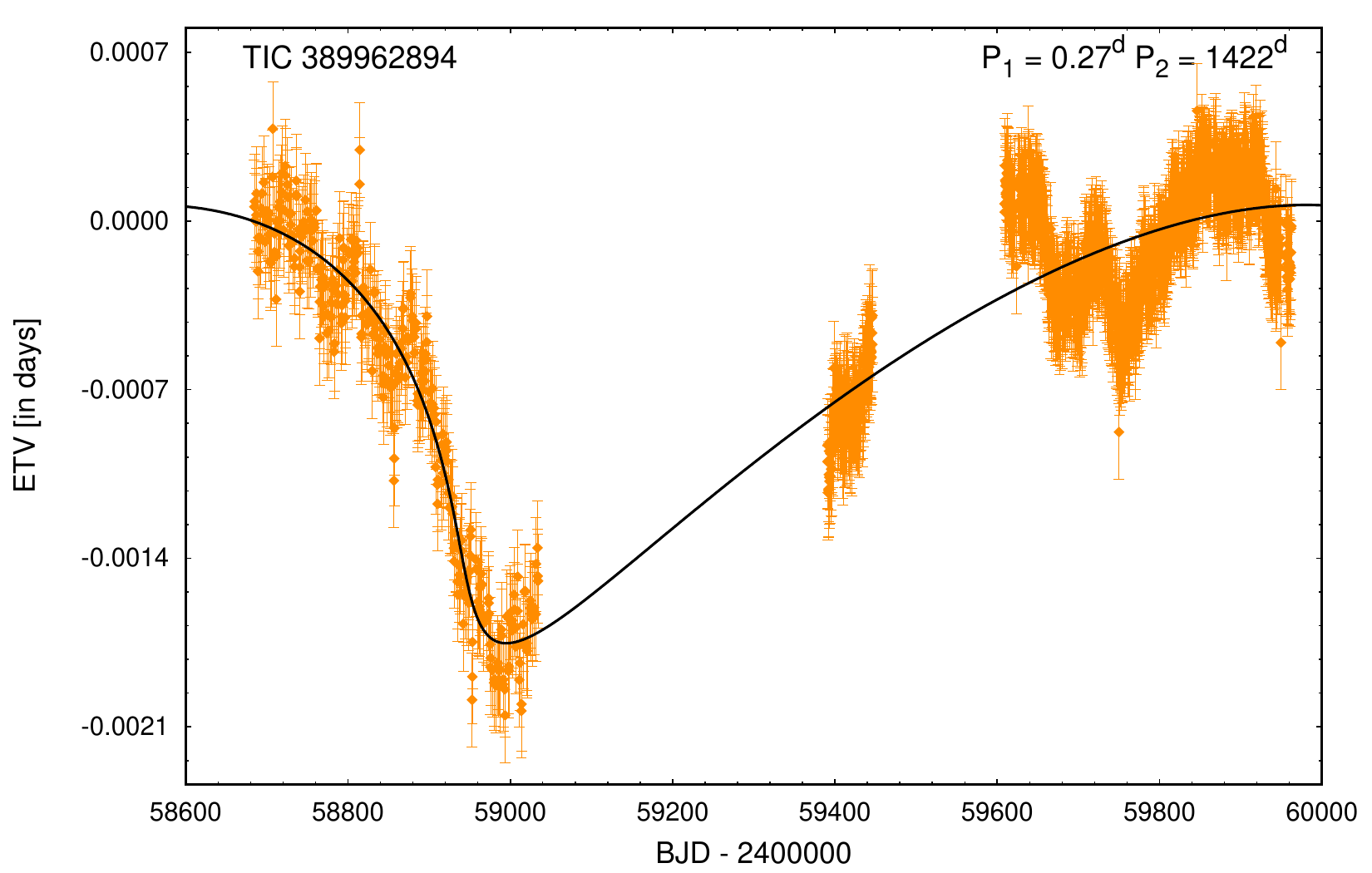}
\includegraphics[width=60mm]{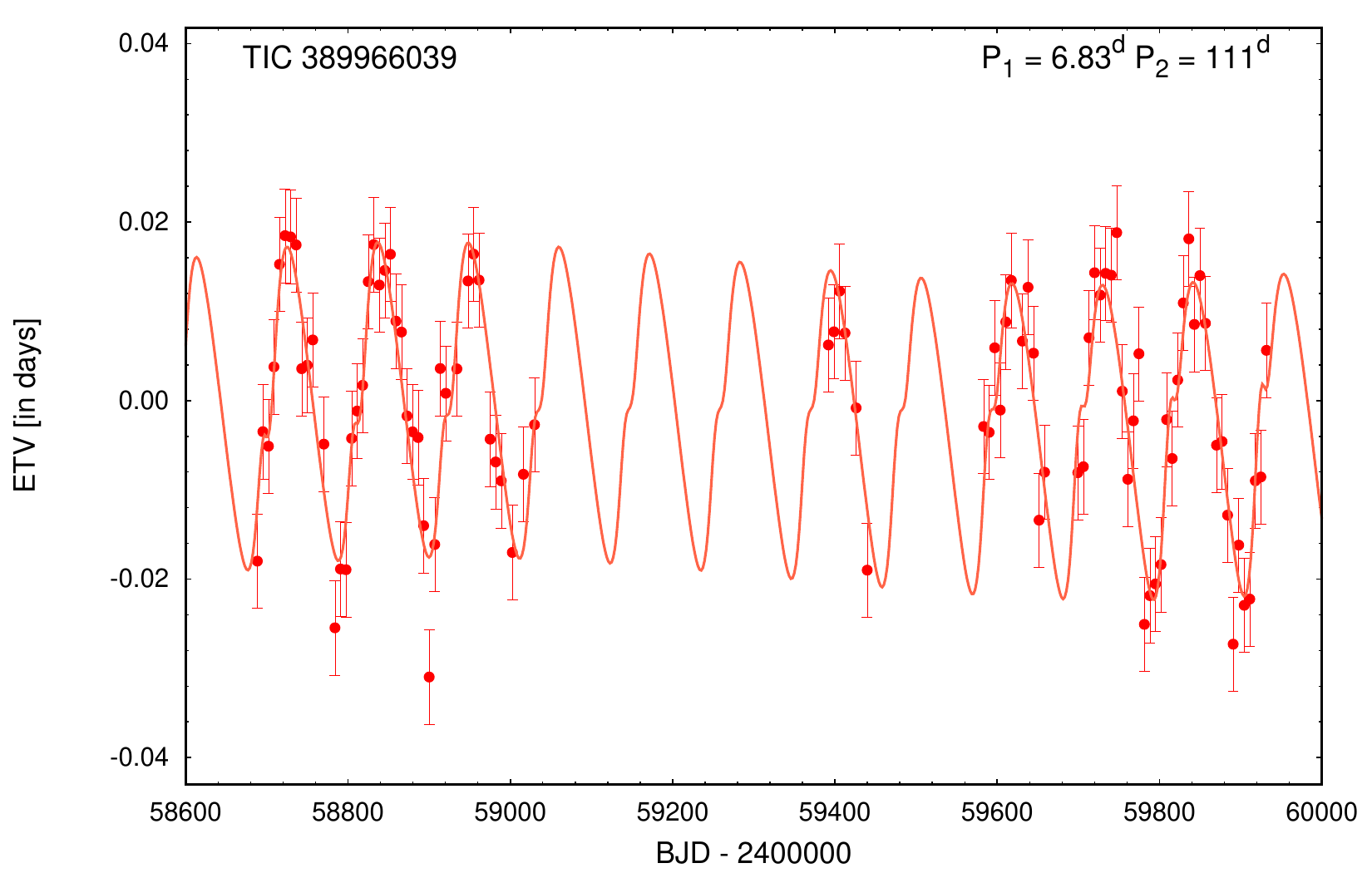}\includegraphics[width=60mm]{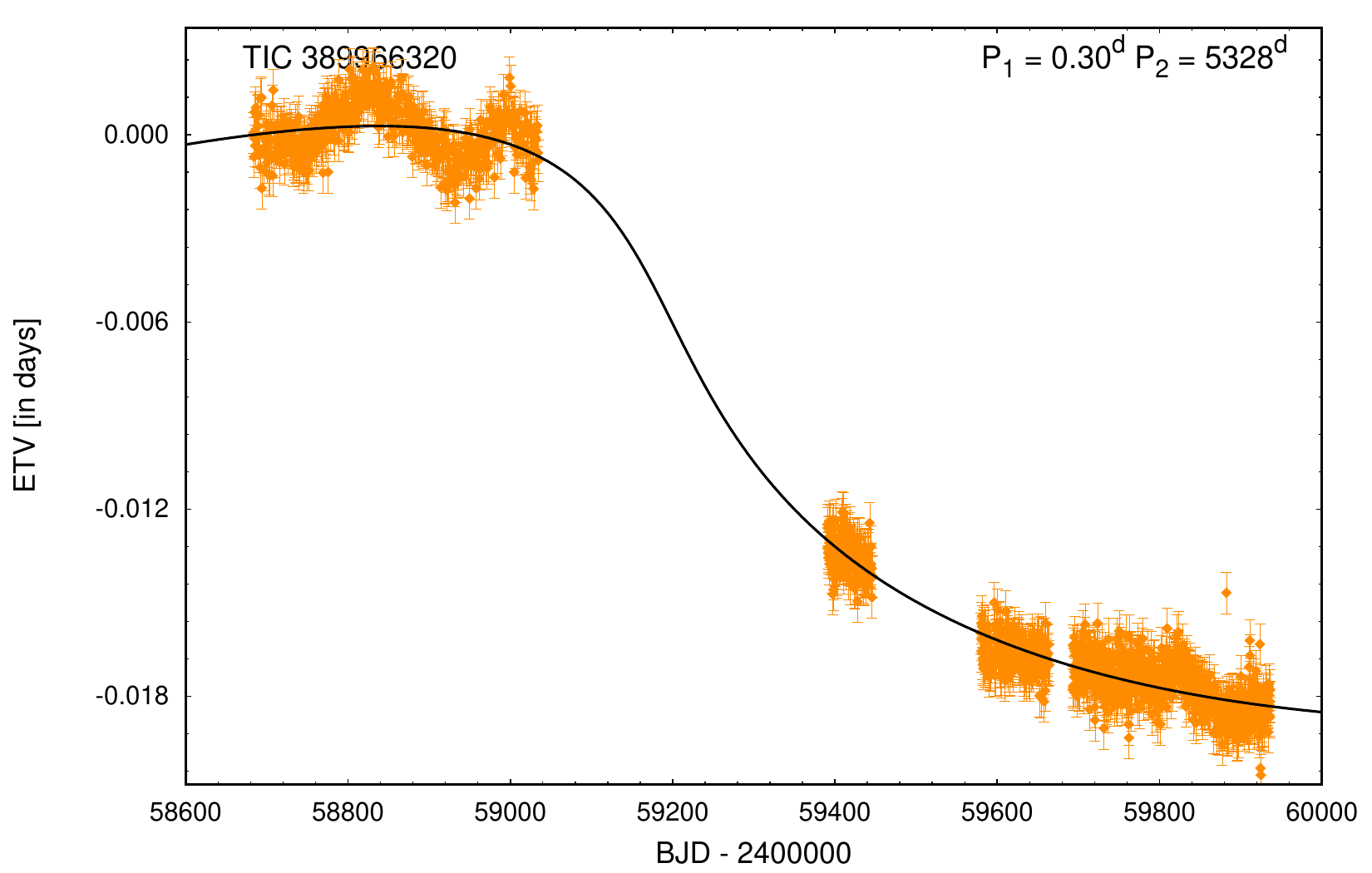}\includegraphics[width=60mm]{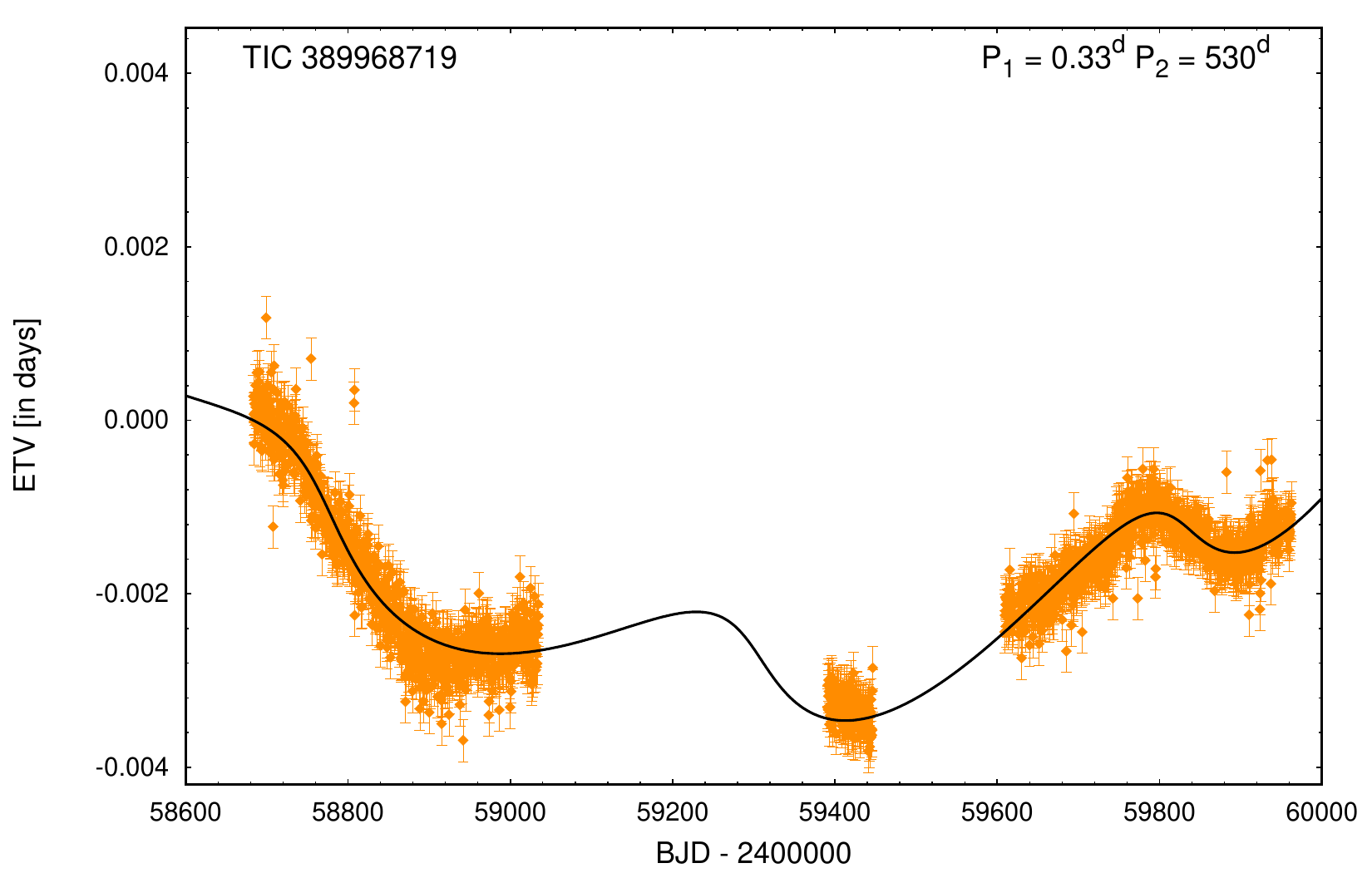}
\includegraphics[width=60mm]{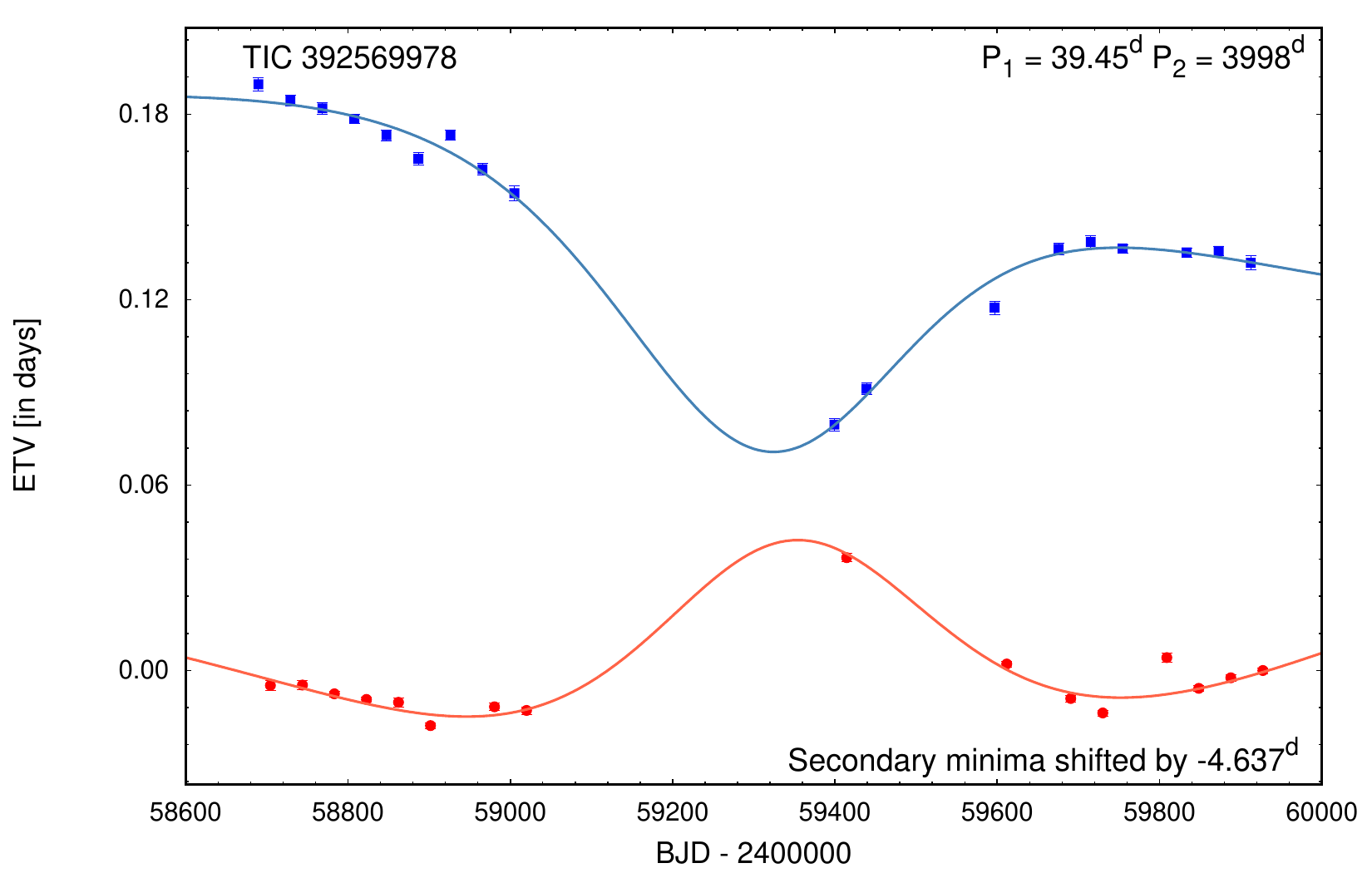}\includegraphics[width=60mm]{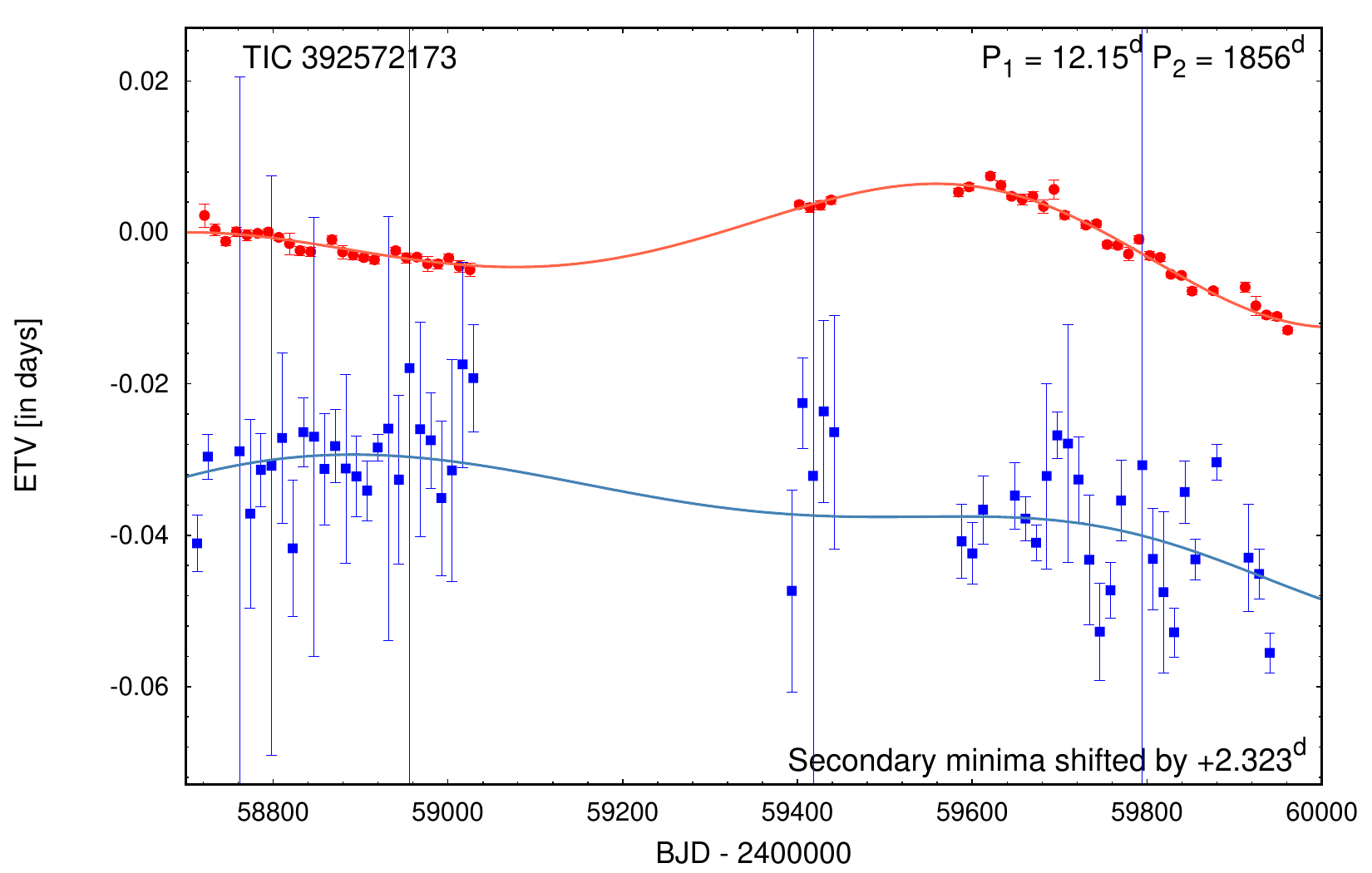}\includegraphics[width=60mm]{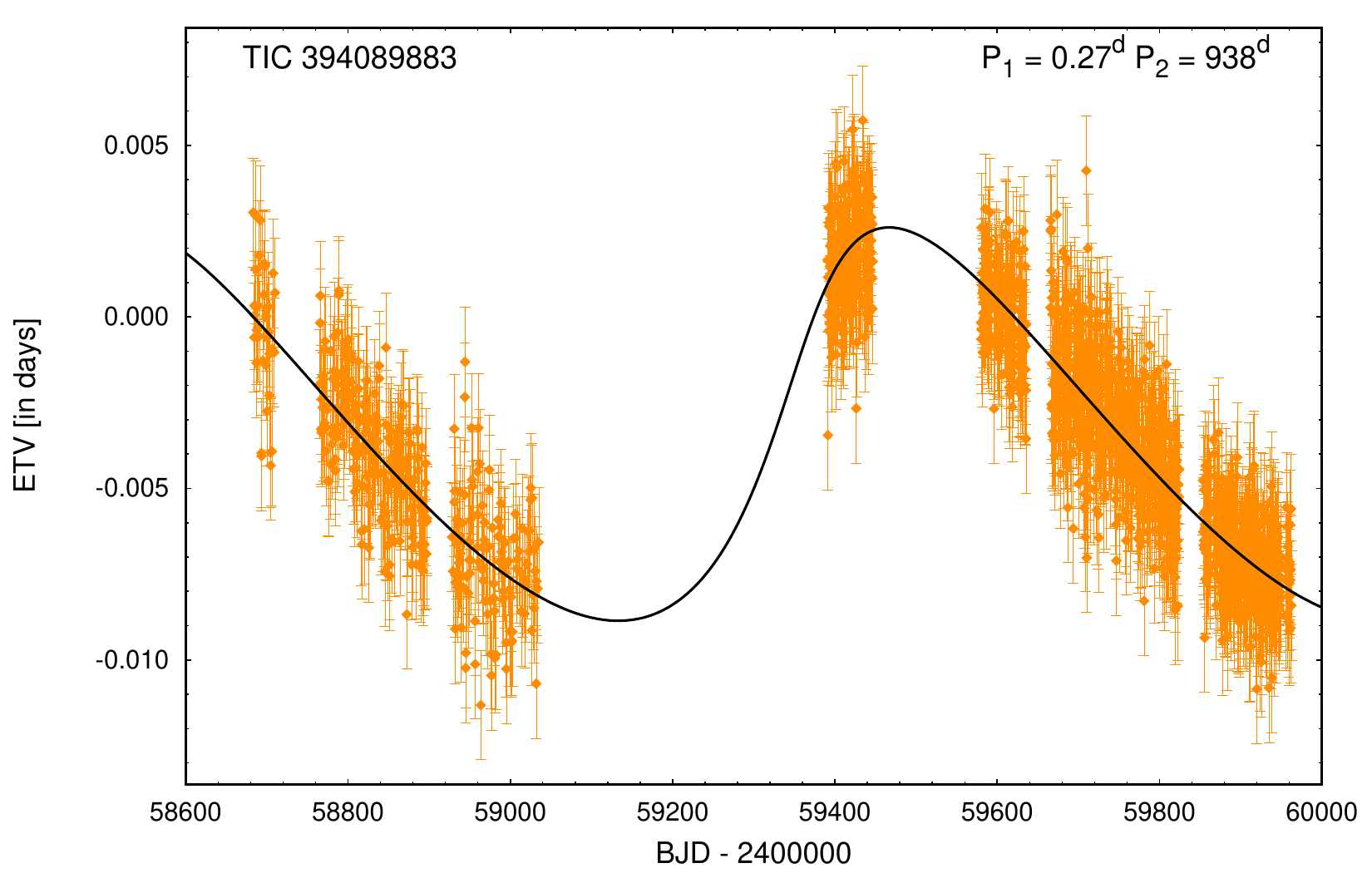}
\caption{(continued)}
\end{figure*}

\addtocounter{figure}{-1}

\begin{figure*}
\includegraphics[width=60mm]{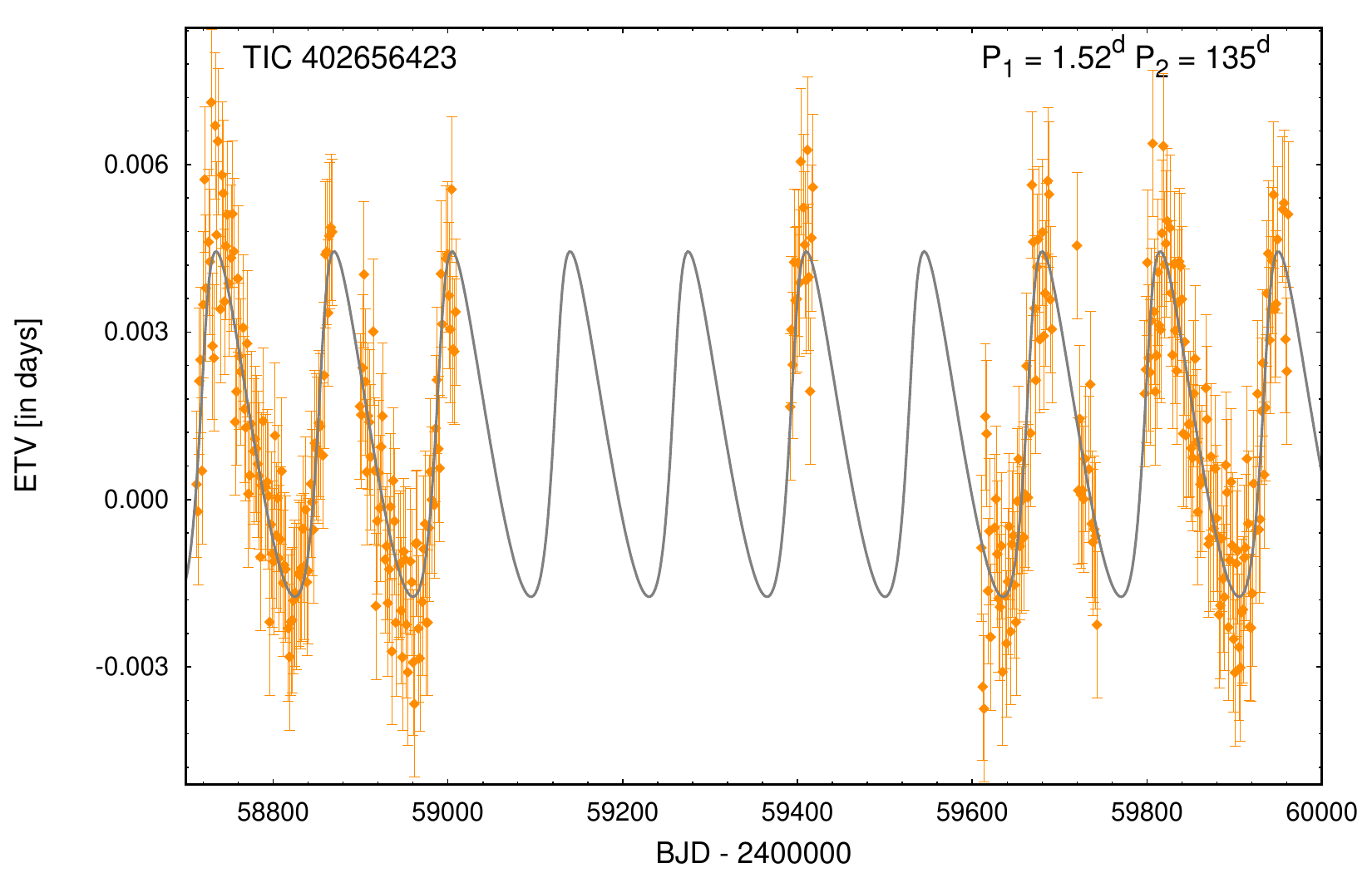}\includegraphics[width=60mm]{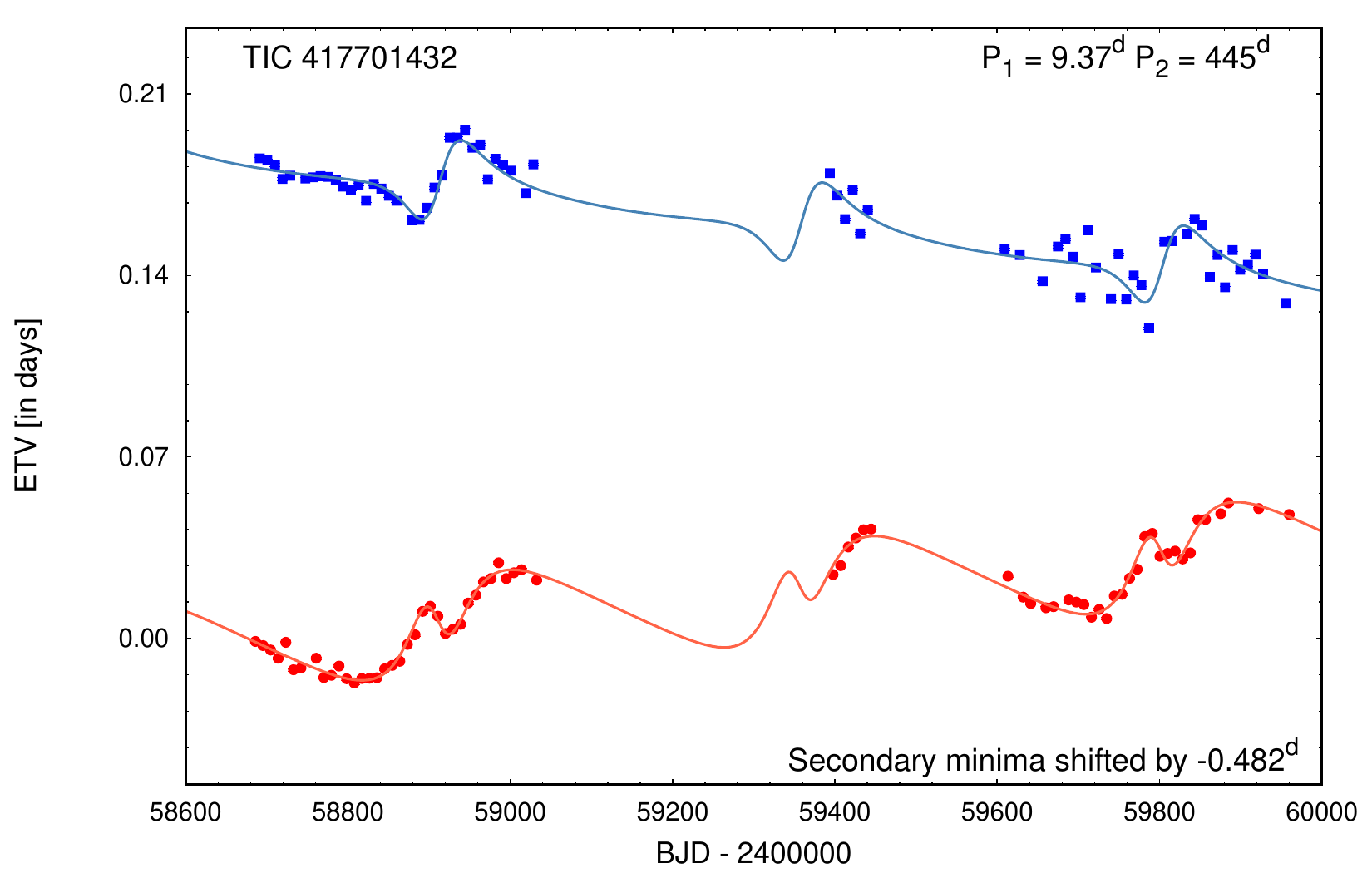}\includegraphics[width=60mm]{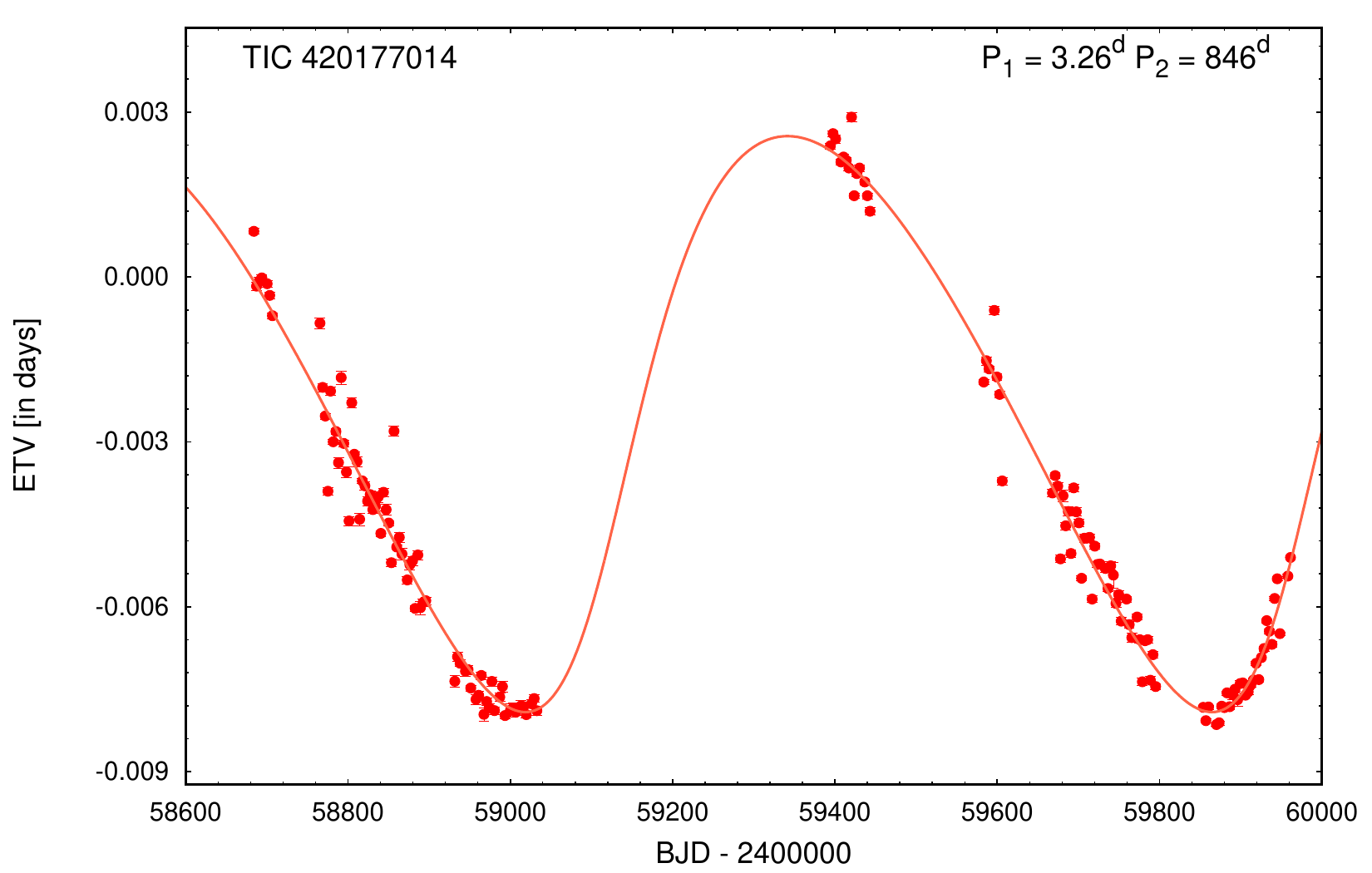}
\includegraphics[width=60mm]{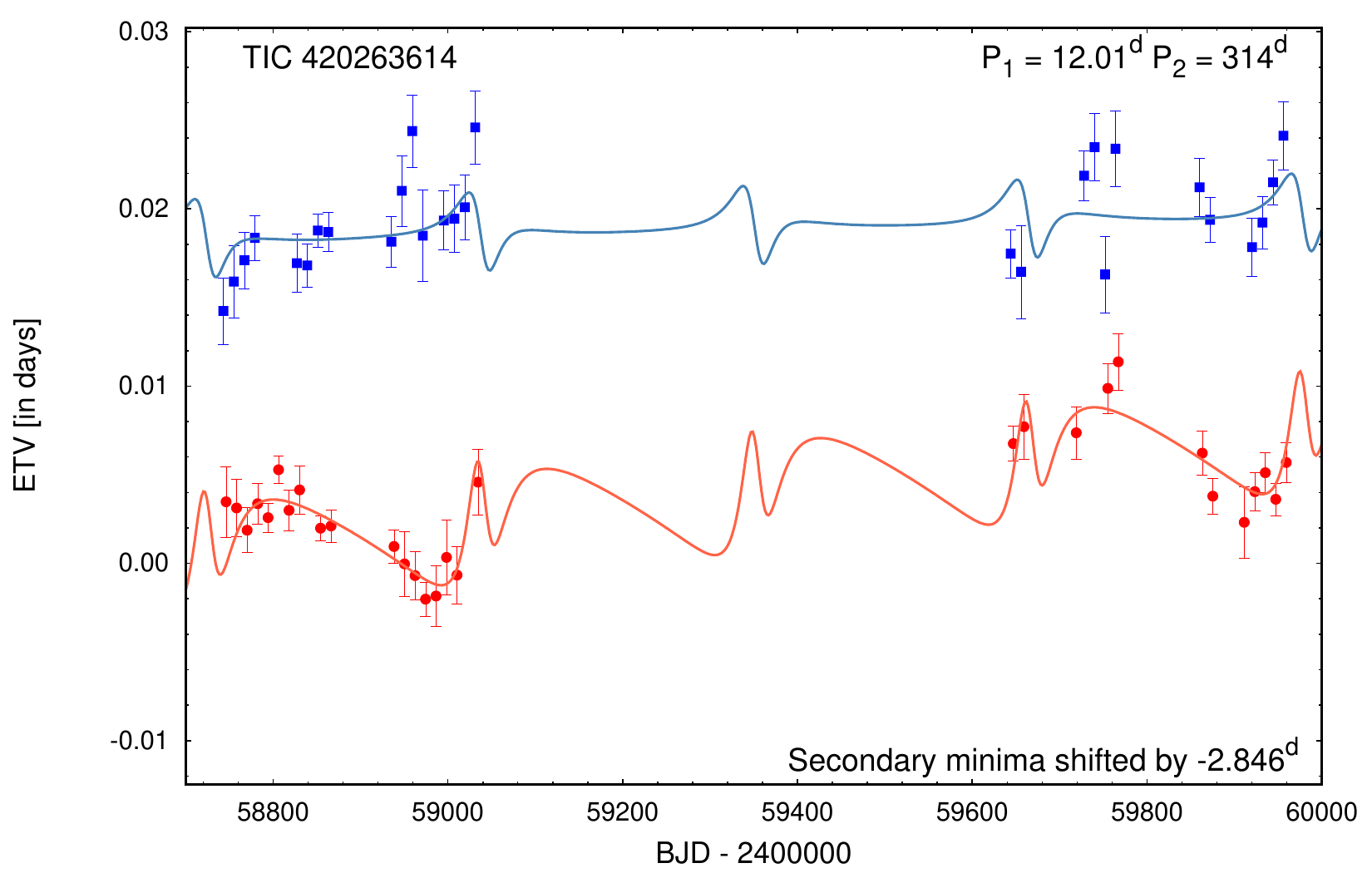}\includegraphics[width=60mm]{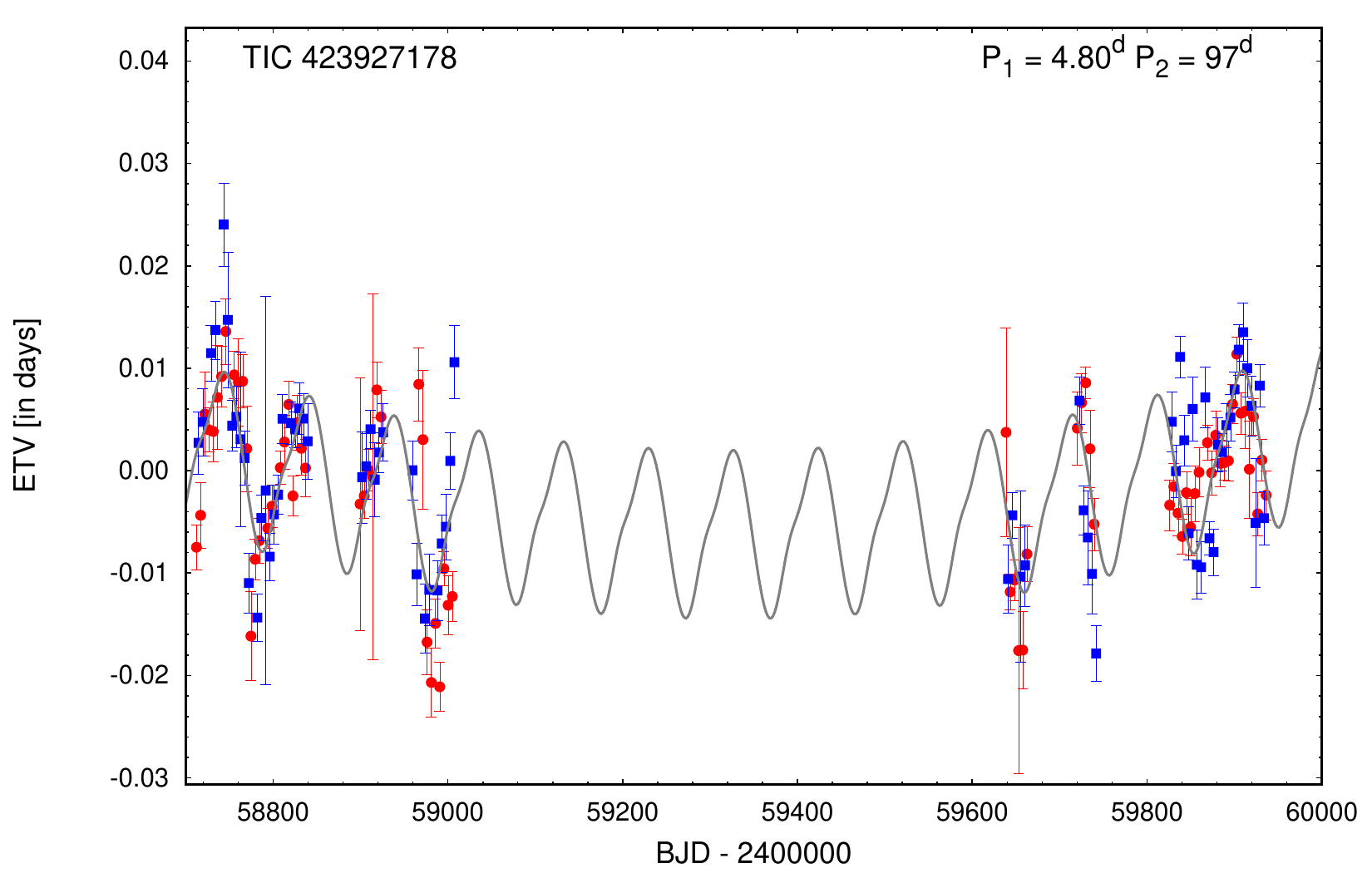}\includegraphics[width=60mm]{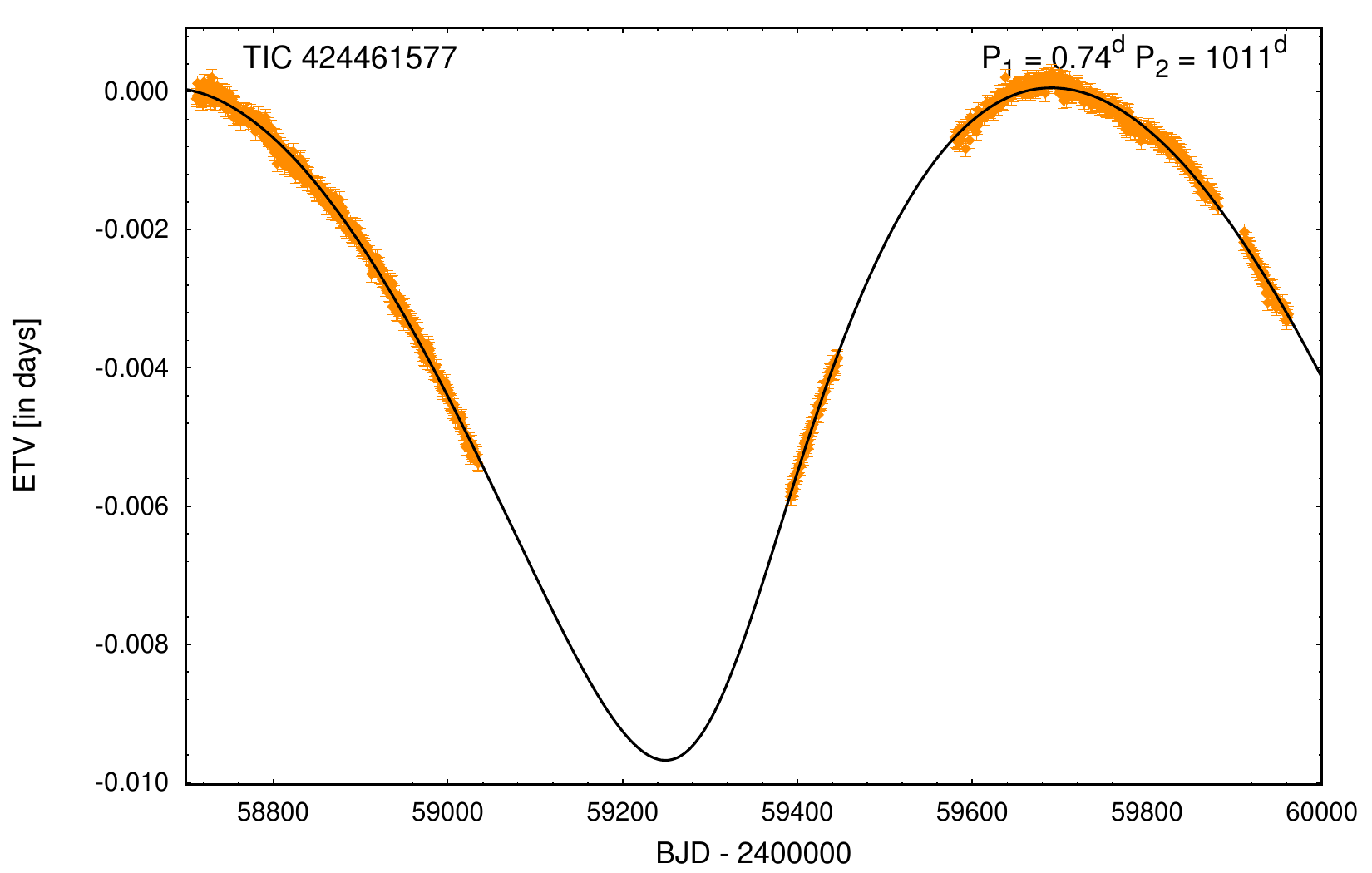}
\includegraphics[width=60mm]{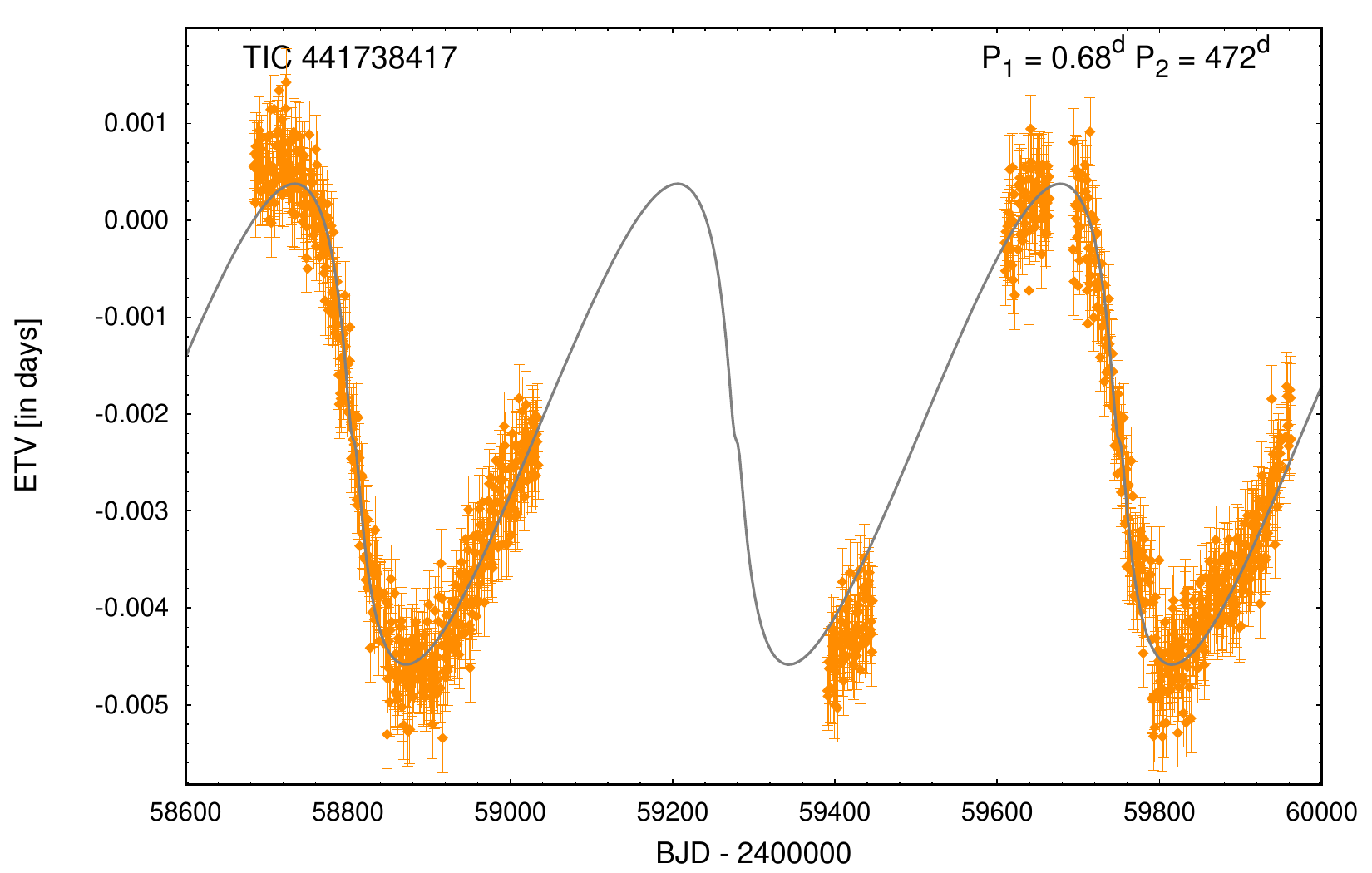}\includegraphics[width=60mm]{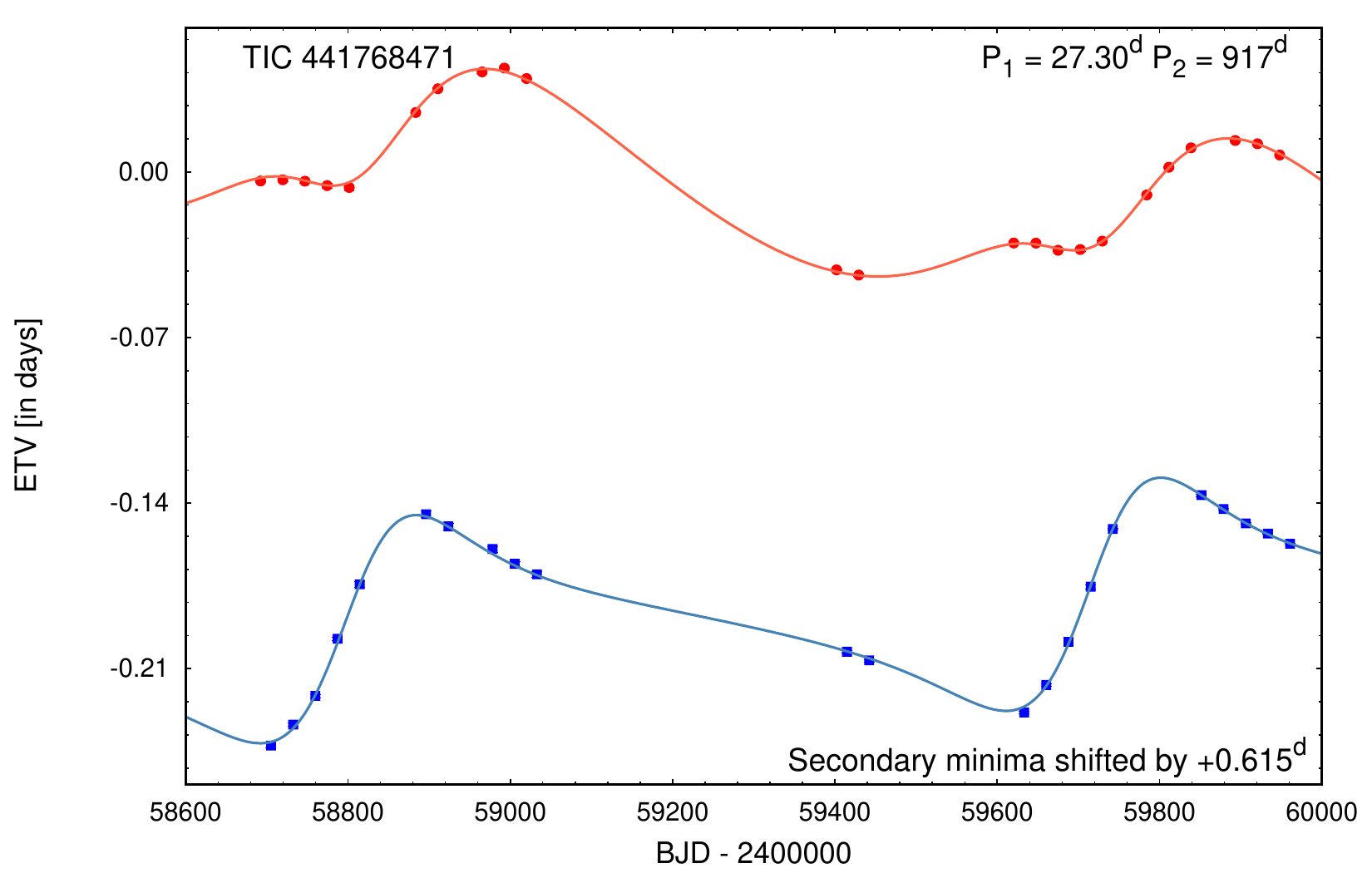}\includegraphics[width=60mm]{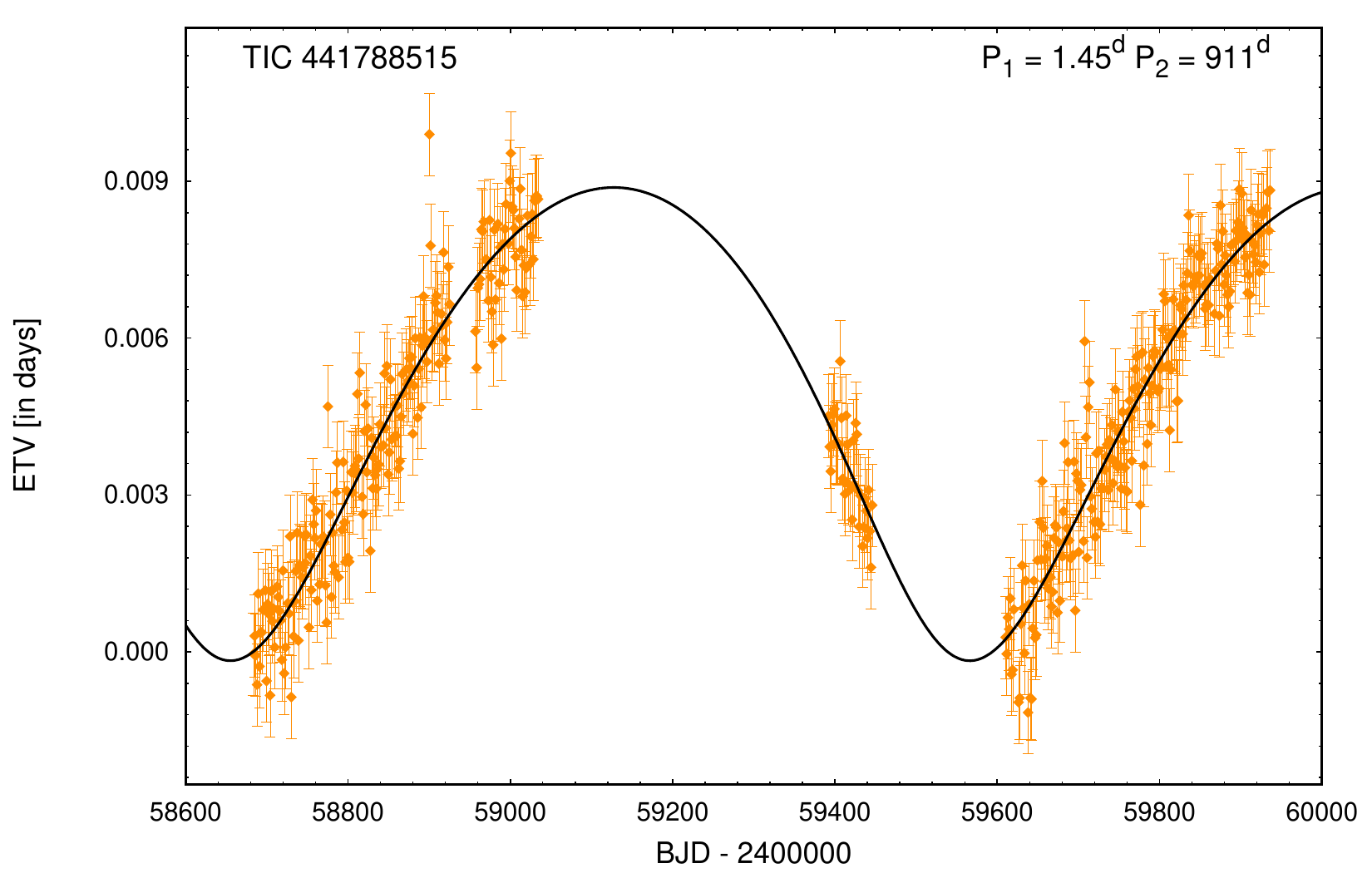}
\includegraphics[width=60mm]{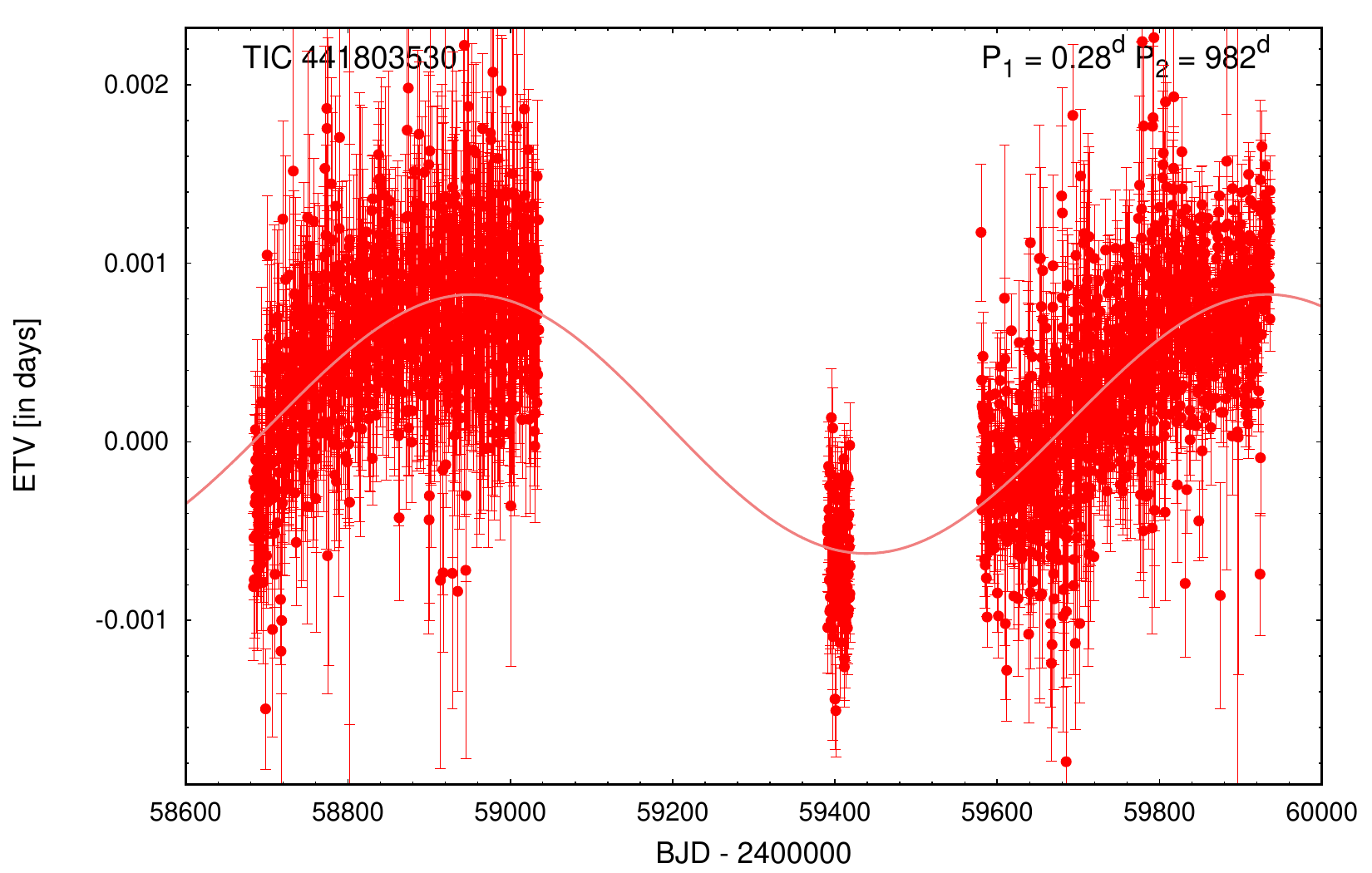}\includegraphics[width=60mm]{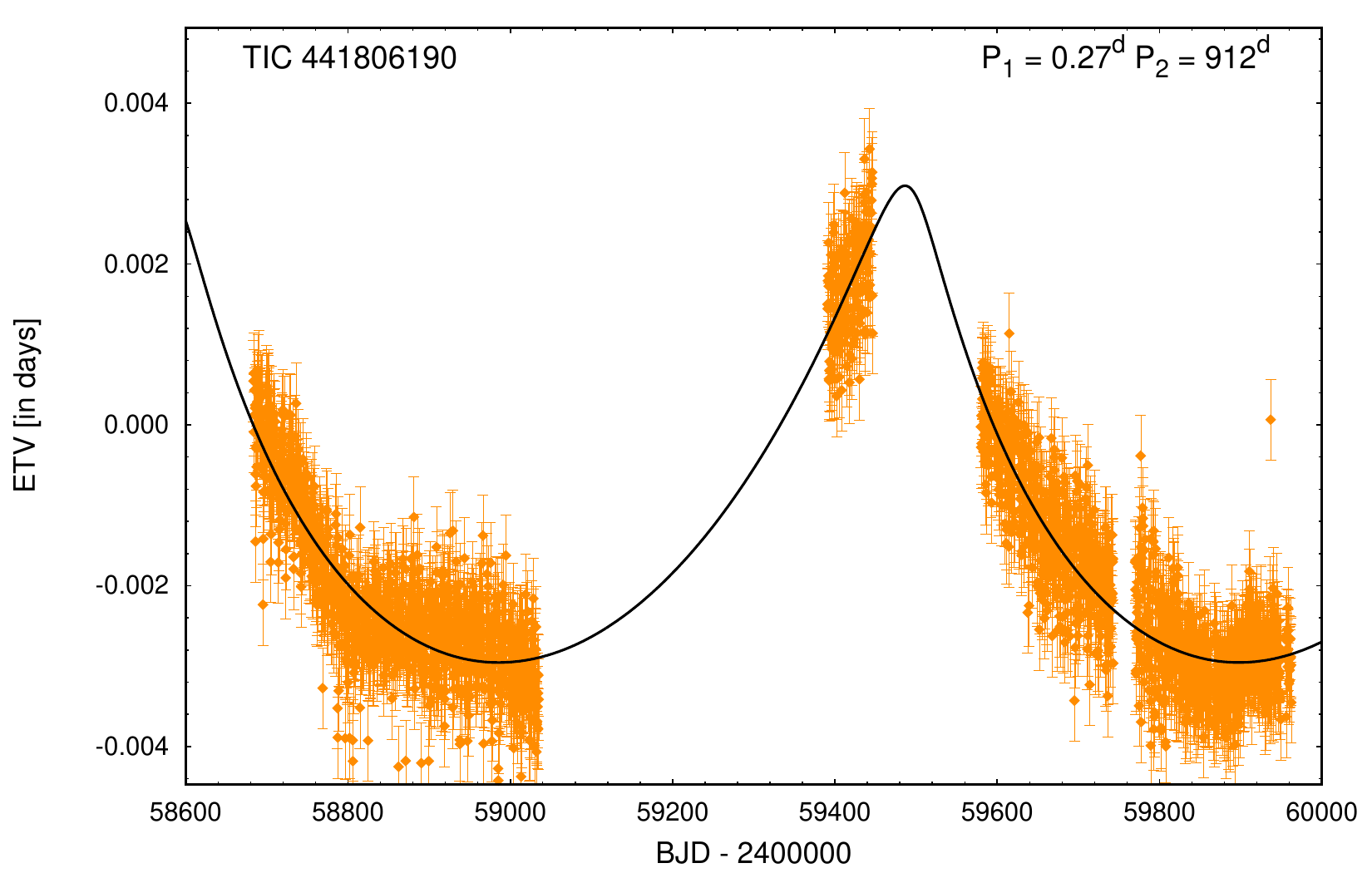}\includegraphics[width=60mm]{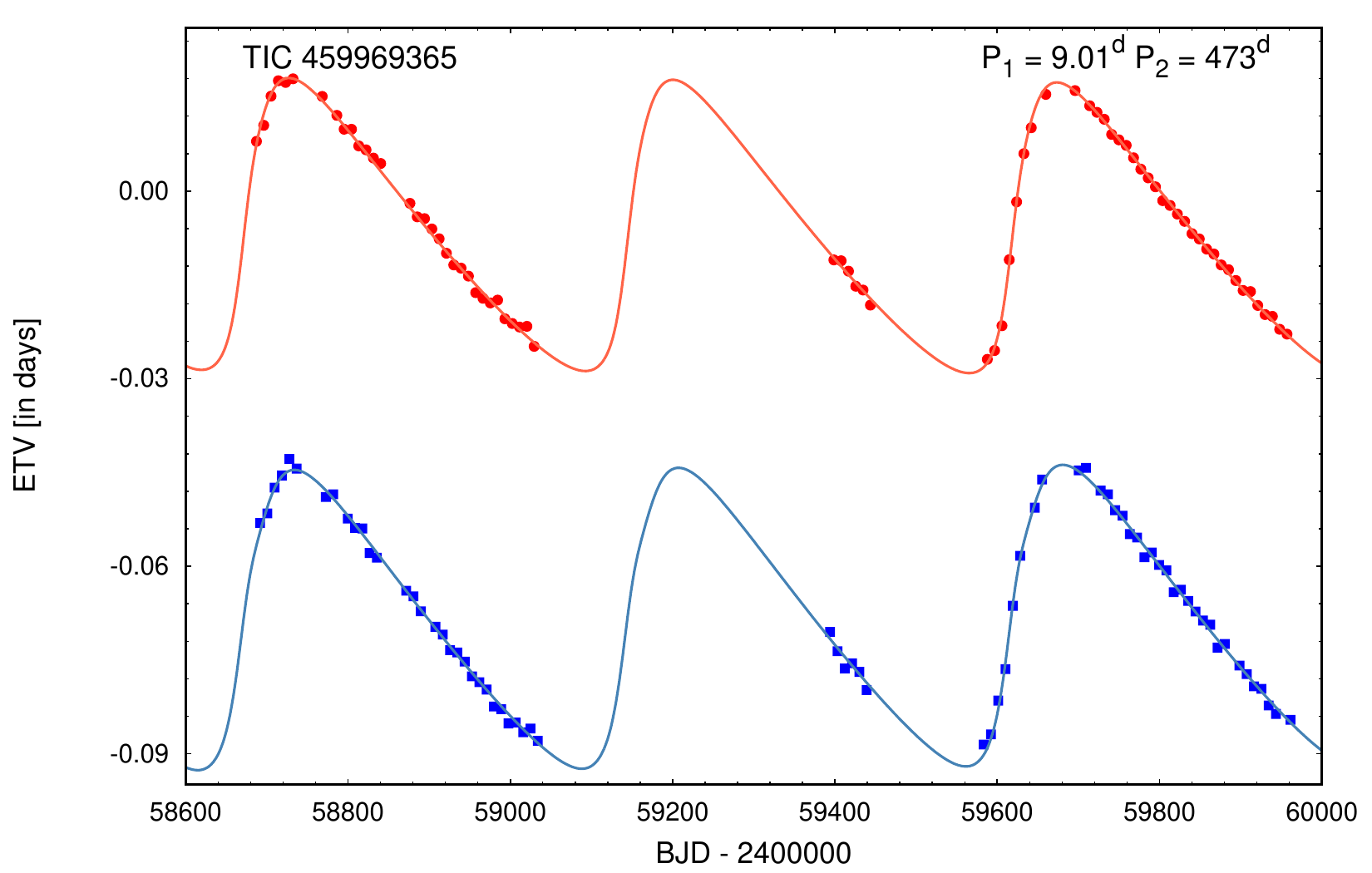}
\caption{(continued)}
\end{figure*}

\end{appendix}

\end{document}